\documentclass[12pt,preprint]{aastex}
\begin{document}

\title{Examining High Energy Radiation Mechanisms of Knots and Hotspots in Active Galactic Nucleus Jets}
\author{ Jin Zhang\altaffilmark{1}, Shen-shi Du\altaffilmark{2}, Sheng-Chu Guo\altaffilmark{2}, Hai-Ming Zhang\altaffilmark{2}, Liang Chen\altaffilmark{3}, En-Wei Liang\altaffilmark{2}, Shuang-Nan Zhang\altaffilmark{1,4}}

\altaffiltext{1}{Key Laboratory of Space Astronomy and Technology, National Astronomical Observatories, Chinese Academy of Sciences, Beijing 100012, China; jinzhang@bao.ac.cn}
\altaffiltext{2}{Guangxi Key Laboratory for Relativistic Astrophysics, Department of Physics, Guangxi University, Nanning 530004, China}
\altaffiltext{3}{Key Laboratory for Research in Galaxies and Cosmology, Shanghai Astronomical Observatory, Chinese Academy of Sciences, 80 Nandan Road, Shanghai 200030, China}
\altaffiltext{4}{Key Laboratory of Particle Astrophysics, Institute of High Energy Physics, Chinese Academy of Sciences, Beijing 100049, China}
\begin{abstract}
We compile the radio-optical-X-ray spectral energy distributions (SEDs) of 65 knots and 29 hotspots in 41 active galactic nucleus jets to examine their high energy radiation mechanisms. Their SEDs can be fitted with the single-zone leptonic models, except for the hotspot of Pictor A and six knots of 3C 273. The X-ray emission of one hotspot and 22 knots is well explained as synchrotron radiations under the equipartition condition; they usually have lower X-ray and radio luminosities than the others, which may be due to a lower beaming factor. An inverse Compton (IC) process is involved for explaining the X-ray emission of the other SEDs. Without considering the equipartition condition, their X-ray emission can be attributed to the synchrotron-self-Compton (SSC) process, but the derived jet power ($P_{\rm jet}$) are not correlated with $L_{\rm k}$ and most of them are larger than $L_{\rm k}$ with more than three orders of magnitude, where $L_{\rm k}$ is the jet kinetic power estimated with their radio emission. Under the equipartition condition, the X-ray emission is well interpreted with the IC process to the cosmic microwave background photons (IC/CMB). In this scenario, the derived $P_{\rm jet}$ of knots and hotspots are correlated with and comparable to $L_{\rm k}$. These results suggest that the IC/CMB model may be the promising interpretation of their X-ray emission. In addition, a tentative knot-hotspot sequence in the synchrotron peak-energy--peak-luminosity plane is observed, similar to the blazar sequence, which may be attributed to their different cooling mechanisms of electrons.
\end{abstract}


%%%%%%%%%%%%%%%%%%%%%%%%%%%%%%%%%%%%%%%%%%%%%%%%%%%%%%%%%%%%%%%%
\section{Introduction}           %% first-level sections will be auto-capitalized
\label{sect:intro}

Knots and hotspots, which have been detected in radio, optical, and X-ray bands (e.g., Harris \& Krawczynski 2006), are substructures of kpc--Mpc jets in radio-loud active galactic nuclei (AGNs). Their emission in the radio-optical band usually can be fitted with the synchrotron radiation of electrons accelerated locally in these substructures (e.g., Kataoka \& Stawarz 2005; Harris \& Krawczynski 2006; Zhang et al. 2009, 2010). The X-ray emission of some substructures can be well fit with the synchrotron radiation extended from the radio-optical band (e.g., the substructure in 3C 371, Sambruna et al. 2007). However, the observed X-ray spectra for most knots and hotspots are harder than the extrapolation of the radio-to-optical components (Kataoka \& Stawarz 2005; Zhang et al. 2010), and then the inverse-Compton (IC) scattering process is involved to model their X-ray emission. The synchrotron-self-Compton (SSC) scattering model can explain the X-ray emission, but the derived  magnetic field strength of the radiation region is extremely low, which significantly deviates the equipartition condition (Harris \& Krawczynski 2006; Zhang et al. 2010). The IC scattering of cosmic microwave background (IC/CMB) is the most favorable model to explain the X-ray emission of most knots and hotspots (Tavecchio et al. 2000; Kataoka \& Stawarz 2005;
Harris \& Krawczynski 2006; Zhang et al. 2010), but it is still being debated.

The IC/CMB scenario requires that these substructures are mildly relativistic with a Lorentz factor of about 10, in order to make the predicted X-ray flux matching with the observations (e.g., Tavecchio et al. 2000; Kataoka \& Stawarz 2005; Zhang et al. 2010). Superluminal proper motions, which suggest that the jets start out highly relativistic with an Lorentz factor of tens, have been detected with the very long baseline interferometry (VLBI) in multi-epoch measurements of parsec (pc) scale jets of AGNs (Jorstad et al. 2005; Lister et al. 2009, 2013). It was proposed that the jets are decelerated, but are still mildly relativistic at kpc-scale (e.g., Arshakian \& Longair 2004; Mullin \& Hardcastle 2009; Uchiyama et al. 2006; Zhang et al. 2010). Meyer et al. (2017a) presented measurements of the kpc-scale proper motions of eleven nearby optical jets with Hubble Space Telescope (HST) and mapped the full velocity profile of these jets from near the black hole to the kpc-scale together with VLBI proper motion measurements. They showed convincing evidence for jet deceleration and transverse motions  in M87, and the apparent velocity is still superluminal in its outer jet. McKeough et al. (2016) analyzed \emph{Chandra} X-ray images of eleven quasars with high redshifts and found that the X-ray to radio flux ratios of these high-redshift jets are marginally inconsistent with those from lower redshifts, and they suggested that the X-ray emission may be due to the IC/CMB rather than the synchrotron process, since the IC/CMB radiation depends on the redshift, but the synchrotron model does not. These facts in some way indicate that the IC/CMB emission should be responsible for the X-rays of the substructures in the large-scale jets.

Observations in the $\gamma$-ray band, if the emission region can be resolved, may be helpful for evaluating the IC/CMB model since the IC/CMB model sometimes predicts prominent $\gamma$-ray emission in the \emph{Fermi}/LAT energy band (e.g., Zhang et al. 2010). As shown in Zhang et al. (2010), the predicted $\gamma$-ray emission from the IC/CMB process strongly depends on the Doppler factors of these substructures, though most of them may be undetectable with \emph{Fermi}/LAT. In addition, the $\gamma$-ray emission region cannot be resolved with \emph{Fermi}/LAT for most AGN jets, and the $\gamma$-ray emission from hotspots and knots may be highly contaminated by the emission from core region. So far, only the radio lobes of radio galaxies (RGs) Cen A (Abdo et al. 2010; Sun et al. 2016) and Fornax A (McKinley et al. 2015; Ackermann et al. 2016) are convincingly detected and resolved with \emph{Fermi}/LAT. Using the monitoring data of \emph{Fermi}/LAT over six years, Meyer et al. (2014, 2015) showed that the expected hard, steady $\gamma$-ray emission from the IC/CMB model is not seen in PKS 0637-752 and 3C 273. Therefore, it is still difficult to use \emph{Fermi}/LAT data to draw a conclusion on the IC/CMB model, albeit they can play strong constraints (e.g., Zhang et al. 2010; Meyer et al. 2015).

In this work, we examine the high energy radiation mechanisms for the knots and hotspots by using their spectral energy distributions (SEDs) in the radio-optical-X-ray band together with the $\gamma$-ray data observed by \emph{Fermi}/LAT. Our sample and data are presented in Section 2. SED modeling is shown in Section 3. We explore the possible similarity of the synchrotron radiation peak between the core regions and the substructures in Section 4, in order to find out some possible clues to the high energy radiation mechanisms. In Section 5, we evaluate these radiation mechanisms by comparing their total jet powers ($P_{\rm jet}$) to the kinetic powers ($L_{\rm k}$). Conclusions and discussion are given in Section 6. Throughout, $H_0=71$ km s$^{-1}$ Mpc$^{-1}$, $\Omega_{m}=0.27$, and $\Omega_{\Lambda}=0.73$ are adopted, which are also used by the XJET Home Page\footnote{http://hea-www.harvard.edu/XJET/}.

\section{Sample and Data}
We compile the radio-optical-X-ray SEDs for a sample of 65 knots and 29 hotspots in 41 AGNs found in literature, as listed in Table 1, and their observed SEDs are given in the appendix. Zhang et al. (2010) presented a comprehensive analysis for a sample of 22 hotspots and 45 knots. Most of them are also included in our sample and are updated with new observations. Among the 41 AGNs, 16 sources were detected by \emph{Fermi}/LAT and the $\gamma$-ray data are also shown in the SEDs of substructures (see the appendix).

In the leptonic models, a broadband SED is composed of two bumps, a synchrotron radiation and an IC process. If $\alpha_{\rm x}<1$ or $\alpha_{\rm ox}<\alpha_{\rm ro}$,  where $\alpha_{\rm x}$ is the spectral index at the X-ray band, $\alpha_{\rm ox}$ and $\alpha_{\rm ro}$ are broadband spectral indices\footnote{A broadband spectral index in $i$ and $j$ bands is defined as $\alpha_{ij} =-\frac{\log(S_{i}/S_{j})}{\log(\nu_{i}/\nu_{j})}$, where $i$ and $j$ stand for the radio (5 GHz), optical (available data in the observed SEDs in this paper), and X-ray (1 keV) bands, and $S_{i}$ and $S_{j}$ are the flux densities in the $i$ and $j$ bands.} in the radio, optical, and X-ray bands, the X-ray emission is attributed to the IC process and cannot be explained by the synchrotron component extended from the radio-optical band. Among the substructure sample, 27 hotspots and 37 knots are classified into this group (IC Group, as displayed in the appendix). If $\alpha_{\rm x}>1$ or $\alpha_{\rm ox}>\alpha_{\rm ro}$, the X-ray emission is likely the high energy end of the synchrotron radiation at the radio and optical bands. One hotspot (3C 111 H-S) and 22 knots are in this group (Syn Group, as displayed in the appendix). Note that the X-ray spectra of the hotspot in Pictor A and six knots in 3C 273 are also soft with $\alpha_{\rm X}>1$, but their SEDs from radio to X-ray bands cannot be represented by one synchrotron radiation component, and thus their X-ray emission may be produced by the synchrotron radiation of a second electron population with very high energy (e.g., Zhang et al. 2009; Zargaryan et al. 2017; Sun et al. 2017). We thus ignore the hotspot of Pictor A and the six knots of 3C 273 in our following analysis.

\begin{figure*}
\includegraphics[angle=0,scale=0.4]{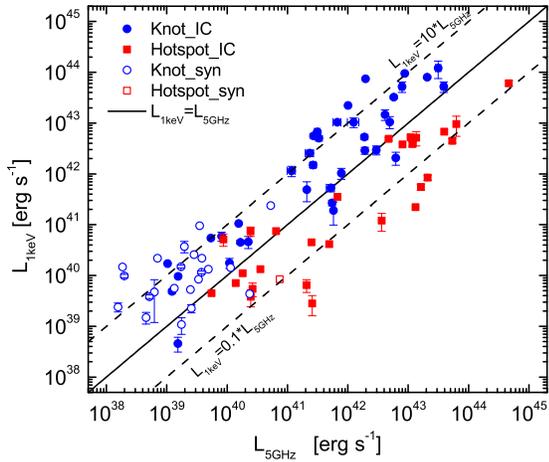}
\caption{Observed luminosity at 1 keV as a function of that at 5 GHz for the 59 knots and 28 hotspots. The solid and opened symbols indicate that their X-ray emission can be explained with the IC process and synchrotron radiation, respectively.}\label{Lr-Lx}
\end{figure*}

We finally have a sample of 59 knots and 28 hotspots for our analysis. We plot their observed luminosity at 1 keV ($L_{\rm 1keV}$) against their observed luminosity at 5 GHz ($L_{\rm 5GHz}$) in Figure \ref{Lr-Lx}. It is found that $L_{\rm 1keV}$ is strongly correlated with $L_{\rm 5GHz}$. The Pearson correlation analysis yields a coefficient of $r=0.85$ with chance probability of $p\sim0$. Hotspots and knots can be roughly separated with a division line of $L_{\rm 1keV}$=$L_{\rm 5GHz}$ (see also Zhang et al. 2010). The substructures of the Syn Group are located in the low luminosity end.

\section{SED Modeling}

The single-zone leptonic models are used to reproduce the SEDs of these substructures. The radiation region is assumed to be a sphere with radius $R$, magnetic field strength $B$, and bulk Lorentz factor $\Gamma$. The radius is derived from the angular radius, which is obtained from the optical or the X-ray observations. The values of $R$ are reported in Table 2 (see the appendix). The beaming factor is calculated with $\delta=1/\Gamma(1-\beta \cos\theta)$, where $\theta$ is the viewing angle. The radiation electron distribution is taken as a broken power-law, and this distribution is characterized by an electron density parameter ($N_0$), a break energy ($\gamma_{\rm b}$), and two slope indices ($p_1$ and $p_2$) below and above the break energy in the energy range of $\gamma_{\rm e}\in[\gamma_{\rm min},\gamma_{\rm max}]$. The Klein-Nashina (KN) effects and the absorption of high energy $\gamma$-ray photons by extragalactic background light (Franceschini et al. 2008) are also taken into account in our model calculations.

As discussed above, an IC component is required to represent the X-ray emission of the substructures in IC Group. Assuming that these substructures are at rest and without considering the equipartition condition, the X-ray emission of these SEDs can be reproduced by the SSC process (syn+SSC case), as reported in Zhang et al. (2010). In this case the derived $B$ is severely smaller than that under the equipartition condition. The fitting results with the syn+SSC model and the derived parameters are given in the appendix.

Considering the beaming effect, we then fit the SEDs of the substructures in IC Group by considering the IC/CMB process (syn+IC/CMB case) under the equipartition condition. Although the contribution of SSC is negligible compared with the IC/CMB component, we still take SSC process into account in syn+IC/CMB case. The CMB peak frequency at $z=0$ is $\nu_{\rm CMB}=1.6\times10^{11}$ Hz and the CMB energy density in the comoving frame is $U^{'}_{\rm CMB}=\frac{4}{3}\Gamma^2U_{\rm CMB}(1+z)^4$ (Dermer \& Schlickeiser 1994), where $U_{\rm CMB}=4.2\times10^{-13}$ erg cm$^{-3}$. The equipartition magnetic field is estimated by equation (A1) in Zhang et al. (2010) in the comoving frame. We assume $\delta=\Gamma$ in the SED modeling in syn+IC/CMB case. The values of $\gamma_{\rm min}$ are constrained with the observed SEDs via a method reported by Tavecchio et al. (2000, see also Zargaryan et al. 2017). The fitting results are also presented in the appendix\footnote{The fitting results are slightly different from that in Zhang et al. (2010) owing to the new data added and the different assumptions, e.g., the calculation methods of the equipartition magnetic field strength and the value of $U^{'}_{\rm CMB}$.}.

We fit the SEDs from radio to X-ray bands of the substructures in Syn Group with the synchrotron radiation under the equipartition condition. Without the detection of an IC component, the constraints on $\delta$ and $\gamma_{\rm min}$ are lost. We take $\delta=\Gamma=1$ and $\gamma_{\rm min}=100$ to model the SEDs. The fitting results in this scenario are also displayed in the appendix.

Note that all substructures in the Syn Group are less luminous than that in the IC Group as shown in Figure \ref{Lr-Lx}. We suspect that these substructures may have a smaller beaming factor than others, hence the cooling of their relativistic electrons by the IC/CMB process is not effective and their relativistic electrons can be accelerated to higher energy to produce the X-ray emission by the synchrotron radiation. A smaller beaming factor then may result in the lower luminosity.

Observations in a higher energy band than the X-ray band are useful for discriminating these models. As described in Section 2, there are 16 sources among the 41 AGNs that have been detected in the $\gamma$-ray band by \emph{Fermi}/LAT. However, the spatial resolution of \emph{Fermi}/LAT cannot resolve the $\gamma$-ray emission location that is either from the core jets or the substructures in large-scale jets. Therefore, we do not model the $\gamma$-ray emission with our models. We show the LAT observation data in their SEDs (see the appendix) in order to compare the observed $\gamma$-ray emission with the predictions of the model fits. It is found that, except for two RGs 3C 120 and 3C 207, the observed fluxes in the GeV band are much higher than the model predictions, likely indicating that the observed $\gamma$-ray fluxes of these AGNs are dominated by their core emission.

\section{Knot-Hotspot Sequence in the $\nu_{\rm s}$--$L_{\rm s}$ Plane}

\begin{figure*}
\includegraphics[angle=0,scale=0.42]{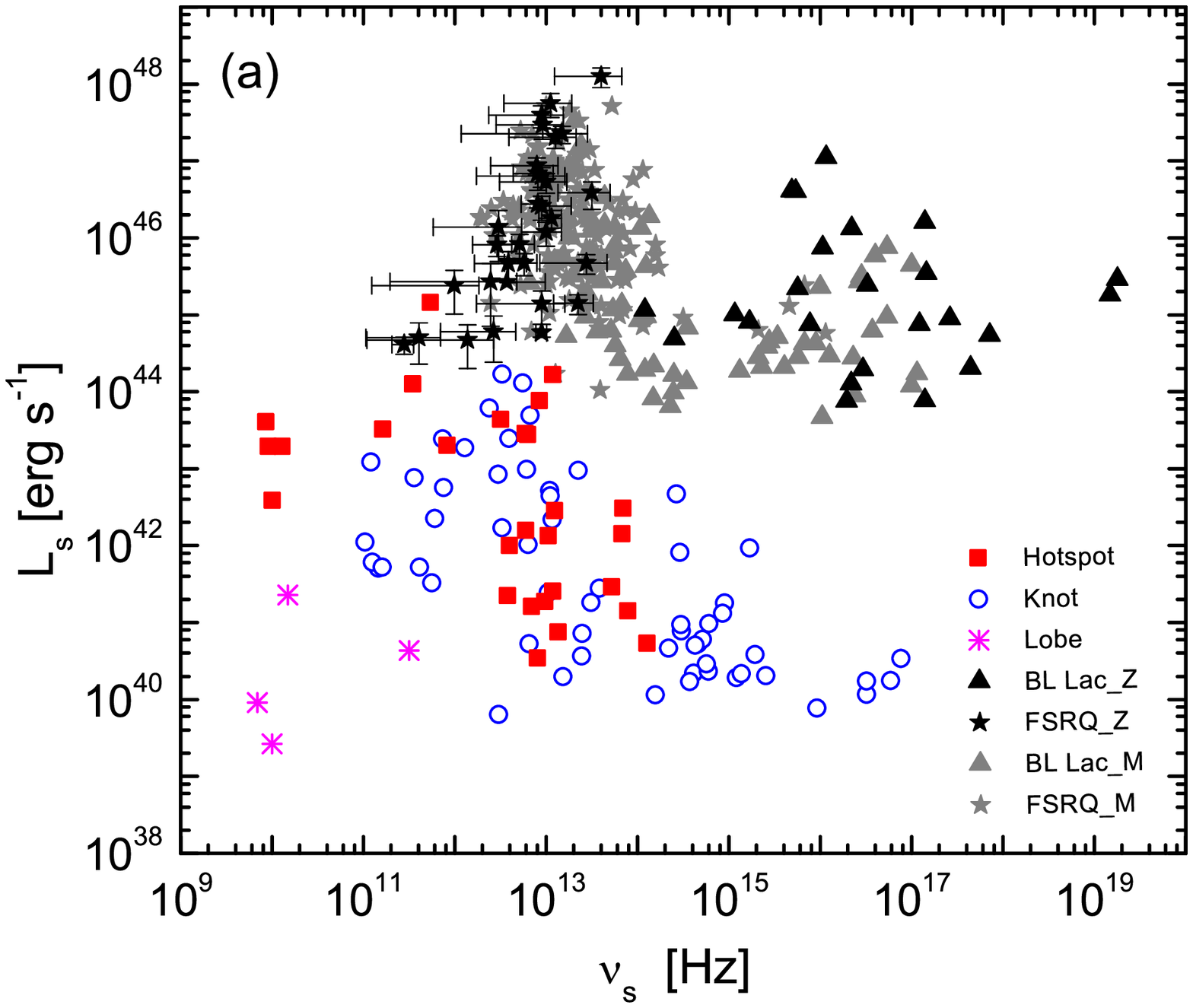}
\includegraphics[angle=0,scale=0.42]{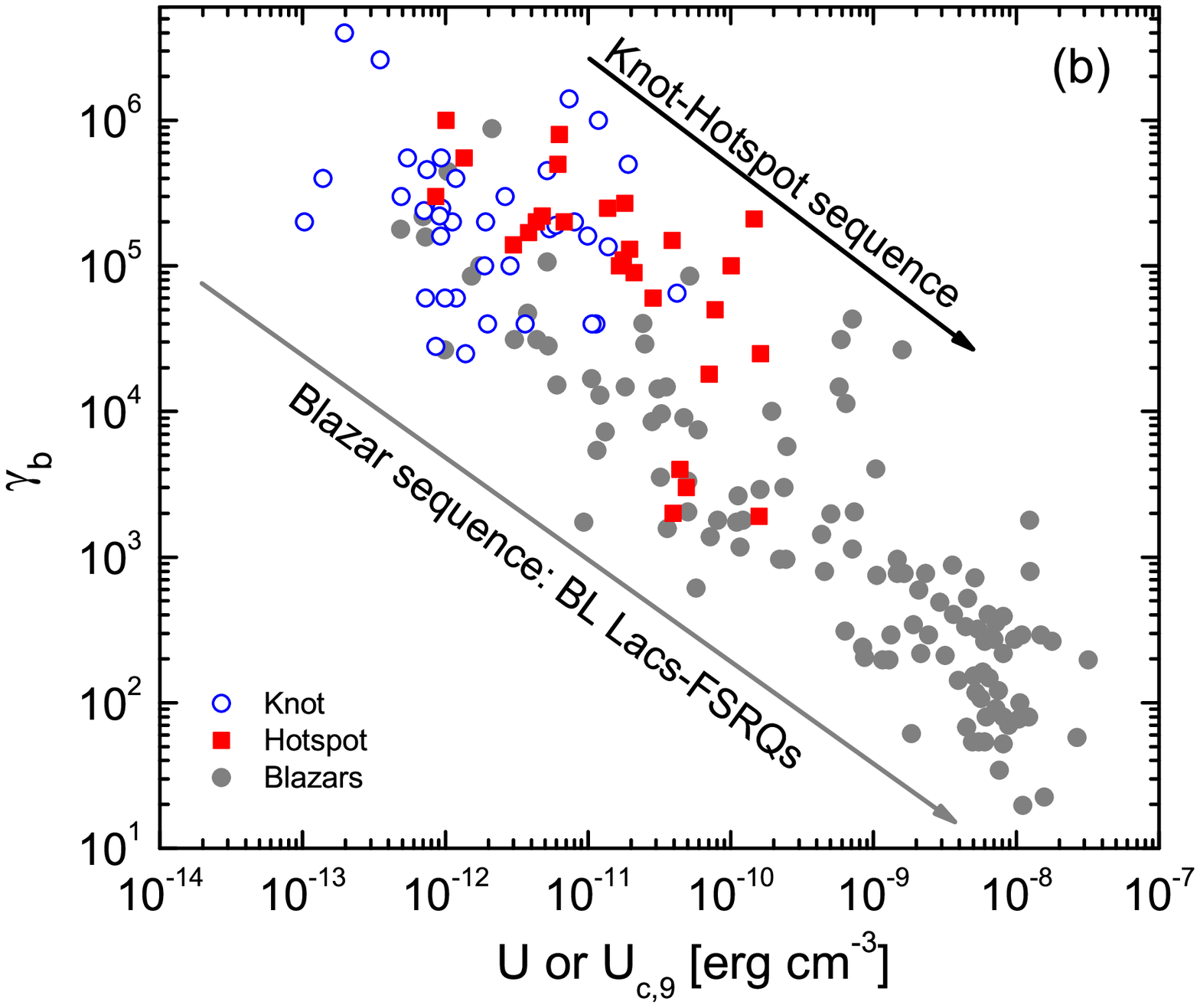}
\caption{\emph{panel (a)}---Synchrotron peak luminosity ($L_{\rm s}$) as a function of the synchrotron peak frequency ($\nu_{\rm s}$) for the 28 hotspots and 59 knots. The data of blazars are taken from Meyer et al. (2011; grey symbols) and Zhang et al. (2012, 2014, 2015; black symbols). The data of the four radio lobes are taken from literature, Cen A in  Abdo et al. (2010a), Fornax A in McKinley et al. (2015), and NGC 6251 in Takeuchi et al. (2012). \emph{panel (b)}---$\gamma_{\rm b}$ (derived in the syn+IC/CMB case) is plotted against $U$ for the 27 hotspots and 37 knots, where $U$ is the energy density of radiation region and $\gamma_{\rm b}$ is the Lorentz factor of the electrons emitting at the peaks of the synchrotron and IC components. The data of a blazar sample from Ghisellini et al. (2010) are also shown in the figure, where the energy densities of blazars in the core regions ($U_{\rm c}$) are re-scaled by $U_{\rm c}=U/10^9$ for illustration. }\label{Ls-nu}
\end{figure*}

Fossati et al. (1998) reported a spectral sequence of FSRQ--LBL--HBL (FSRQ: flat spectrum radio quasar, LBL and HBL: low- and high-frequency-peaked BL Lacs), and along with this sequence an increase in the peak frequency corresponds to the decrease of bolometric luminosity, i.e., FSRQs tend to have a higher $L_{\rm s}$ and lower $\nu_{\rm s}$ than BL Lacs (see also Ghisellini et al. 1998), where $\nu_{\rm s}$ and $L_{\rm s}$ are the synchrotron peak frequency and luminosity, respectively. This phenomenon is the so called ``blazar sequence", which aims to unify the different blazars with the bolometric luminosity. Figure \ref{Ls-nu} (a) illustrates the substructures in the $\nu_{\rm s}$--$L_{\rm s}$ plane, where $\nu_{\rm s}$ and $L_{\rm s}$ are derived from our SED fits with the syn+IC/CMB model for 64 substructures and with the synchrotron model for 23 substructures, respectively. For comparison, the blazar data from Meyer et al. (2011) and Zhang et al. (2012, 2014, 2015) and the data of four radio lobes\footnote{Normally the scales of radio lobes are much larger than hotspot and knots, and no observation data at the optical band are available for them, hence we do not take this kind of large-scale jet substructures into account in this work. The values of $\nu_{\rm s}$ and $L_{\rm s}$ for the four radio lobes are collected from literature, i.e., two radio lobes of Cen A in Abdo et al. (2010), one radio lobe of Fornax A in McKinley et al. 2015, and one radio lobe of NGC 6251 in Takeuchi et al. (2012). } are also shown in Figure \ref{Ls-nu} (a). It is found that on average the hotspots have a higher $L_{\rm s}$ and lower $\nu_{\rm s}$ than knots, showing up as an anti-correlation between $\nu_{\rm s}$ and $L_{\rm s}$. Many hotspots overlap with knots, but the tentative knot-hotspot sequence seems similar to the ``blazar sequence", i.e., many BL Lacs overlap with FSRQs in the $\nu_{\rm s}$--$L_{\rm s}$ plane. However, the four radio lobes seem not follow the $\nu_{\rm s}$--$L_{\rm s}$ anti-correlation of hotspots and knots.

Note that RGs are the parent populations of blazars, according to the unification models for radio-loud AGNs. Hotspots generally appear on boundaries of radio lobes in FR II RGs (the parent population of FSRQs), and knots are usually thought to be a part of a well-collimated jet in FR I RGs (the parent population of BL Lacs). In this respect, the knot-hotspot sequence is likely related to blazar sequence. This may shed light on the radiation mechanisms of these substructures.

Ghisellini et al. (1998) showed an anti-correlation of $\gamma_{\rm b}-U$ for blazars (see also Zhang et al. 2012), where $U$ is the energy density of radiation region including the energy of the synchrotron radiation photon field and the magnetic field, and $\gamma_{\rm b}$ is the Lorentz factor of the electrons emitting at the peaks of the synchrotron and IC components. They proposed that the blazar sequence may physically result from the different dominant cooling mechanisms of relativistic electrons in blazars; the relativistic electrons in FSRQs are mainly cooled down by the external Compton scattering process and cannot be accelerated to high energy, and thus FSRQs have higher luminosity and lower peak frequency than BL Lacs. We suspect that knots and hotspots may also have the different dominant cooling mechanisms of relativistic electrons. We therefore calculate the energy density ($U$) of these substructures in the syn+IC/CMB case\footnote{We only demonstrate the result in the syn+IC/CMB case, but this anti-correlation also exists for the syn+SSC case. We do not take the synchrotron sub-sample of 23 substructures into account here because their $\delta$ cannot be constrained when the synchrotron radiation is used to explain the X-ray emission. } and show they in the $\gamma_{\rm b}-U$ plane in Figure \ref{Ls-nu} (b). The data of the blazar sample from Ghisellini et al. (2010) are also shown in Figure \ref{Ls-nu} (b), where the energy densities of blazars in the core regions ($U_{\rm c}$) are re-scaled by $U_{\rm c}=U/10^9$ for illustration. The tentative knot-hotspot sequence is apparently similar to the FSRQ-BL Lac sequence. We suggest that the knot-hotspot sequence tendency is also likely due to the different cooling effect of their relativistic electrons.

\begin{figure*}
\includegraphics[angle=0,scale=0.4]{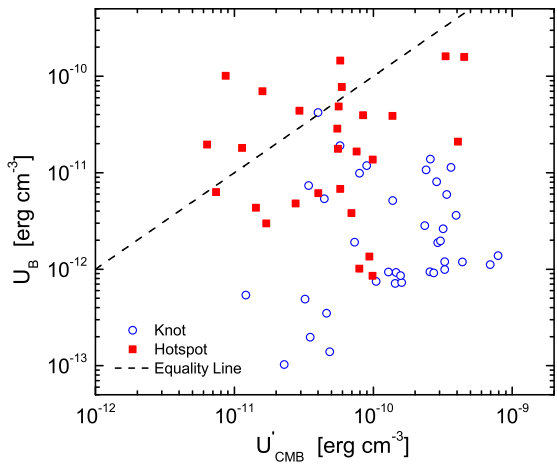}
\caption{$U_B$ is plotted against $U^{'}_{\rm CMB}$ in the comoving frame for the 27 hotspots and 37 knots in the syn+IC/CMB case.}\label{Ucmb-UB}
\end{figure*}

We compare the energy densities between the magnetic field ($U_B$) and the CMB photon field ($U^{'}_{\rm CMB}$) in the jet-comoving frame in Figure \ref{Ucmb-UB}. One can observe $U^{'}_{\rm CMB}>U_B$ for the knots with only one outlier. Therefore, the IC/CMB process may dominate the cooling of the relativistic electrons in the knots. Most hotspots also have $U^{'}_{\rm CMB}>U_{\rm B}$, similar to the knots. The IC/CMB process thus may also dominate the cooling of the relativistic electrons in these hotspots. For the rest hotspots with $U^{'}_{\rm CMB}<U_{\rm B}$, the cooling by the magnetic field via synchrotron radiation may dominate.

The $\gamma_b$ values of knots vary from $2.5\times 10^4$ to $4\times 10^6$, and most of them narrowly cluster at $\sim10^5$. This is comparable to that in BL Lacs (clustered at $10^4\sim 10^5$, see Figure 3 in Zhang et al. 2012), implying that their electrons can be accelerated effectively. The $\nu_{\rm s}$ values of knots range from $\sim10^{11}$ to $\sim10^{16}$ Hz, but the $\nu_{\rm s}$ values of BL Lacs are from $10^{14}$ to $10^{20}$ Hz (e.g., Zhang et al. 2012). The superluminal proper motions in pc-scale jets of BL Lacs observed with the VLBI have been widely reported (e.g., Lister et al. 2009, 2013), but the apparent velocity of knots in kpc-scale is still not very clear, or much lower than that in pc-scale jets (Harris \& Krawczynski 2006). Hence, the difference of the $\nu_{\rm s}$ distribution in knots and BL Lacs should be due to no (or smaller) amplification effect of relativistic jets in knots.

The distributions of the $\nu_{\rm s}$ and $\gamma_{\rm b}$ values in hotspots are much broader than that of FSRQs. It is found $\gamma_b\in [2\times 10^{3},\ 1\times 10^6]$ and $\nu_s\in [8\times10^9, \ 2\times10^{14}]$ for hotspots, while $\gamma_{\rm b} \in [10^2,\ 10^3]$ and $\nu_s\in [10^{12}, \ 10^{14}]$ for FSRQs. We do not find a narrow $\gamma_{\rm b}$ distribution in hotspots as seen in FSRQs. This would be due to that the relativistic electrons in FSRQs are mainly cooled by the photon field of the broad line region and could not be efficiently accelerated (e.g., Ghisellini et al. 1998, 2010; Zhang et al. 2015), but the cooling mechanisms of electrons in hotspots may be not universal; as shown in Figure \ref{Ucmb-UB}, the cooling of the relativistic electrons in some hotspots is dominated by the IC/CMB process, while for others their relativistic electrons are cooled by the synchrotron radiation.

\section{Testing Radiation Models with Jet Kinetic Powers}

All the substructures in our sample are spatially-resolved in the radio band of 5 GHz. In order to explain physically the observed correlations between the narrow-line luminosity and the radio luminosity of RGs and steep-spectrum quasars, Willott et al. (1999) reported that the bulk kinetic power of the jet can be estimated with the extended radio luminosity at 151 MHz ($L_{151}$) under the assumption that the minimum stored energy is required in the extended regions to produce the observed synchrotron luminosity, i.e.,
\begin{equation}
L_{\rm k}=3\times10^{45}f^{3/2}L_{151}^{6/7} \quad\rm erg\ s^{-1},
\end{equation}
where $L_{151}$ is in units of $10^{28}$ W Hz$^{-1}$ sr$^{-1}$, and $f$ incorporates many unknown factors related to the jet composition and environment, which is taken as  $f=10$ in our analysis (Cattaneo \& Best 2009). We derive the 151 MHz radio fluxes of these substructures from their 5 GHz radio fluxes by adopting a power-law radio spectrum\footnote{Notation $f_\nu=K\nu^{-\beta}$ is used. The radio spectral index ($\beta$) is taken as 0.75 for the substructures whose radio spectral indices are not available in Table 1, which is the median of the other substructures.}, then calculate their $L_{151}$ and $L_k$ values (see also Godfrey et al. 2012). Our results are reported in Table 2 (see the appendix).

We also calculate the total jet power ($P_{\rm jet}$) of these substructures on the basis of our SED fits. Following Ghisellini et al. (2010), we assume that the jet powers are carried by electrons and cold protons with a one-to-one ratio. We then calculate $P_{\rm jet}^{\rm ep} = \sum_i \pi R^2 \Gamma^2 c U_{\rm i}$, where $U_{\rm i}$ (i=e, p, $B$) is measured in the comoving frame, which are given by
\begin{eqnarray}
U_{\rm e}=m_{\rm e}c^2\int N(\gamma)\gamma d\gamma,\\
U_{\rm p}=m_{\rm p}c^2\int N(\gamma)d\gamma,\\
U_B=B^2/8\pi.
\end{eqnarray}
The derived $P^{\rm ep}_{\rm jet}$ values for the syn+IC/CMB case and the syn+SSC case are also presented in Table 2 (see the appendix).

\begin{figure*}
\includegraphics[angle=0,scale=0.42]{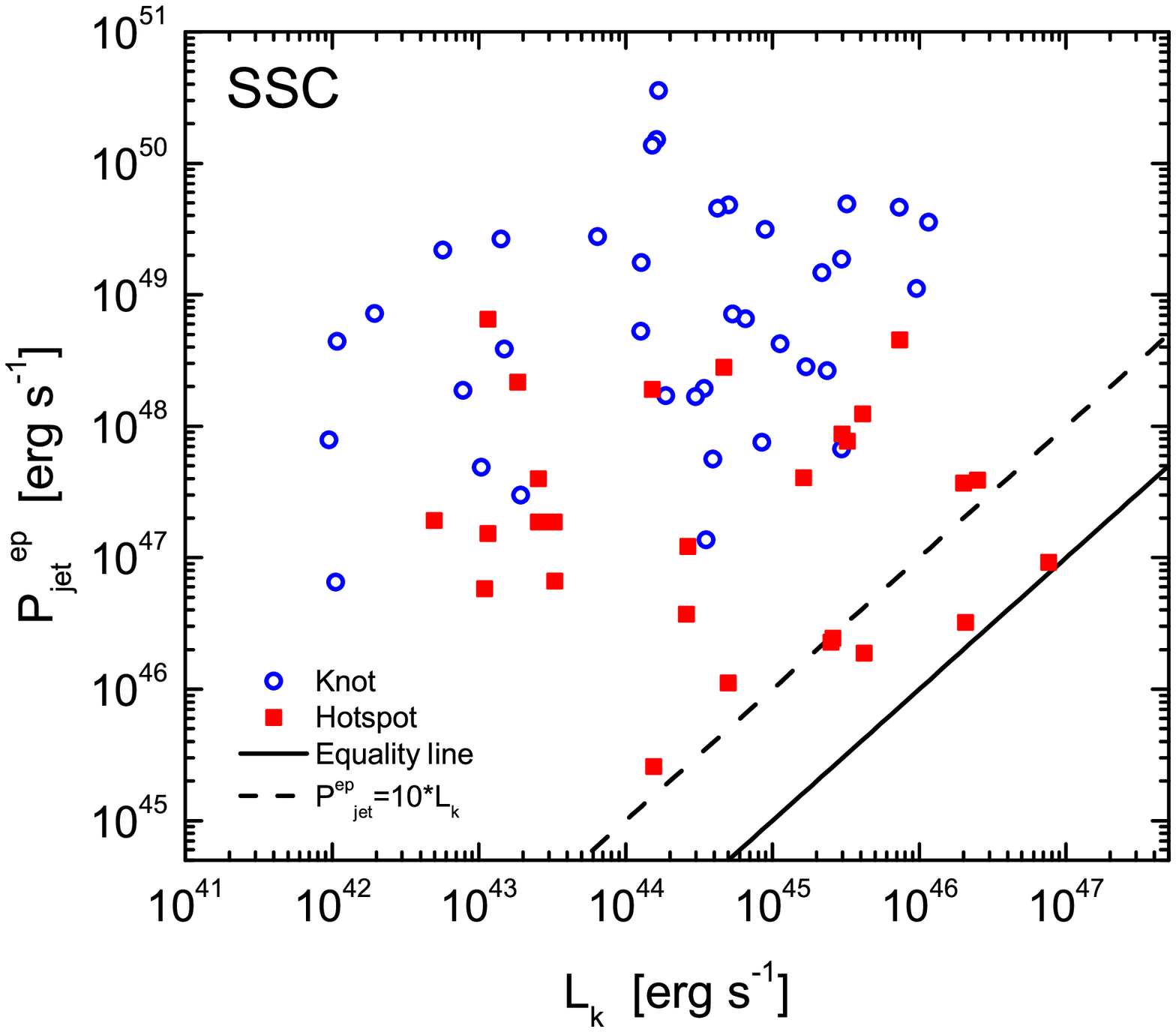}
\includegraphics[angle=0,scale=0.42]{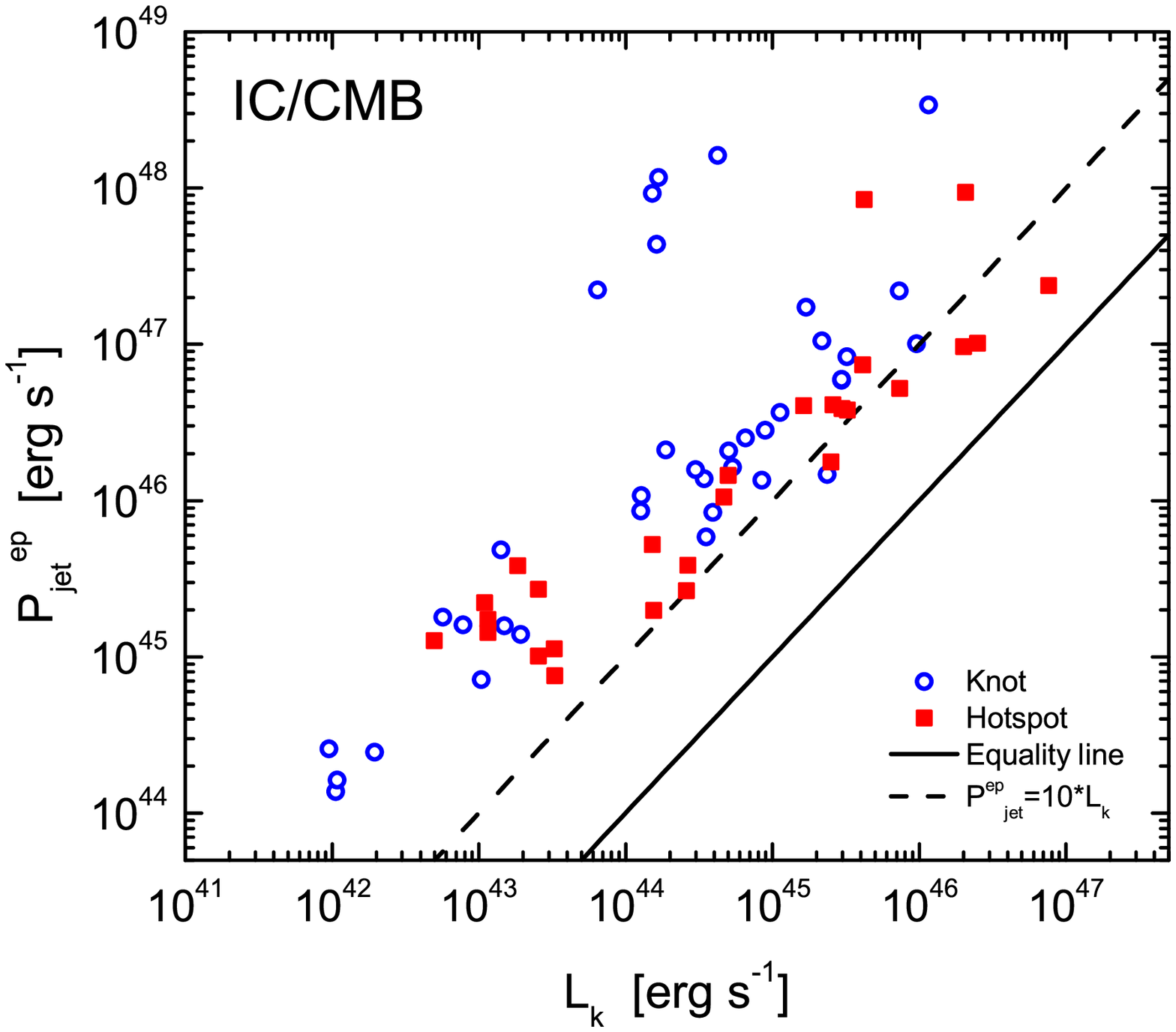}\\
\includegraphics[angle=0,scale=0.42]{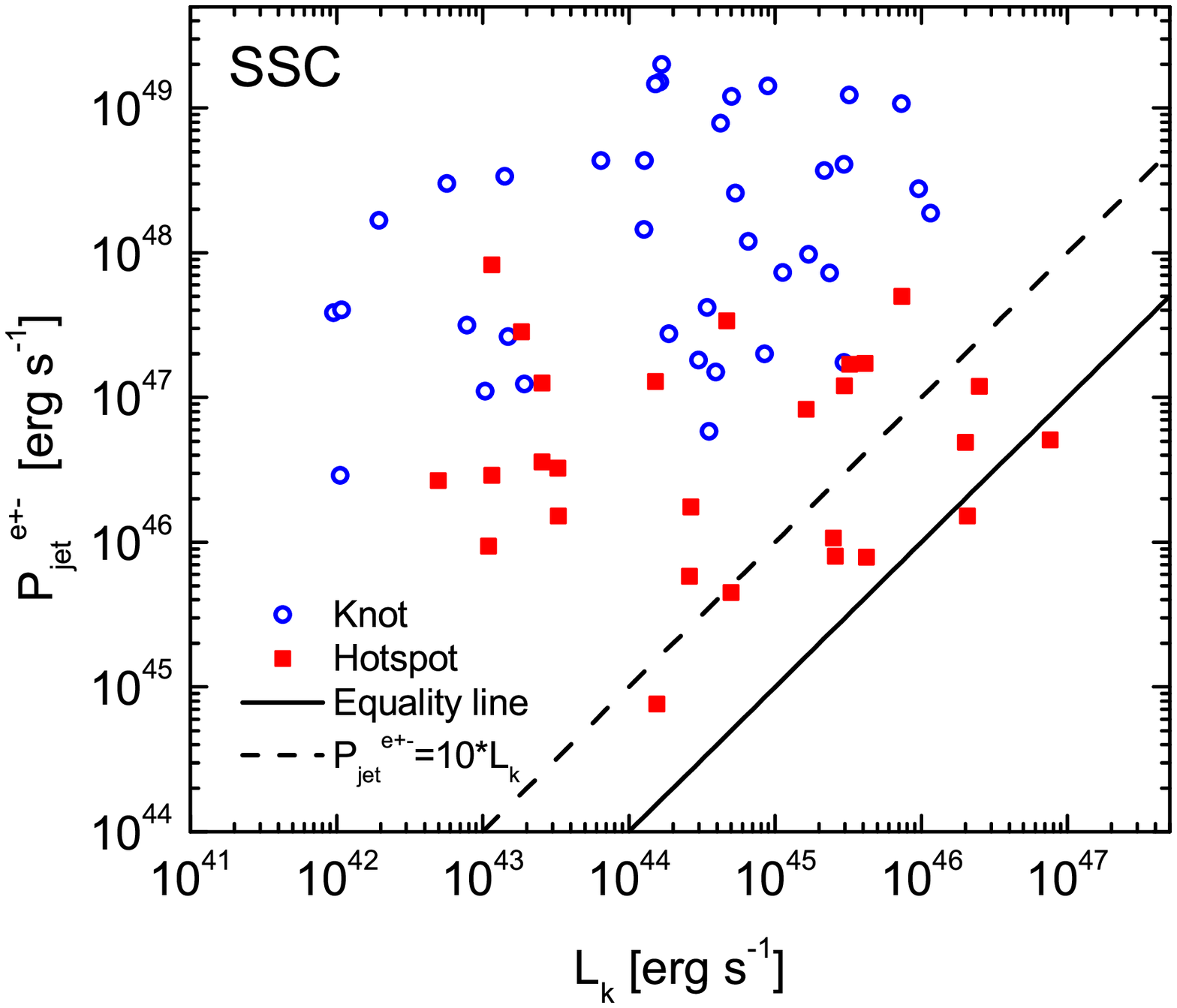}
\includegraphics[angle=0,scale=0.42]{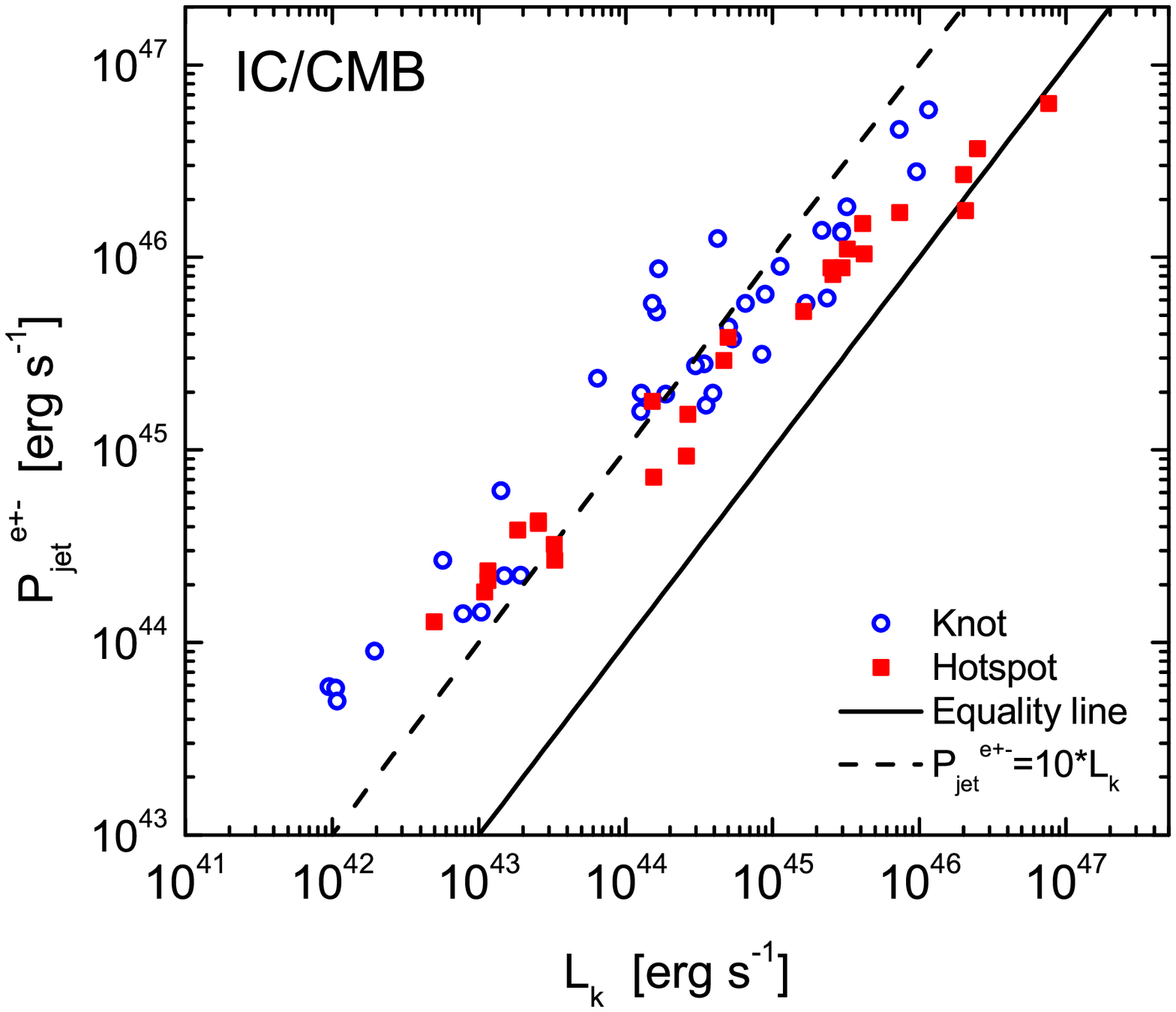}
\caption{Jet powers derived under the syn+SSC and syn+IC/CMB cases as a function of $L_{\rm k}$ for the 27 hotspots and 37 knots, where $L_{\rm k}$ is derived with the radio radiation power. Two cases are considered to calculate the jet powers, i.e., the proton-electron pair jets ($P_{\rm jet}^{\rm ep}$, \emph{the top panels}) and the positron-electron pair jets ($P_{\rm jet}^{\pm e}$, \emph{the below panels}). }\label{Pjet-Lmech}
\end{figure*}

$P_{\rm jet}^{\rm ep}$ of these substructures (27 hotspots and 37 knots) as a function of $L_{\rm k}$ for the sync+IC/CMB and syn+SSC cases are shown in Figure \ref{Pjet-Lmech}. It is found that $L_{\rm k}$ sets a lower limit to $P_{\rm jet}^{\rm ep}$ in both model cases. We find $P_{\rm jet}^{\rm ep}$ in the syn+IC/CMB case is strongly correlated with $L_{\rm k}$. The Pearson correlation analysis yields the coefficient $r=0.785$ and chance probability $p=1.55\times10^{-14}$ for all these substructures, $r=0.906$ and $p=7.63\times10^{-11}$ for the 27 hotspots and $r=0.767$ and $p=3.1\times10^{-8}$ for the 37 knots, respectively. Note that there are five knots (K-$\alpha$, K-A and K-B in PKS 1136--135, K-B and K-D in PKS 1150+497) with much higher $P_{\rm jet}^{\rm ep}$ than $L_{\rm k}$ since their proton powers may be overestimated with very small $\gamma_{\min}$ values, i.e., $\gamma_{\min}<5$.

The $P_{\rm jet}^{\rm ep}$ values of the knots in the syn+SSC scenario are at least three orders of magnitude higher than $L_{\rm k}$. Except for five hotspots, the derived $P_{\rm jet}^{\rm ep}$ values from the syn+SSC model for the other hotspots are also larger than $L_{\rm k}$ with at least one order of magnitude. We do not find any statistical correlation between $L_{\rm k}$ and $P_{\rm jet}^{\rm ep}$ with chance probability $p<10^{-4}$ in this scenario.

The above analysis is on based the assumption of equal numbers of cold protons to relativistic electrons. However, the jet composition is still a debatable issue. We thus also consider the case of the electron-positron pair jets, in which the jet powers are carried by e$^{\pm}$ pairs and magnetic fields, i.e., $P_{\rm jet}^{e^{\pm}}$ (see also Zargaryan et al. 2017). $P_{\rm jet}^{e^{\pm}}$ as a function of $L_{\rm k}$ is also presented in Figure \ref{Pjet-Lmech}. Similar results as that in the proton-electron pair case can be observed. In the syn+SSC case, $P_{\rm jet}^{e^{\pm}}$ is still much higher than $L_{\rm k}$ for all the knots and most of the hotspots, and is not correlated with $L_{\rm k}$. In the syn+IC/CMB case, the correlation between $P_{\rm jet}^{e^{\pm}}$ and  $L_{\rm k}$ is even tighter than that in the proton-electron pair case, i.e., $r=0.961$ and $p\sim0$.

\section{Conclusions and Discussion}

Based on our leptonic model fits to the SEDs in the radio-optical-X-ray band for a sample of 29 hotspots and 65 knots from the large-scale jets of 41 AGNs, we found a tentative knot-hotspot sequence in the $\nu_{\rm s}$--$L_{\rm s}$ plane, i.e., substructures with a higher synchrotron peak luminosity tend to have a lower peak frequency, similar to the ``blazar sequence". This may be physically due to the different dominant cooling mechanisms of relativistic electrons in these substructures. We estimated their kinetic powers ($L_{\rm k}$) with their radio data and calculated their jet powers on the basis of our SED fits by considering the electron-proton pair jets and the electron-positron pair jets. We show that $L_{\rm k}$ is strongly correlated with $P_{\rm jet}^{\rm ep}$ in the syn+IC/CMB case. But in the syn+SSC case, no similar correlation is presented, $P_{\rm jet}^{\rm ep}$ of knots are at least three orders of magnitude higher than $L_{\rm k}$, and only five hotspots have a $P_{\rm jet}^{\rm ep}$ comparable to $L_{\rm k}$. Considering the electron-positron pair jets, similar results are also observed. We argue that the syn+IC/CMB may be the favorable model for the high energy emission above the X-ray band in large-scale jets. Interestingly the substructures, for which their SEDs from radio to X-ray bands can be well explained by one synchrotron radiation component, have lower observed luminosity than the rest of substructures that need a IC process to reproduce their X-ray emission. This may be due to the smaller beaming factors of these synchrotron radiation substructures than others.

The tentative knot-hotspot sequence may shed light on the radiation physics of knots and hotspots. Different from the core jets, the hotspots and knots are in large-scale jets and their environments should be similar. The tentative knot-hotspot sequence then may be the results of cooling competition between the magnetic field and the CMB photon field. Future observations of polarization in the UV and X-ray bands would examine and distinguish the radiation models (Cara et al. 2013).

Strong correlation between $L_{\rm k}$ and $P_{\rm jet}^{\rm ep}$ in the syn+IC/CMB scenario is interesting. It was suggested that the syn+IC/CMB model may result in the so-called ``super-Eddington" jet powers (e.g., Dermer \& Atoyan 2004; Uchiyama et al. 2006; Meyer et al. 2015) since the syn+IC/CMB model needs that the derived electron energy distribution
extends down to lower-energy band, i.e., $\gamma_{\min}$ being very small. The $\gamma_{\min}$ values derived from our fits for most substructures are much larger than 1. This may avoid the issue of ``super-Eddington" jet power. By assuming that the jet powers are carried by the electron-positron pairs and magnetic fields, we found that $P_{\rm jet}^{\pm e}$ is more comparable to $L_{\rm k}$ in the syn+IC/CMB scenario and the correlation between them becomes tighter than that in the proton-electron pair jet case.

\acknowledgments

We thank the anonymous referee for his/her valuable suggestions. This work is supported by the National Natural Science Foundation of China (grants 11573034, 11533003, and U1731239), the National Basic Research Program (973 Programme) of China (grant 2014CB845800), and En-Wei Liang acknowledges support from the special funding from the Guangxi Science Foundation for Guangxi Distinguished Professors (Bagui Yingcai \& Bagui Xuezh; 2017AD22006).

\begin{deluxetable}{lccccccc}
\tabletypesize{\scriptsize}\tablecolumns{8}\tablewidth{40pc} \tablecaption{List of Large-scale Jet Substructures in AGNs Included Our Sample} \tablenum{1}
\tablehead{ \colhead{Source}&\colhead{Comp}&\colhead{\emph{z}\tablenotemark{a}}&\colhead{$\alpha_{\rm r}$\tablenotemark{b}}&\colhead{$\alpha_{\rm x}$\tablenotemark{b}} &\colhead{$F_{\rm 5GHz}$\tablenotemark{c}}&\colhead{$F_{\rm 1keV}$\tablenotemark{c}}&\colhead{Re\tablenotemark{d,e}}\\\colhead{}&\colhead{}&\colhead{}&\colhead{}&\colhead{} &\colhead{[Jy]}&\colhead{[Jy]}&\colhead{}
}
\startdata
3C 15&K-C&0.073& &0.7$\pm$0.4&0.03522&1.5E-9$\pm$4E-10&(1)\\
3C 17&K-S3.7&0.22&0.87& &0.03$\pm$1E-3&1.45E-9$\pm$6.1E-10&(2)\\
&K-S11.3&&0.79& &0.083$\pm$0.002&5.6E-10$\pm$2.7E-10&\\
3C 31&K-R1&0.0167&0.58& &0.04&3.3E-9$\pm$3E-10&(3)\\
&K-R2&&0.55& &0.05&3.1E-10$\pm$1E-10&\\
3C 33&H-S1&0.0597&0.75&0.8$\pm$0.6&0.61327&1.4E-10$\pm$6E-11&(4)\\
&H-S2&&0.98&0.8$\pm$0.6&0.49809&3.2E-10$\pm$9E-11&\\
&H-N1&&0.88&1.2$\pm$0.8&0.0631&2.7E-10$\pm$8E-11&\\
&H-N2&&0.9&1.2$\pm$0.8&0.05912&1.9E-10$\pm$7E-11&\\
PKS 0208-512&K-K0&0.999& &0.69$\pm$0.36&0.00747&5.86E-9&(5)(I)\\
&K-K1&& &0.69$\pm$0.36&0.00385&1.75E-9&\\
&K-K3&& &0.69$\pm$0.36&0.0334&7.48E-9&\\
3C 66B&K-A&0.0215&0.75&0.97$\pm$0.34&0.0039&4E-9$\pm$3E-10&(6)\\
&K-B&&0.6&1.17$\pm$0.14&0.034&6.1E-9$\pm$4E-10&\\
PKS 0405--123&H-N&0.574&0.9$\pm$0.1& &0.204&1.6E-9$\pm$5E-10&(7)\\
3C 105&H-S1&0.089&0.54& &0.067$\pm$0.005&1.6E-9&(8)\\
&H-S3&&0.74& &0.51$\pm$0.015&8.89E-10&\\
3C 111&K-K9&0.0491&&&0.0102&1.93E-9&(9)(II)\\
&K-K14&&&&0.0138&1.64E-9&\\
&K-K22&&&&0.00558&7.21E-10&\\
&K-K30&& &0.76$\pm$0.29&0.0194&4.07E-9&\\
&K-K40&&&&0.01217&6.31E-10&\\
&K-K45&&&&0.01765&9.91E-10&\\
&K-K51&&&&0.0089&3.97E-10&\\
&K-K61&& &1.01$\pm$0.21&0.05627&7.89E-9&\\
&K-K97&&&&0.04114&1.06E-9&\\
&K-K107&&&&0.08559&3.24E-10&\\
&H-N&& &0.83$\pm$0.28&0.9006&3.35E-9&\\
&H-S&&&&0.26933&6.31E-10&\\
3C 120&K-K4&0.033&0.74&0.9$\pm$0.2&0.066$\pm$0.003&1E-8$\pm$2E-9&(10)(II)\\
&K-S2&&0.67&0.2$\pm$0.6&0.0084$\pm$4E-4&2.9E-9&\\
&K-S3&&0.69& &0.0051$\pm$3E-4&8E-10$\pm$6E-10&\\
&K-K7&&0.68&2.4$\pm$0.6&0.016$\pm$1E-3&6.3E-9$\pm$1.6E-9&\\
Pictor A&H-W&0.035&0.74&1.07$\pm$0.11&2&4.5E-8&(11)(III)\\
PKS 0521--365&K-K&0.055&0.89&1.3$\pm$0.3&0.15&1.4E-8&(12)(IV)\\
PKS 0637--752&K-WK8.9&0.651&0.74& &0.06294&7.34E-9$\pm$4.3E-10&(13)(V)\\
&K-all&&0.78& &0.22477&1.81E-8$\pm$9.4E-10&\\
PKS 0836+299 (4C +29.30)&H-B&0.064&0.85$\pm$0.1& &0.018&2.2E-9$\pm$6E-10&(7)\\
3C 207&K-A&0.68& &0.1$\pm$0.3&0.041&3E-9$\pm$7E-10&(7)(II)\\
3C 228&H-S&0.5524&&&0.132&1.3E-9&(14)(15)\\
3C 245&H-D&1.029&0.87$\pm$0.08& &0.222$\pm$0.022&7E-10$\pm$3E-10&(7)\\
PKS 1136--135&K-$\alpha$&0.554&0.75$\pm$0.1&0.9$\pm$0.4&0.0044&1.9E-9$\pm$2E-10&(16)\\
&K-A&&0.67$\pm$0.11&1.1$^{+0.3}_{-0.6}$&0.0054&1.7E-9$\pm$2E-10&\\
&K-B&&0.81$\pm$0.13&1.1$^{+0.2}_{-0.3}$&0.0109$\pm$0.0016&3.5E-9$\pm$2E-10&\\
&K-C&&0.66$\pm$0.09&0.5$^{+0.3}_{-0.2}$&0.03067&1.8E-9$\pm$2E-10&\\
&K-D&&0.71$\pm$0.08&0.5$\pm$0.5&0.0486$\pm$0.0049&1E-9$\pm$2E-10&\\
&K-E&&0.82$\pm$0.09&1.3$^{+0.6}_{-0.5}$&0.10213&7E-10$\pm$2E-10&\\
3C 263&H-K&0.6563&0.84&1.0$\pm$0.3&0.582&1E-9$\pm$1E-10&(17)\\
PKS 1150+497 (4C +49.22)&K-B&0.334&0.72&0.7$\pm$0.2&0.01705$\pm$0.0017&7.6E-9$\pm$5E-10&(7)(VI)\\
&K-C&&0.71&0.5$\pm$0.3&0.01281$\pm$0.002&2.9E-9$\pm$4E-10&\\
&K-D&&0.68&0.7$\pm$0.5&0.00641$\pm$1E-3&1.3E-9$\pm$3E-10&\\
&K-E&&0.67&0.7$\pm$0.3&0.0146$\pm$0.0015&1.7E-9$\pm$2E-10&\\
&K-IJ&&0.81&1.1$\pm$0.6&0.0285$\pm$0.0043&6E-10$\pm$1E-10&\\
3C 273&K-A&0.1583&0.85&0.83$\pm$0.02&0.13271&4.65E-8$\pm$5.4E-10&(18)(VII)\\
&K-B&&0.82&0.80$\pm$0.03&0.10517&1.09E-8$\pm$2.5E-10&\\
&K-C1&&0.73&1.07$\pm$0.06&0.14735&4.85E-9$\pm$1.6E-10&\\
&K-C2&&0.75&0.96$\pm$0.05&0.30062&6.25E-9$\pm$1.8E-10&\\
&K-D1&&0.77&1.02$\pm$0.05&0.42356&5.16E-9$\pm$1.7E-10&\\
&K-DH&&0.85&1.04$\pm$0.04&1.29672&7.82E-9$\pm$2E-10&\\
M87 &K-A&0.00427& &1.61$\pm$0.07&2.4604&1.56E-7$\pm$8.8E-9&(19)(II)\\
 &K-B&& &1.59$\pm$0.12&1.66927&3.03E-8$\pm$5.5E-9&\\
 &K-C1&& &1.33$\pm$0.06&1.14974&1.46E-8$\pm$5.2E-9&\\
 &K-D&& &1.43$\pm$0.09&0.33725&5.15E-8$\pm$4.2E-9&\\
 &K-E&& &1.48$\pm$0.12&0.10142&3.22E-8$\pm$6.5E-9&\\
 &K-F&& &1.64$\pm$0.15&0.29594&2.01E-8$\pm$5.2E-9&\\
3C 275.1&H-N&0.555&0.69$\pm$0.12& &0.178$\pm$0.027&1.78E-9&(15)(20)(VIII)\\
PKS 1229--021&K-A&1.045&&&0.107$\pm$0.011&8.5E-9$\pm$3.2E-9&(21)\\
PKS 1354+195 (4C +19.44)&K-S4.0&0.72&0.74$\pm$0.05&0.55$\pm$0.13&0.0165$\pm$2E-4&5.13E-10$\pm$7.9E-11&(22)\\
&K-S5.3&&0.74$\pm$0.09&0.90$\pm$0.22&0.00669$\pm$2.1E-4&1.84E-10$\pm$4.5E-11&\\
3C 295&H-NW&0.45&0.94&0.9$\pm$0.5&1.07023&3.8E-9&(23)\\
PKS 1421--490&H-A&0.662&0.67&0.31$\pm$0.32&4.91257$\pm$0.01&1.33E-8$\pm$1.6E-9&(24)\\
3C 303&H-W&0.141&0.84&0.4$\pm$0.2&0.26$\pm$0.03&2.8E-9&(25)\\
3C 345&K-A&0.594&0.55& &0.18163&6.35E-11&(26)(IX)\\
3C 346&K-C&0.161& &1.0$\pm$0.3&0.15798&1.6E-9$\pm$2E-10&(27)\\
3C 351&H-L&0.372&0.75&0.85$\pm$0.1&0.5&3.4E-9$\pm$4E-10&(15)(28)\\
&H-J&&0.7&0.5$\pm$0.1&0.201$\pm$0.01005&4.3E-9$\pm$3E-10&\\
3C 371&K-A&0.051&0.69&1.1$\pm$0.4&0.01172&6.55E-9&(29)(X)\\
3C 390.3&H-B&0.0561& &0.9$\pm$0.15&0.0667&4.2E-9$\pm$8.7E-10&(30)\\
3C 403&K-F1&0.059& &0.75$\pm$0.4&0.02679&9E-10$\pm$2E-10&(15)(31)\\
&K-F6&& &0.7$\pm$0.3&0.04074&2.3E-9$\pm$2E-10&\\
Cygnus A&H-A&0.056&0.5&0.77$\pm$0.13&44.6$\pm$4.46&3.12E-8$\pm$4.3E-9&(32)\\
&H-B&&0.38&0.7$\pm$0.35&9.96$\pm$1&6.8E-9$\pm$2.6E-9&\\
&H-D&&0.59&0.8$\pm$0.11&57.69$\pm$5.77&4.79E-8$\pm$5.9E-9&\\
PKS 2101--490&K-K6&1.04&0.81$\pm$0.01&0.85$\pm$0.2&0.01698&7.5E-10$\pm$2E-10&(33)\\
PKS B2152--699&H-S&0.0283&0.7& &0.15542&1.65E-9&(34)\\
&H-N&&0.7& &0.06173&1.04E-9&\\
&K-D&&&&0.015&1.31E-9&\\
PKS 2201+044&K-A&0.027&0.71$\pm$0.07&1.1$\pm$0.4&0.00875&5.6E-9&(29)(XI)\\
&K-$\beta$&&0.59$\pm$0.07&0.9$\pm$0.5&0.00229&3.8E-9&\\
PKS 2209+080&K-E&0.485&0.71&&0.02813$\pm$0.00605&4.86E-9$\pm$1E-9&(35)\\
3C 445&H-SW&0.056&0.64&&0.04958&6.28E-10&(36)(XII)\\
&H-SE&&0.71&&0.09786&7.57E-10&\\
3C 454.3&K-A&0.859&&&0.044$\pm$0.004&6E-9$\pm$1.4E-9&(21)(VII)\\
&K-B&&&&0.216$\pm$0.022&6E-9$\pm$1.4E-9&\\

\enddata
\tablenotetext{a}{z: redshift.}
\tablenotetext{b}{$\alpha_{\rm r}$ and $\alpha_{\rm x}$ are the radio and X-ray spectral indices, respectively.}
\tablenotetext{c}{$F_{\rm 5GHz}$ and $F_{\rm 1keV}$ are the observed flux densities at 5 GHz and 1 keV, respectively. }
\tablenotetext{d}{References of the substructure data: (1) Dulwich, et al. 2007; (2) Massaro et al. 2009; (3) Lanz et al. 2011; (4) Kraft et al. 2007; (5) Perlman et al. 2011; (6) Kataoka \& Stawarz 2005; (7) Sambruna et al. 2004 (8) Orienti et al. 2012; (9) Clautice et al. 2016; (10) Harris et al. 2004; (11) Wilson et al. 2001; (12) Falomo et al. 2009; (13) Meyer et al. 2017b; (14) Hardcastle et al. 2004; (15) Werner et al. 2012; (16) Sambruna et al. 2006; Cara et al. 2013; (17) Hardcastle et al. 2002; (18) Jester et al. 2007; (19) Liu \& Shen 2007; Perlman et al. 2001; (20) Cheung et al. 2005; (21) Tavecchio et al. 2007; (22) Harris et al. 2017; (23) Harris et al. 2000; (24) Gelbord et al. 2005; (25) Kataoka et al. 2003; (26) Kharb et al. 2012; (27) Worrall \& Birkinshaw 2005; (28) Brunetti et al. 2001; (29) Sambruna et al. 2007; (30) Harris et al. 1998; (31) Kraft et al. 2005; (32) Stawarz et al. 2007; (33) Godfrey et al. 2012; (34) Worrall et al. 2012; (35) Breiding et al. 2017; (36) Orienti et al. 2012.}
\tablenotetext{e}{References for $\gamma$-ray data: (I) Ghisellini et al. 2010; (II) Xue et al. 2017; (III) Brown \& Adams 2012; (IV) D'Ammando et al. 2015; (V) Meyer et al. 2015; (VI) Cutini et al. 2014; (VII) Zhang et al. 2015; (VIII) Liao et al. 2015; (IX) Berton et al. 2017; (X) Ghisellini et al. 2011; (XI) the NASA/IPAC Extragalactic Database (NED); (XII) Kataoka et al. 2011. }
\end{deluxetable}

\clearpage
\appendix
\section{SED Fitting Results}
\begin{figure*}
\includegraphics[angle=0,scale=0.215]{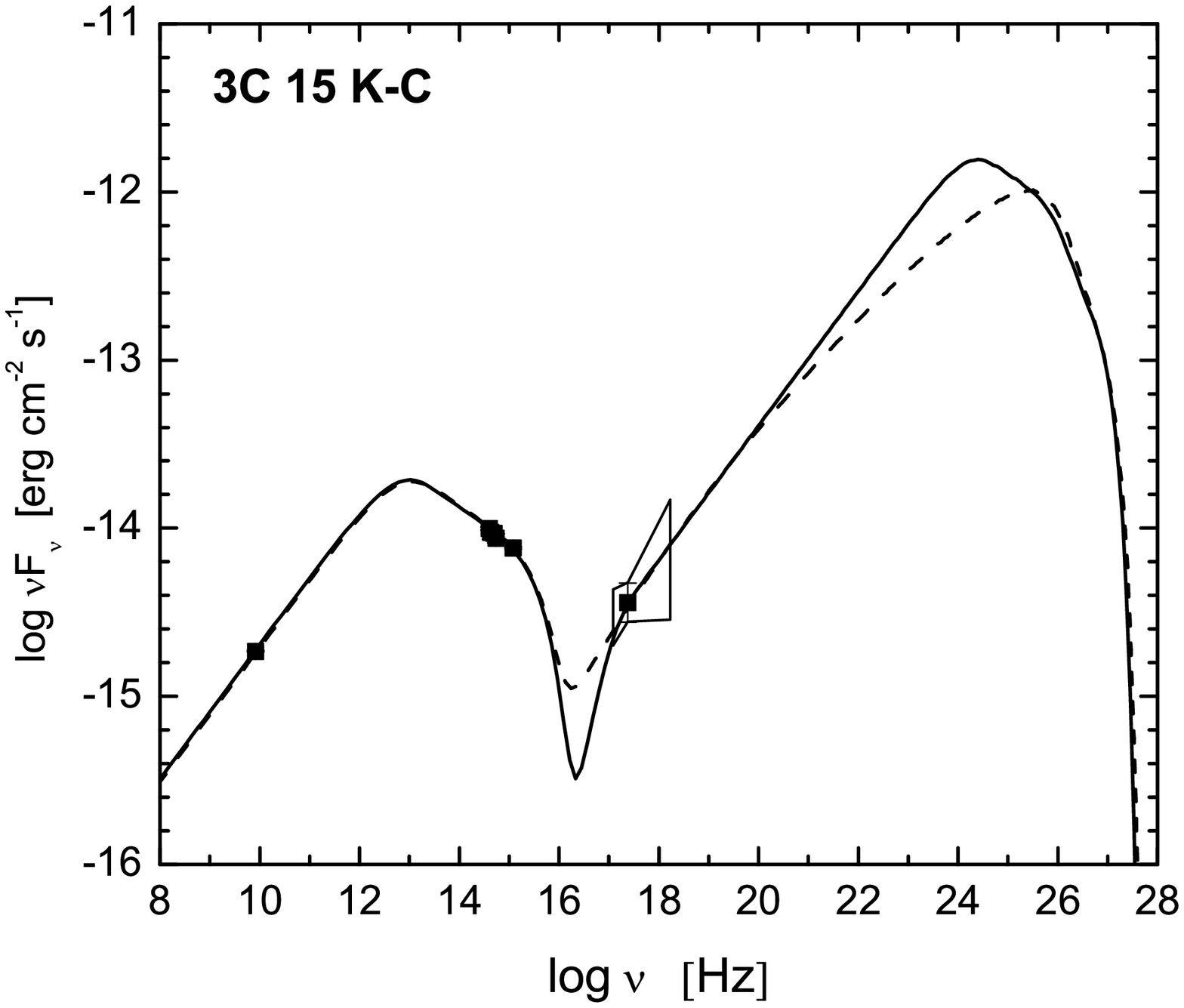}
\includegraphics[angle=0,scale=0.215]{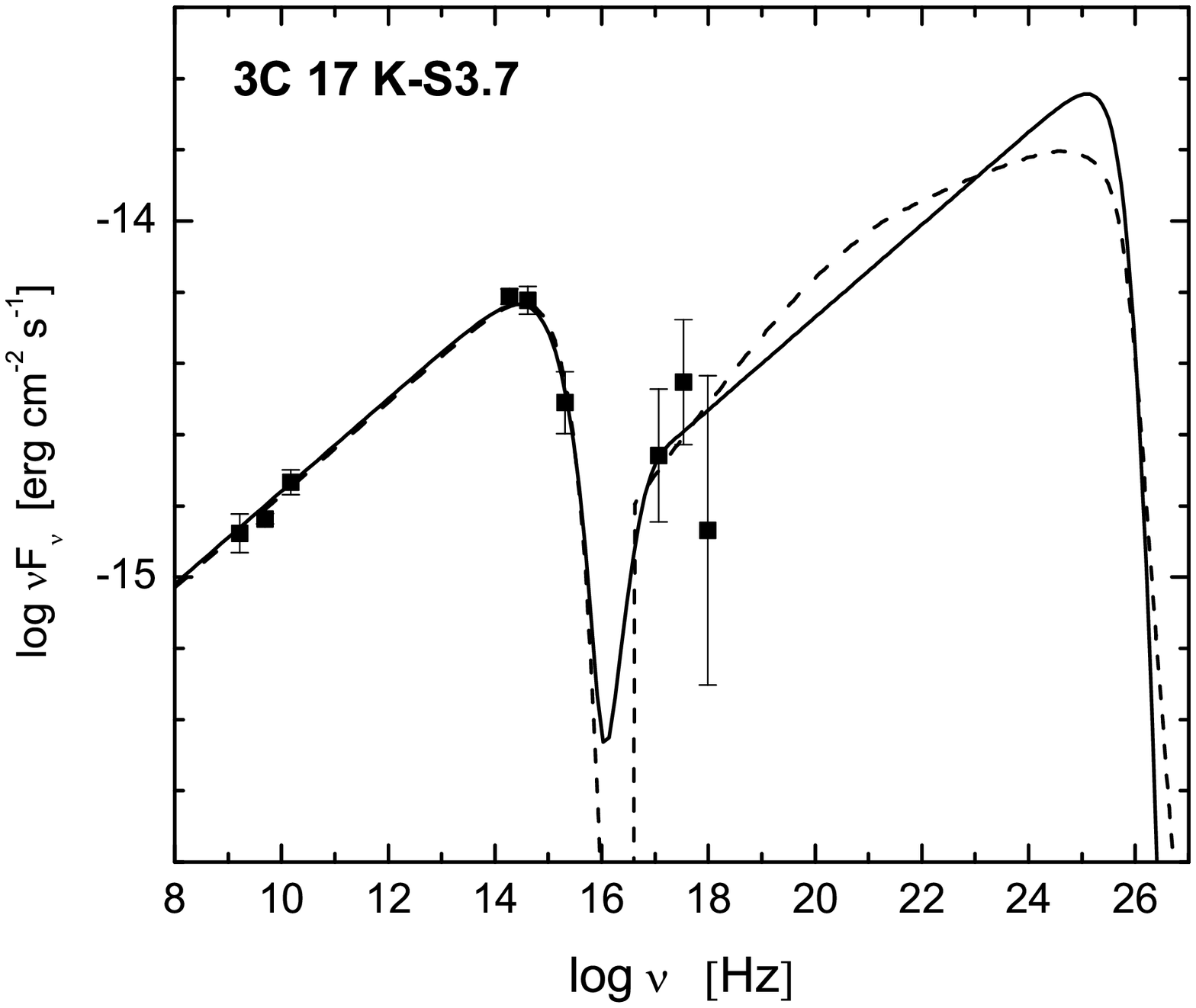}
\includegraphics[angle=0,scale=0.215]{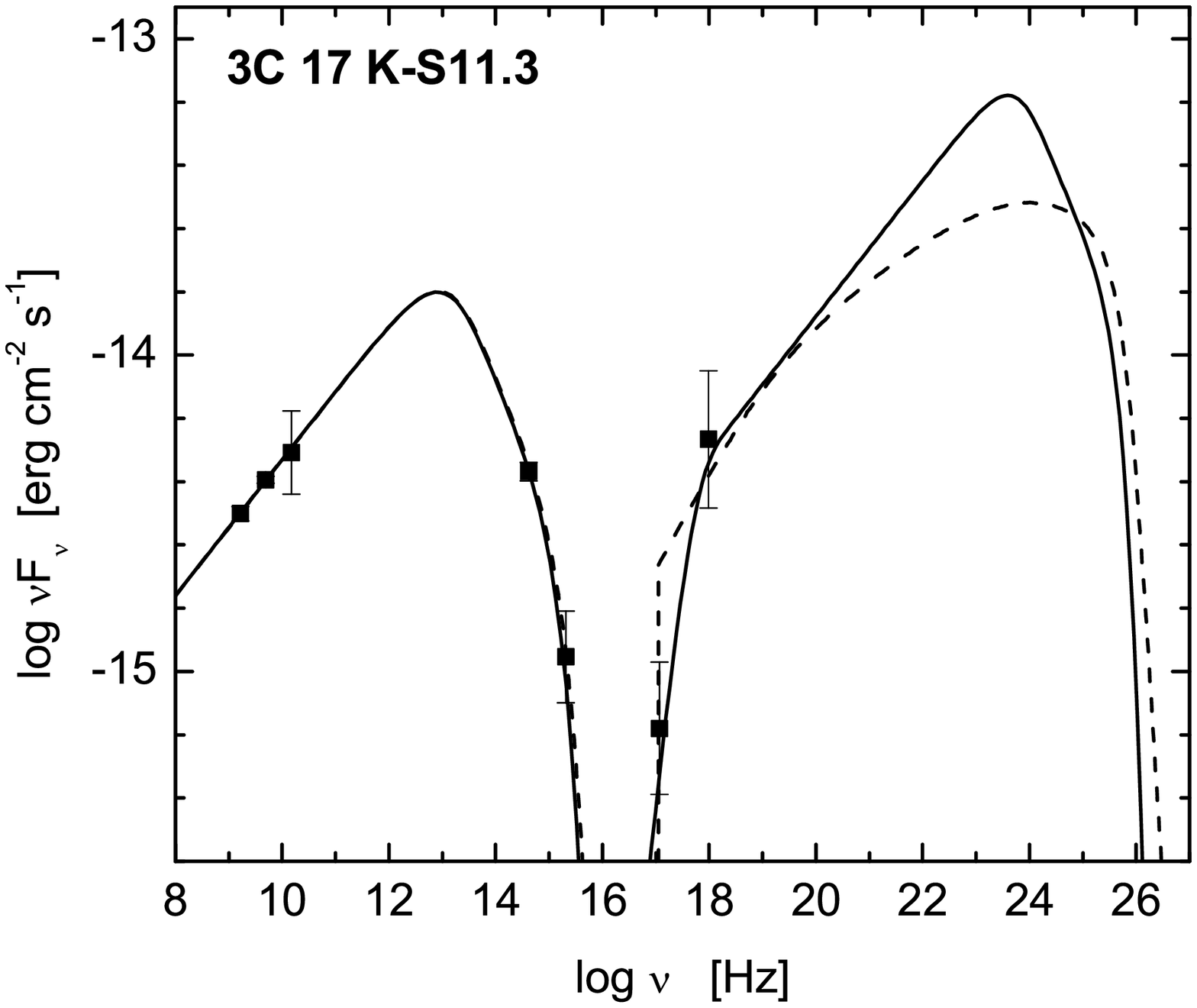}
\includegraphics[angle=0,scale=0.215]{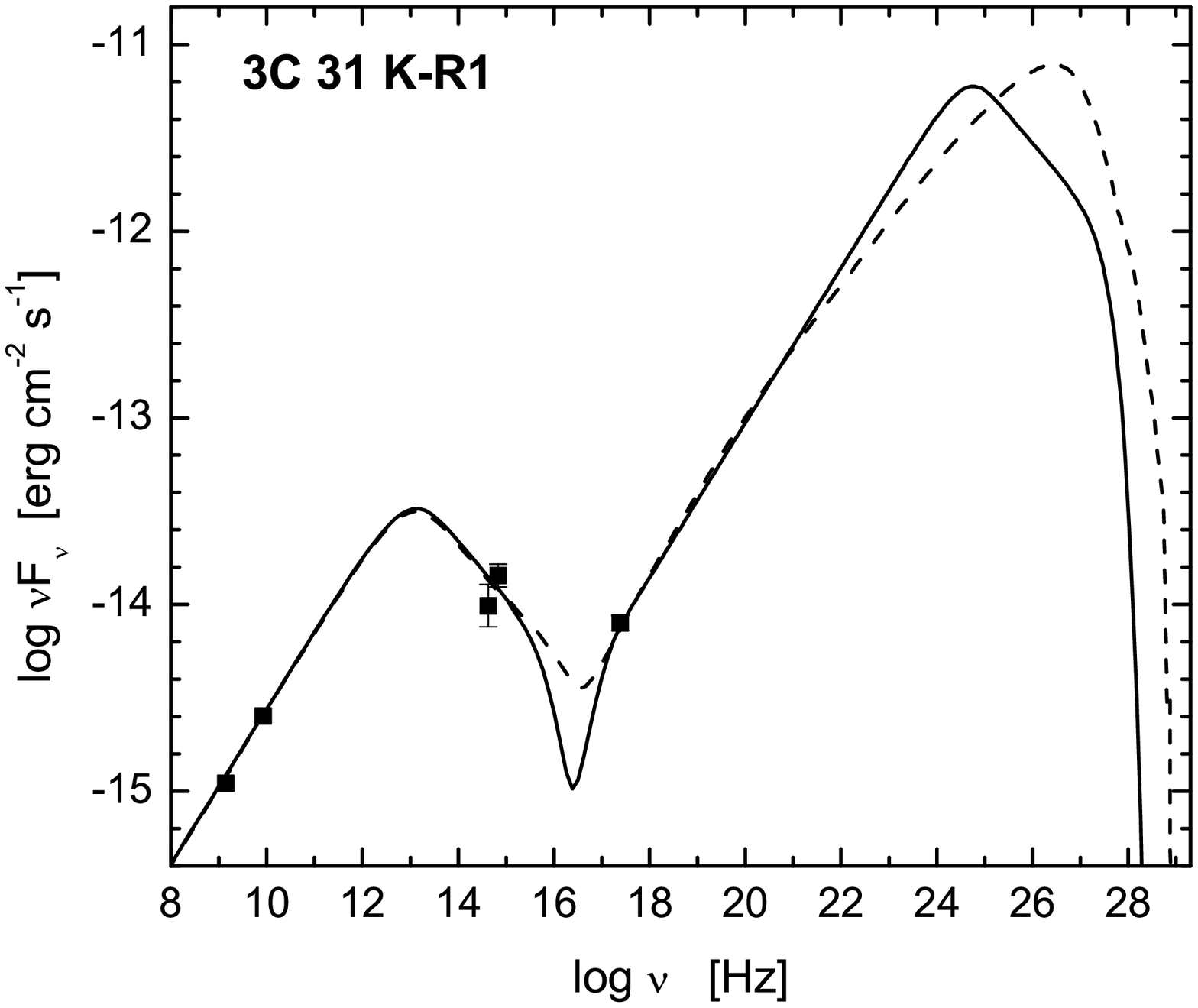}\\
\includegraphics[angle=0,scale=0.215]{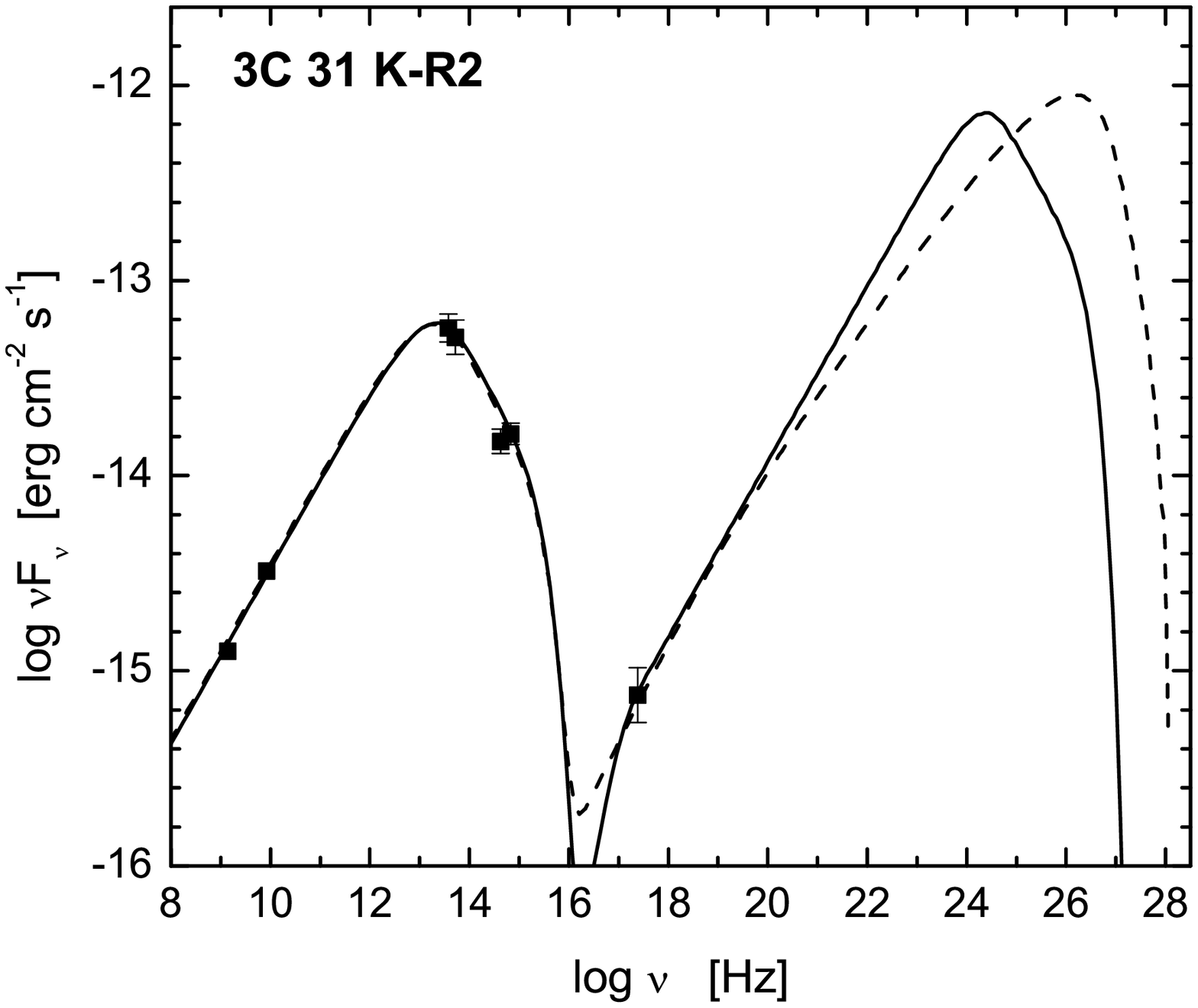}
\includegraphics[angle=0,scale=0.215]{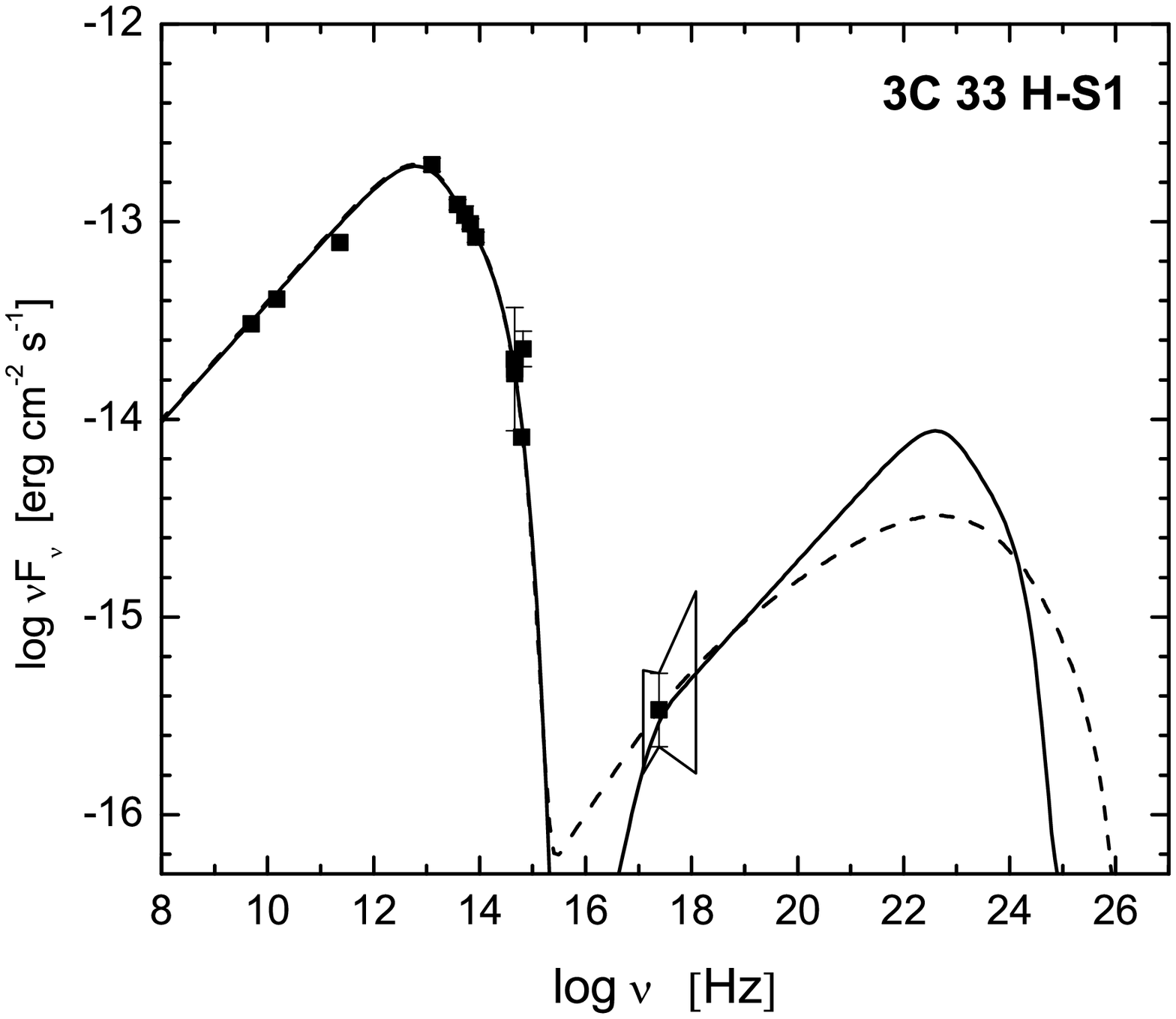}
\includegraphics[angle=0,scale=0.215]{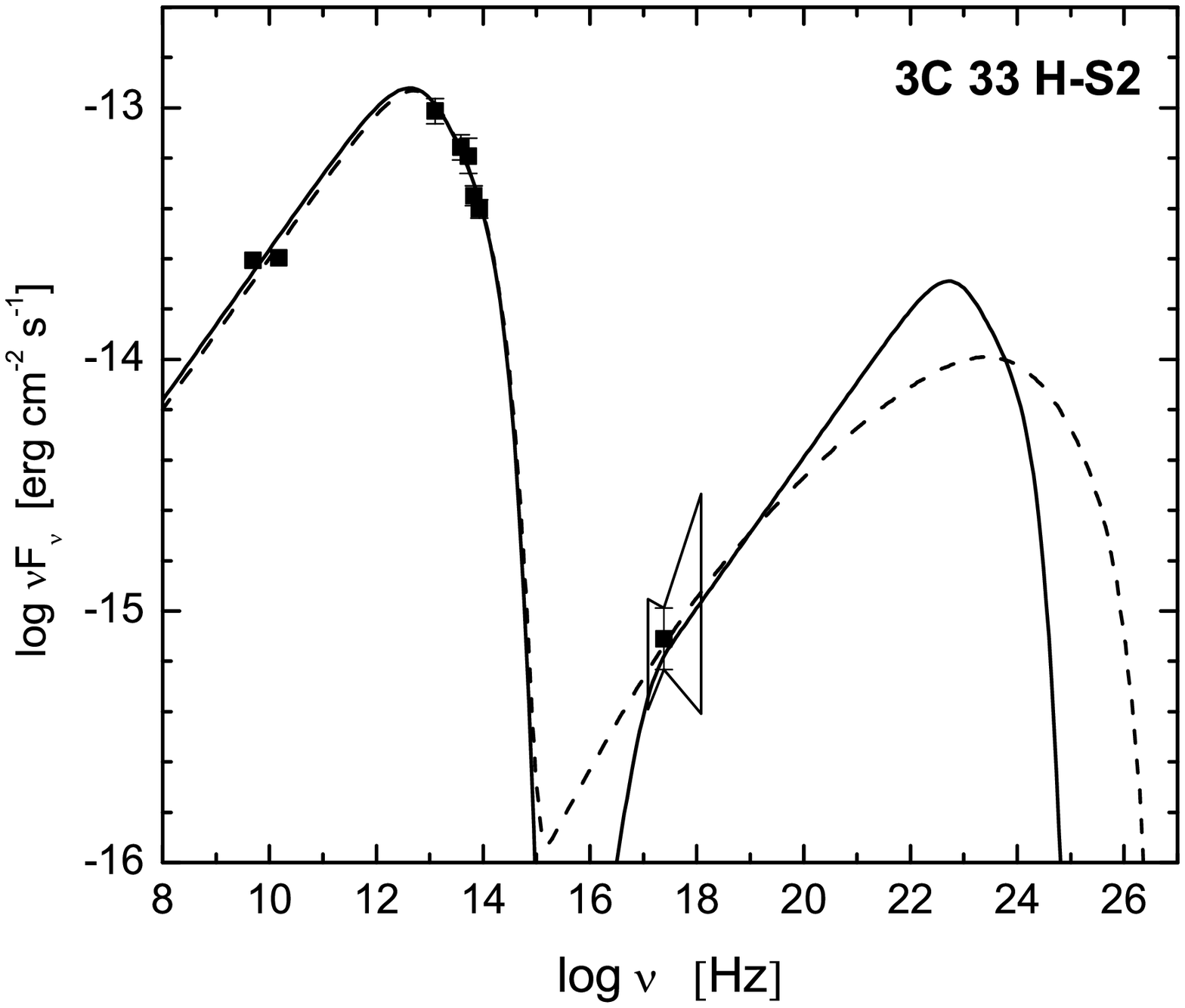}
\includegraphics[angle=0,scale=0.215]{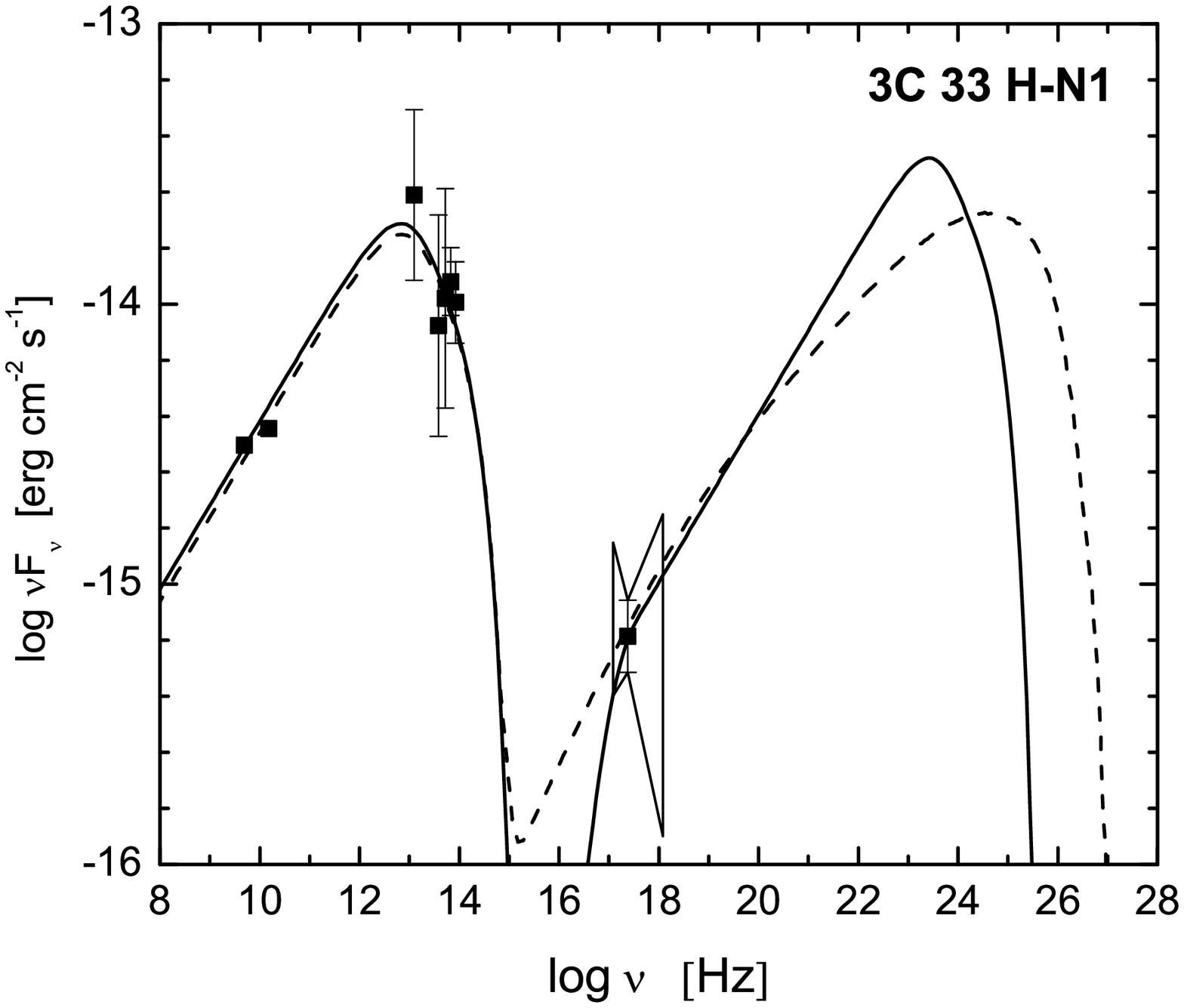}\\
\includegraphics[angle=0,scale=0.215]{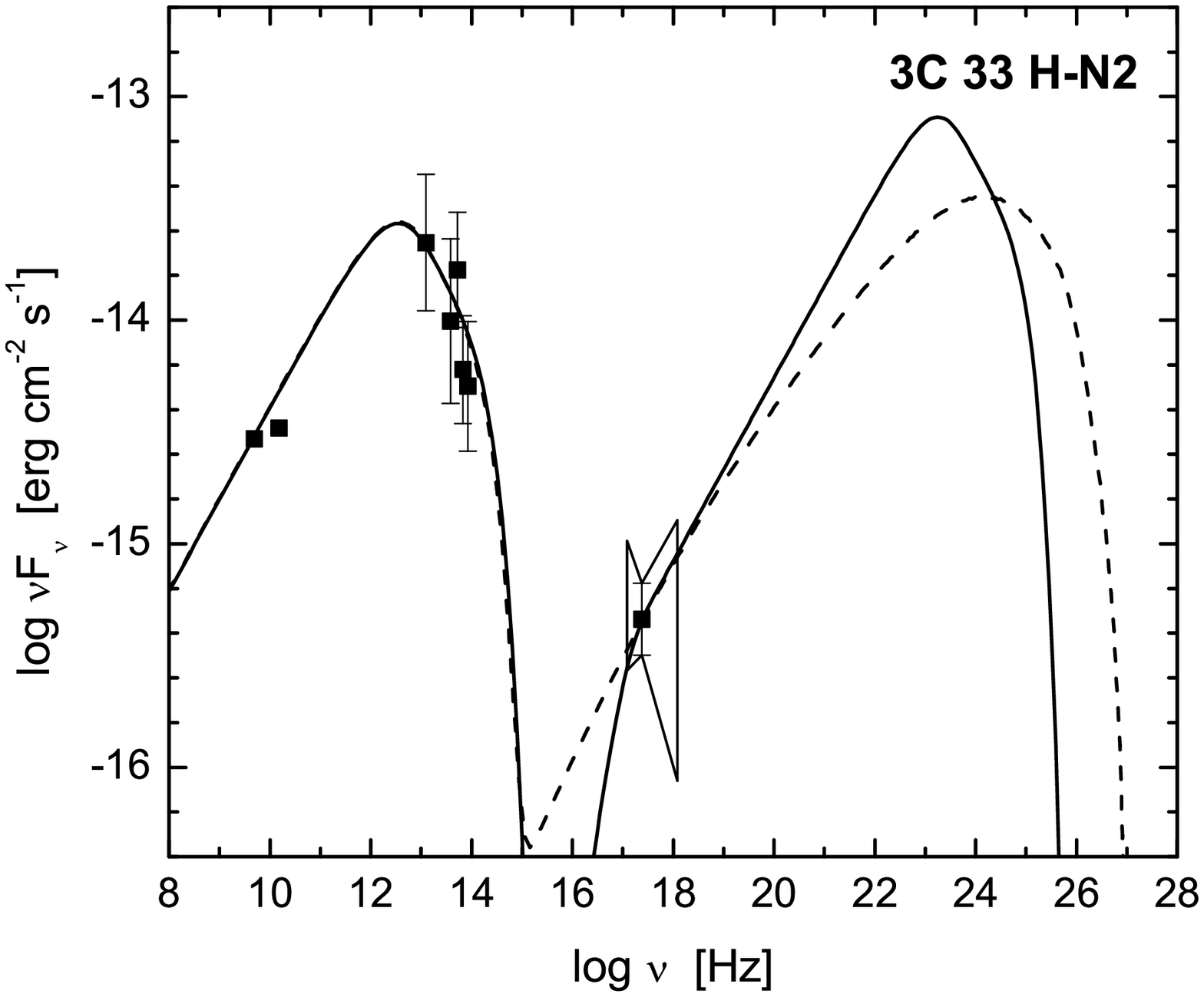}
\includegraphics[angle=0,scale=0.215]{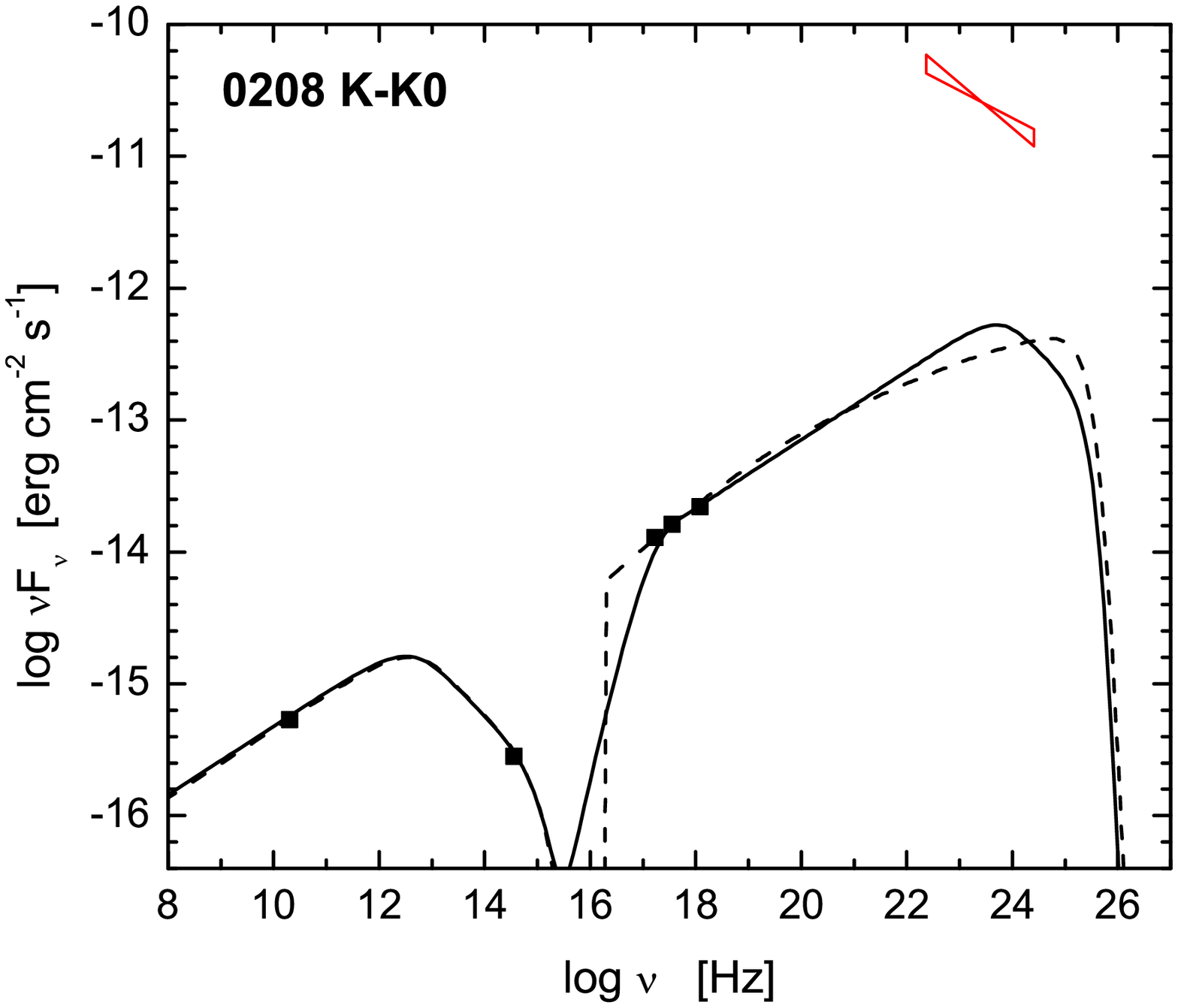}
\includegraphics[angle=0,scale=0.215]{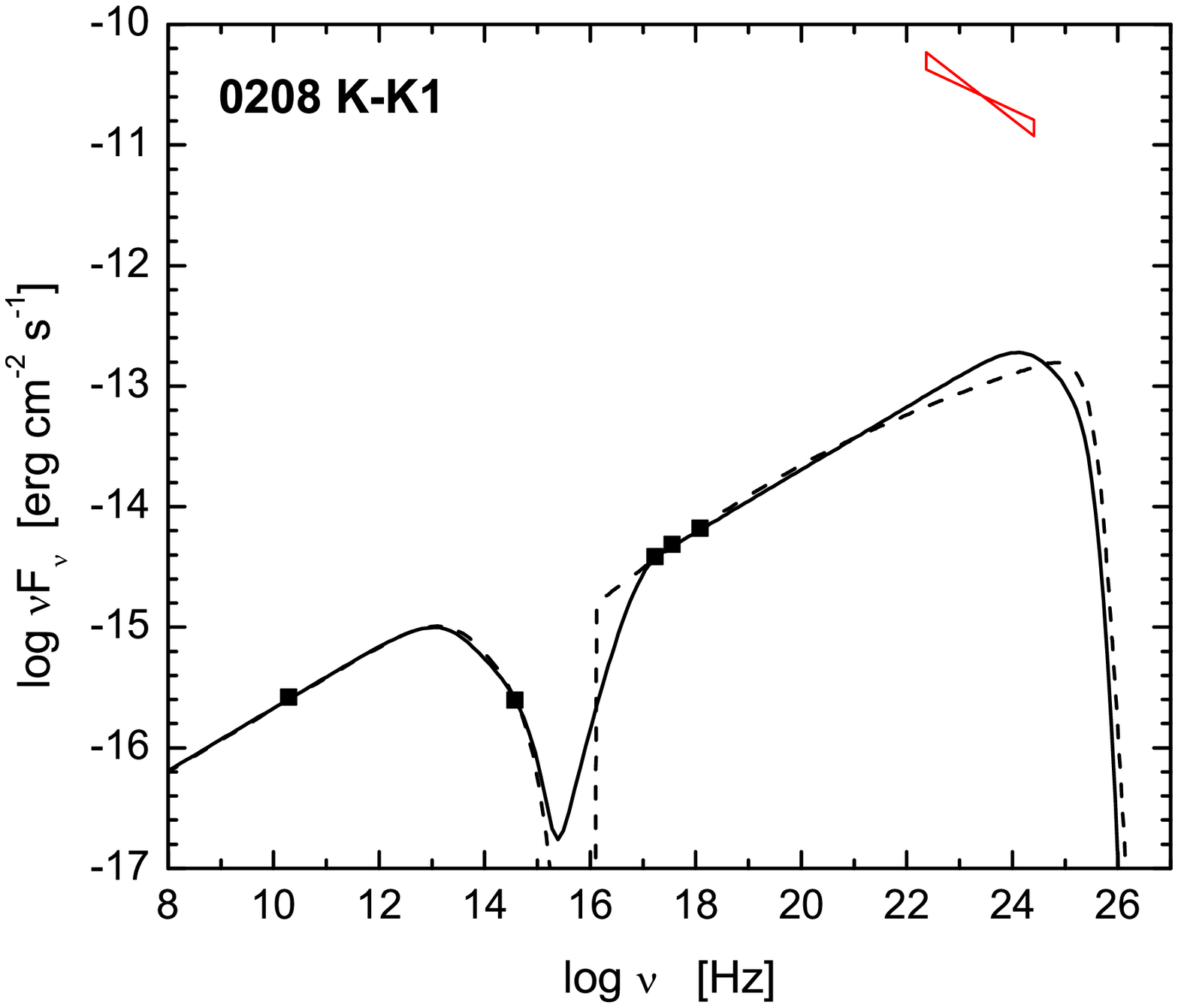}
\includegraphics[angle=0,scale=0.215]{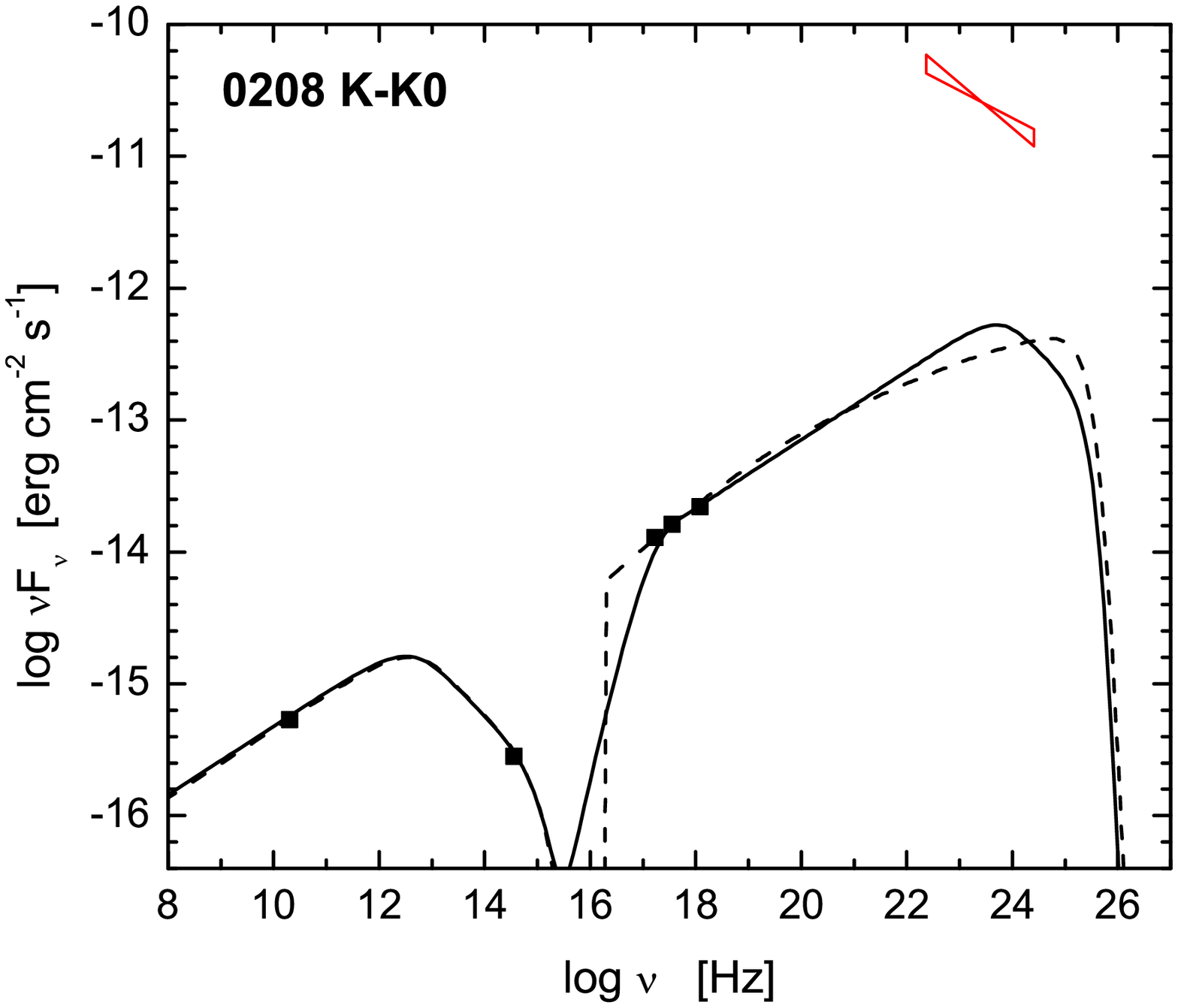}\\
\includegraphics[angle=0,scale=0.215]{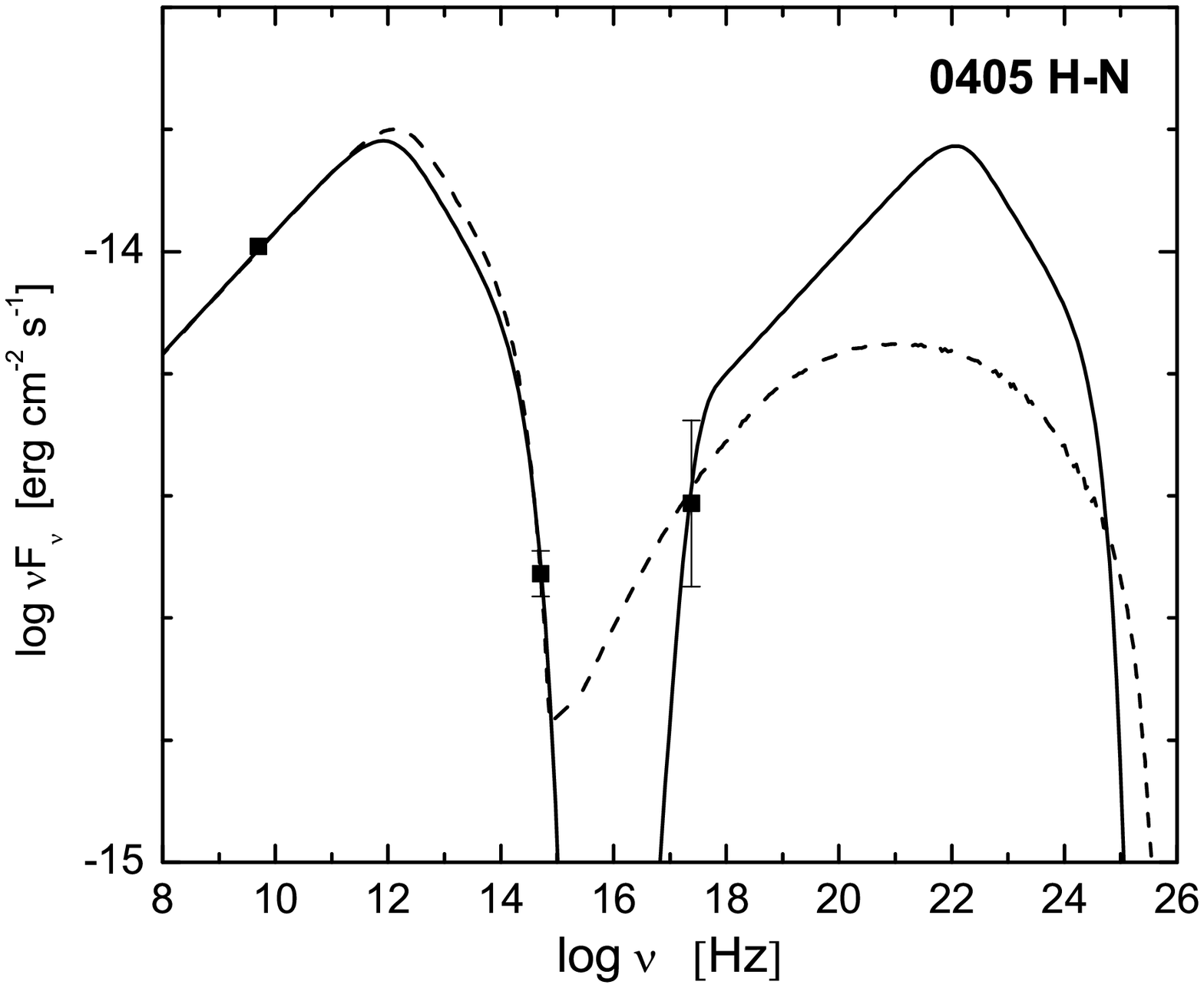}
\includegraphics[angle=0,scale=0.215]{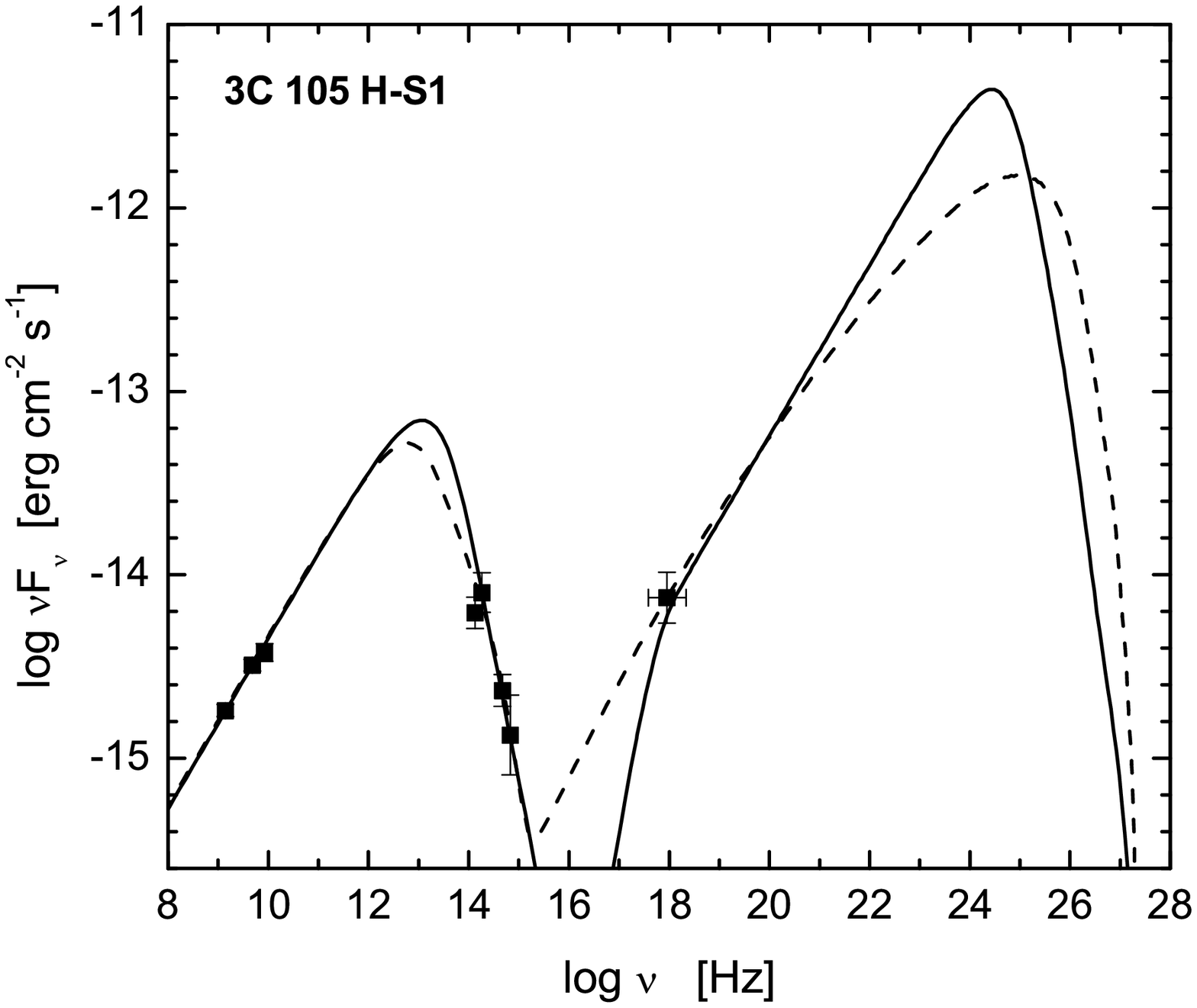}
\includegraphics[angle=0,scale=0.215]{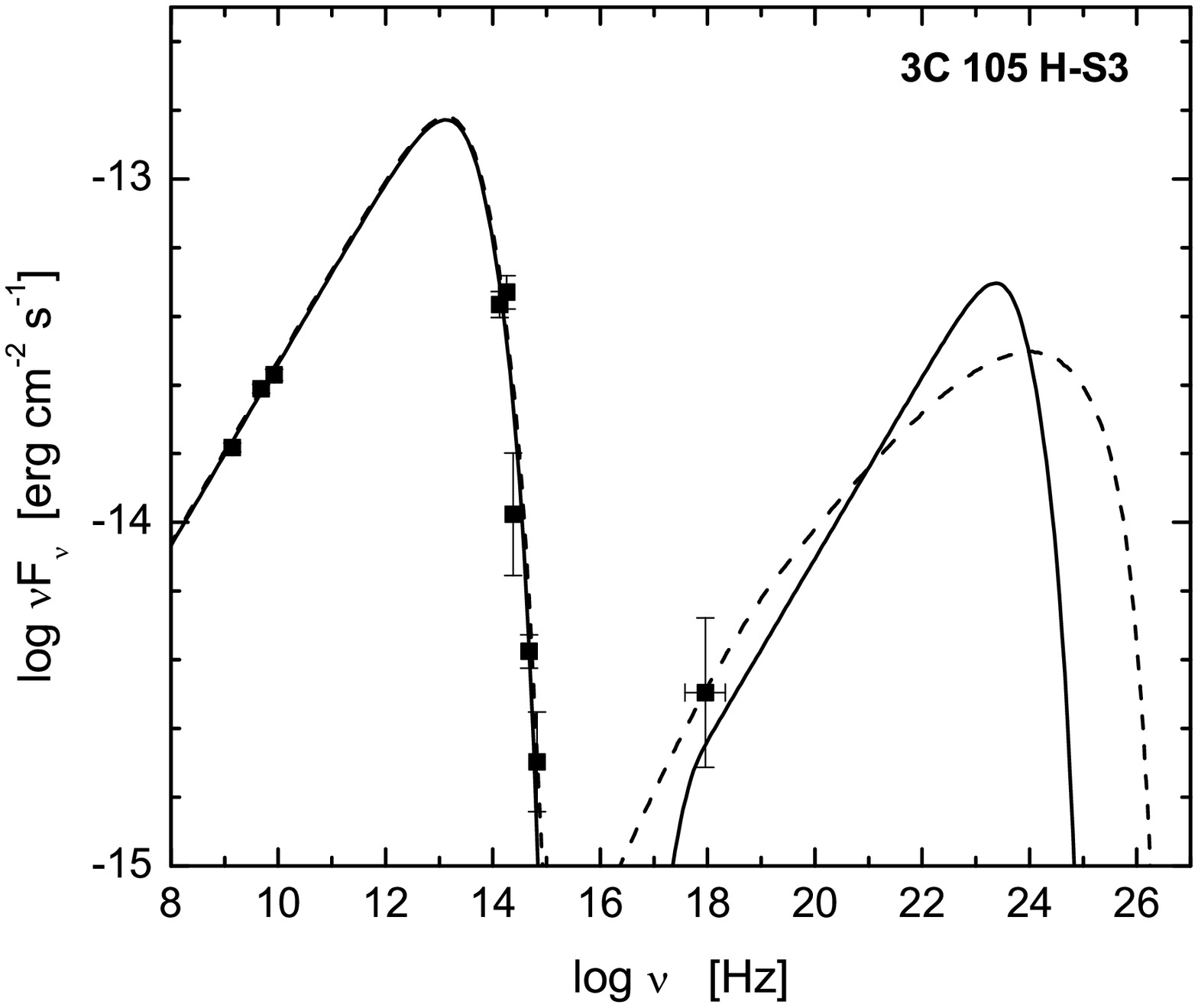}
\includegraphics[angle=0,scale=0.215]{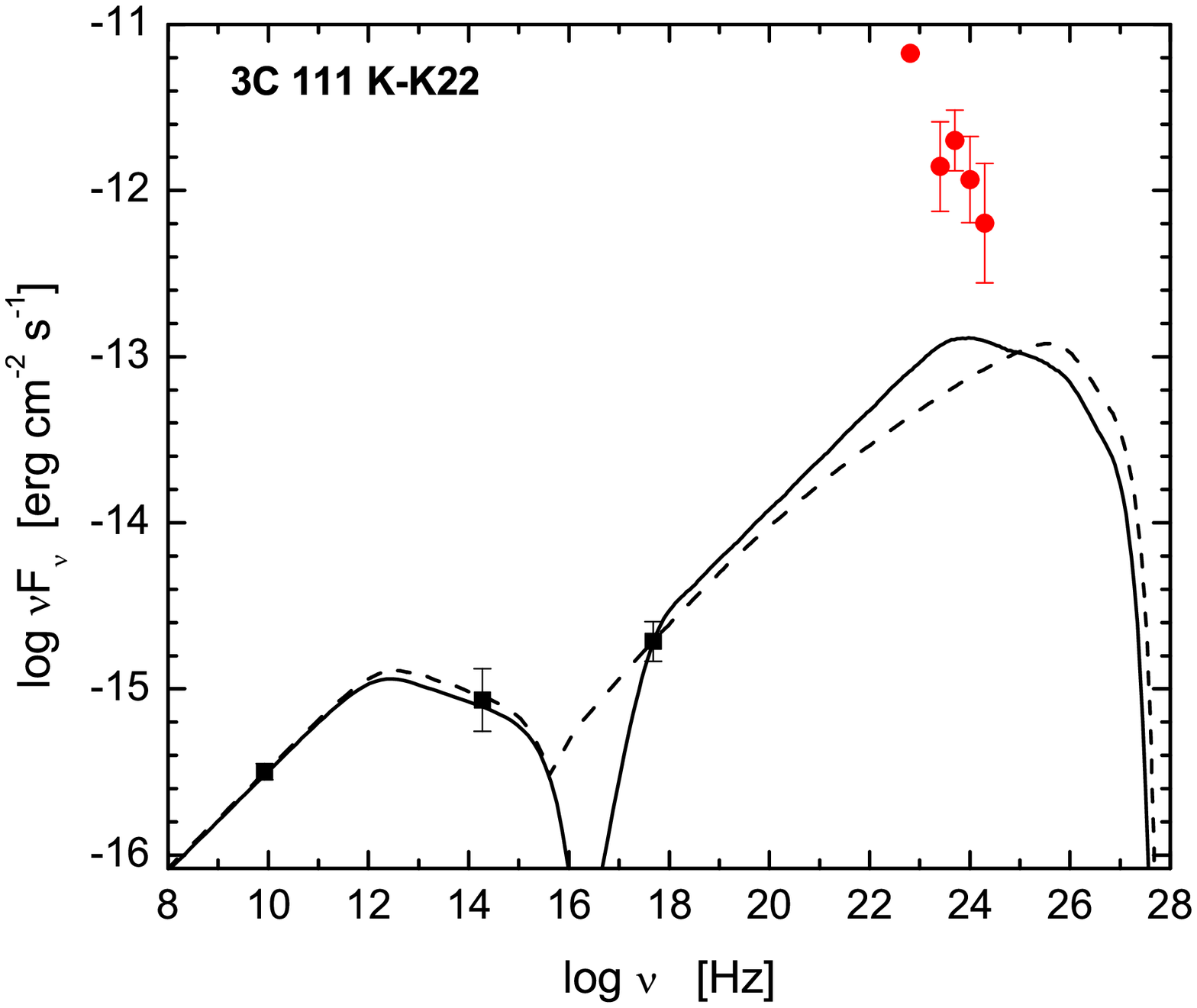}\\
\caption{Observed SEDs of the large-scale jet substructures (\emph{black solid squares and bow-ties}) with the model fitting (\emph{lines}) and the observation data of \emph{Fermi}/LAT (\emph{red solid circles and bow-ties}). The \emph{black solid lines} indicate the fitting results with the syn+IC/CMB model by considering the beaming effect under the equipartition condition, while the \emph{dash lines} are the results of the syn+SSC model without considering the equipartition condition and by assuming $\delta=1$.}
\end{figure*}
\begin{figure*}
\includegraphics[angle=0,scale=0.215]{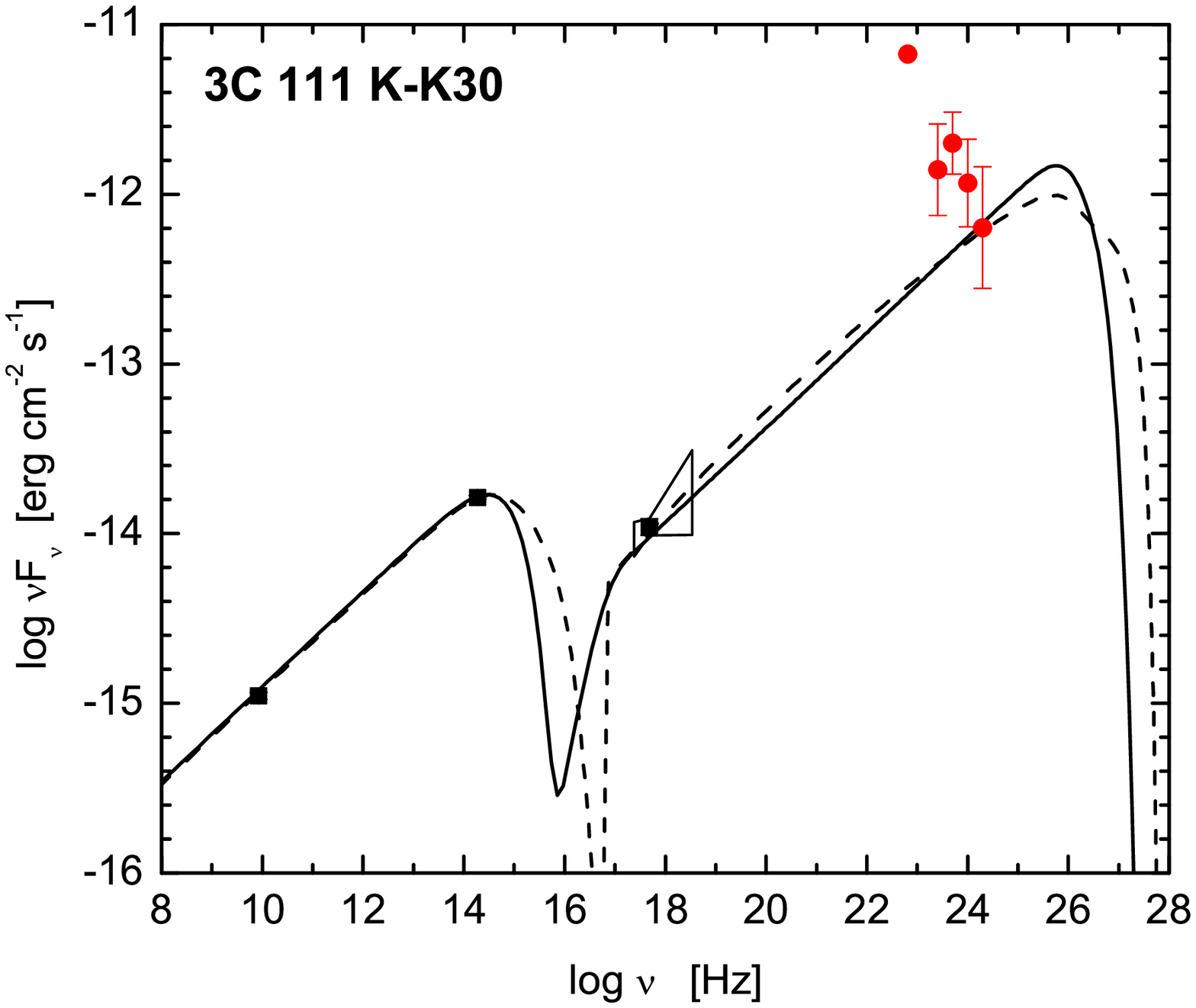}
\includegraphics[angle=0,scale=0.215]{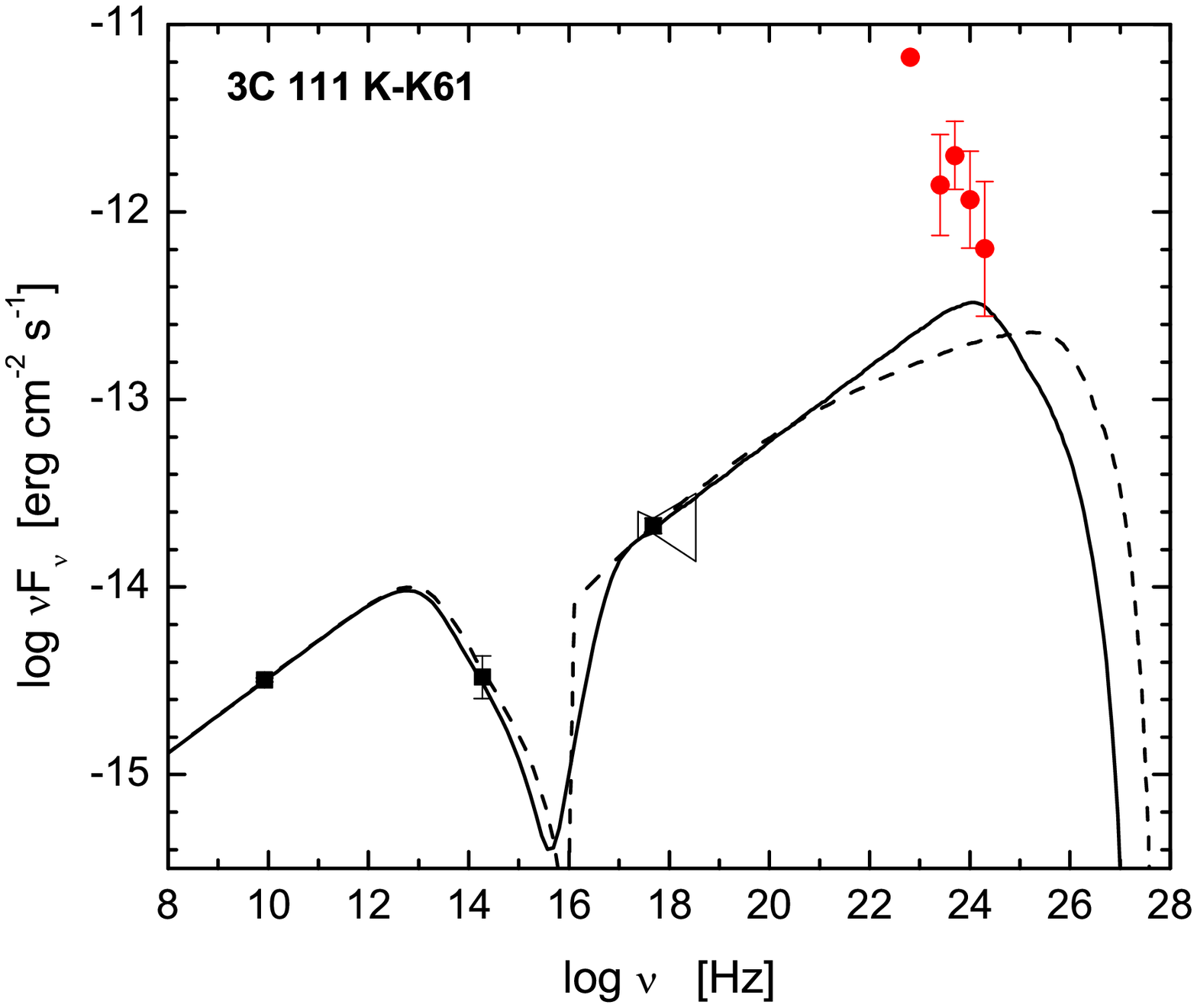}
\includegraphics[angle=0,scale=0.215]{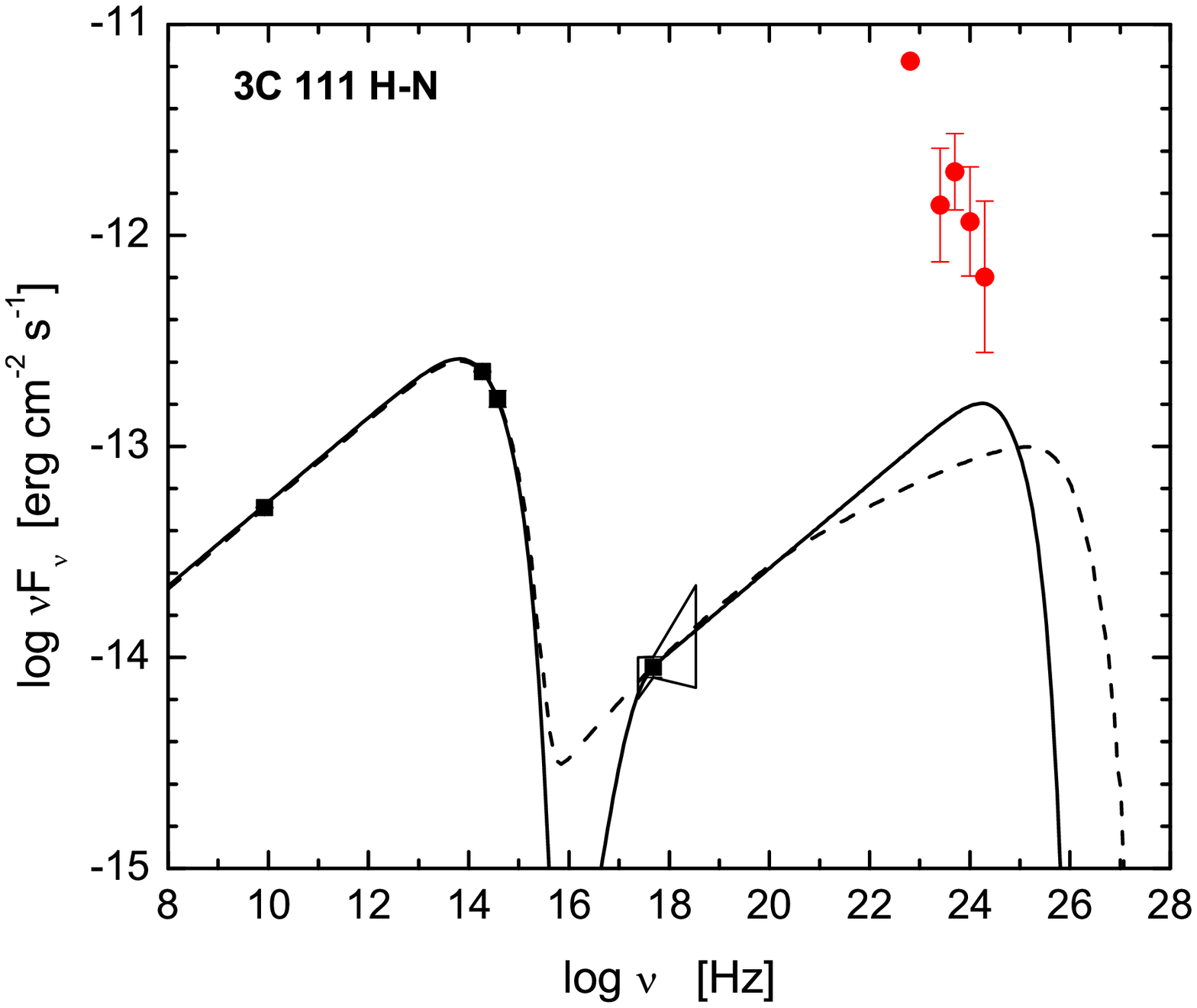}
\includegraphics[angle=0,scale=0.215]{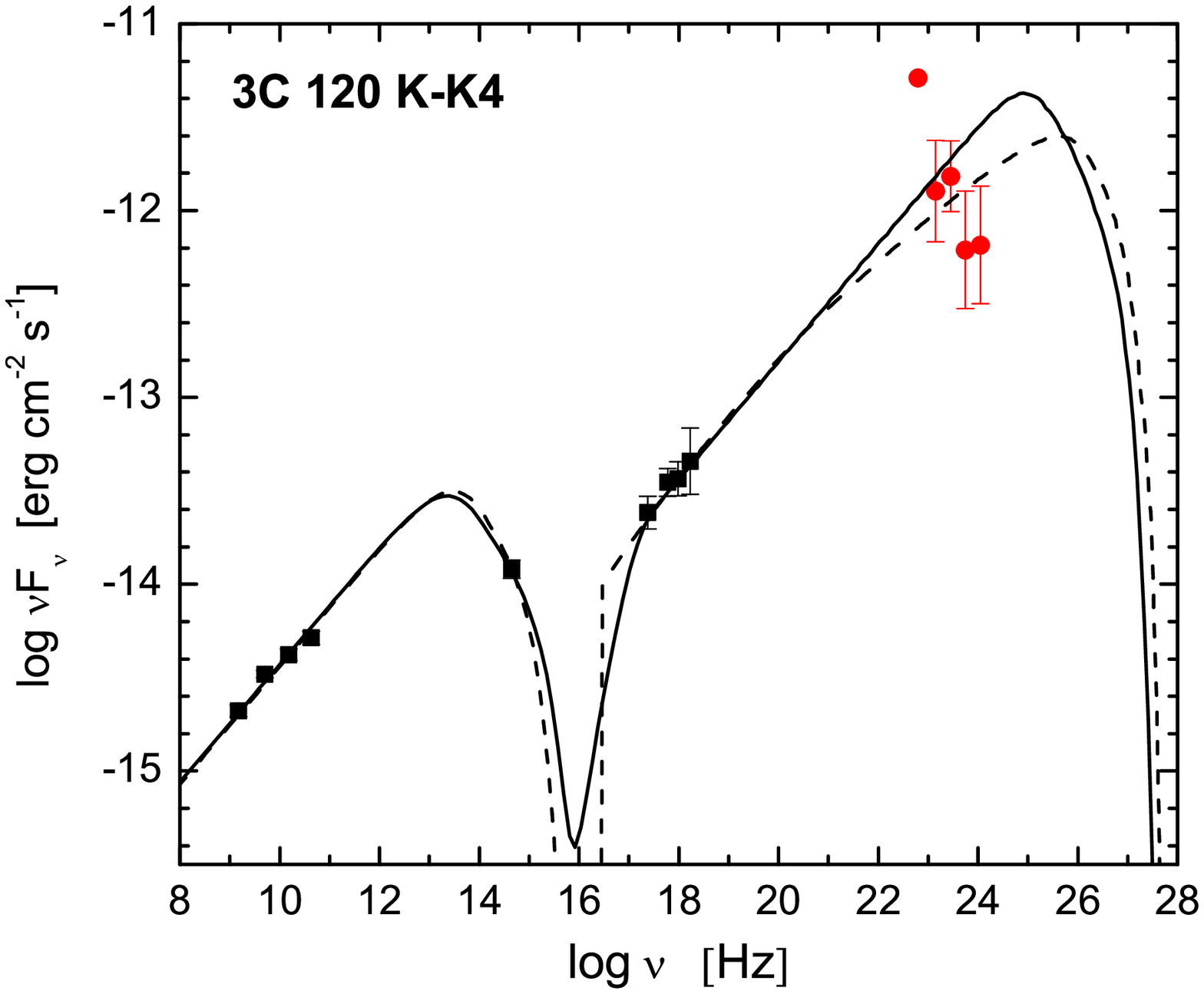}\\
\includegraphics[angle=0,scale=0.215]{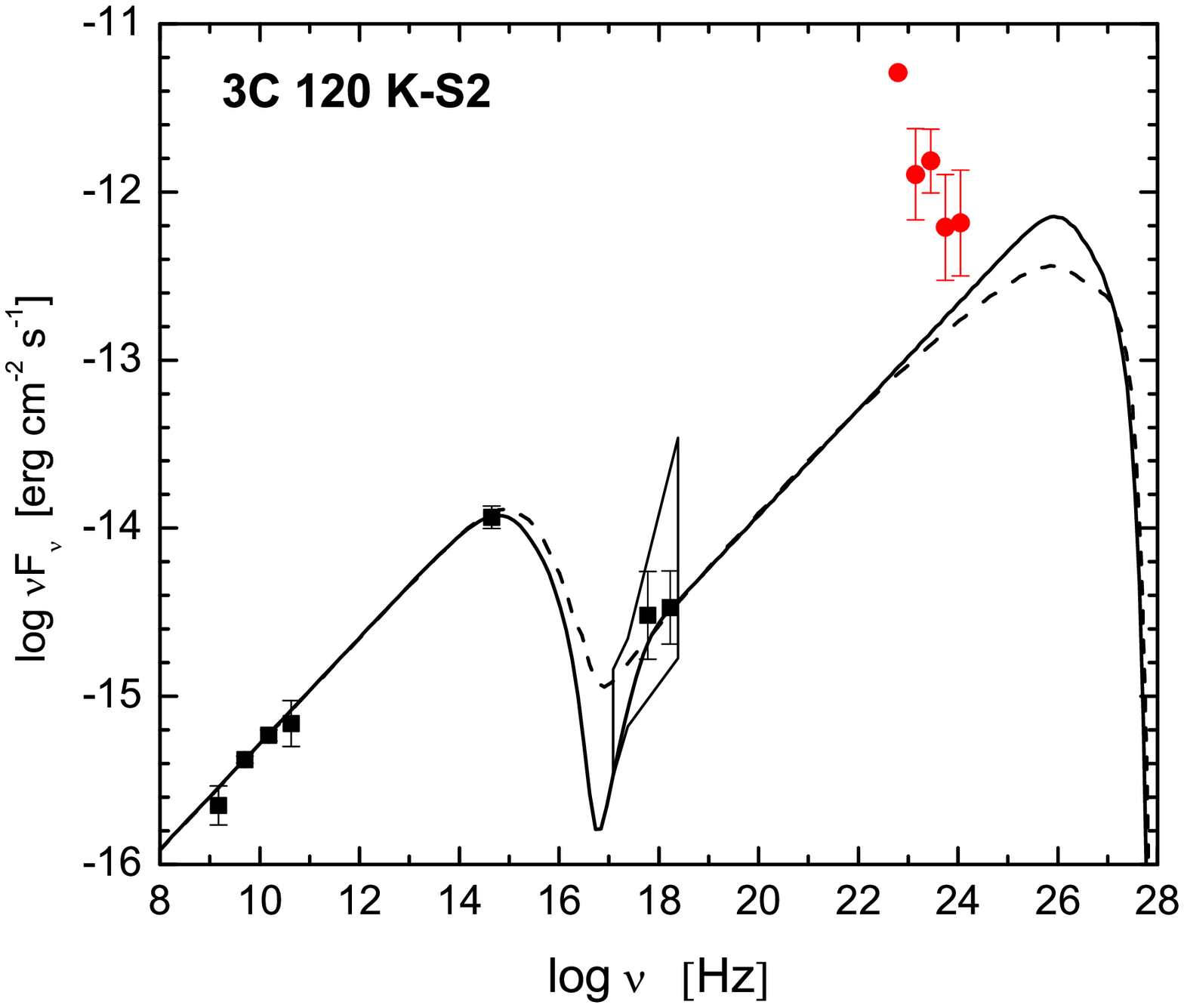}
\includegraphics[angle=0,scale=0.215]{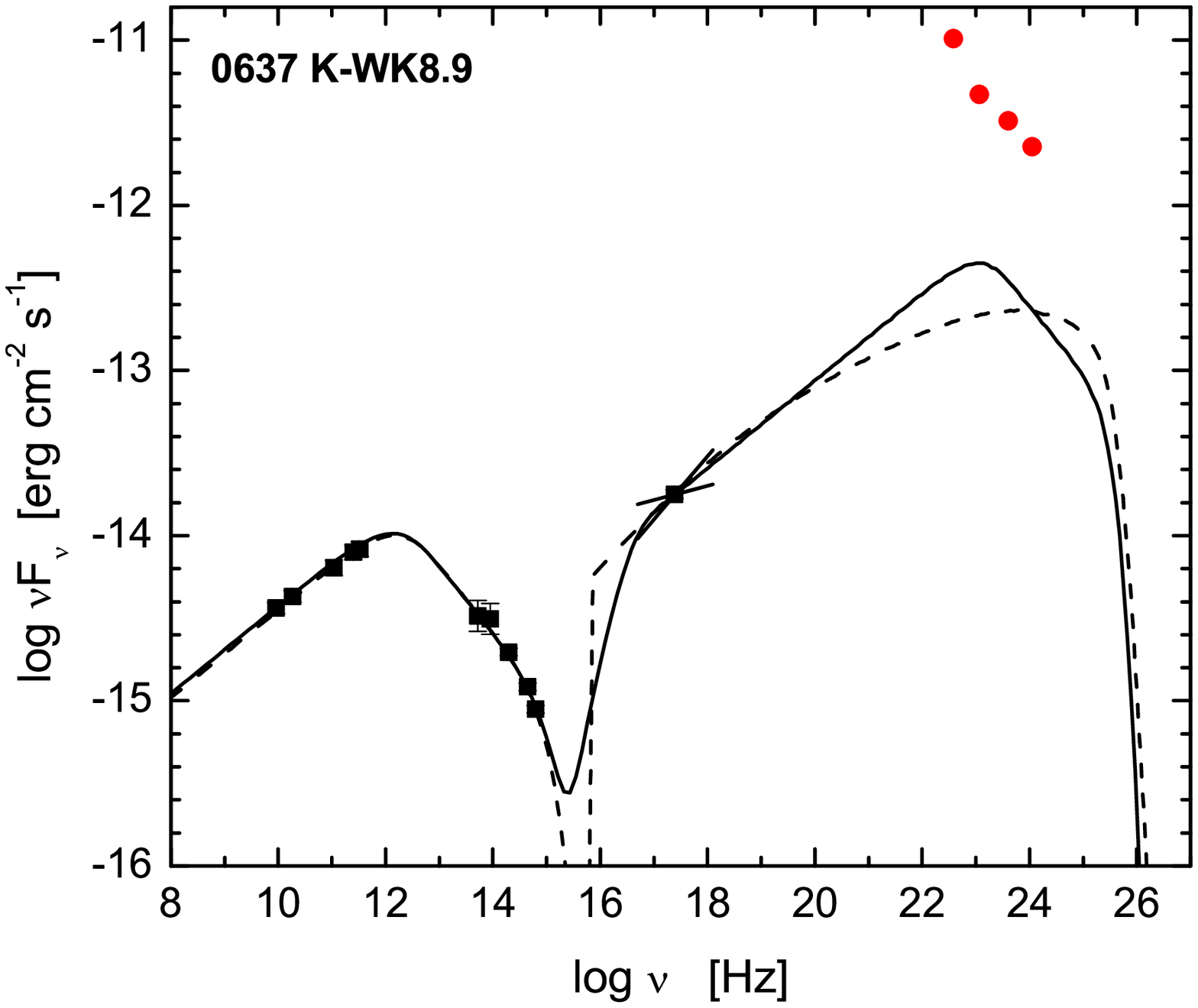}
\includegraphics[angle=0,scale=0.215]{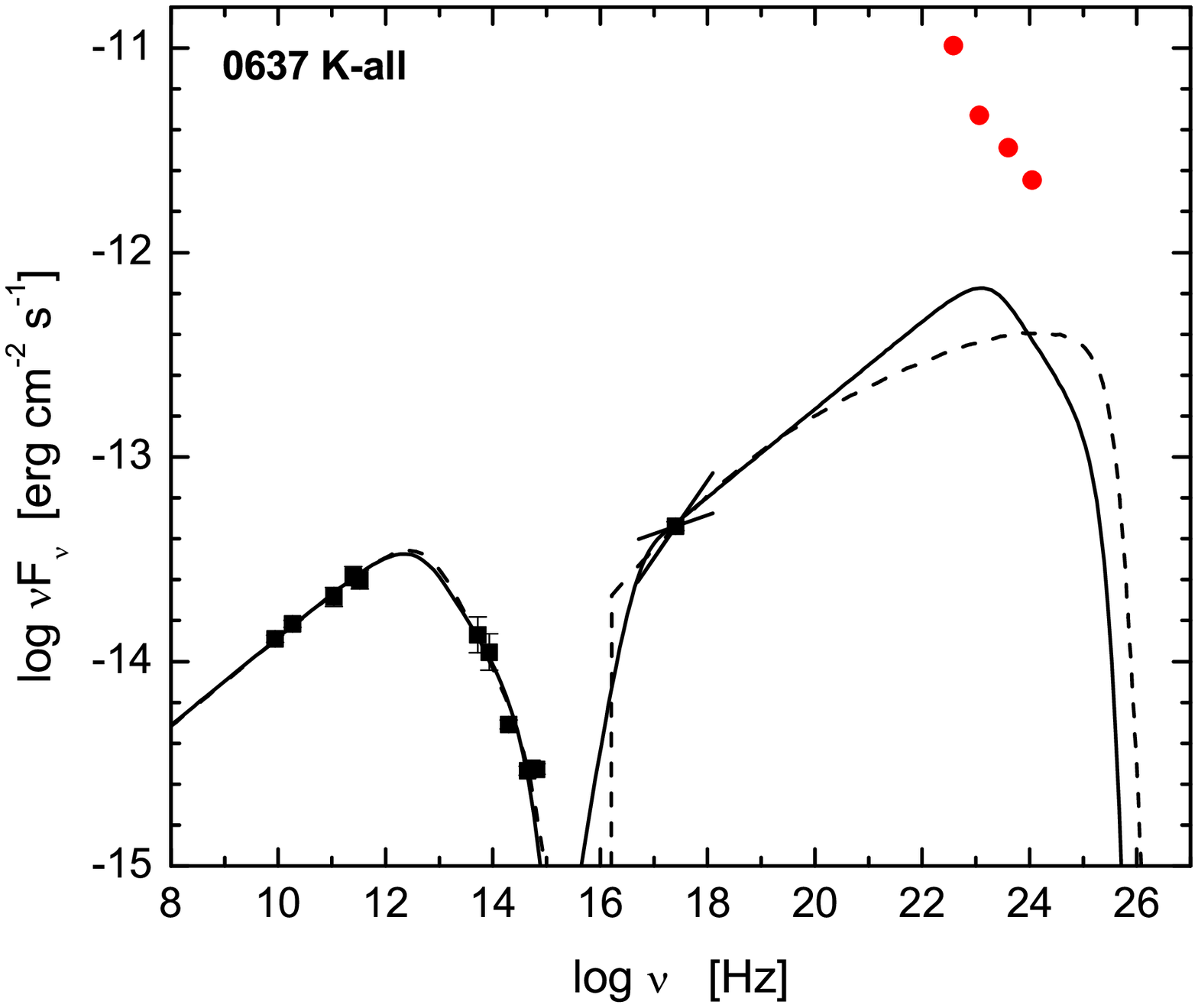}
\includegraphics[angle=0,scale=0.215]{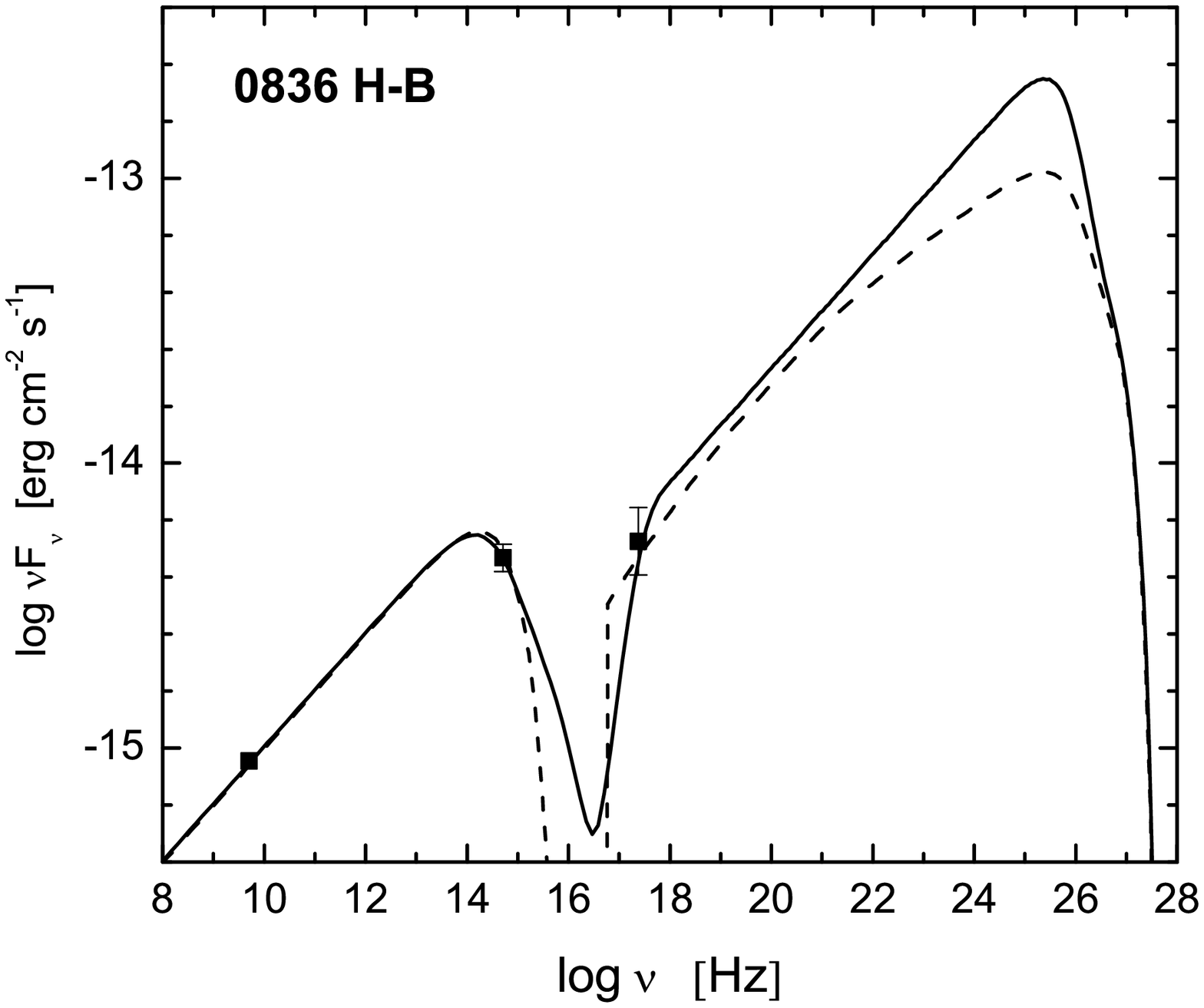}\\
\includegraphics[angle=0,scale=0.215]{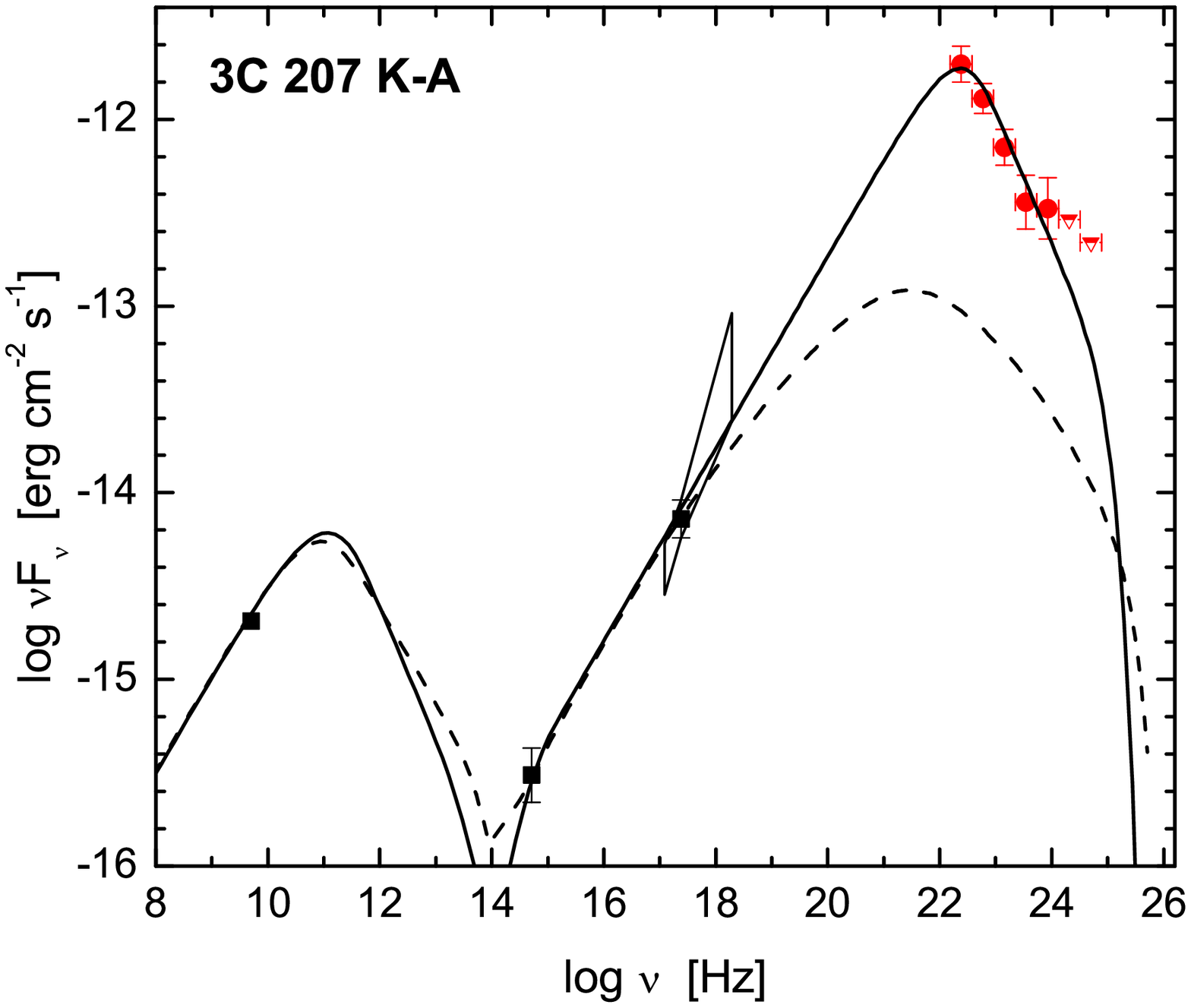}
\includegraphics[angle=0,scale=0.215]{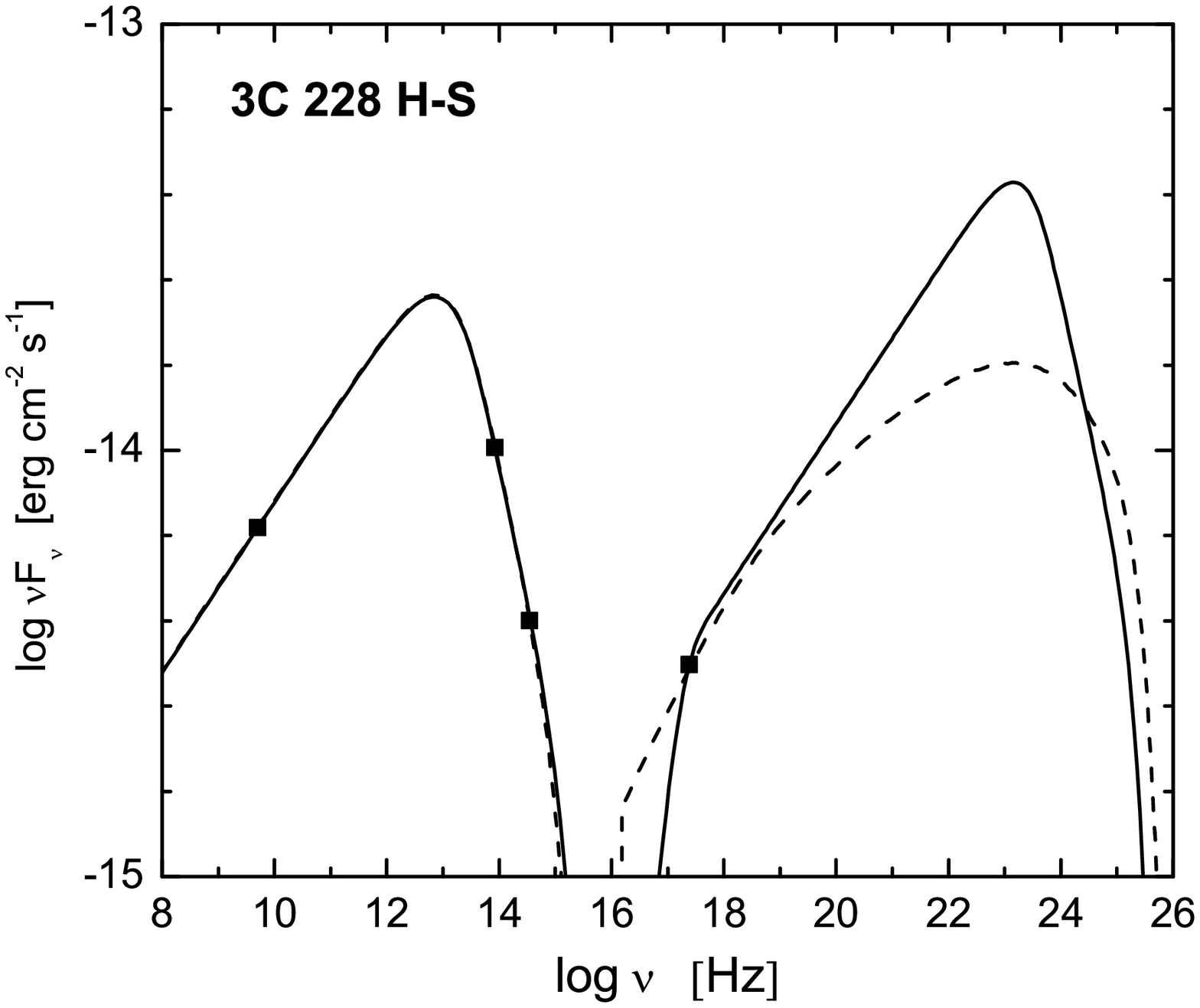}
\includegraphics[angle=0,scale=0.215]{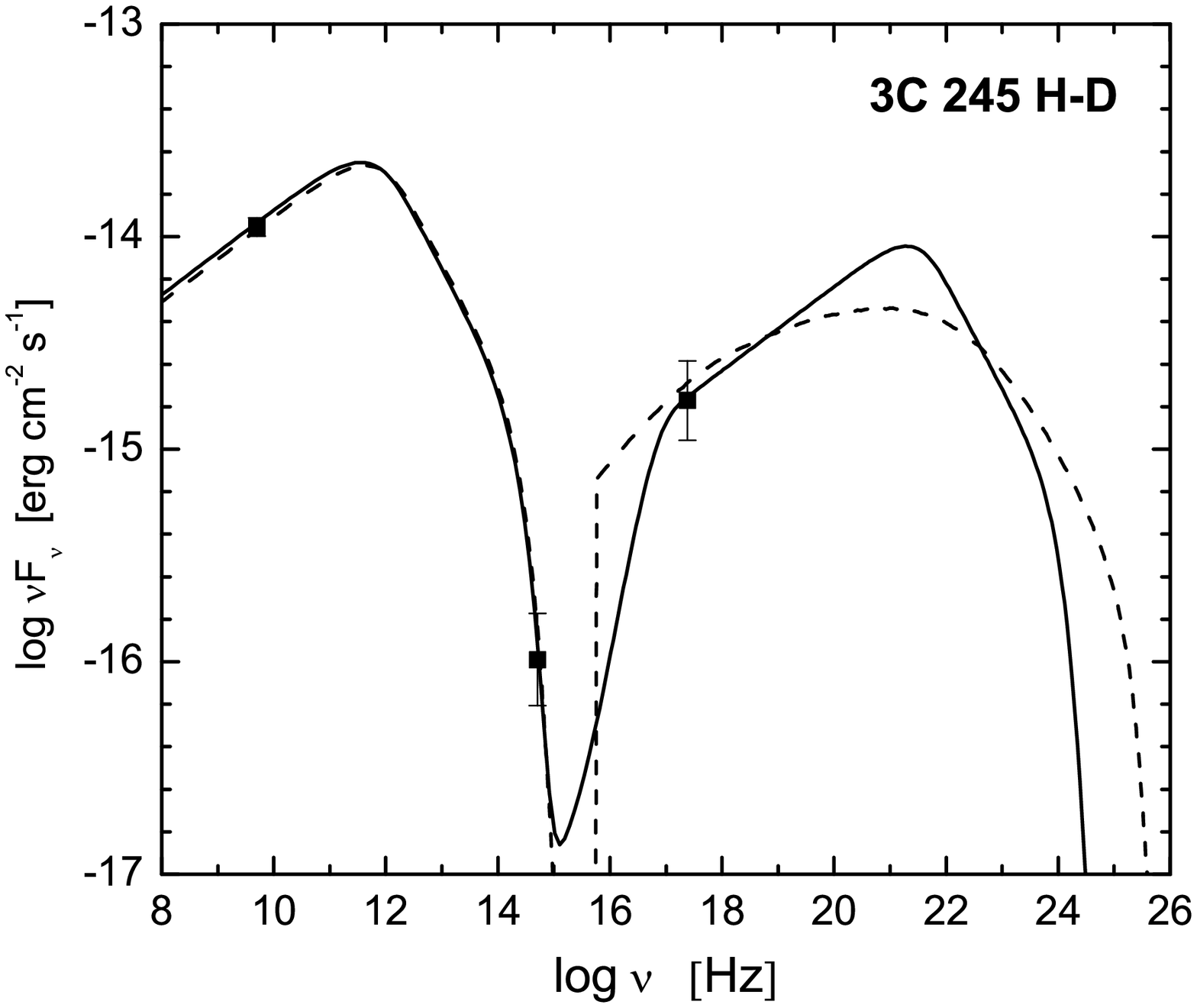}
\includegraphics[angle=0,scale=0.215]{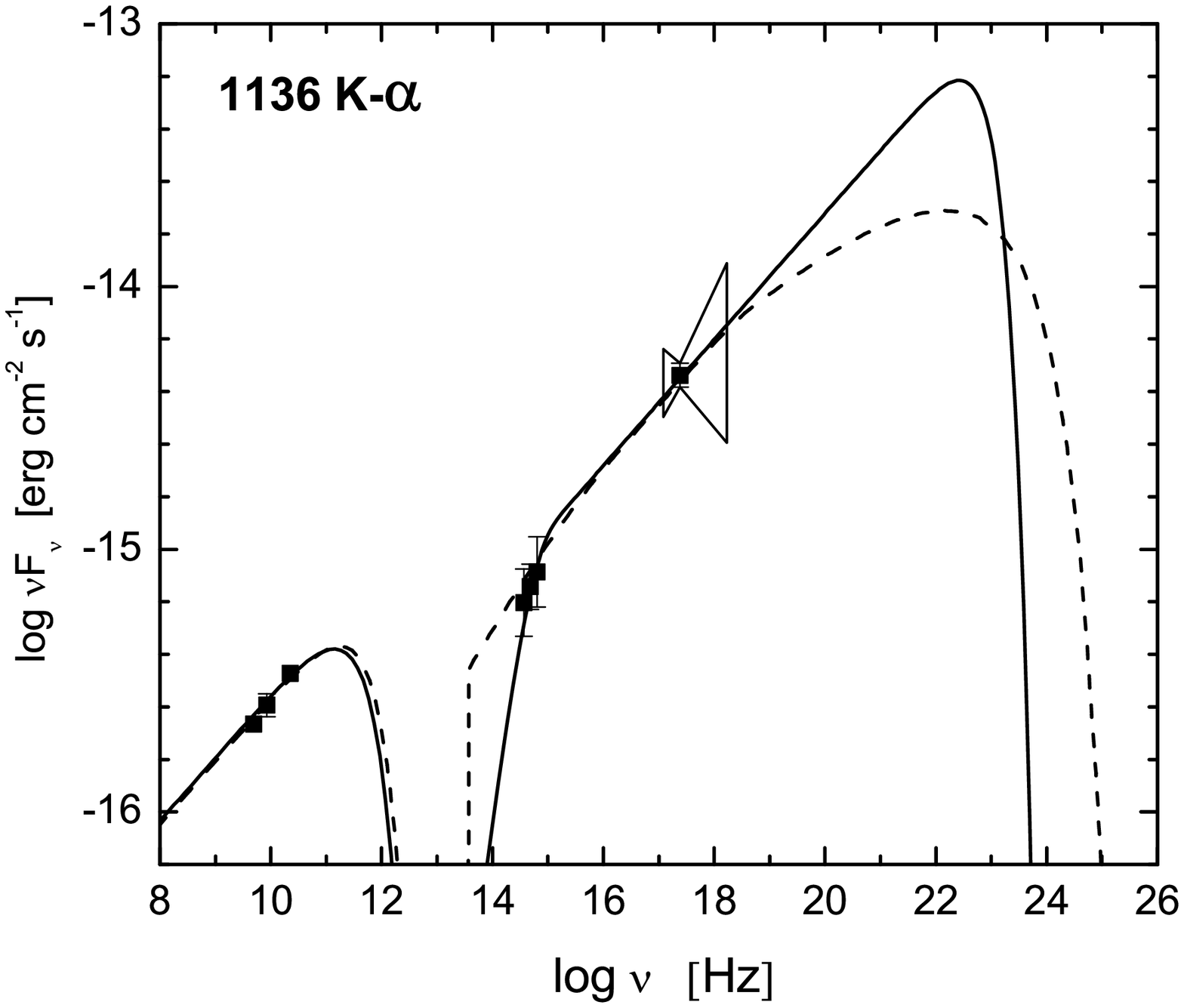}\\
\includegraphics[angle=0,scale=0.215]{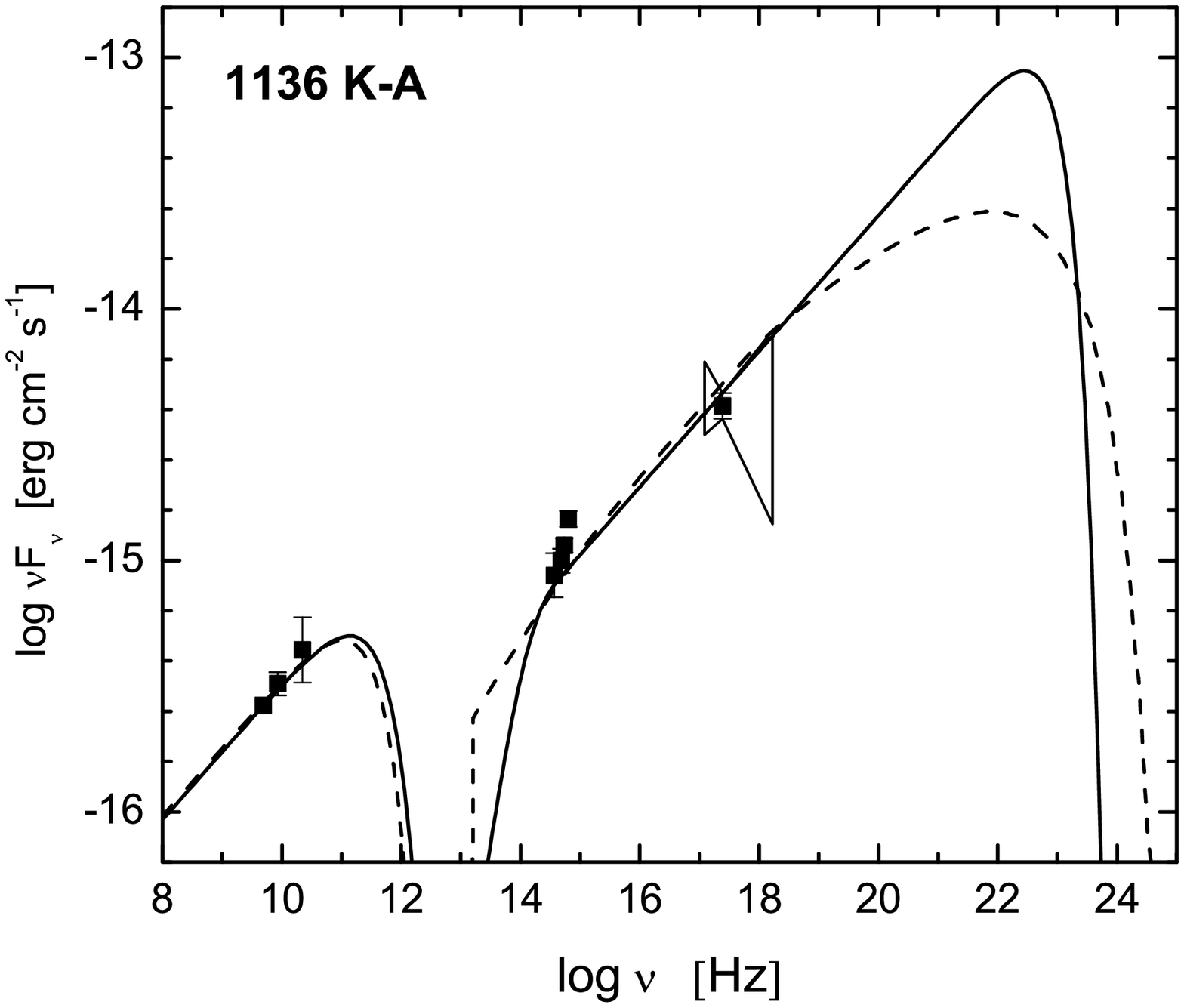}
\includegraphics[angle=0,scale=0.215]{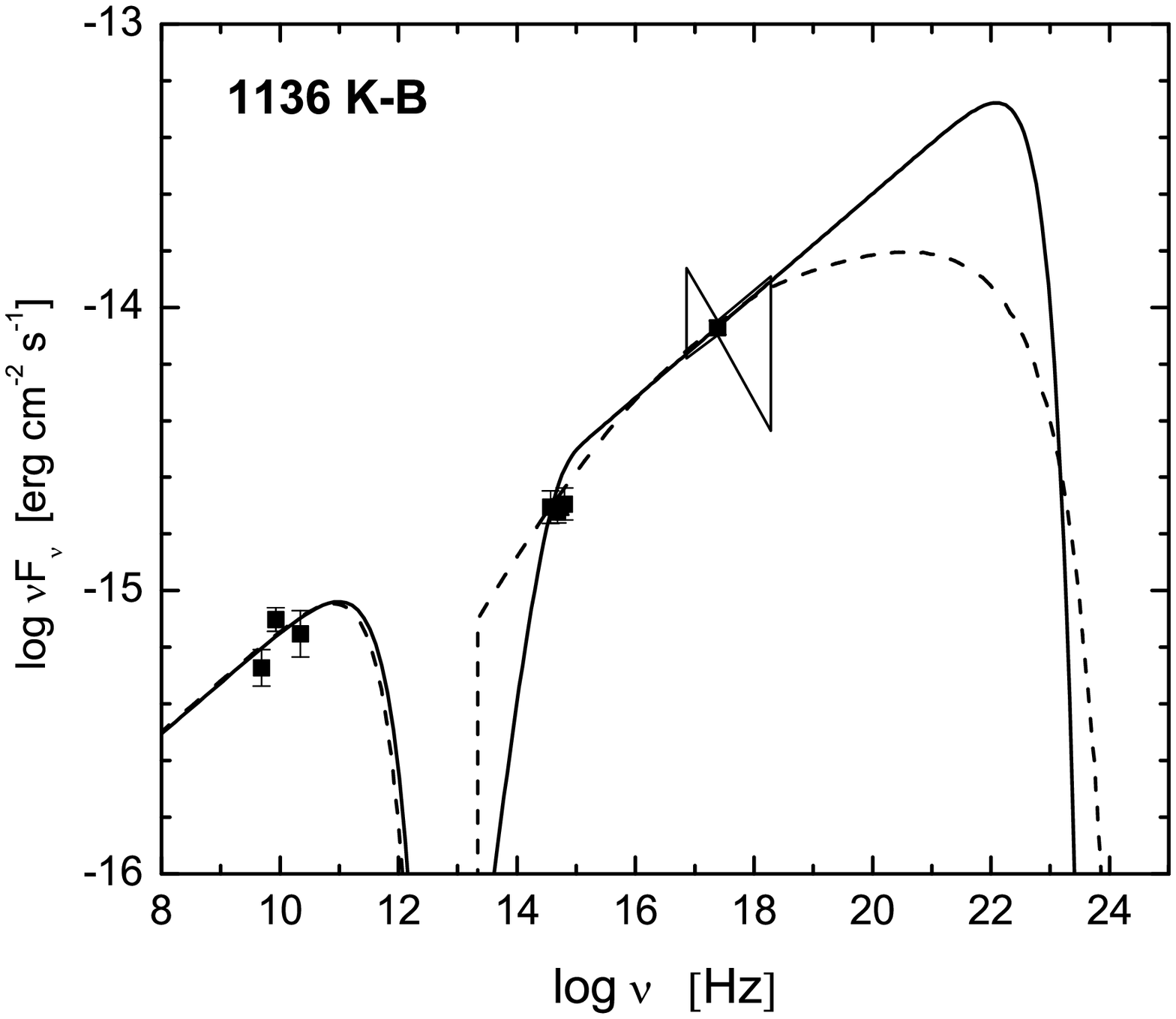}
\includegraphics[angle=0,scale=0.215]{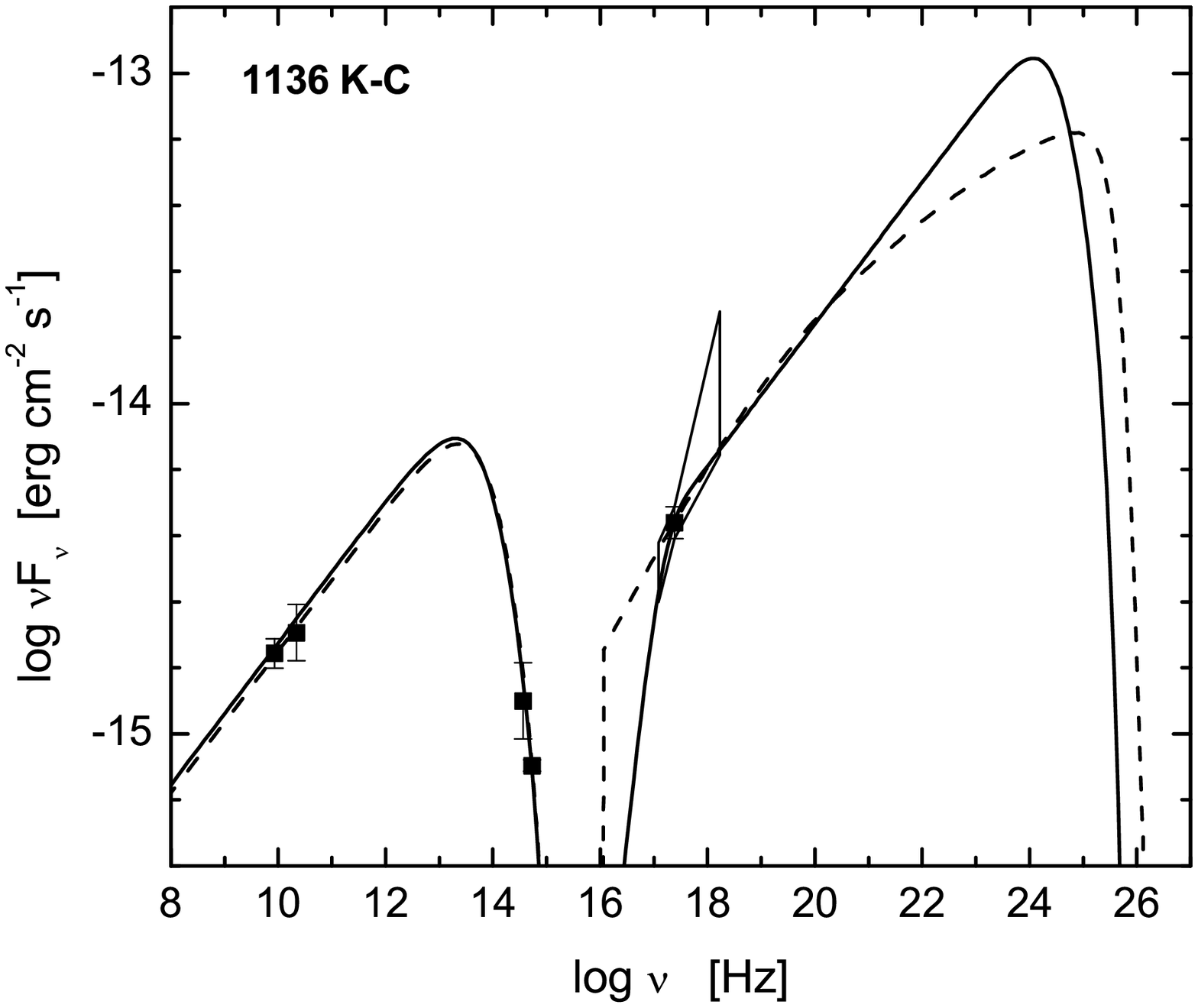}
\includegraphics[angle=0,scale=0.215]{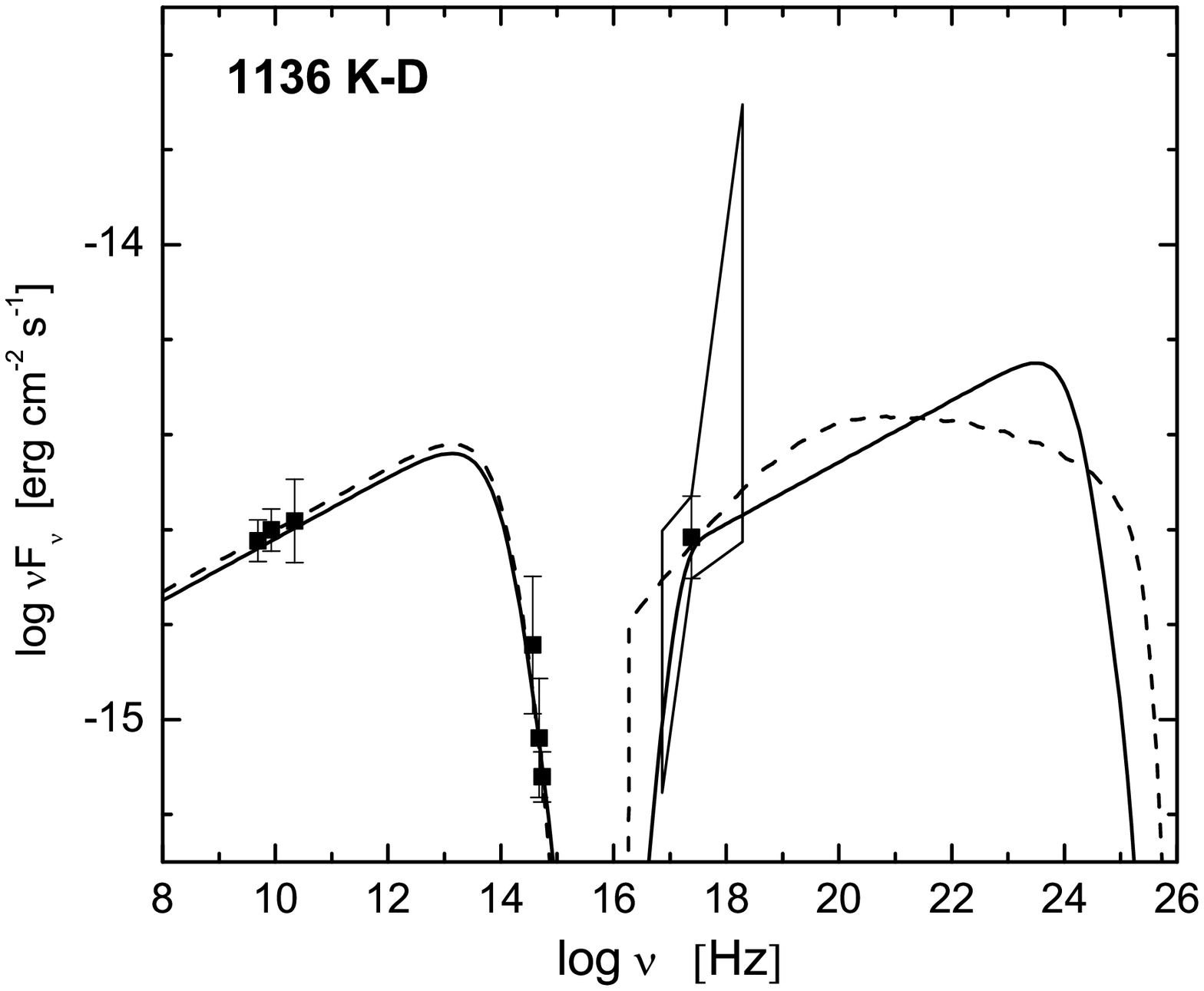}\\
\includegraphics[angle=0,scale=0.215]{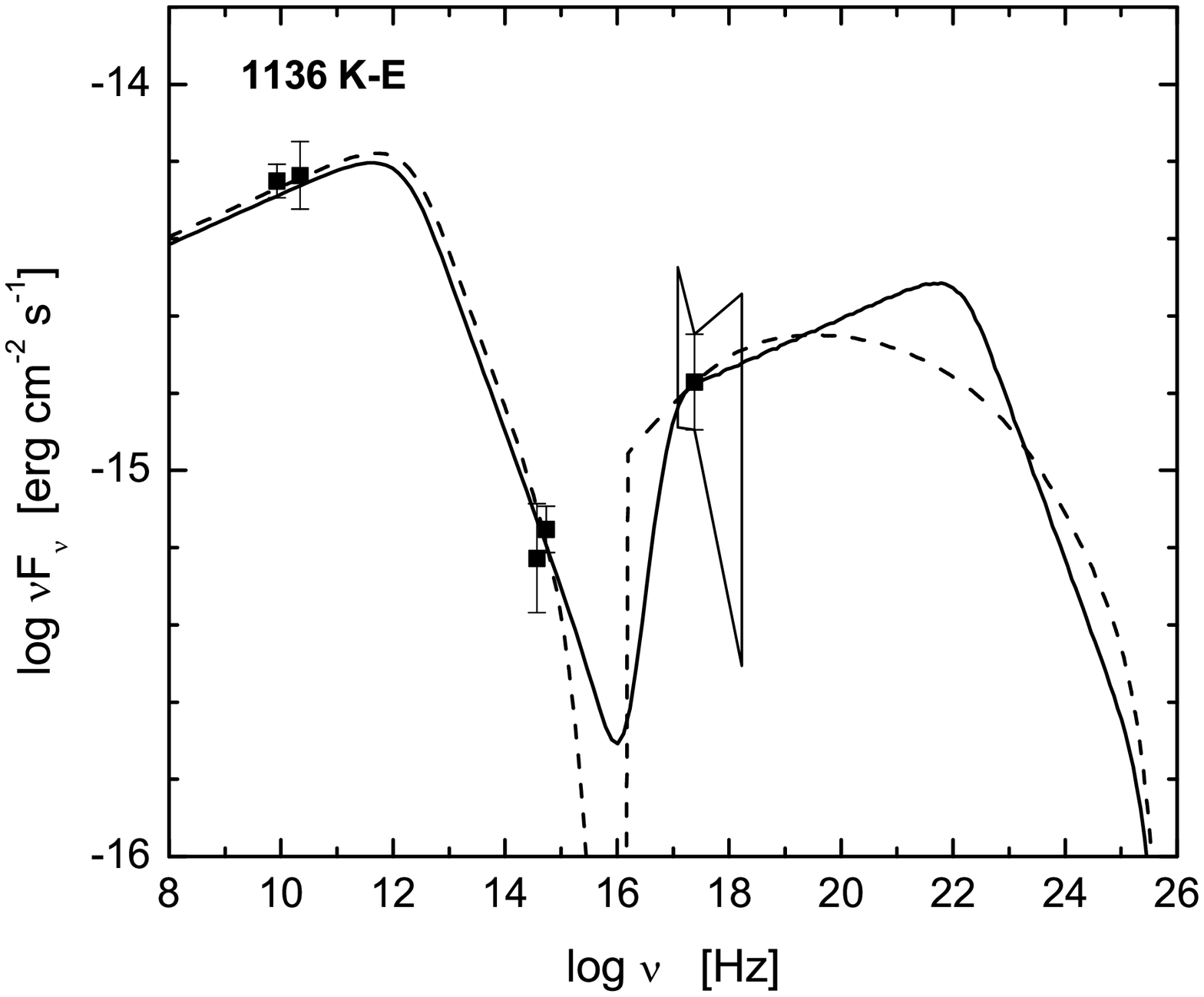}
\includegraphics[angle=0,scale=0.215]{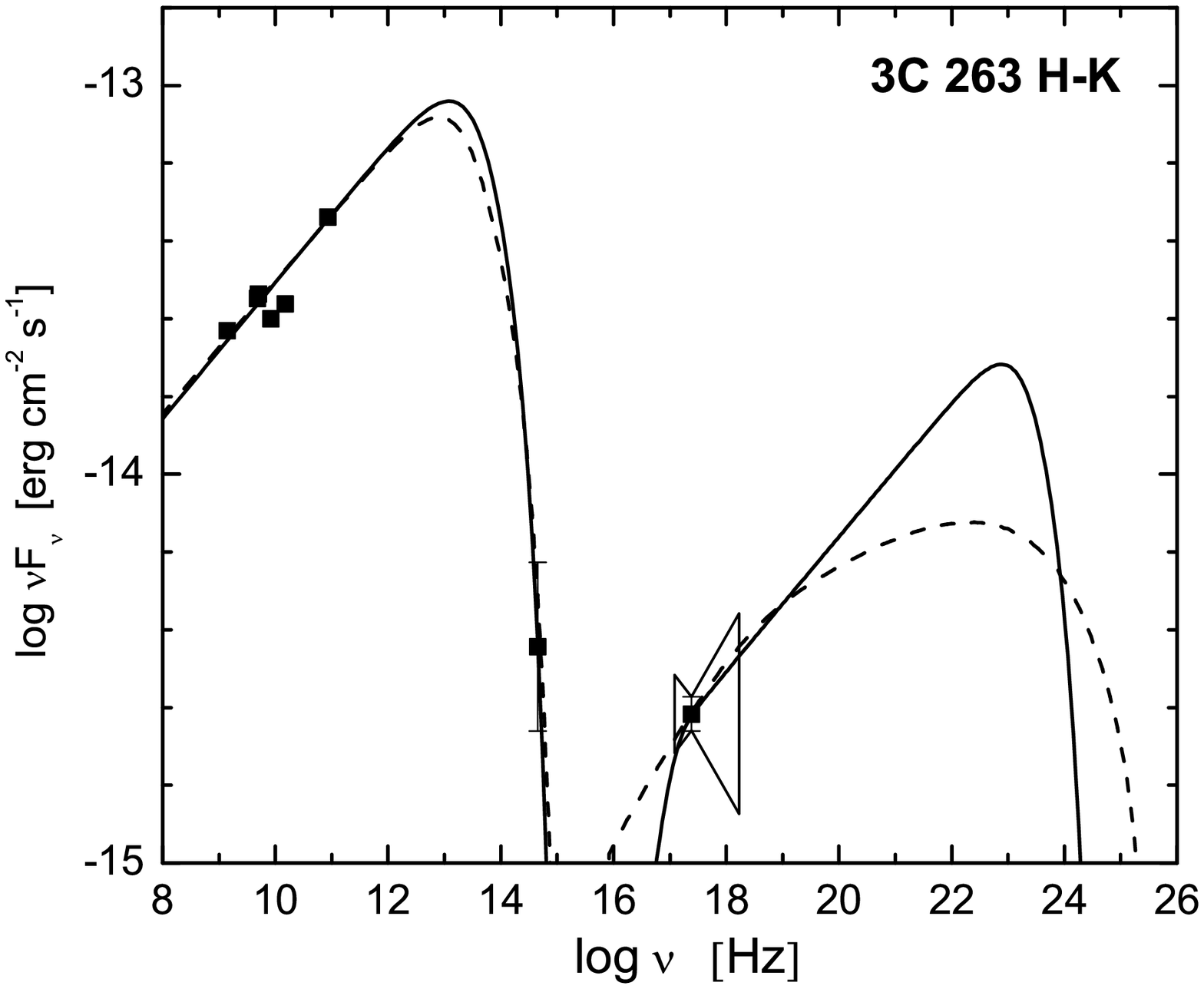}
\includegraphics[angle=0,scale=0.215]{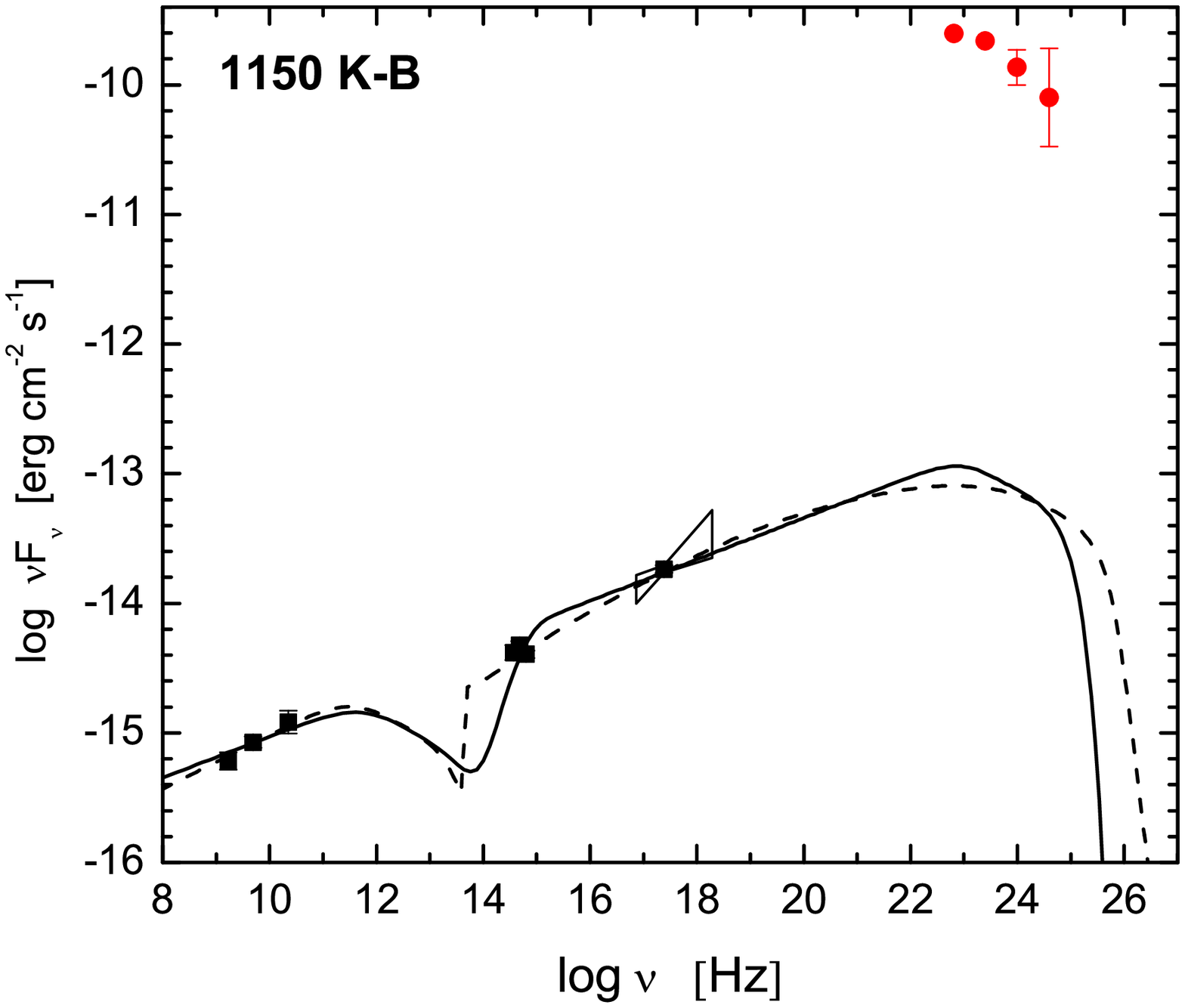}
\includegraphics[angle=0,scale=0.215]{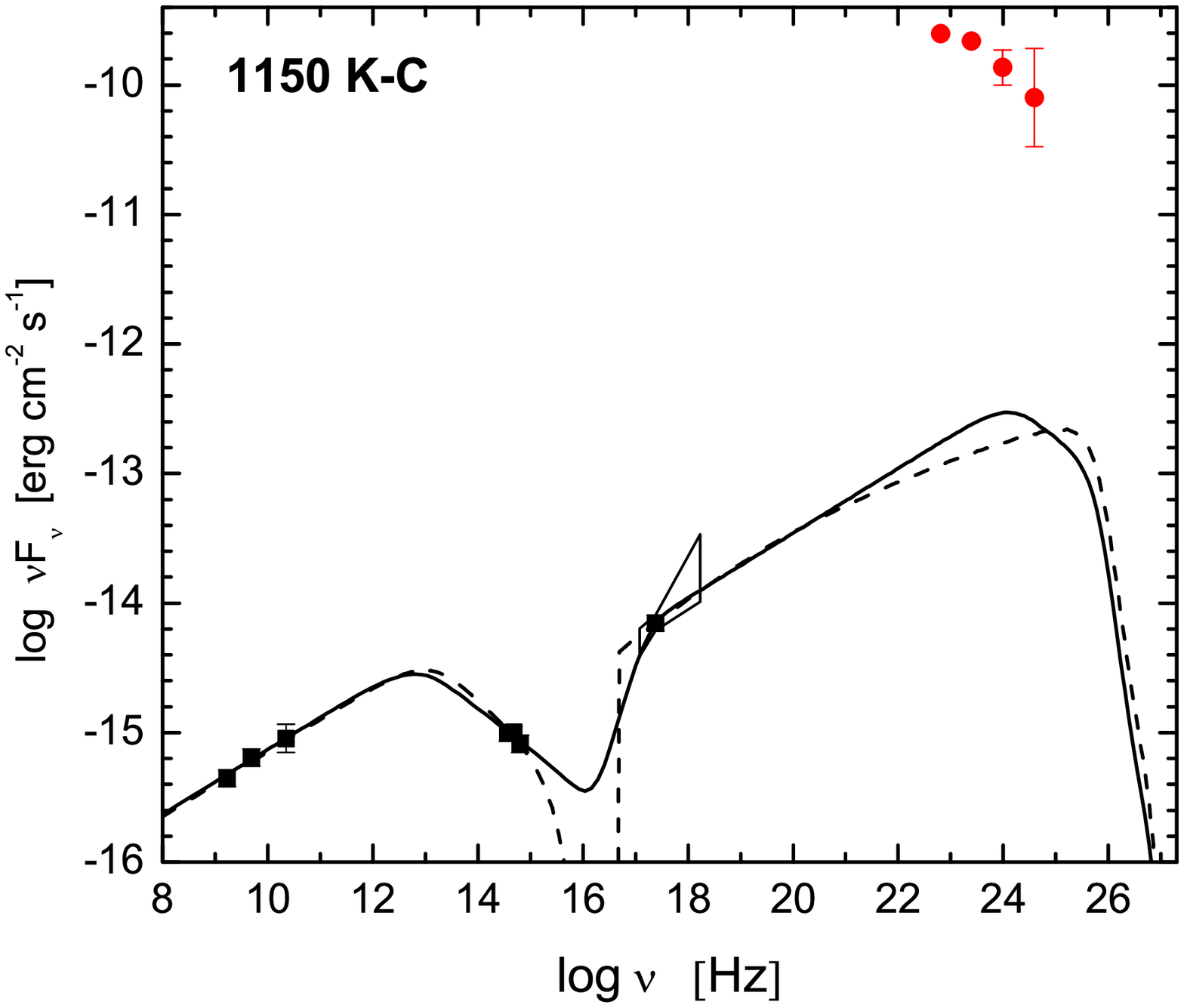}\\
\includegraphics[angle=0,scale=0.215]{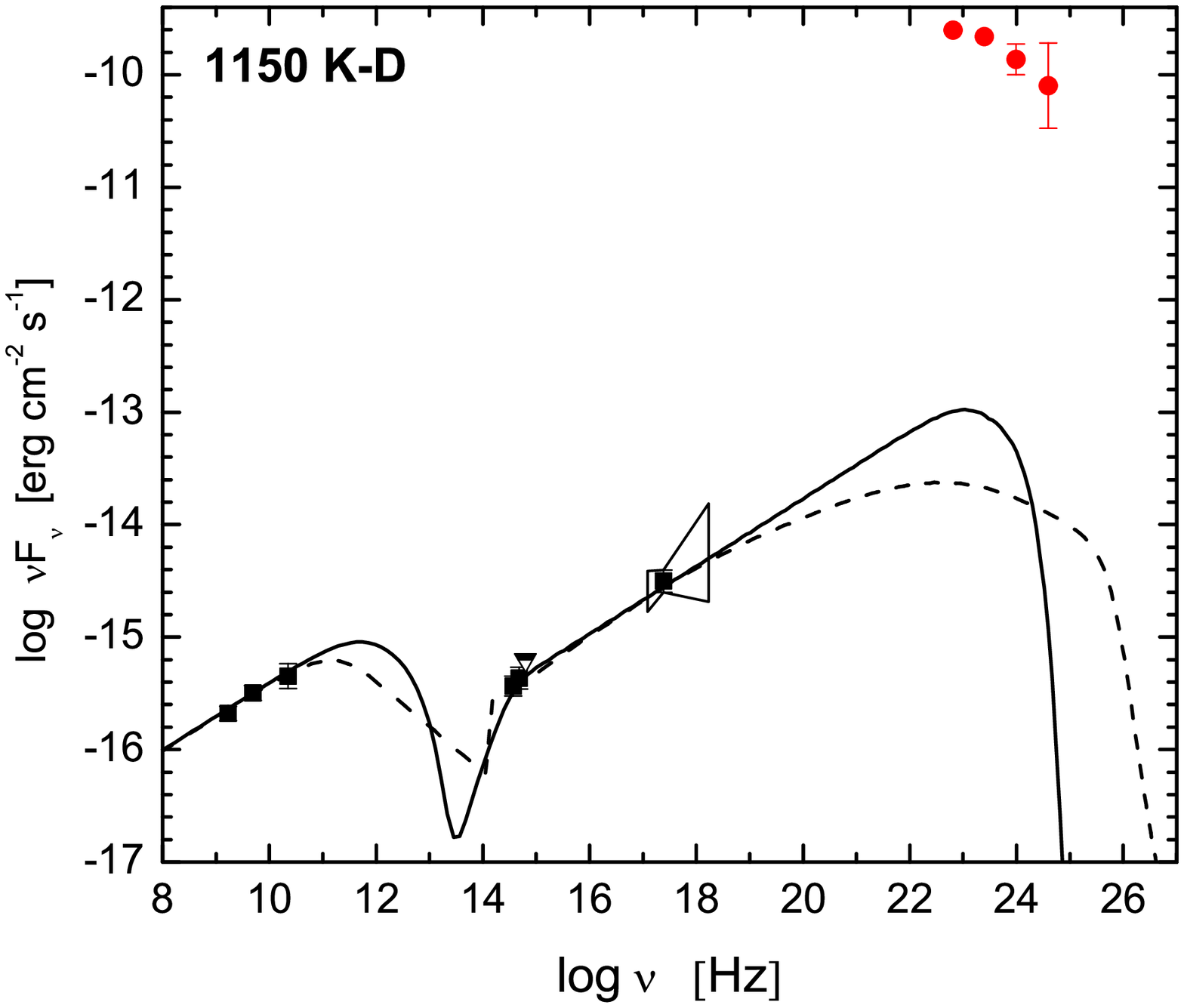}
\includegraphics[angle=0,scale=0.215]{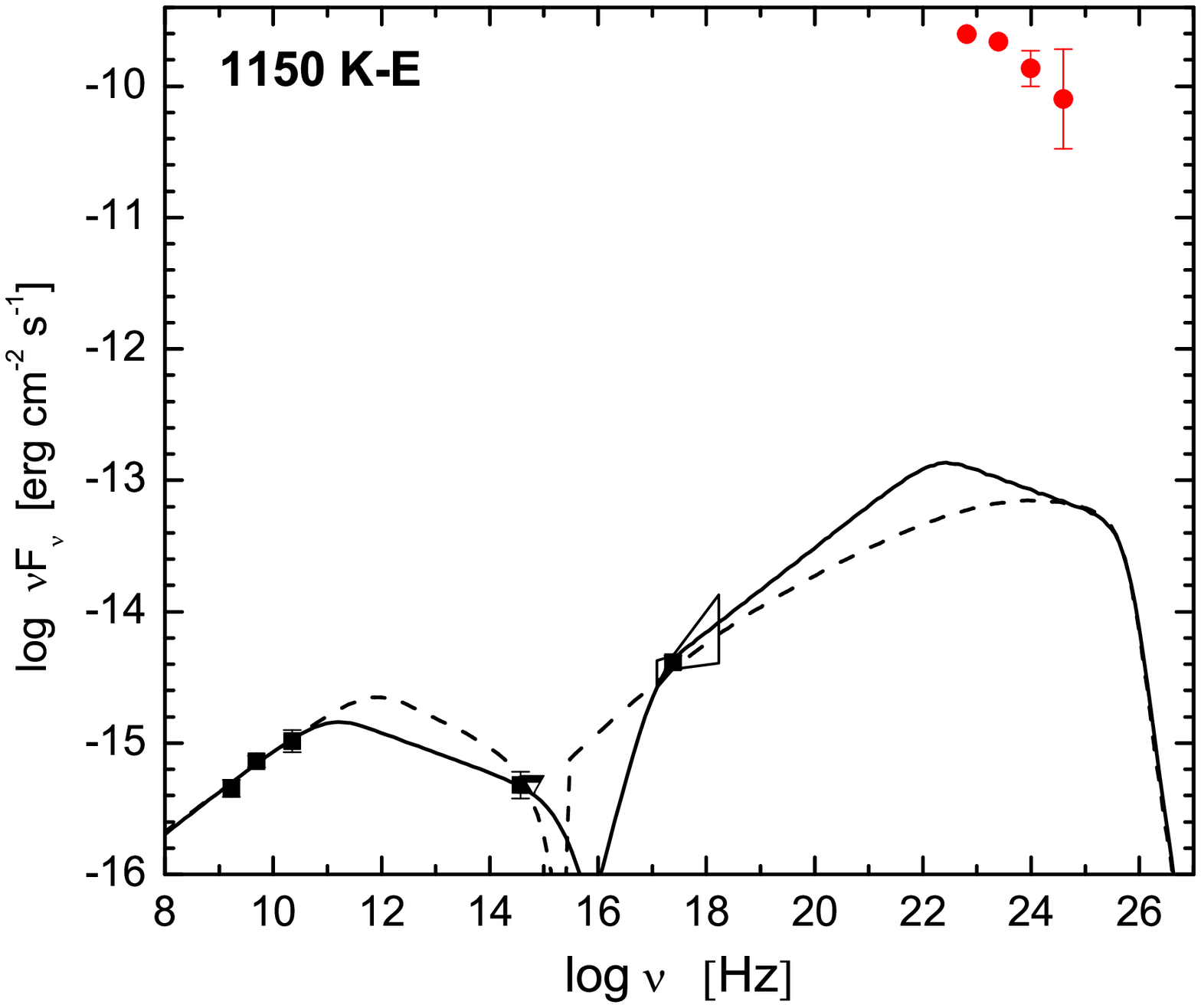}
\includegraphics[angle=0,scale=0.215]{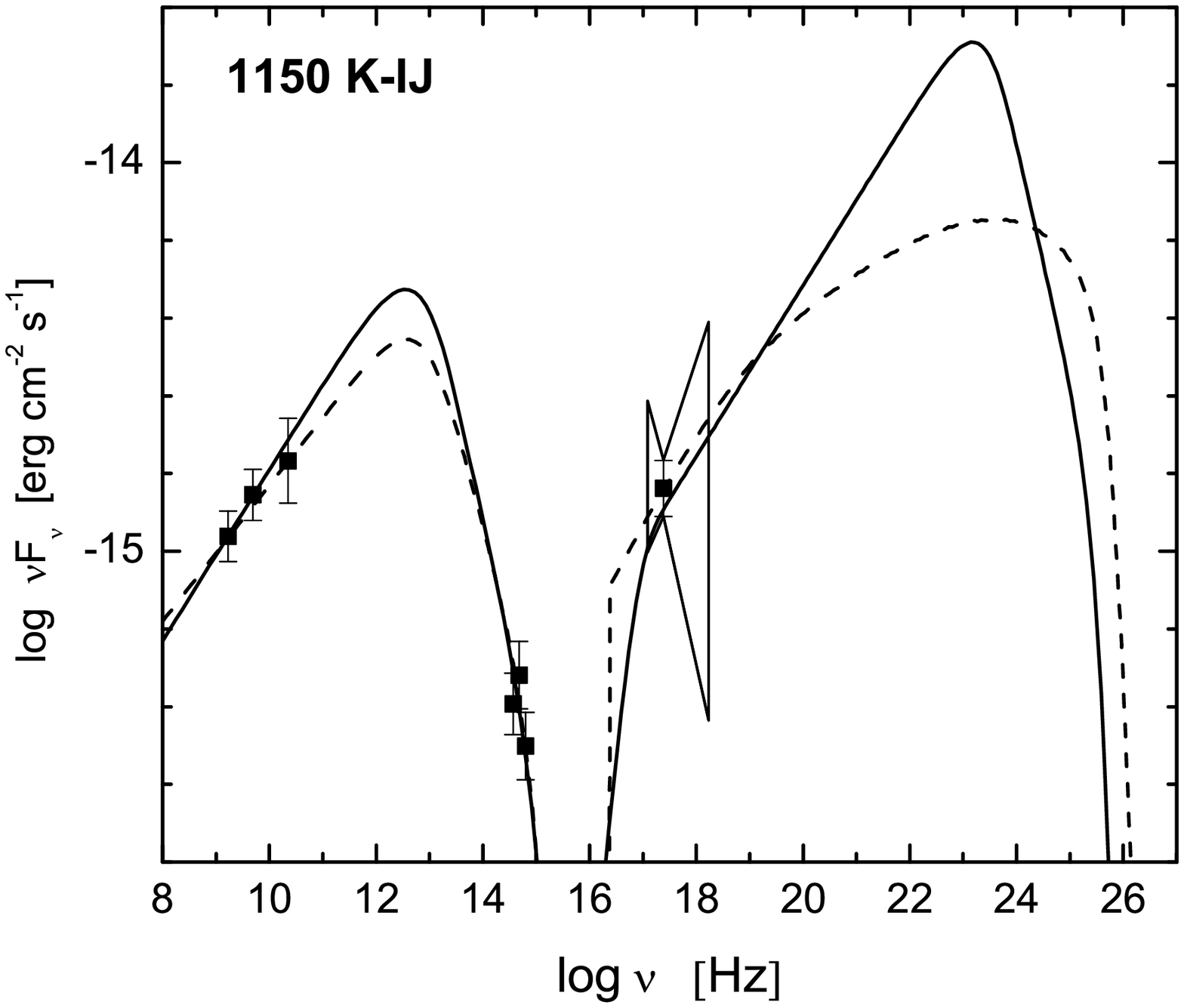}
\includegraphics[angle=0,scale=0.215]{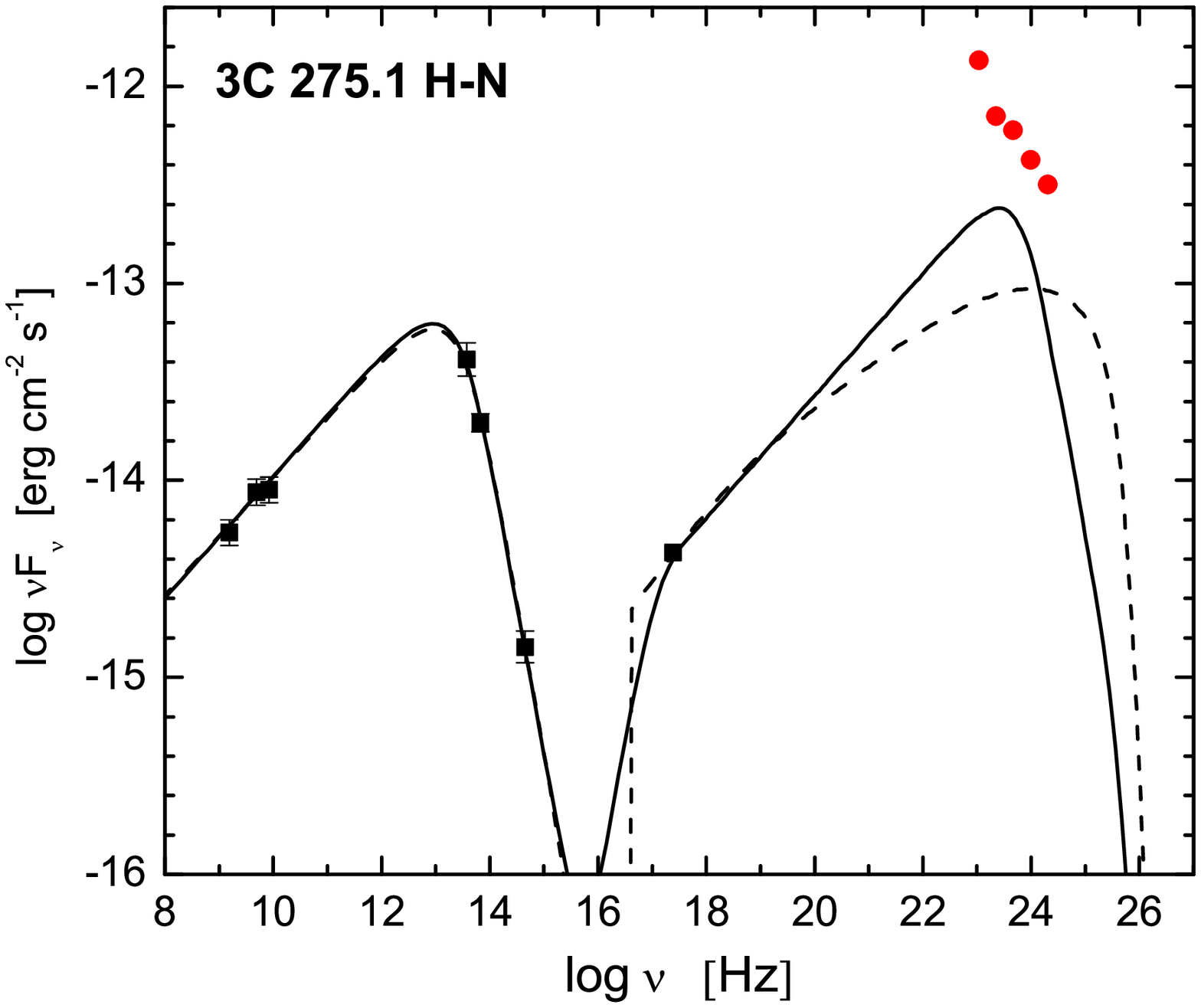}\\

\hfill\center{Fig.5---continued: the results of SED fitting with the IC models.   }
\end{figure*}
\begin{figure*}
\includegraphics[angle=0,scale=0.215]{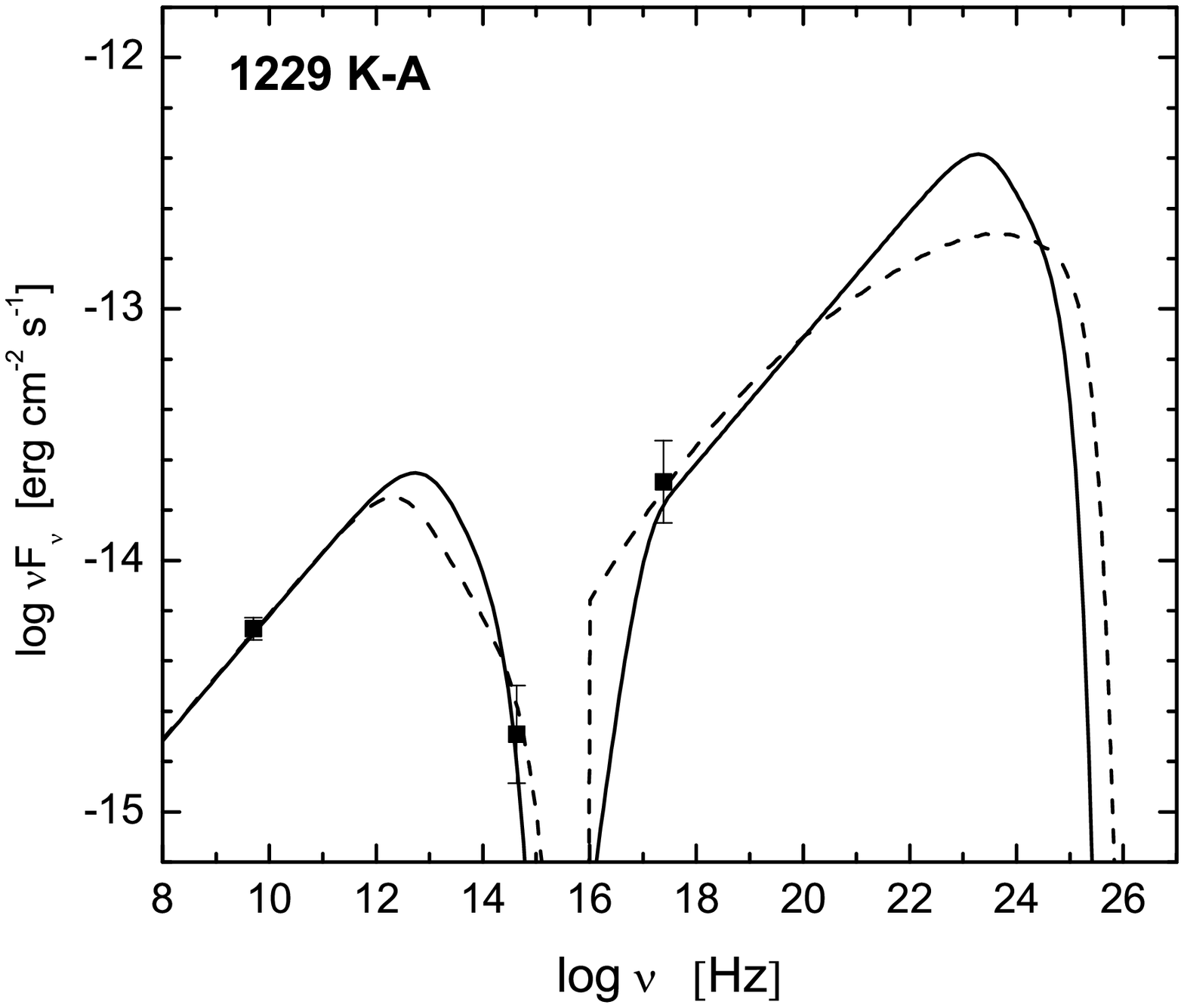}
\includegraphics[angle=0,scale=0.215]{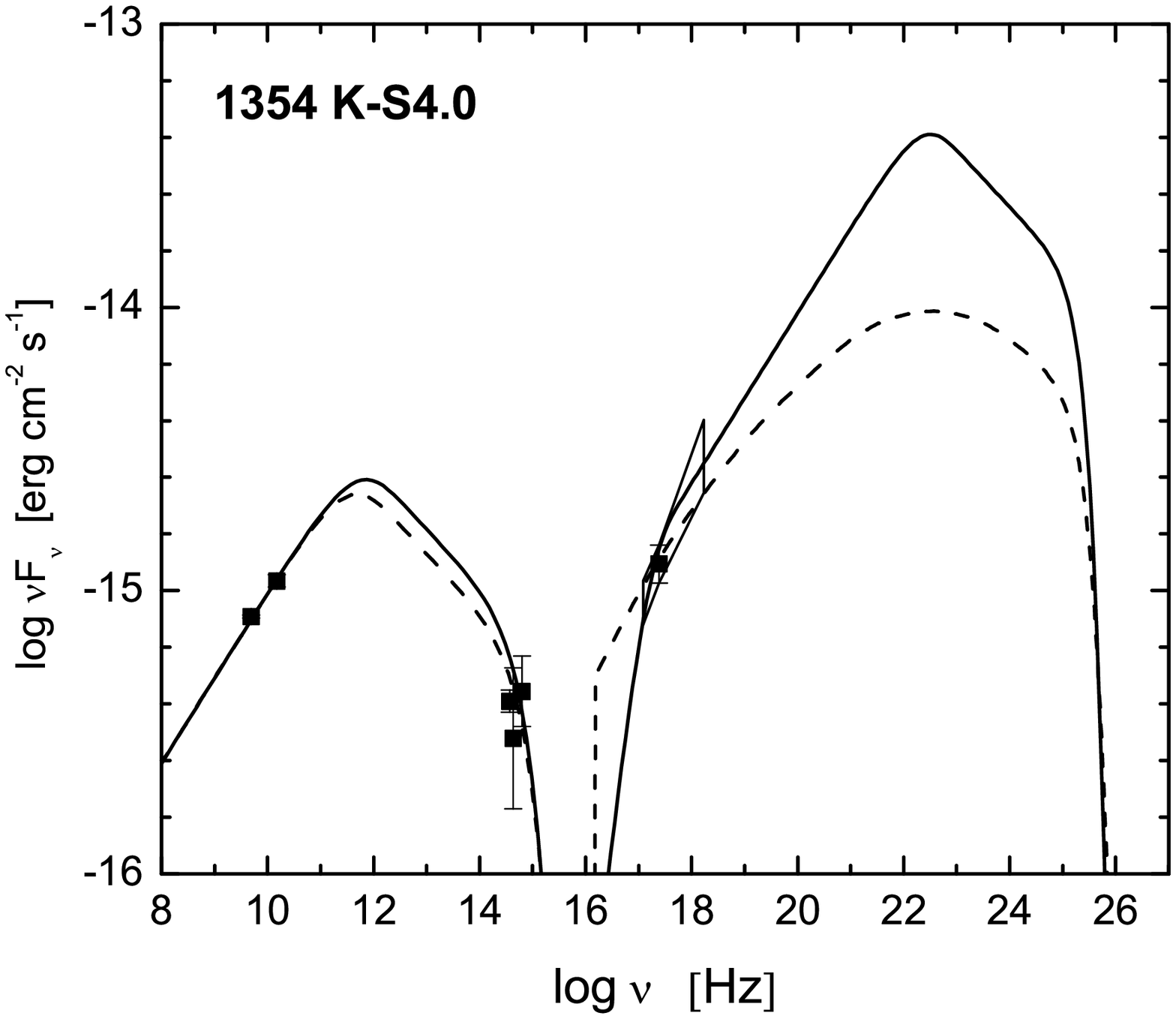}
\includegraphics[angle=0,scale=0.215]{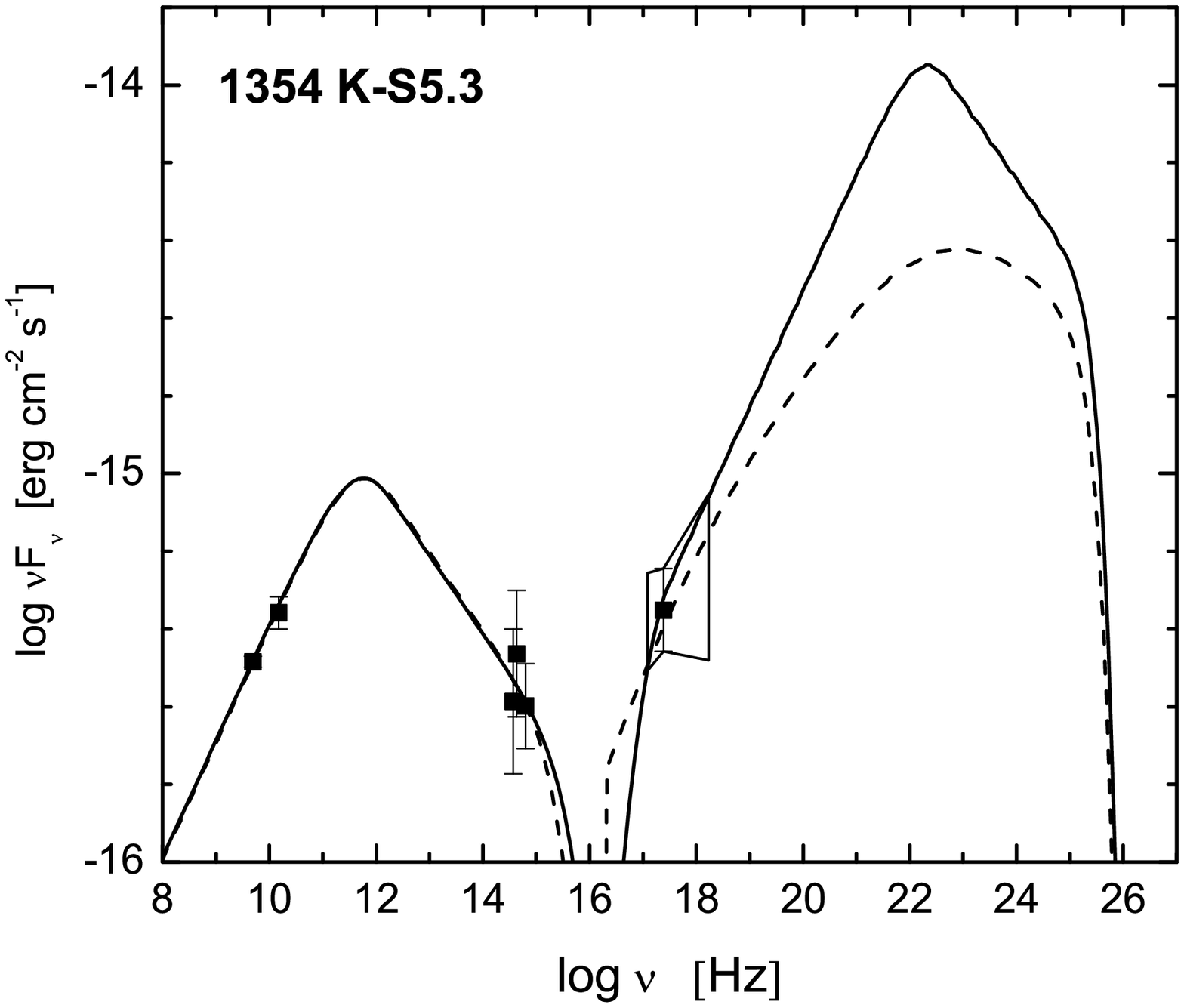}
\includegraphics[angle=0,scale=0.215]{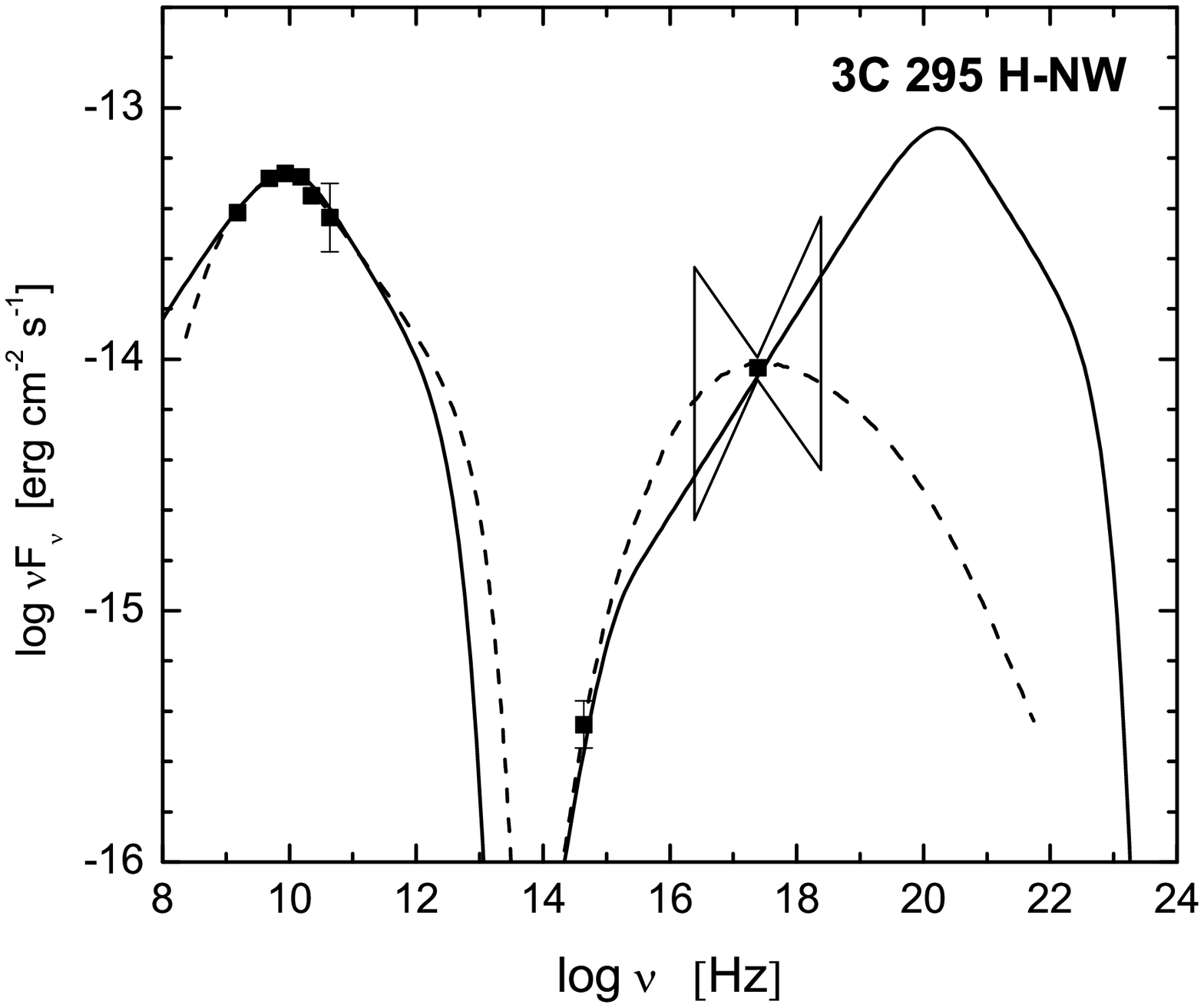}\\
\includegraphics[angle=0,scale=0.215]{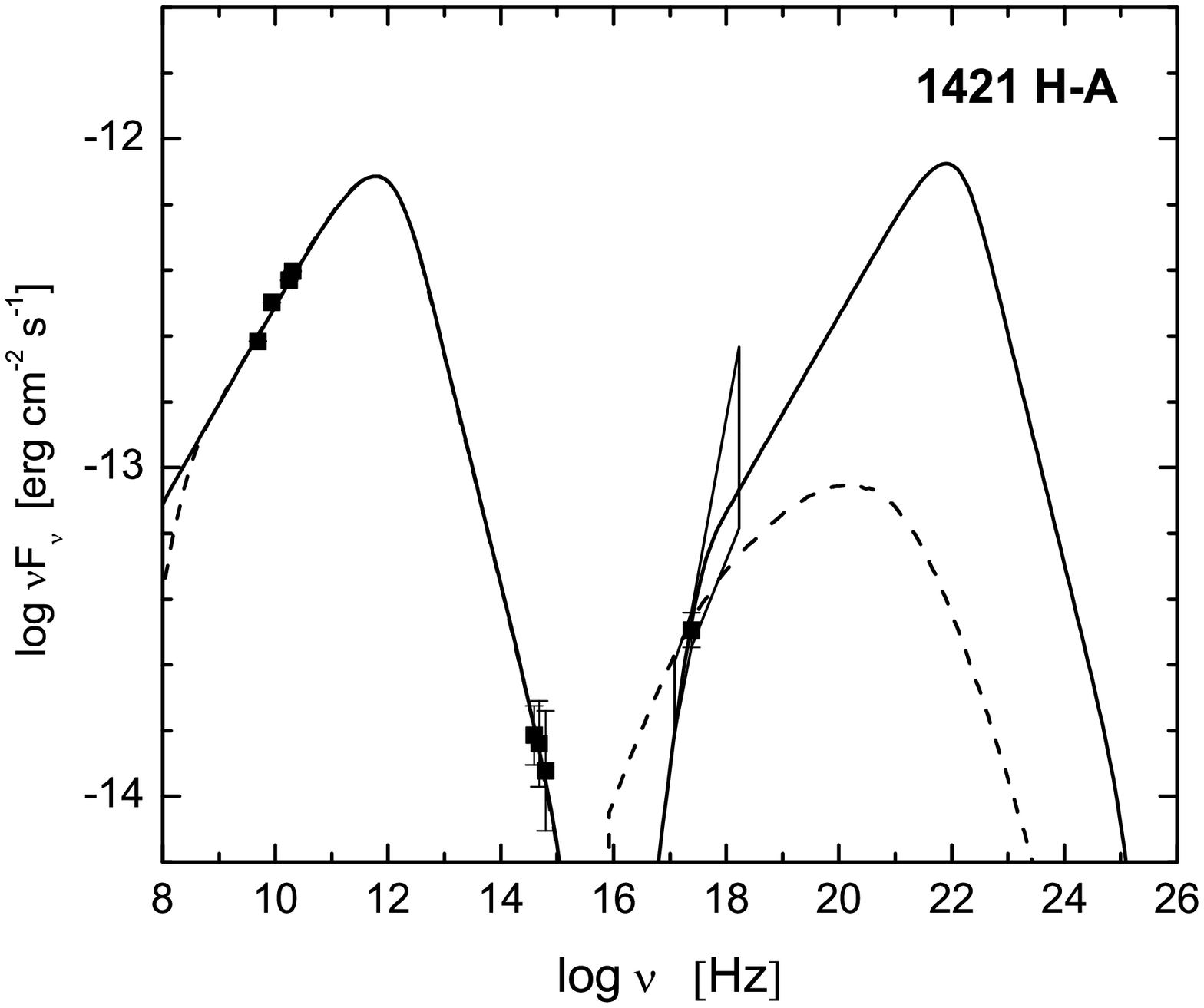}
\includegraphics[angle=0,scale=0.215]{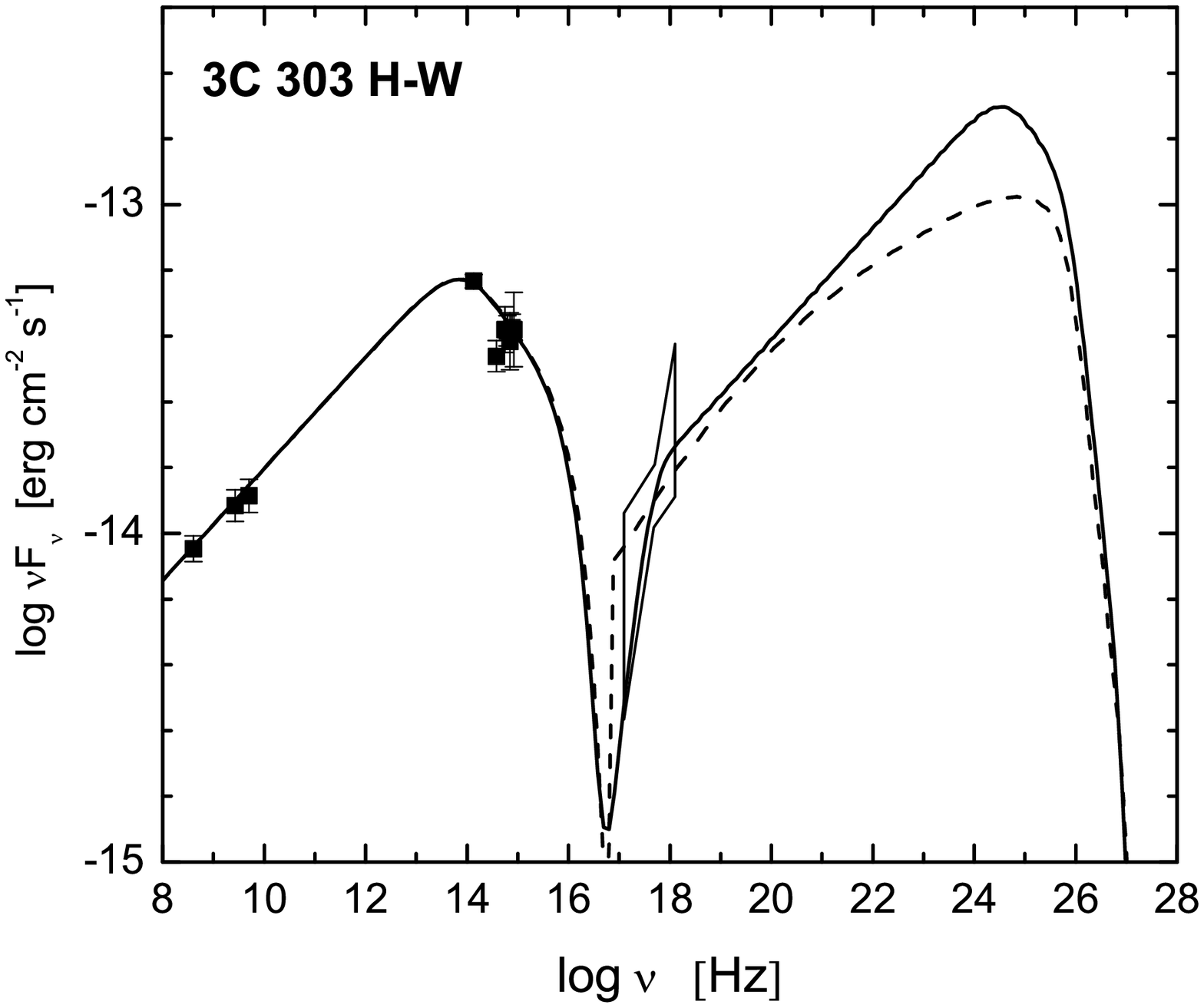}
\includegraphics[angle=0,scale=0.215]{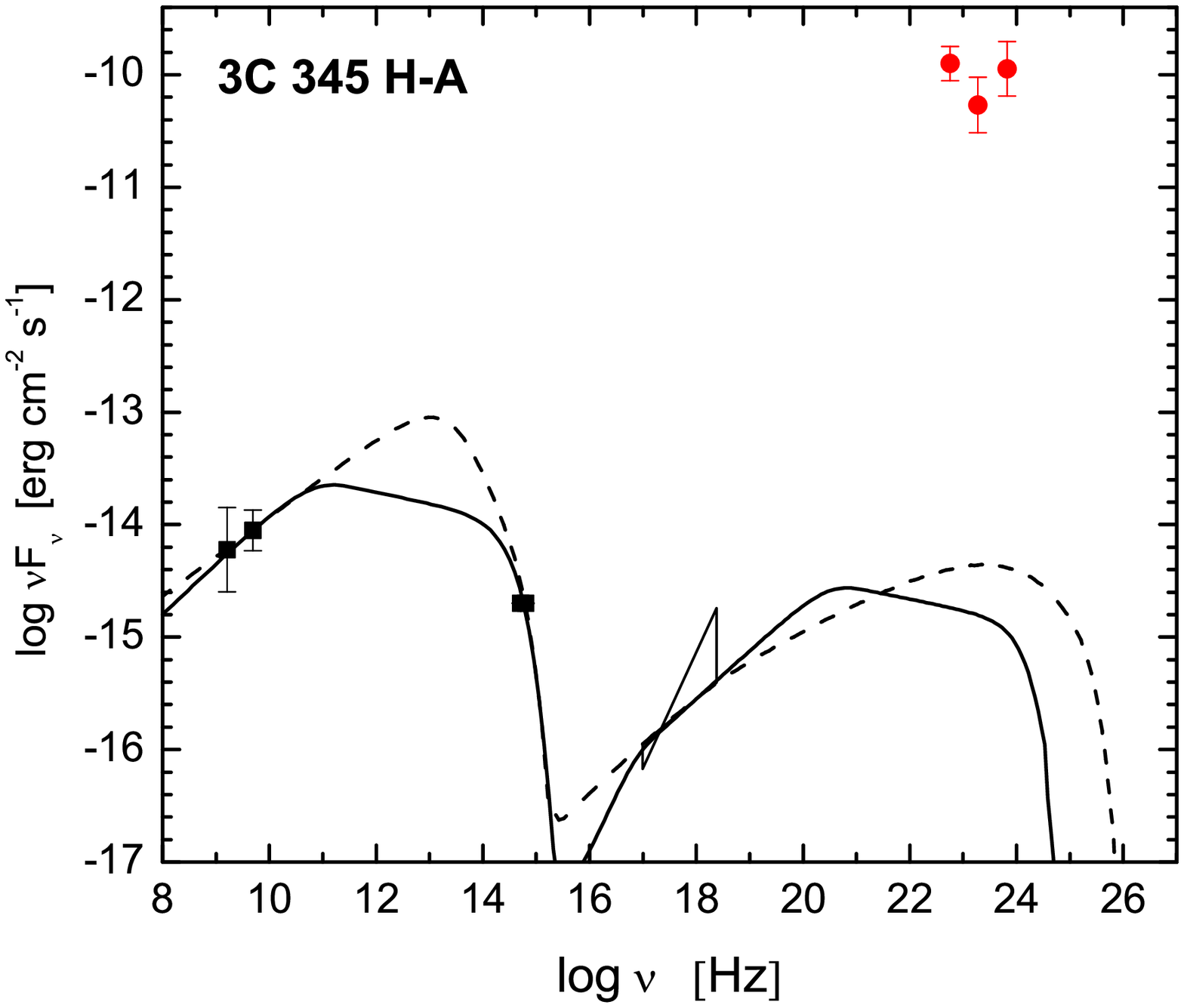}
\includegraphics[angle=0,scale=0.215]{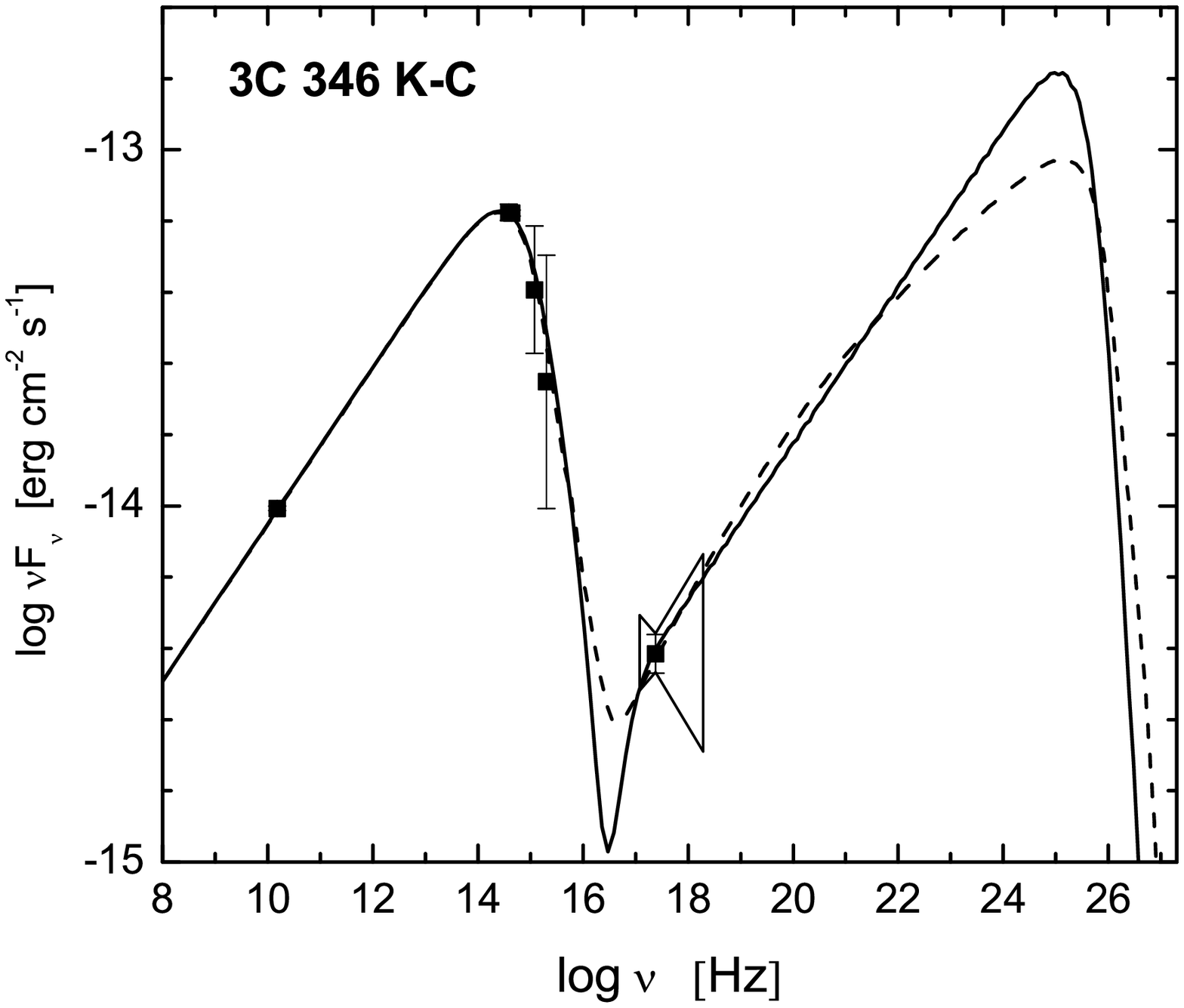}\\
\includegraphics[angle=0,scale=0.215]{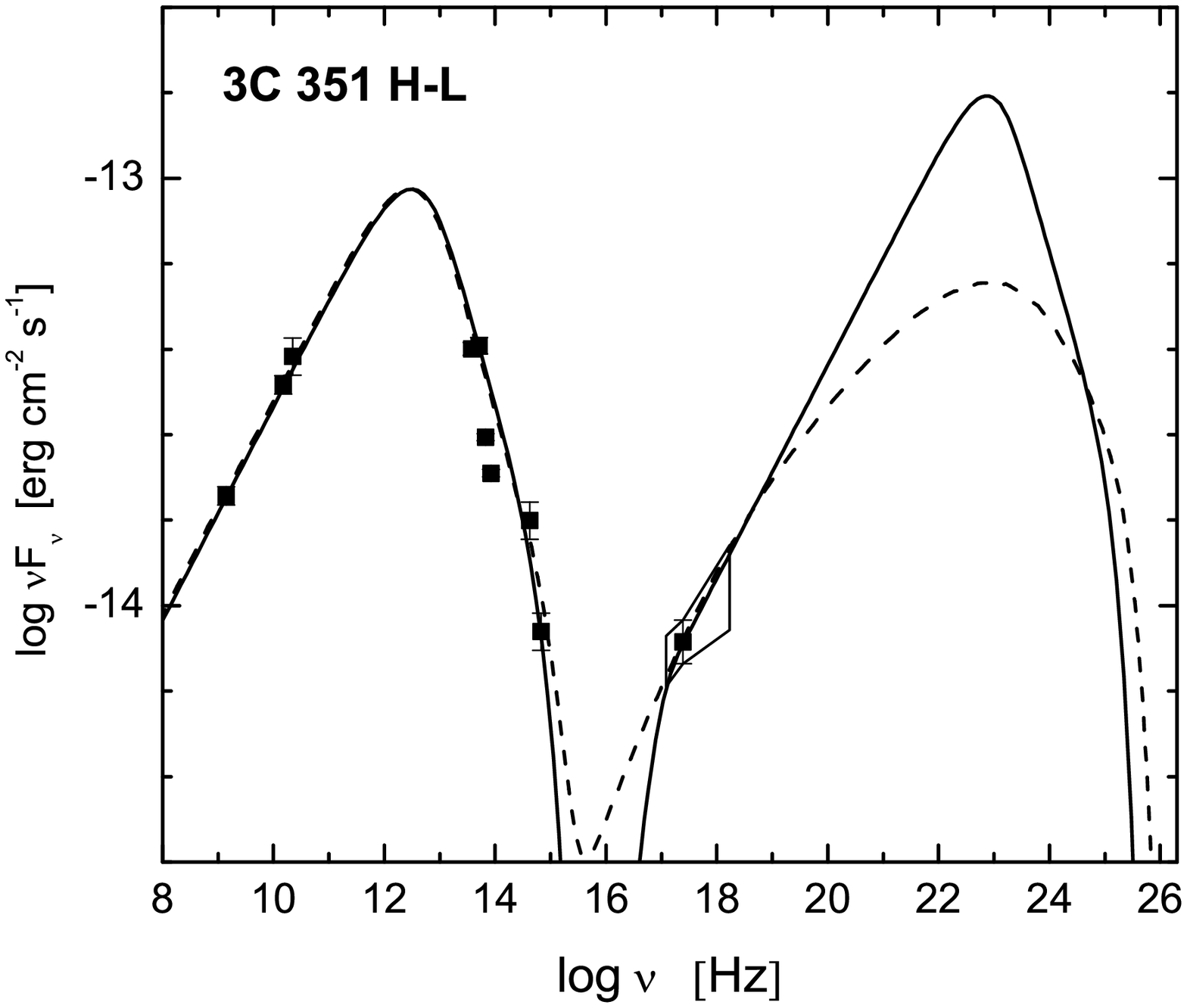}
\includegraphics[angle=0,scale=0.215]{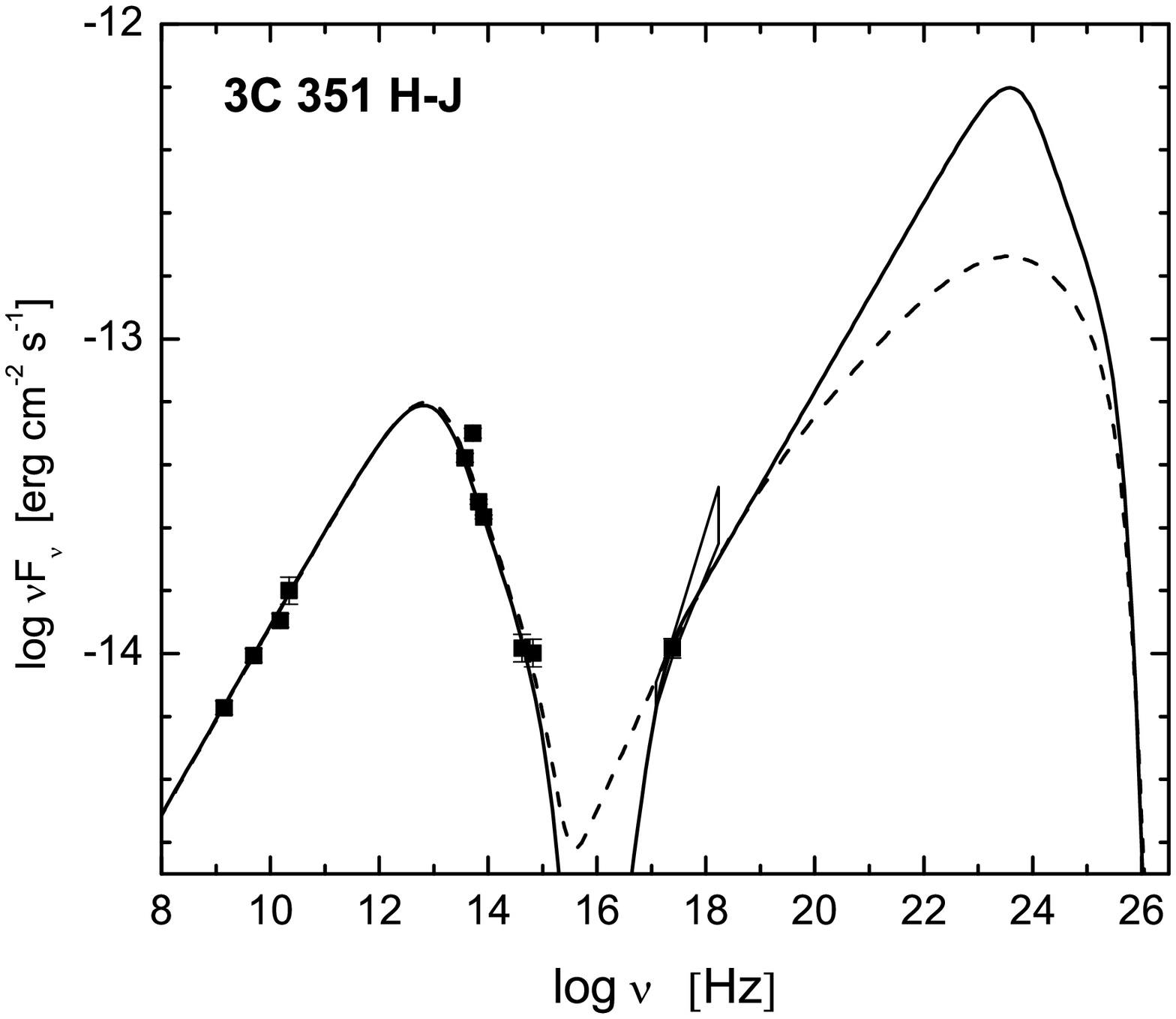}
\includegraphics[angle=0,scale=0.215]{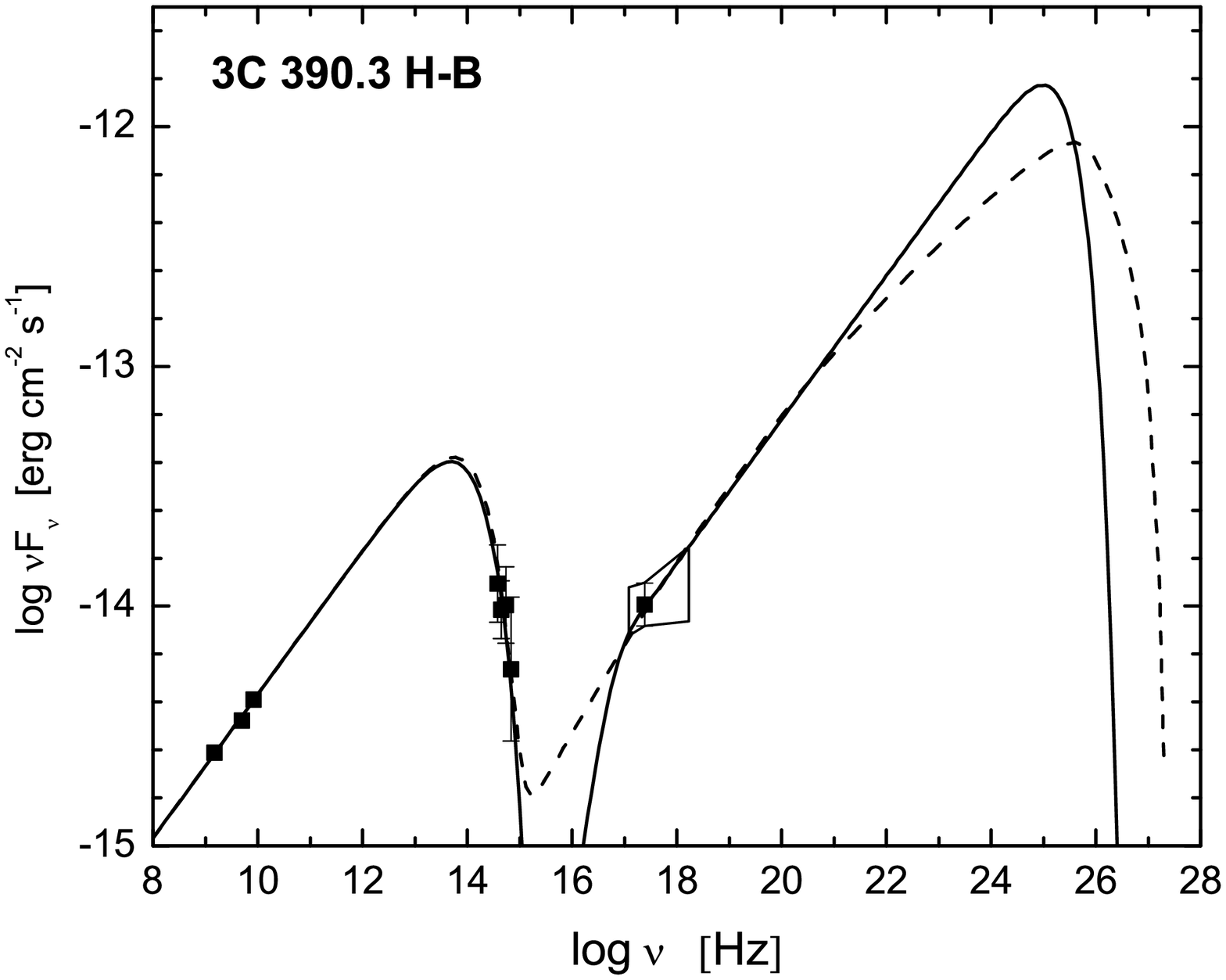}
\includegraphics[angle=0,scale=0.215]{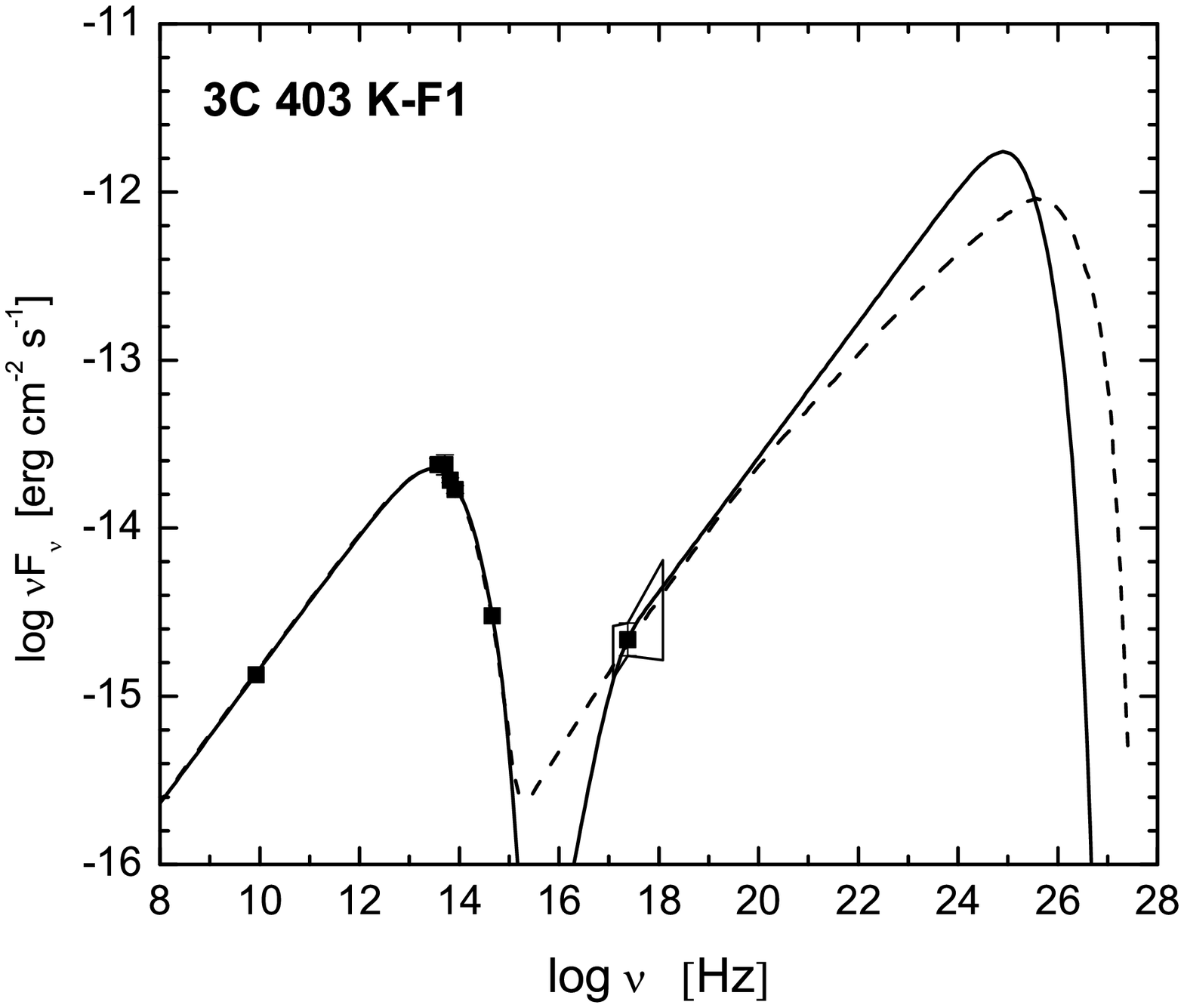}\\
\includegraphics[angle=0,scale=0.215]{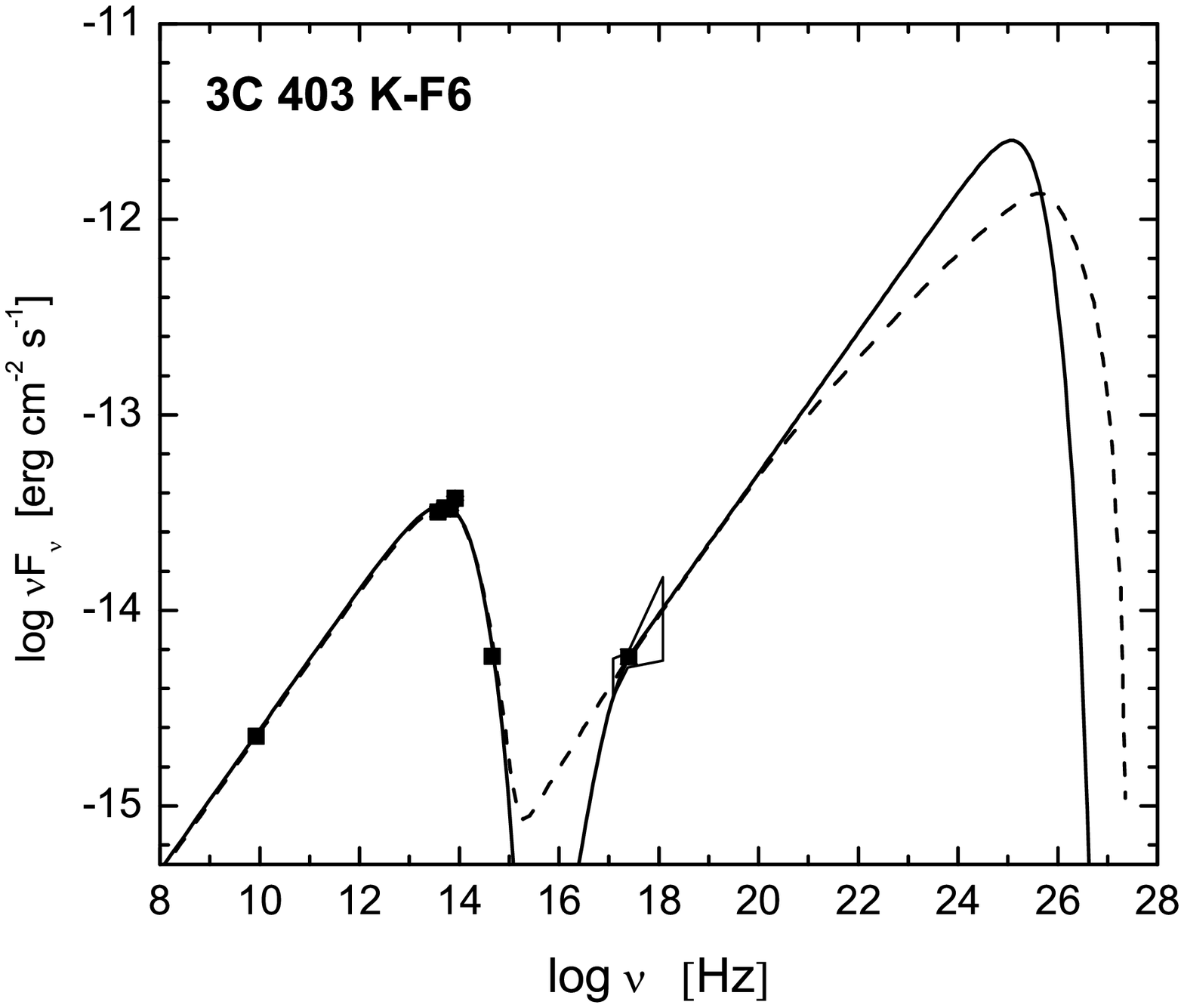}
\includegraphics[angle=0,scale=0.215]{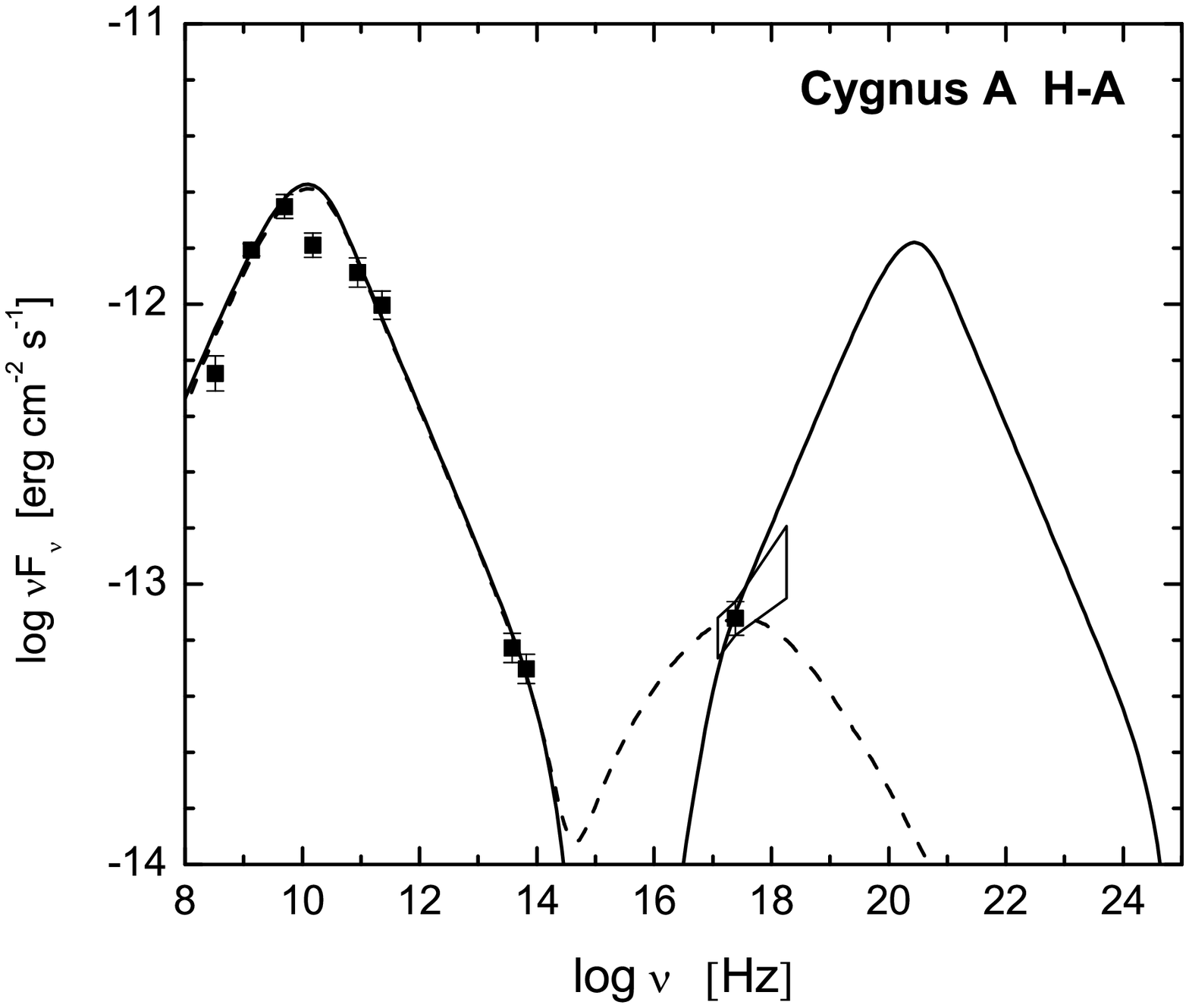}
\includegraphics[angle=0,scale=0.215]{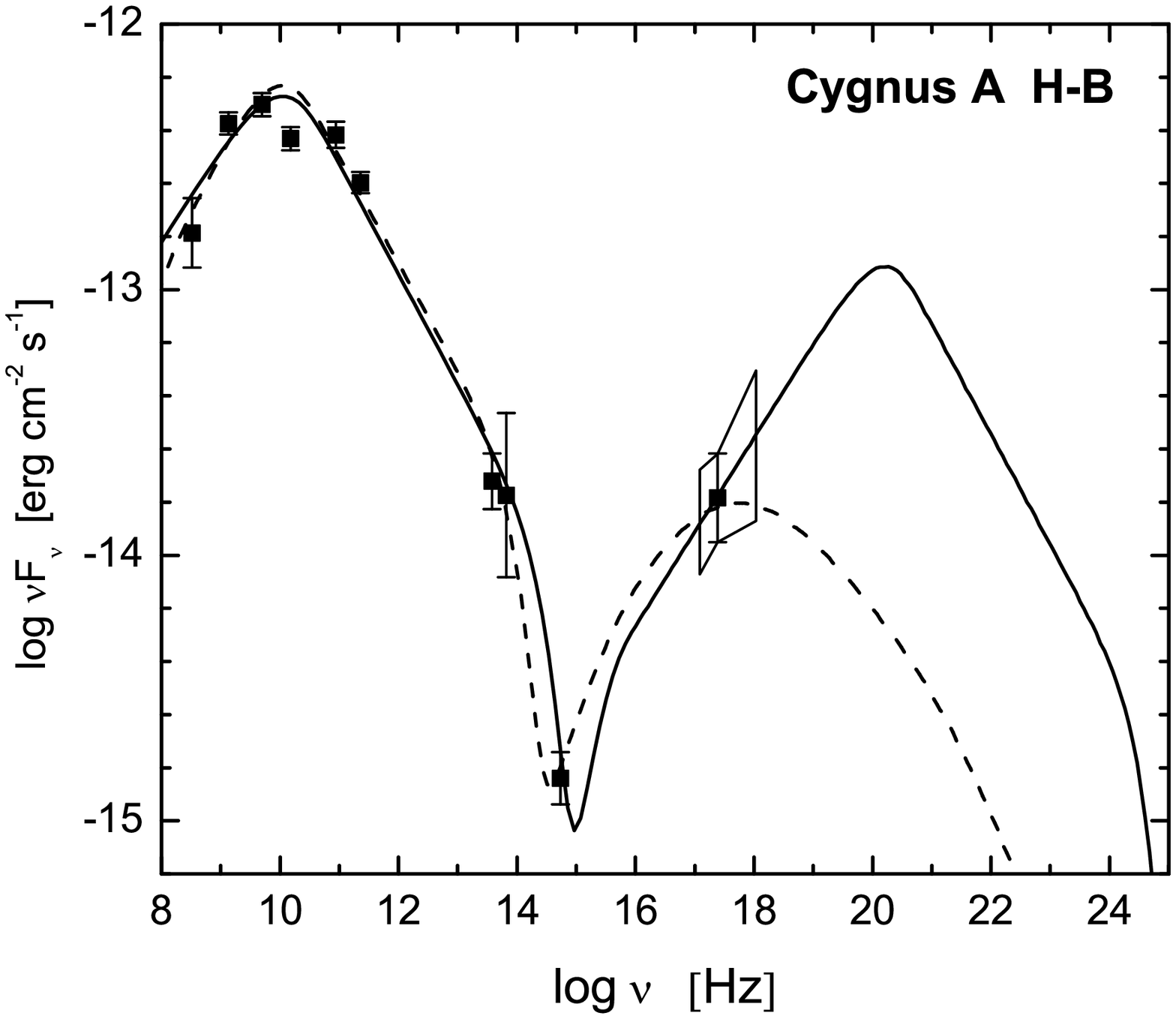}
\includegraphics[angle=0,scale=0.215]{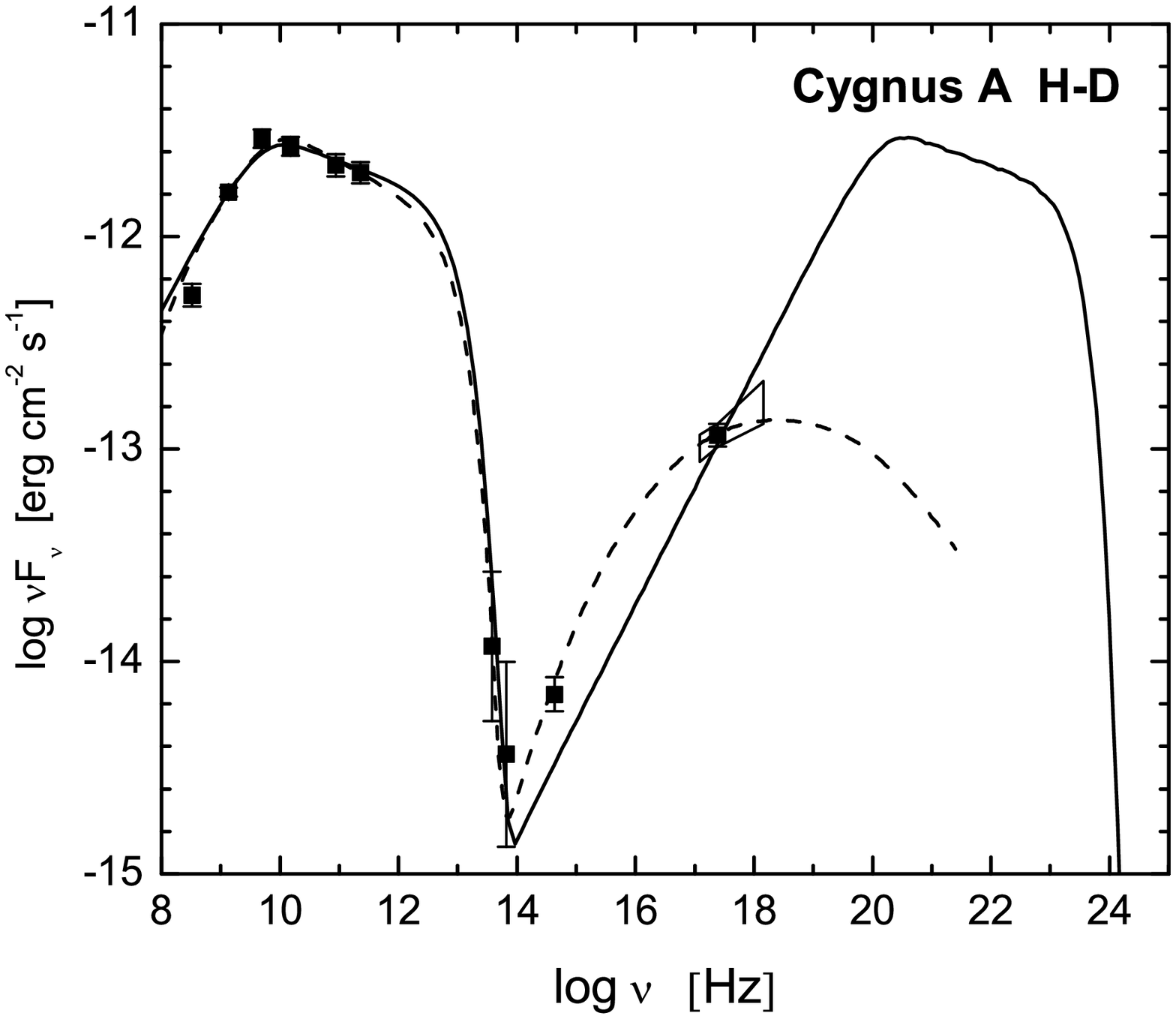}\\
\includegraphics[angle=0,scale=0.215]{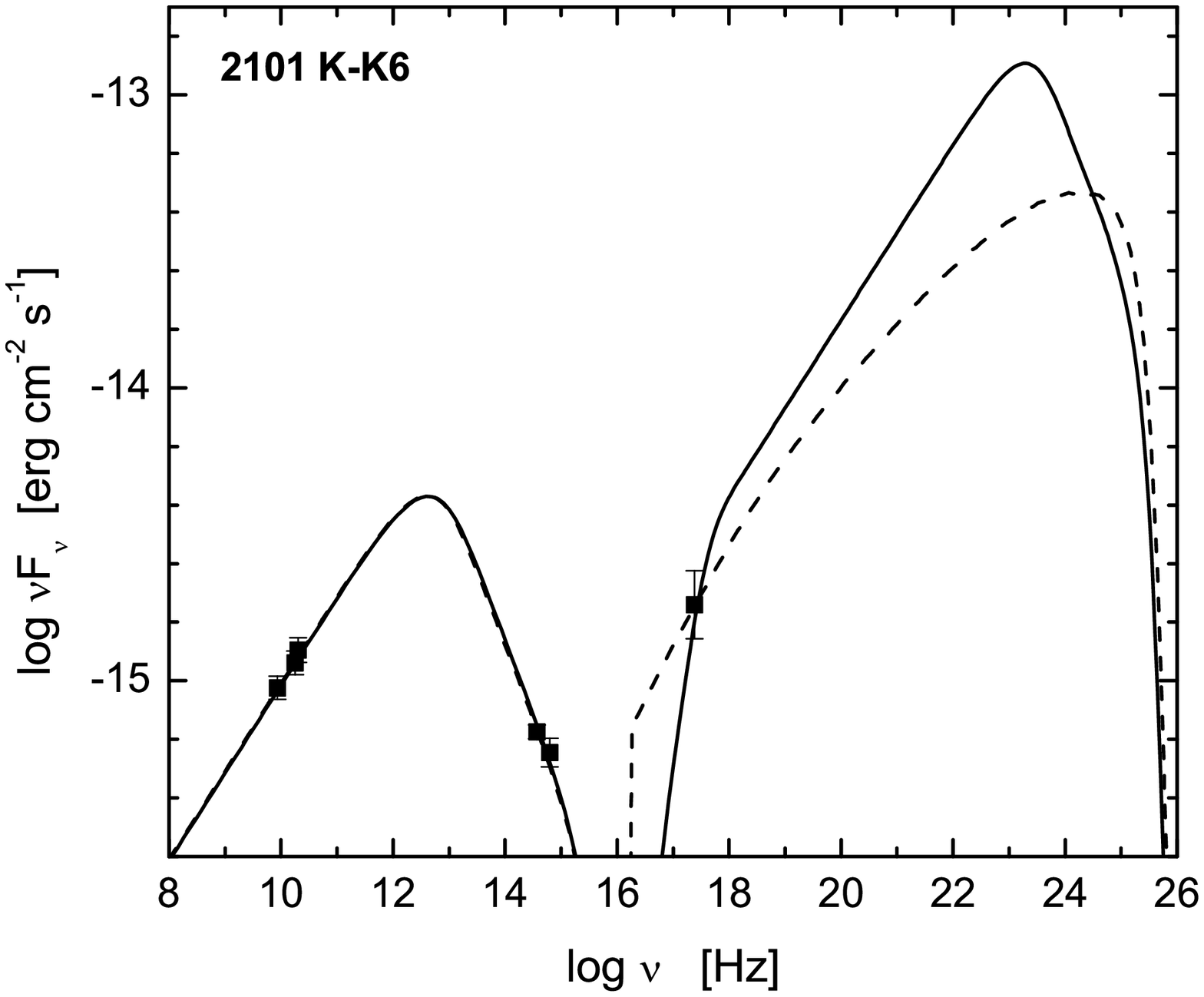}
\includegraphics[angle=0,scale=0.215]{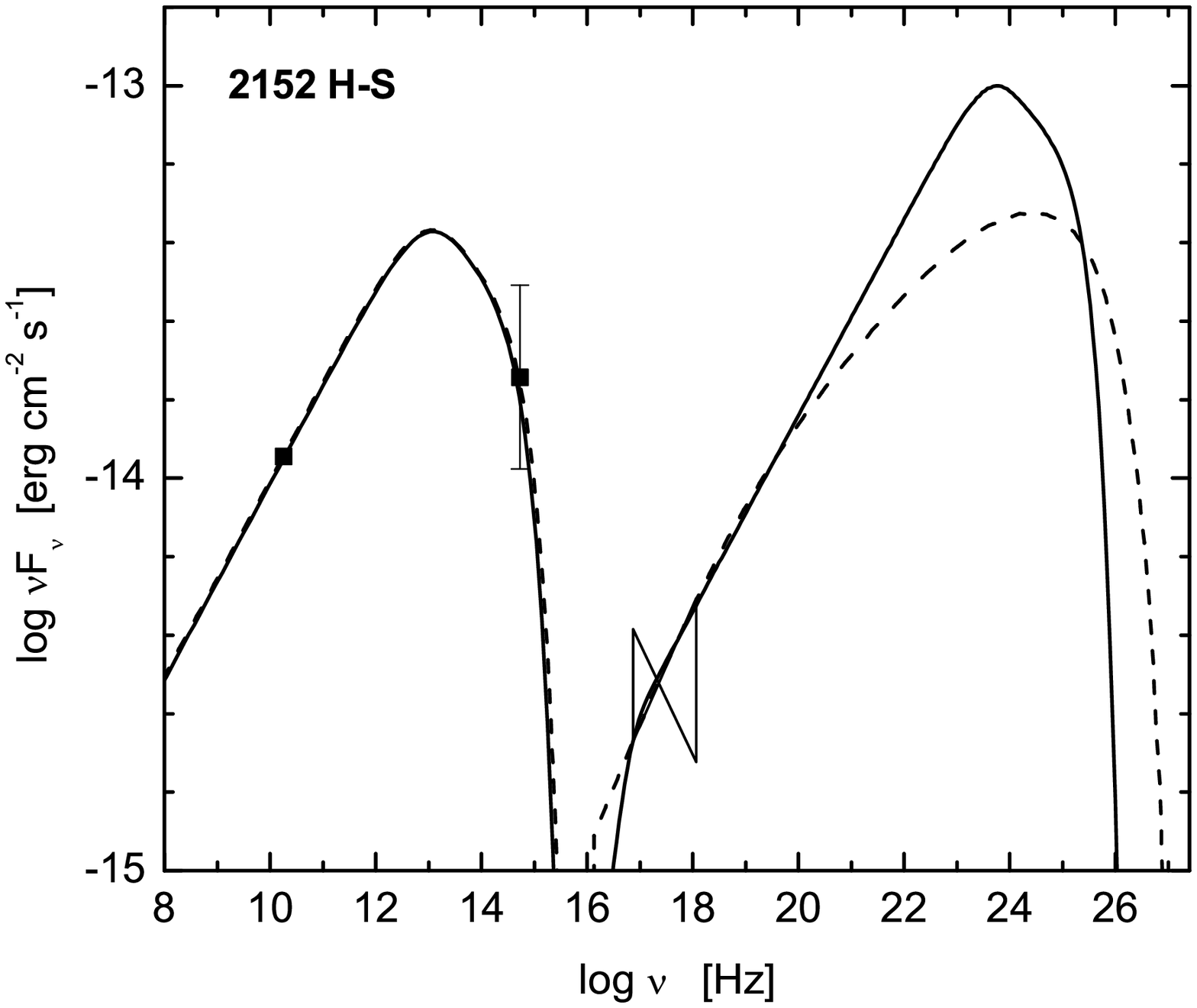}
\includegraphics[angle=0,scale=0.215]{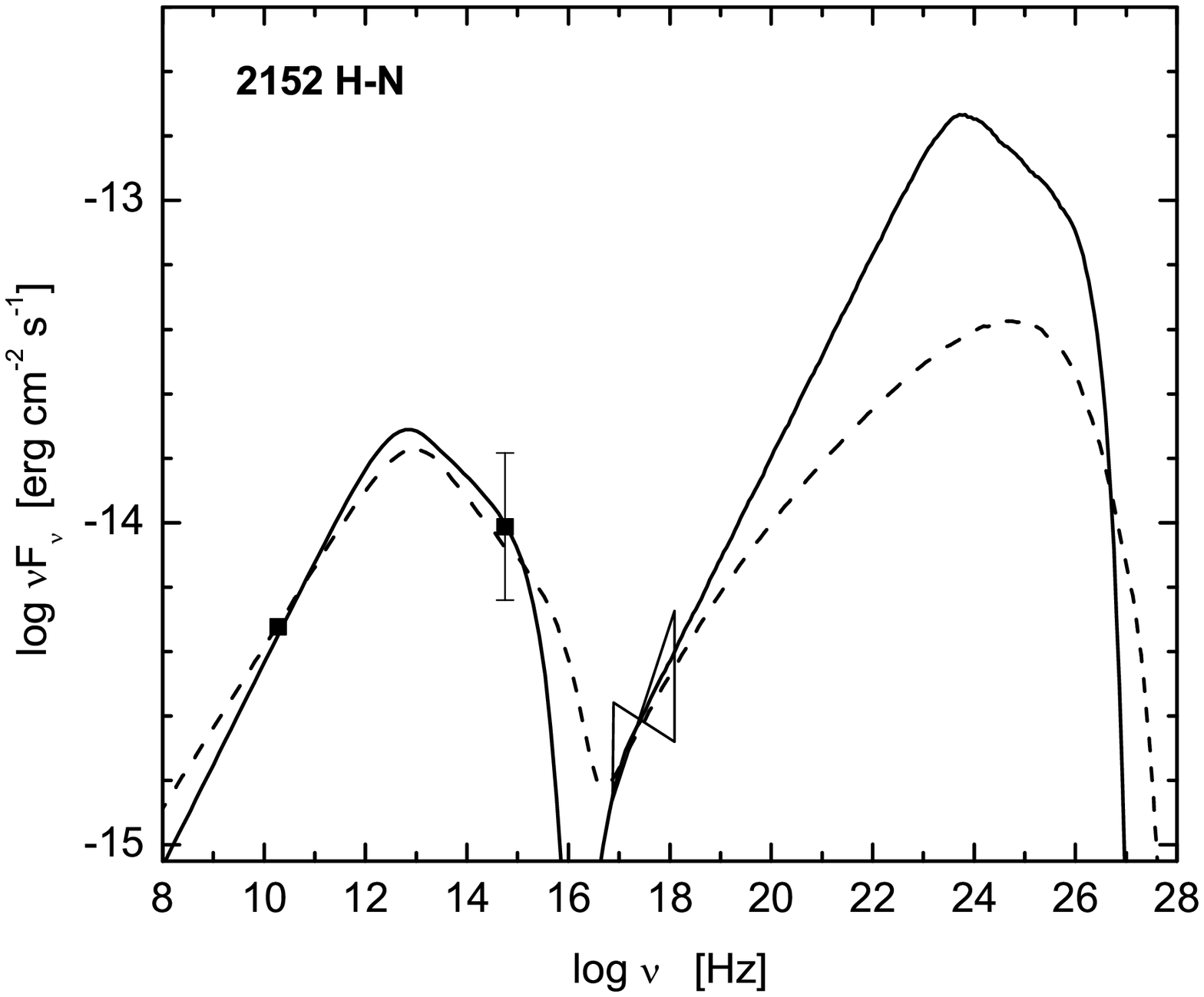}
\includegraphics[angle=0,scale=0.215]{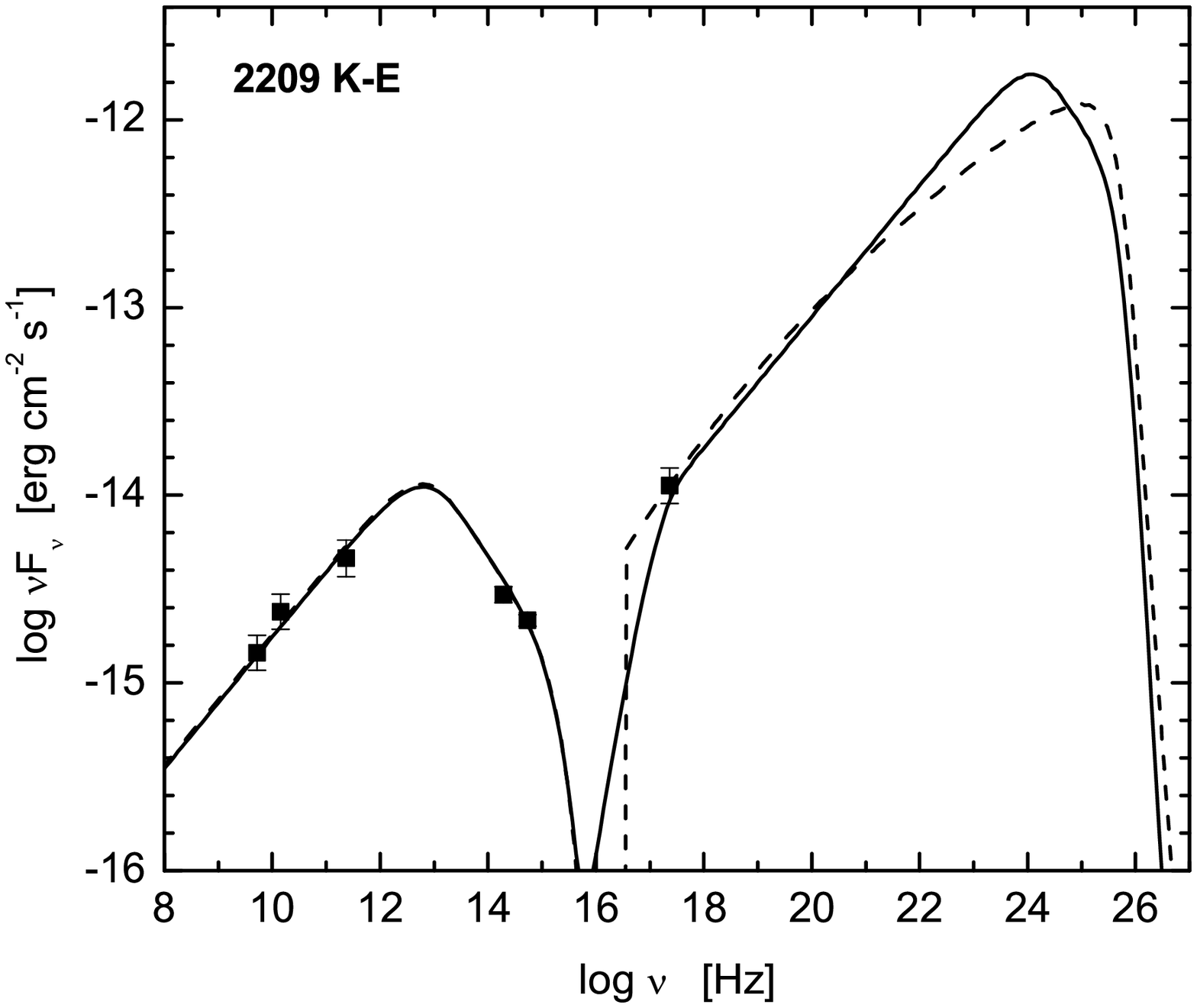}\\
\includegraphics[angle=0,scale=0.215]{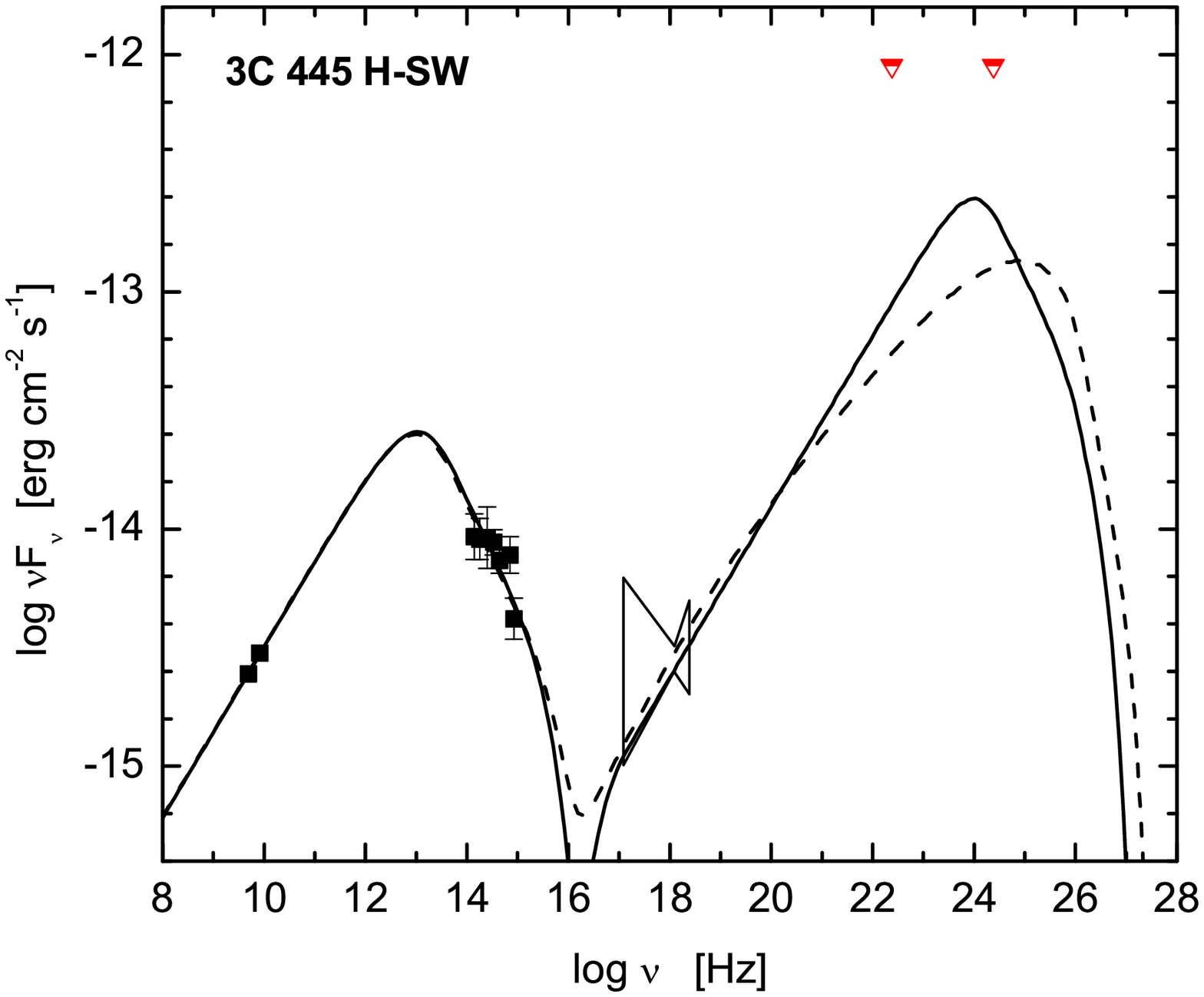}
\includegraphics[angle=0,scale=0.215]{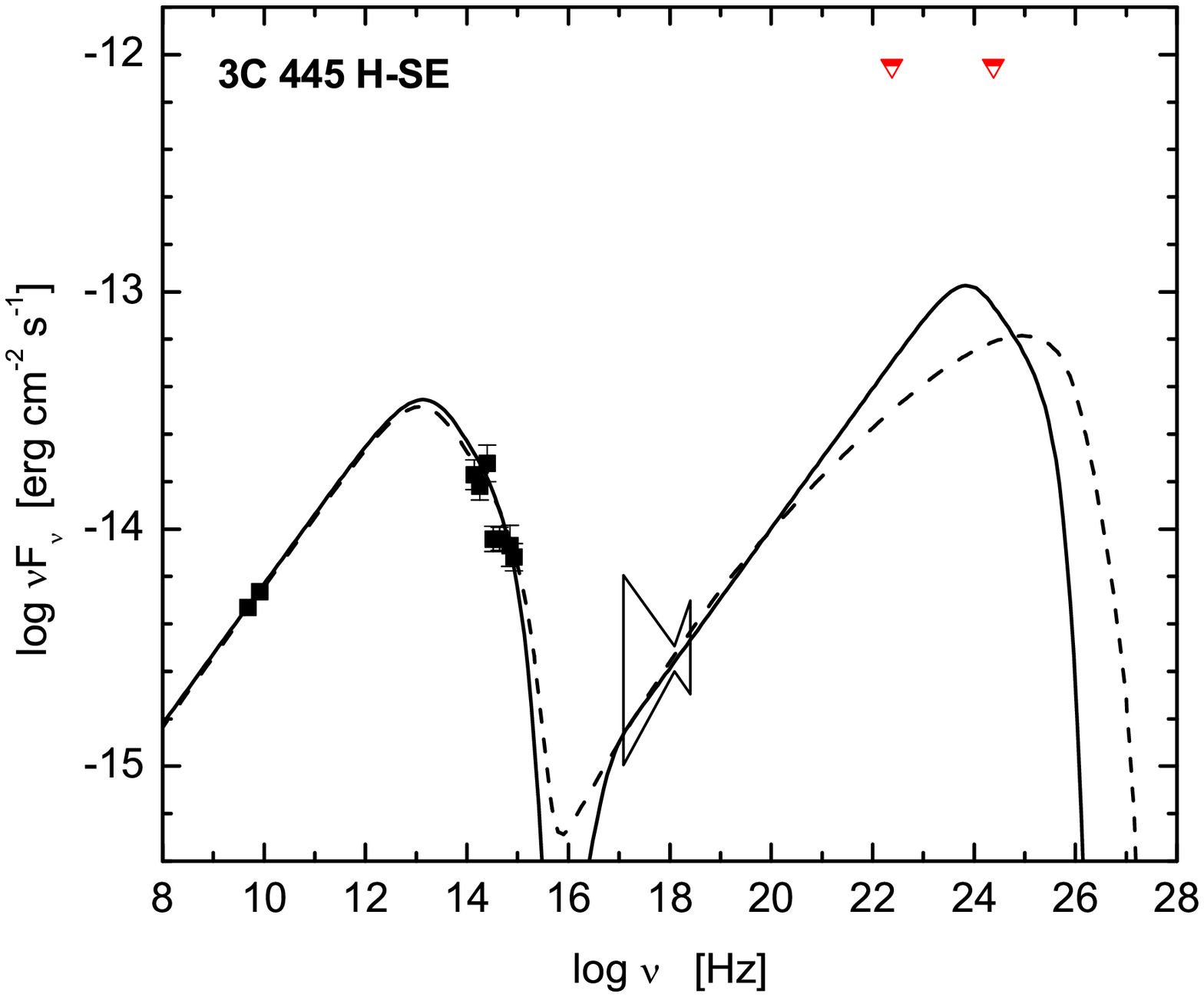}
\includegraphics[angle=0,scale=0.215]{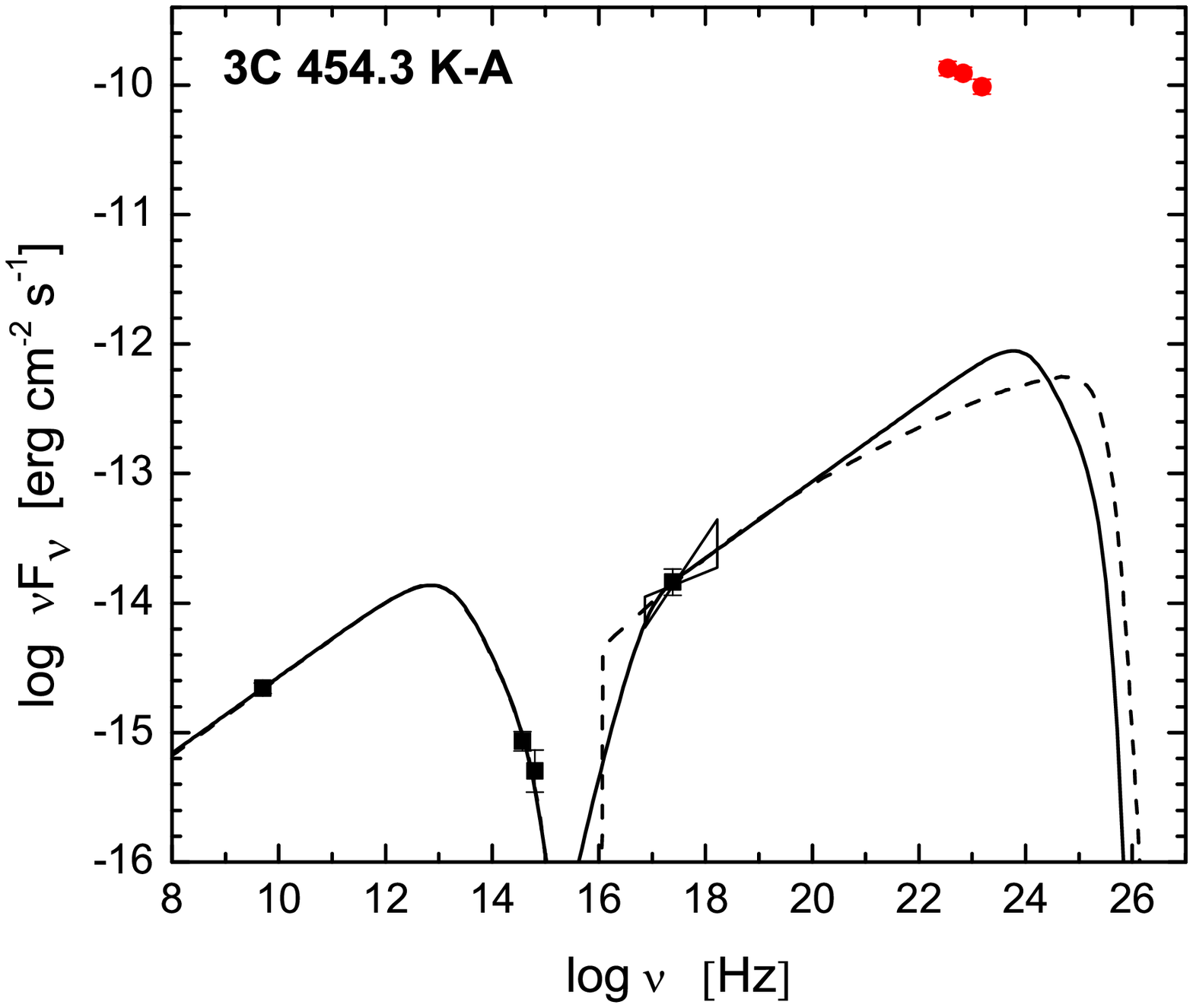}
\includegraphics[angle=0,scale=0.215]{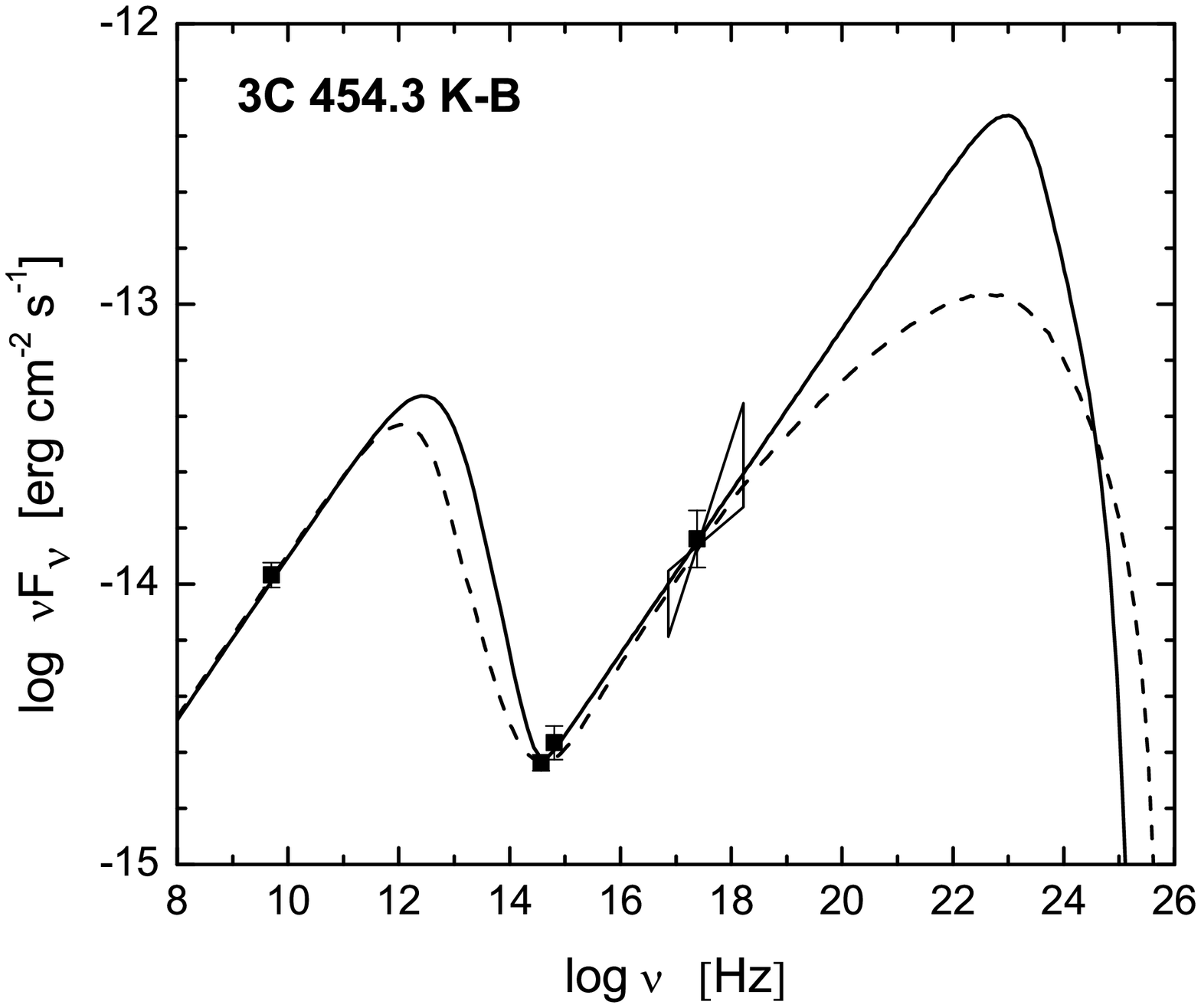}\\
\hfill\center{Fig.5---continued: the results of SED fitting with the IC models.   }
\end{figure*}
\begin{figure*}
\includegraphics[angle=0,scale=0.215]{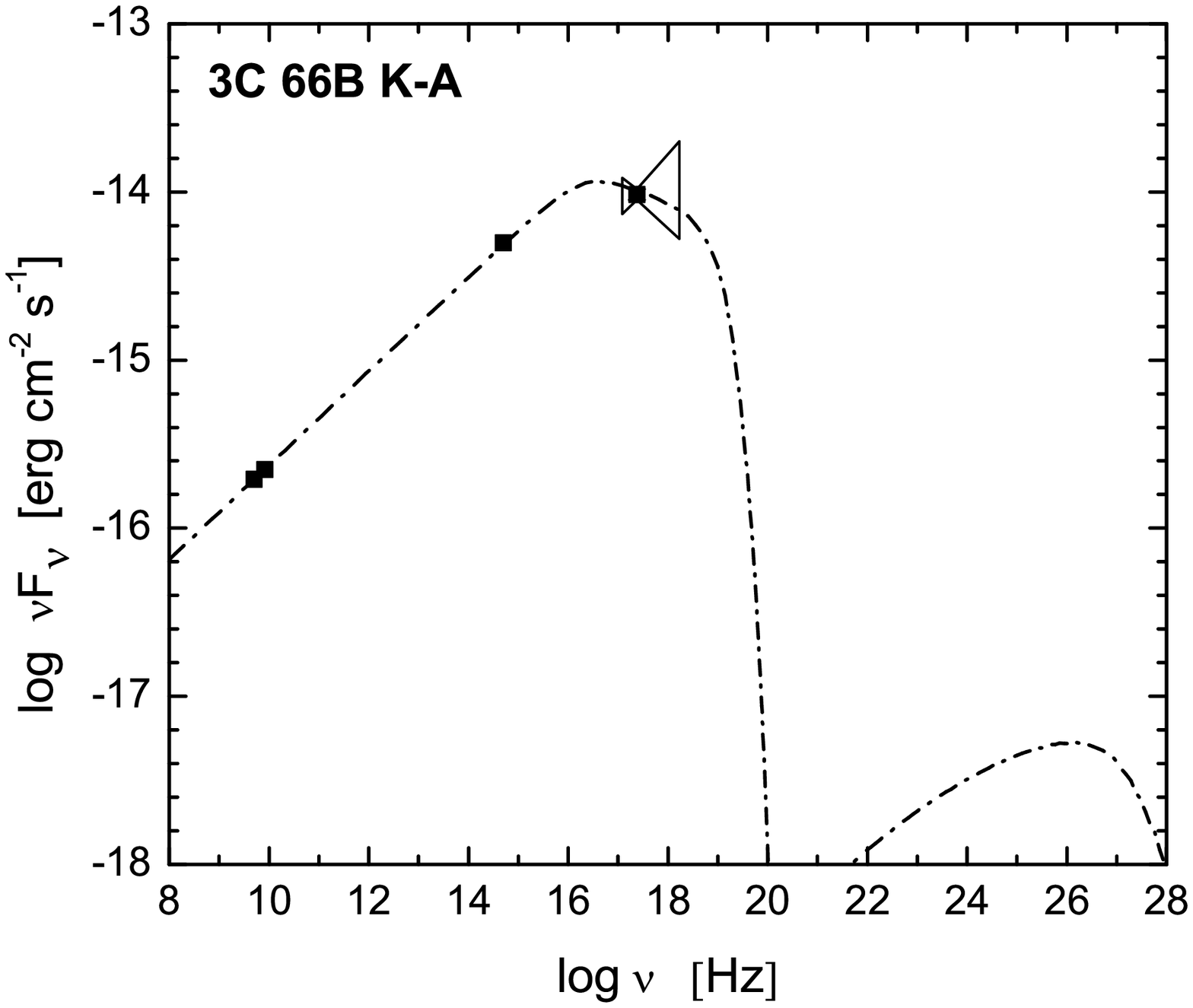}
\includegraphics[angle=0,scale=0.215]{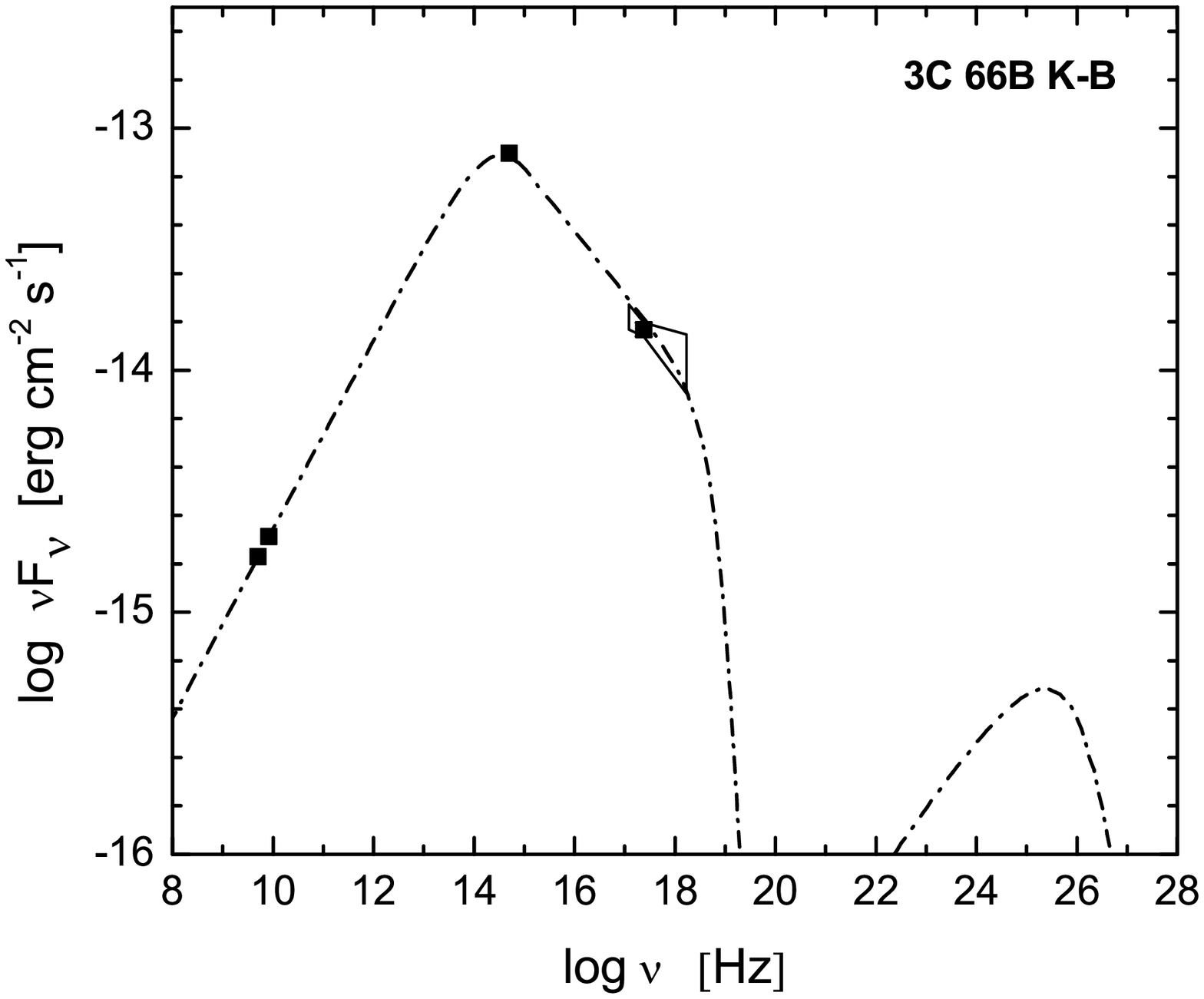}
\includegraphics[angle=0,scale=0.215]{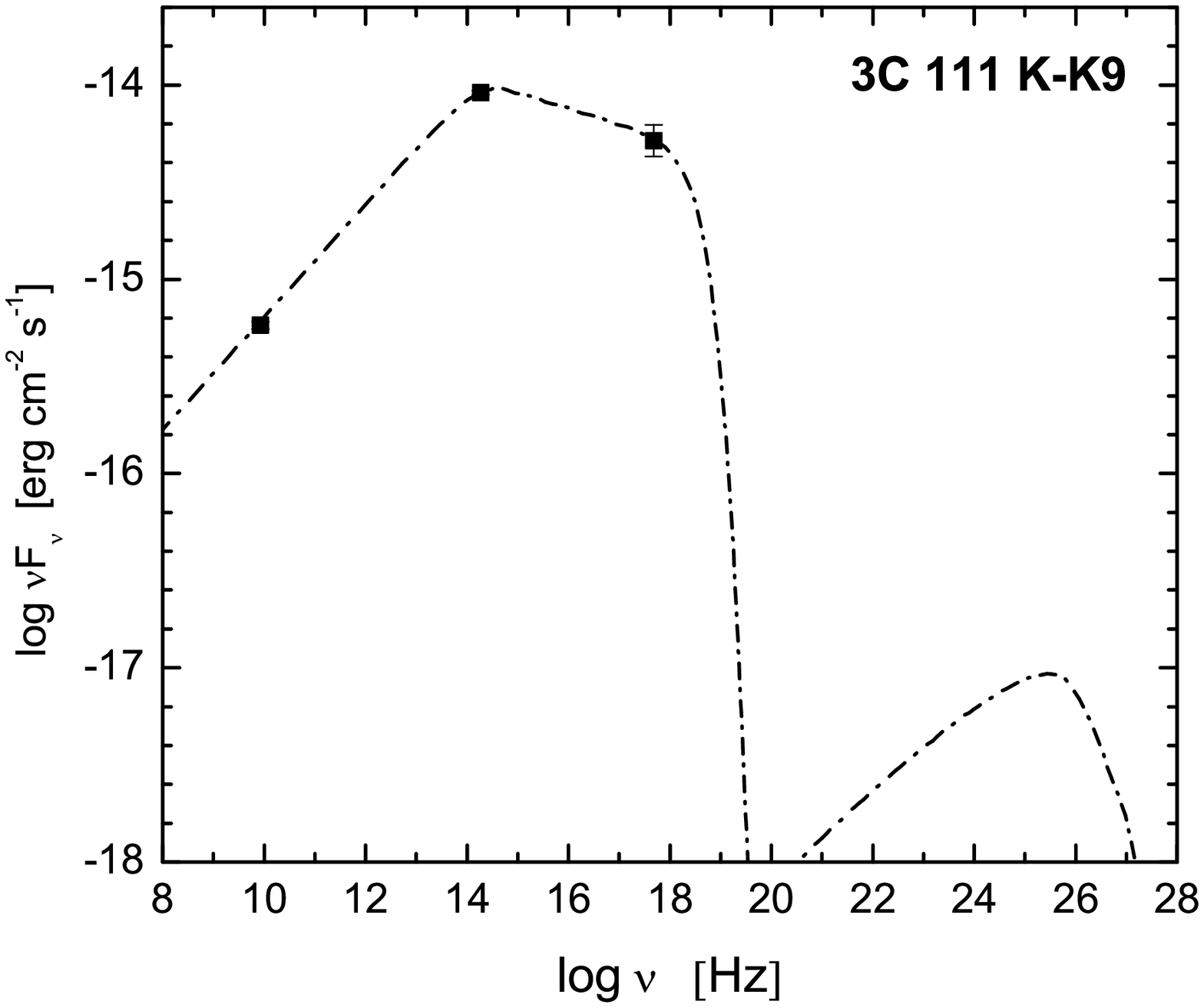}
\includegraphics[angle=0,scale=0.215]{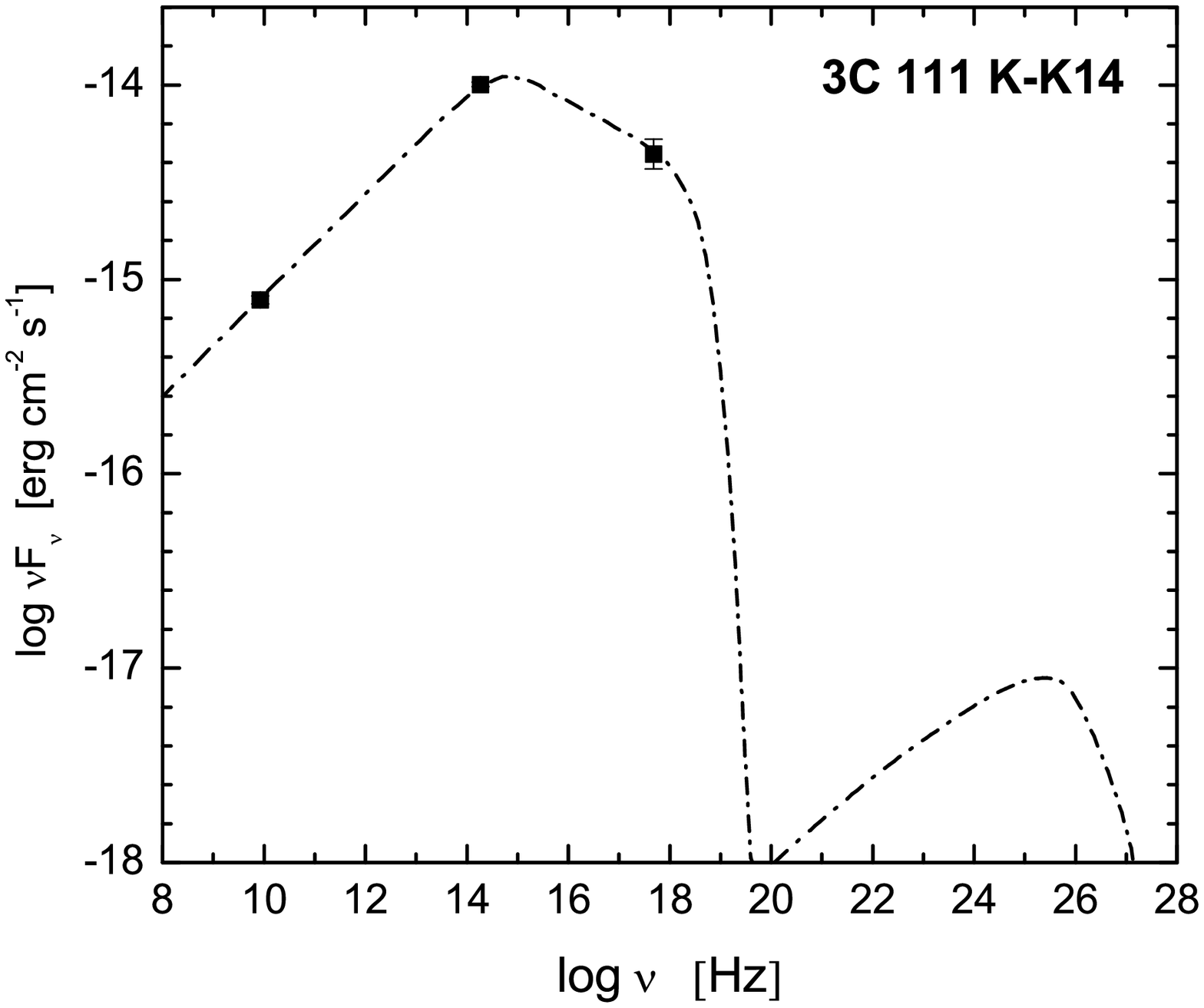}\\
\includegraphics[angle=0,scale=0.215]{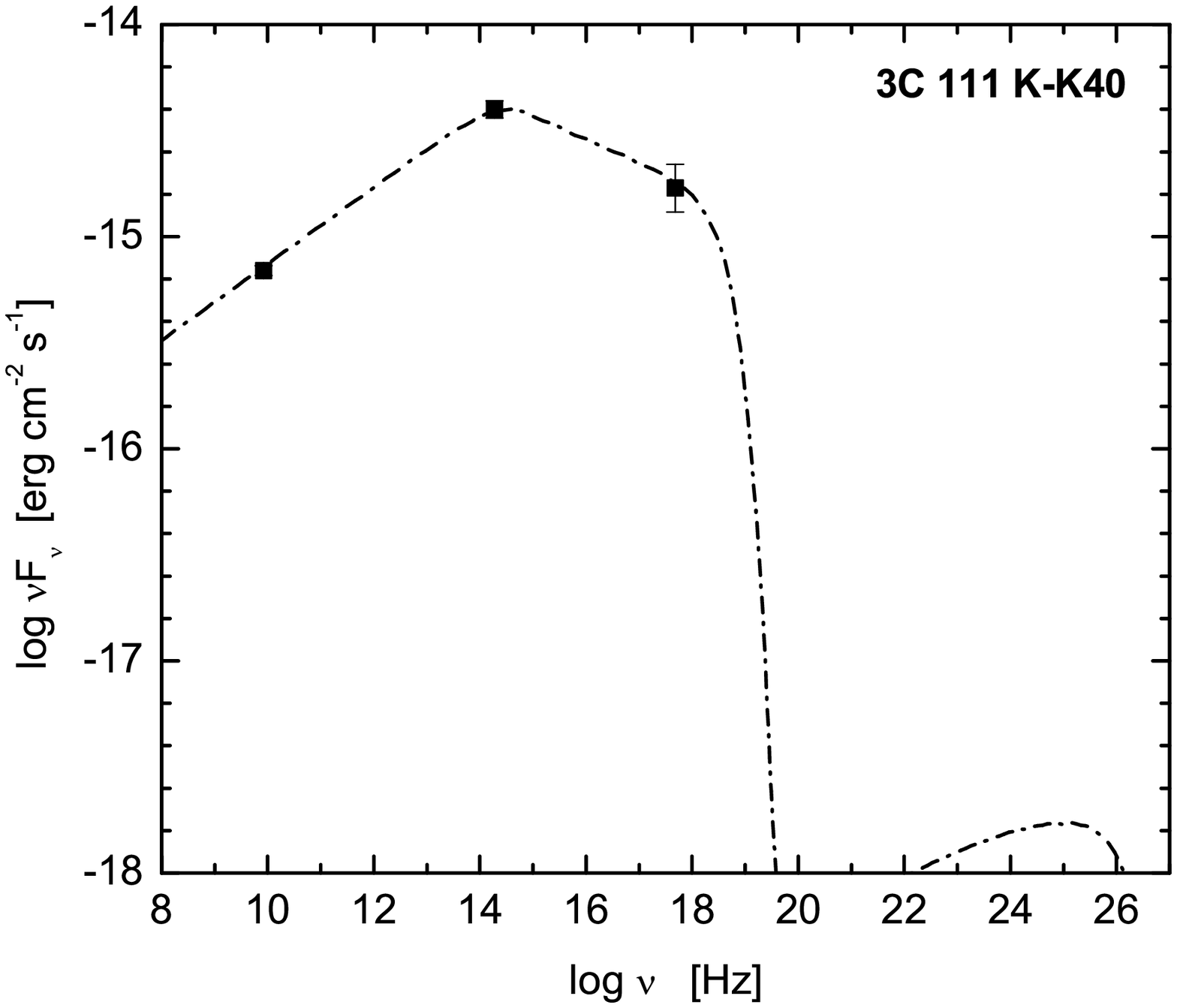}
\includegraphics[angle=0,scale=0.215]{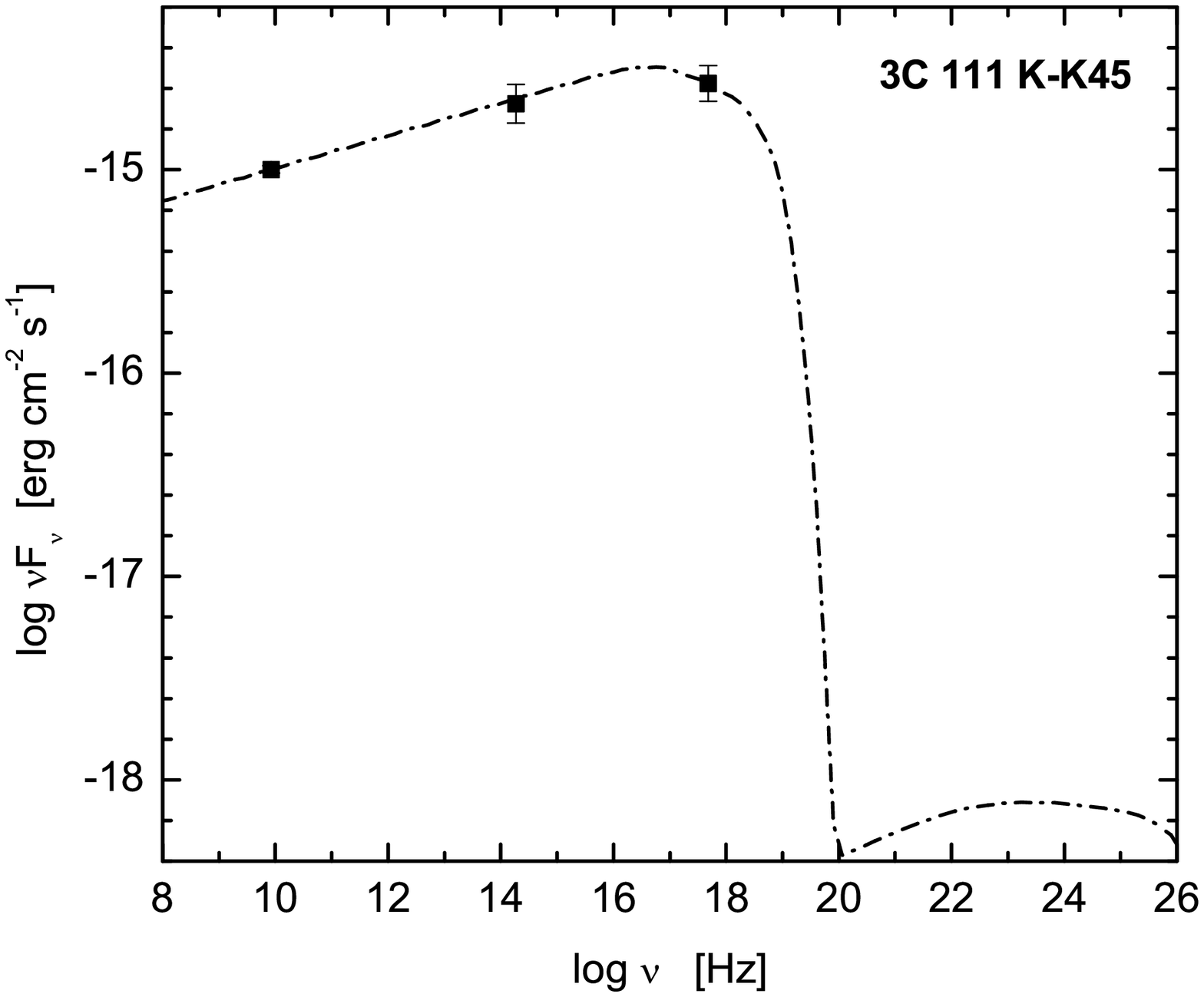}
\includegraphics[angle=0,scale=0.215]{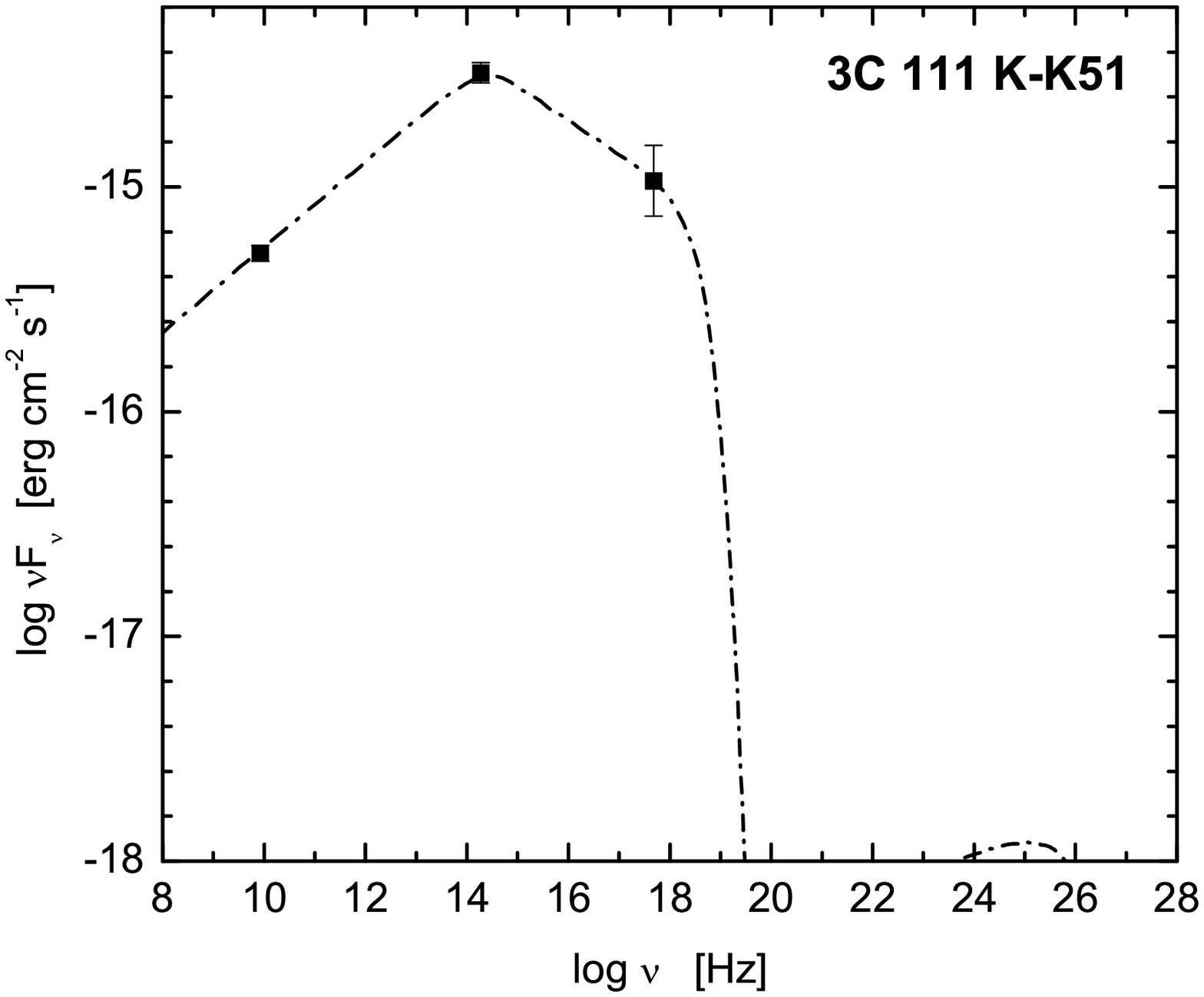}
\includegraphics[angle=0,scale=0.215]{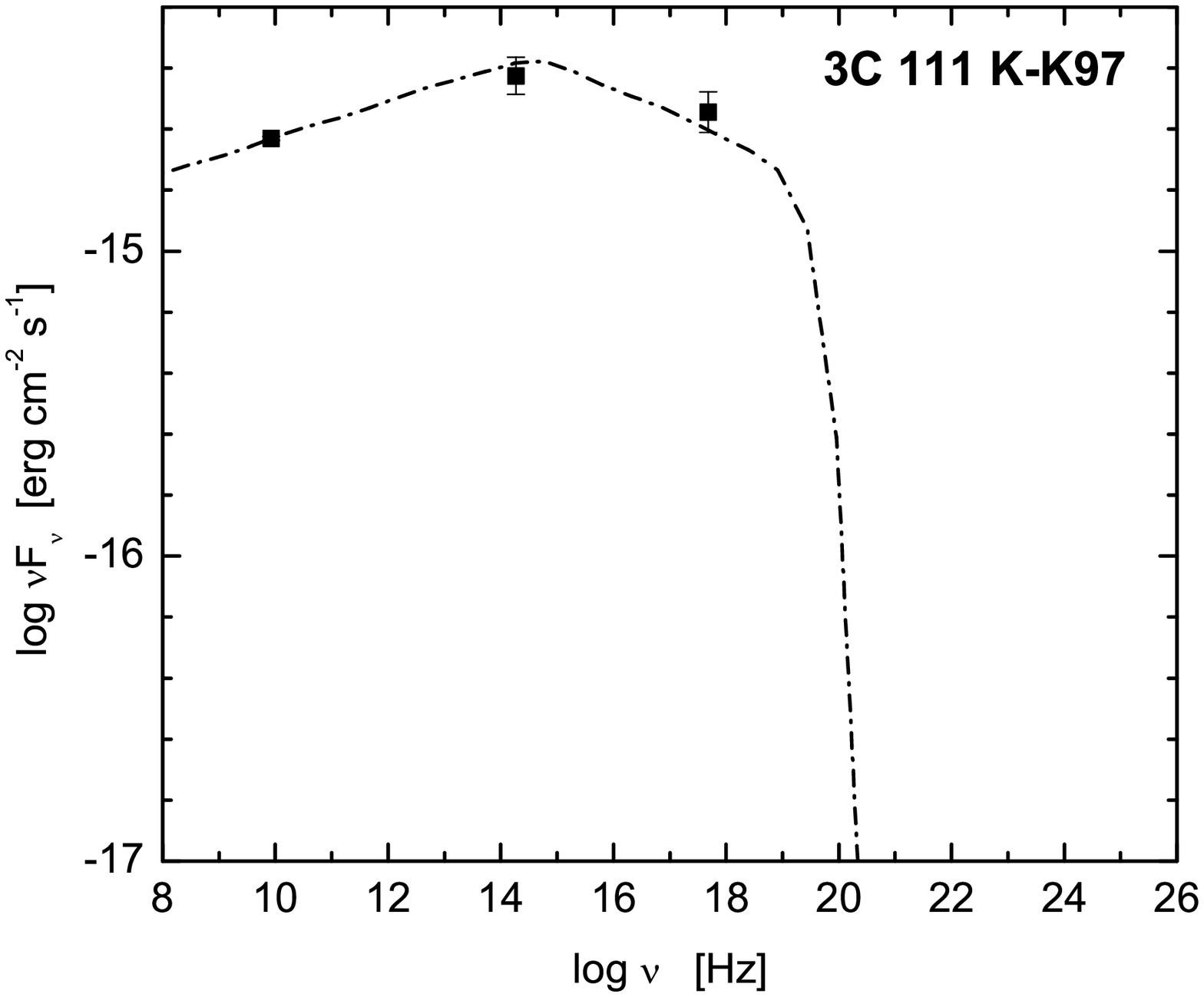}\\
\includegraphics[angle=0,scale=0.215]{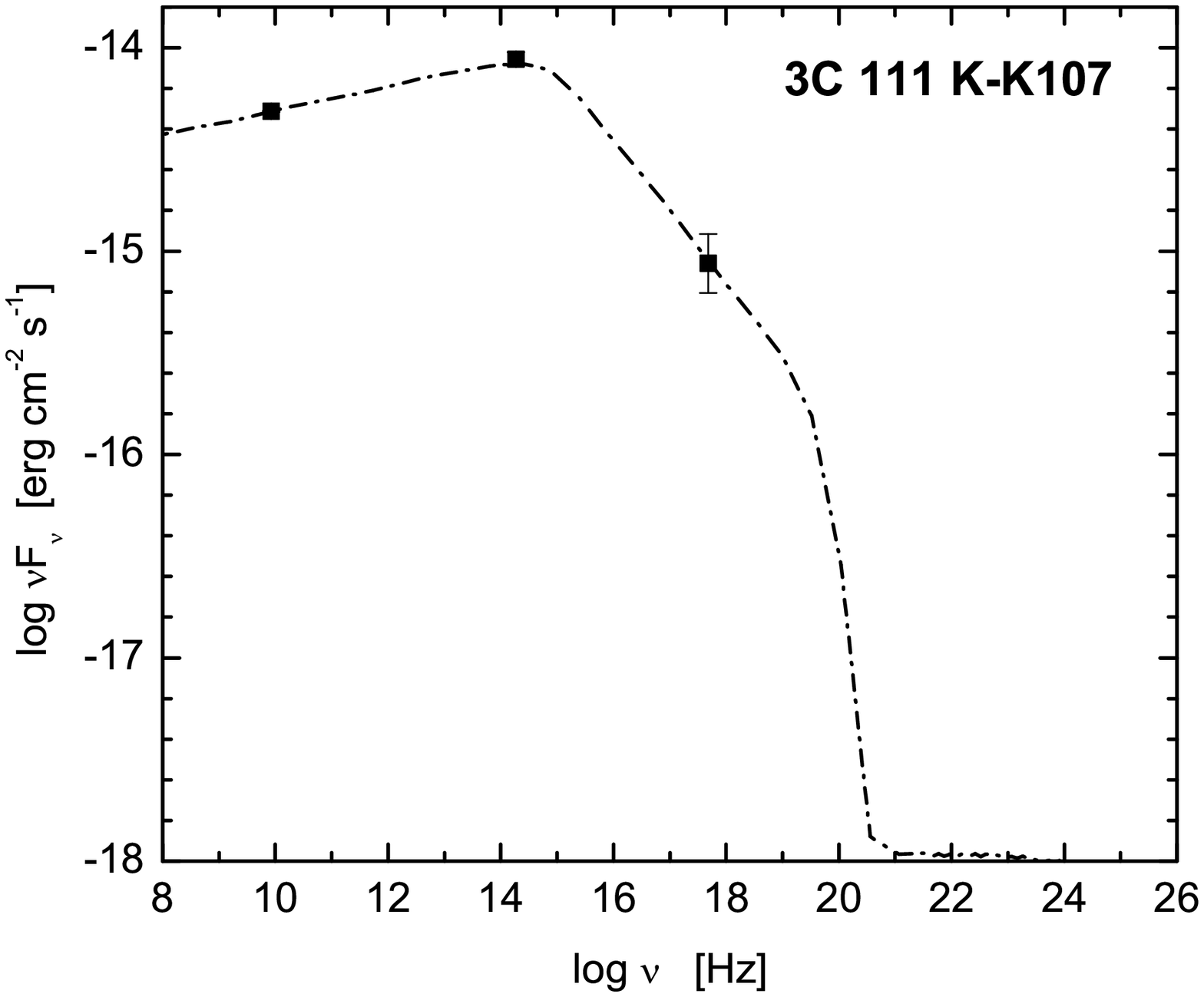}
\includegraphics[angle=0,scale=0.215]{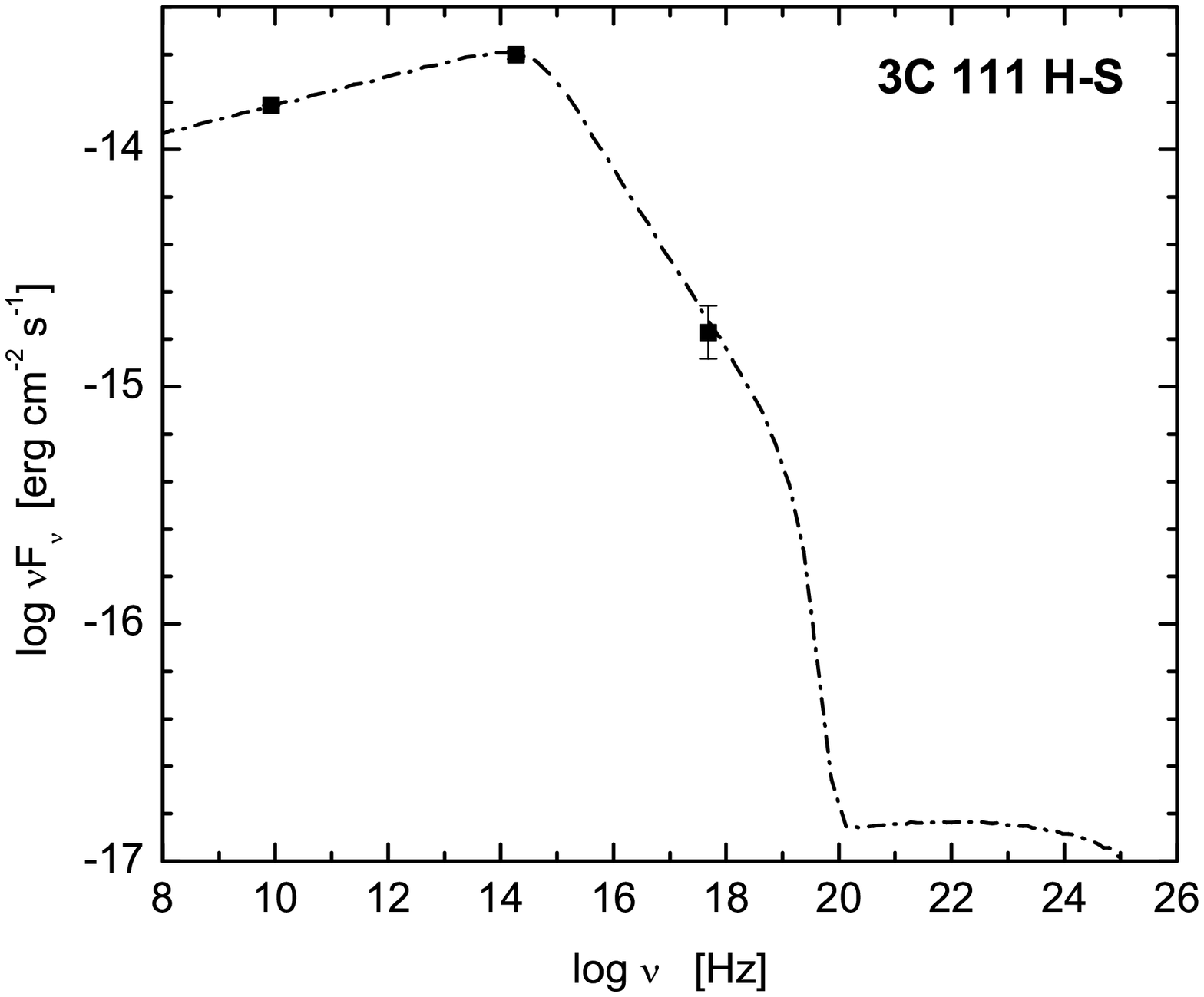}
\includegraphics[angle=0,scale=0.215]{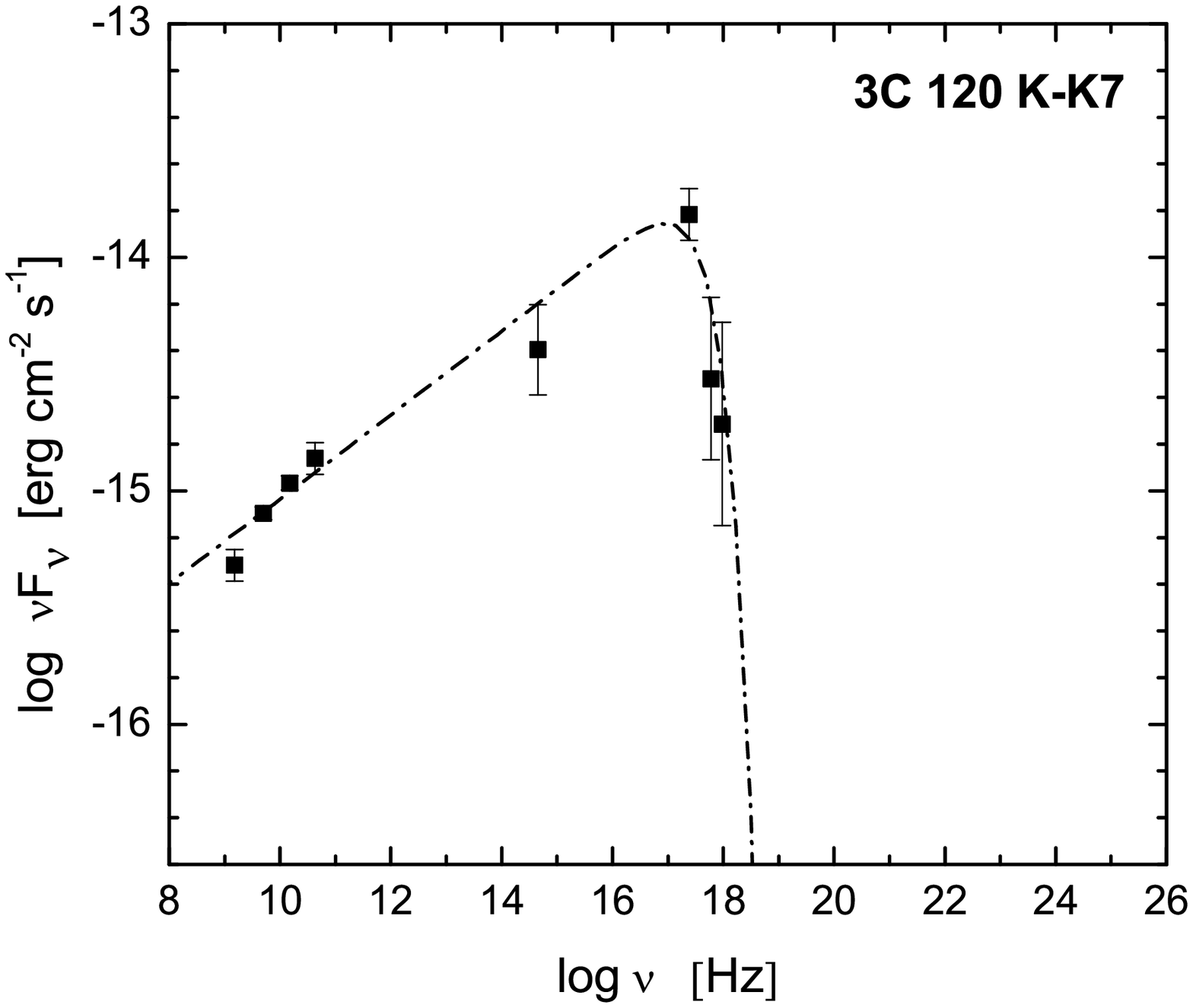}
\includegraphics[angle=0,scale=0.215]{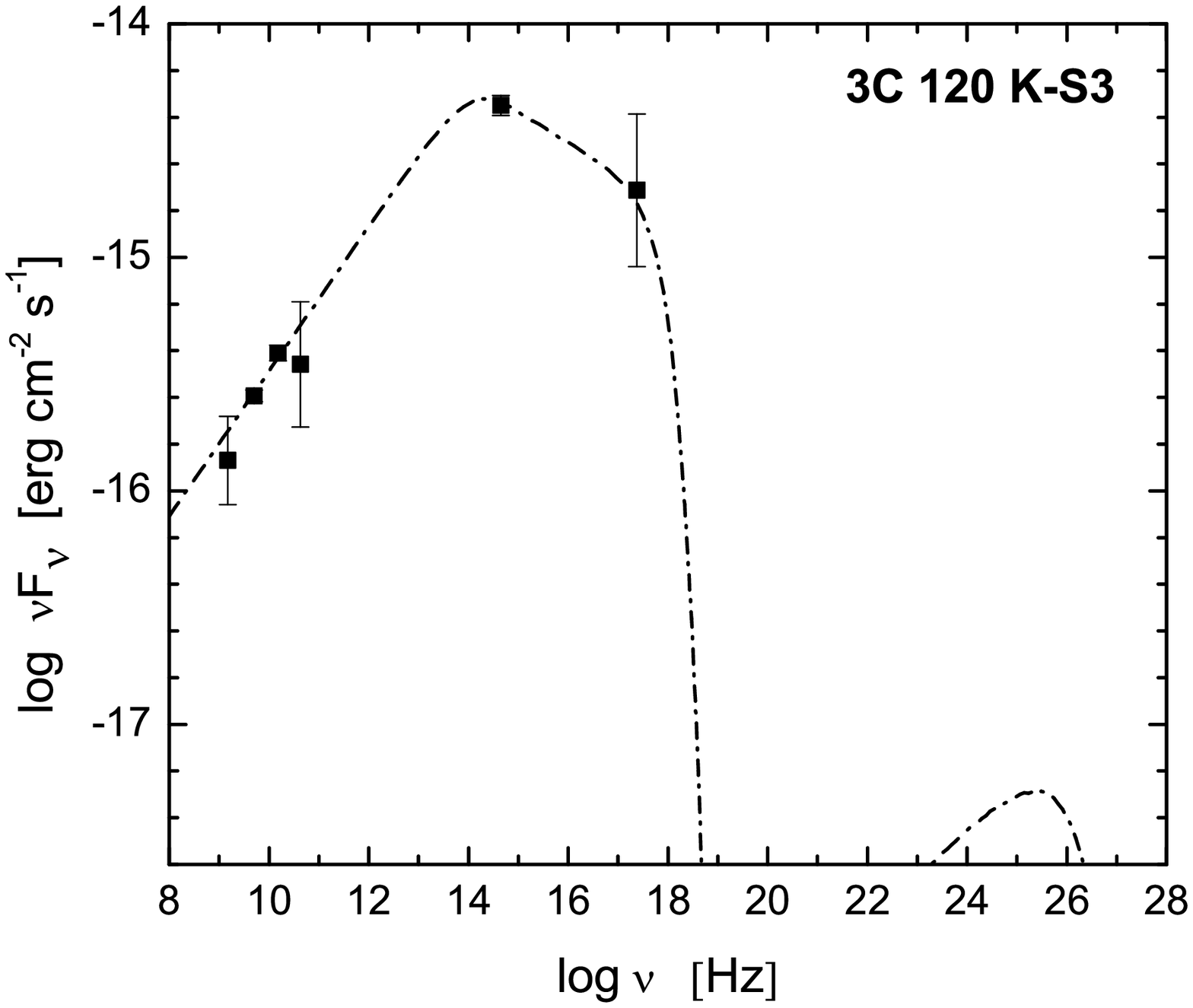}\\
\includegraphics[angle=0,scale=0.215]{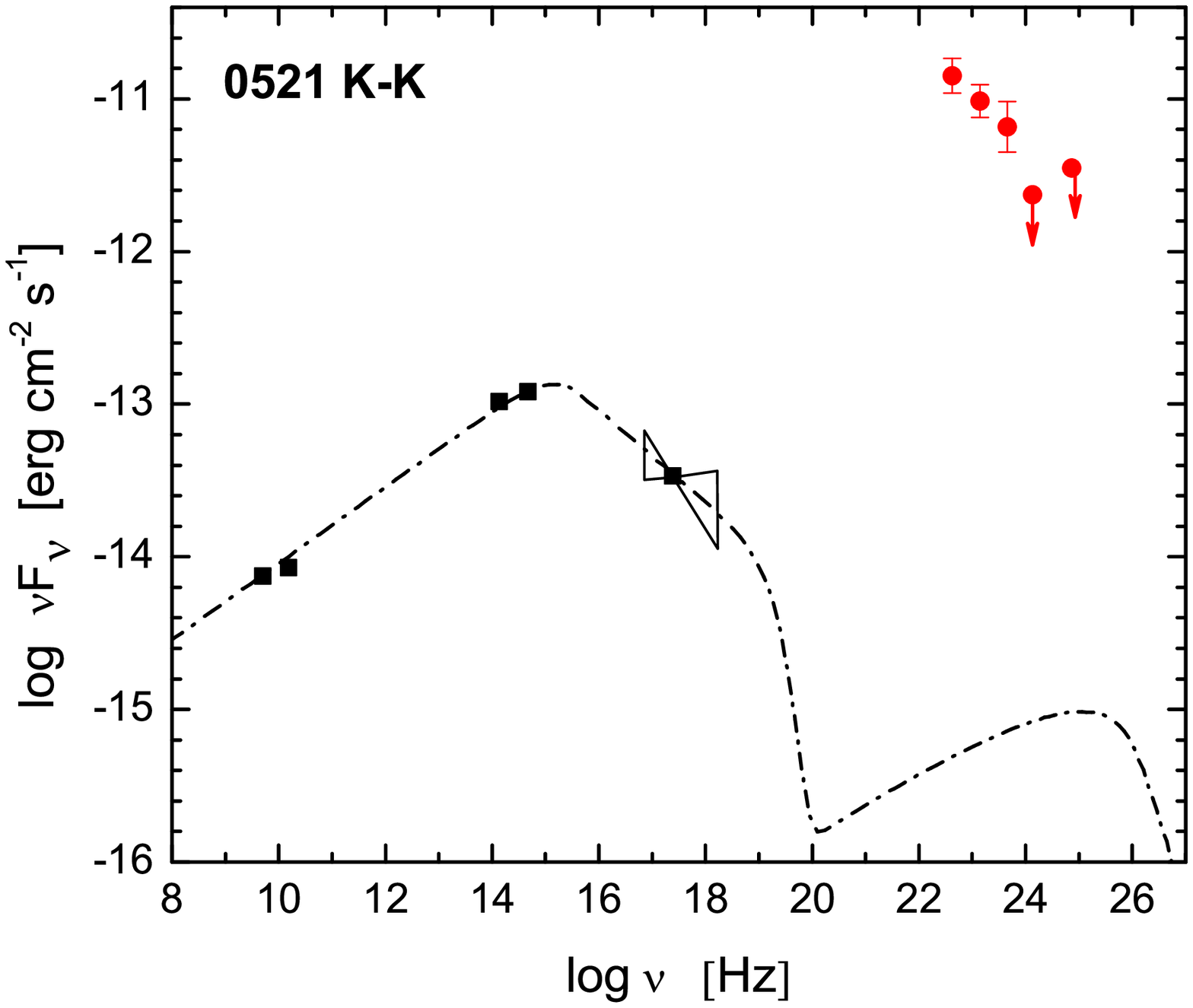}
\includegraphics[angle=0,scale=0.215]{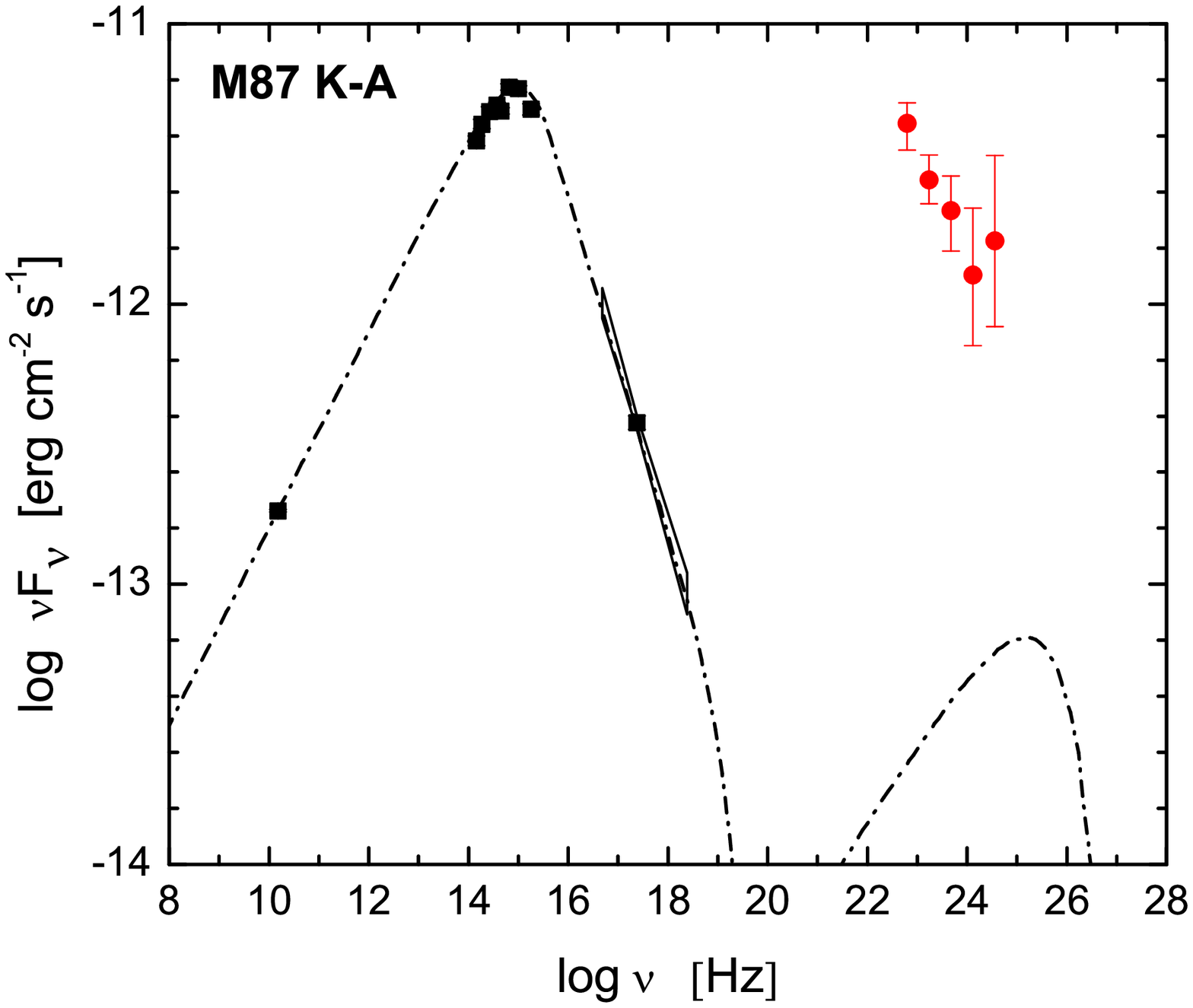}
\includegraphics[angle=0,scale=0.215]{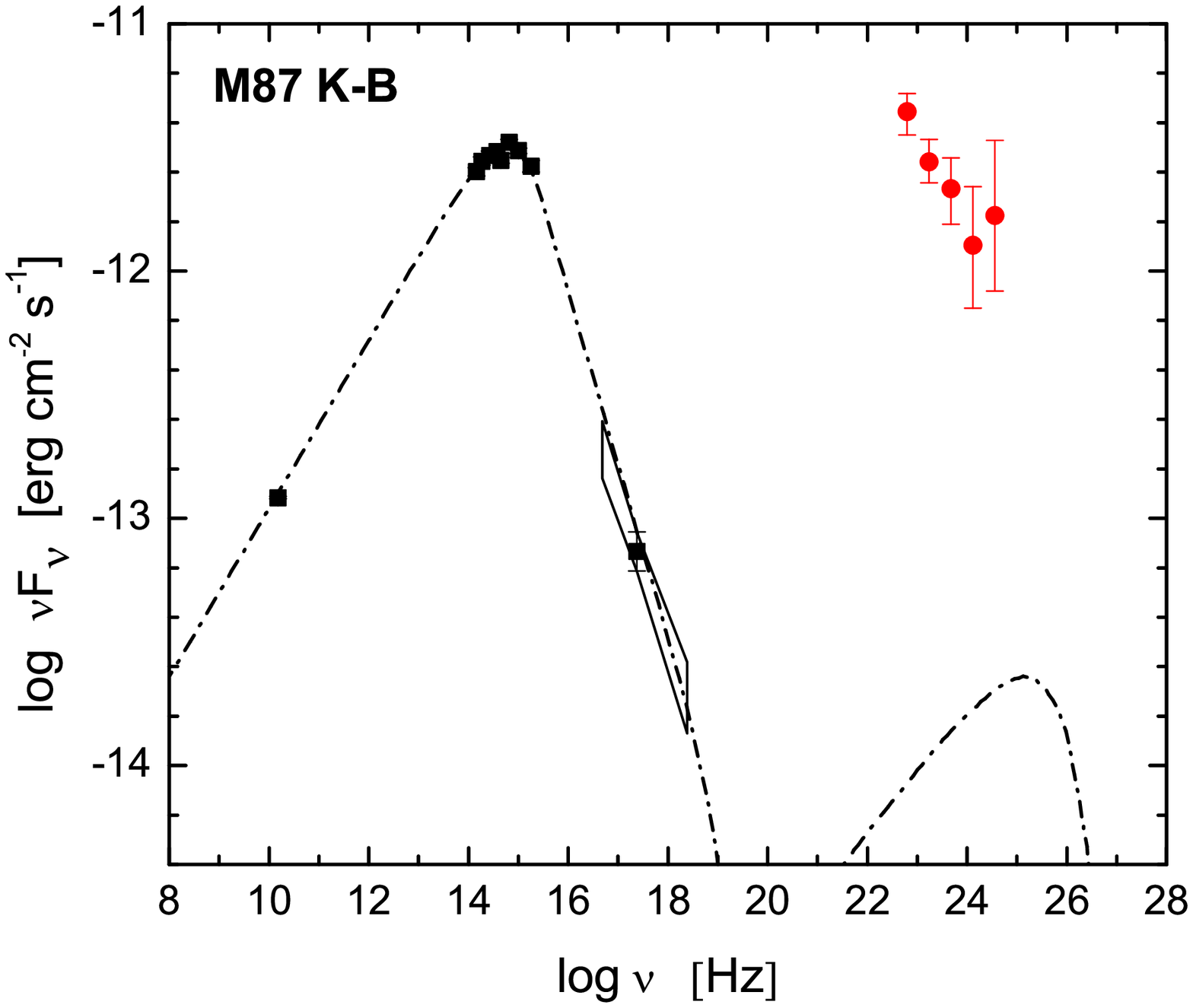}
\includegraphics[angle=0,scale=0.215]{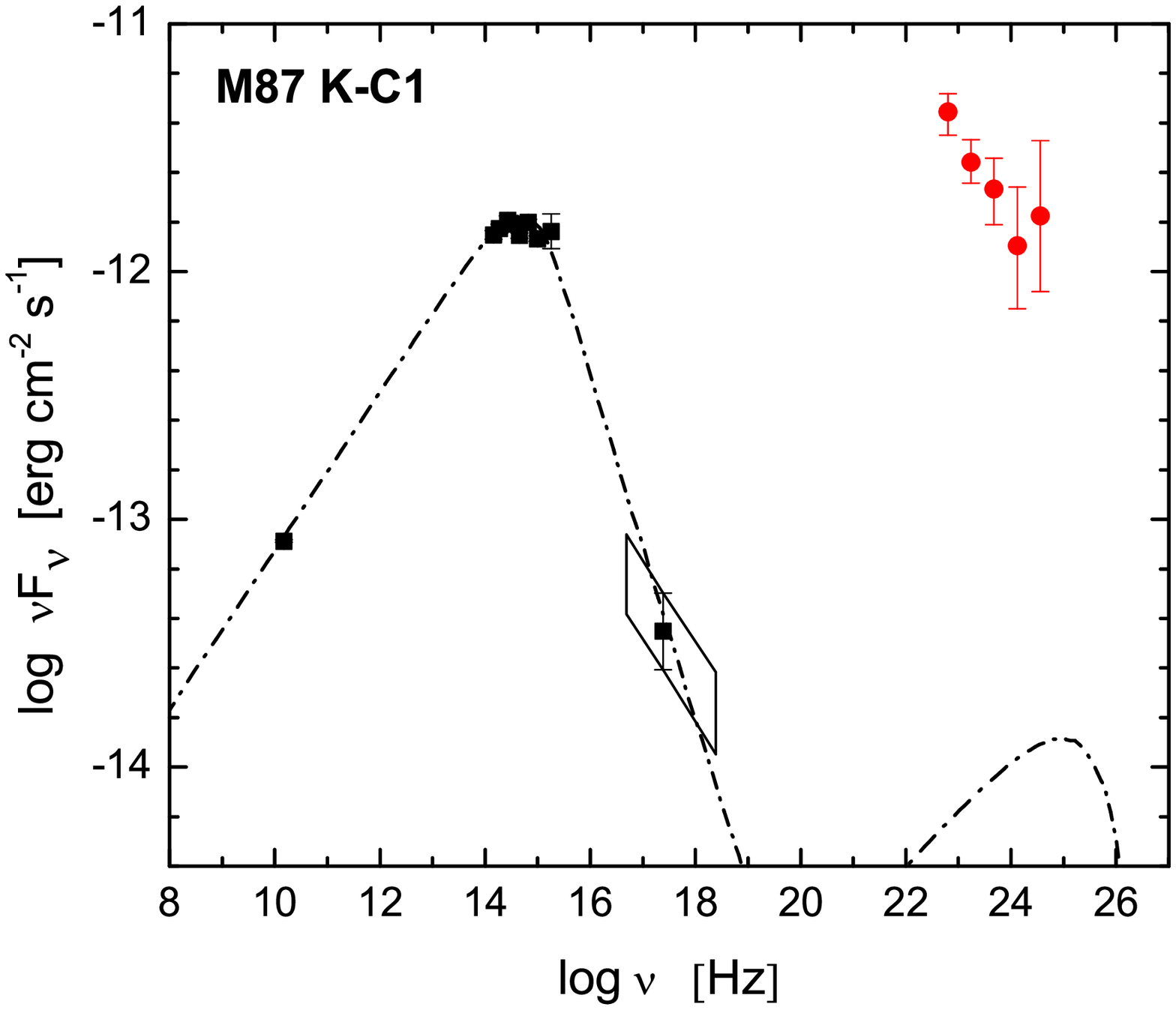}\\
\includegraphics[angle=0,scale=0.215]{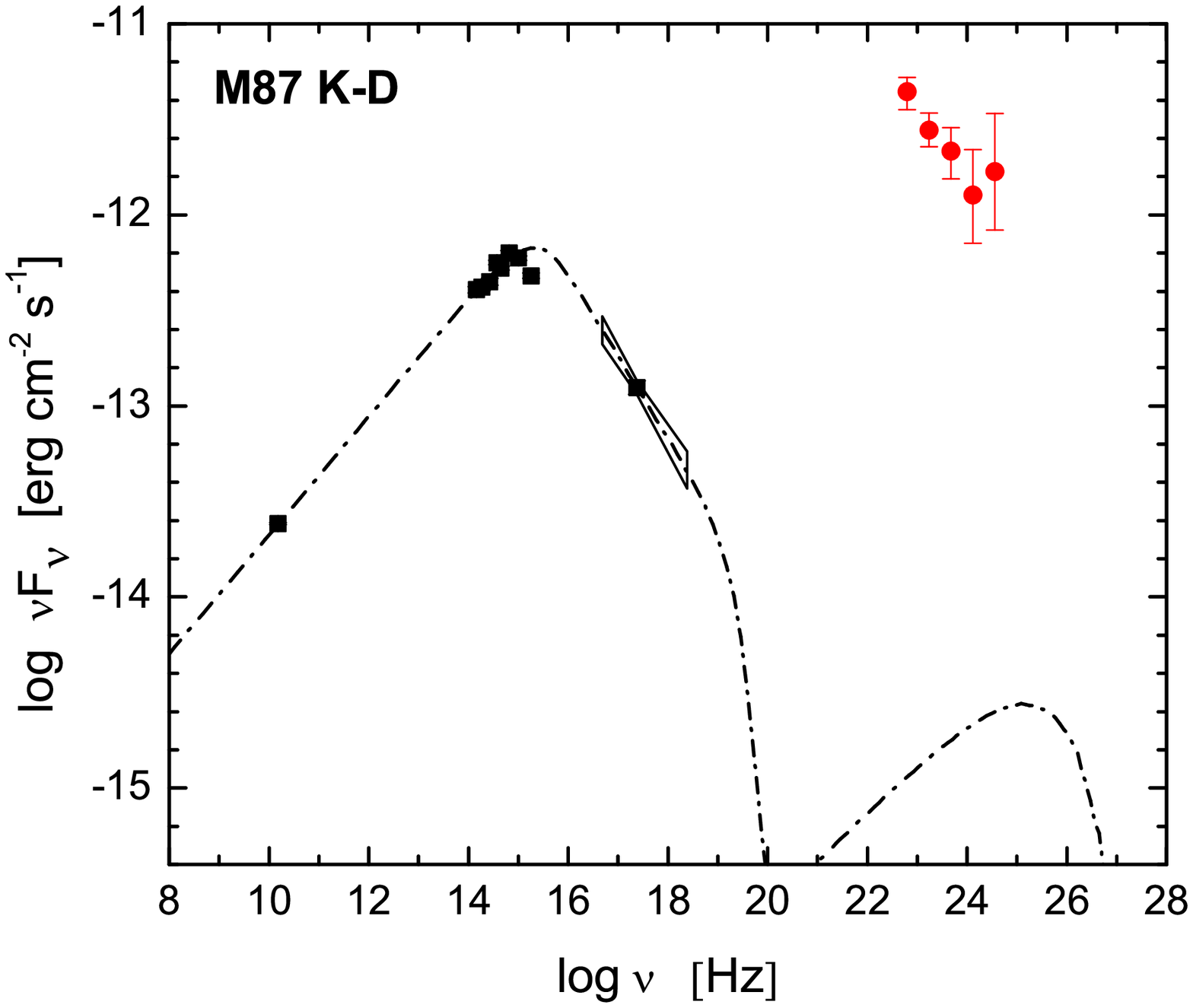}
\includegraphics[angle=0,scale=0.215]{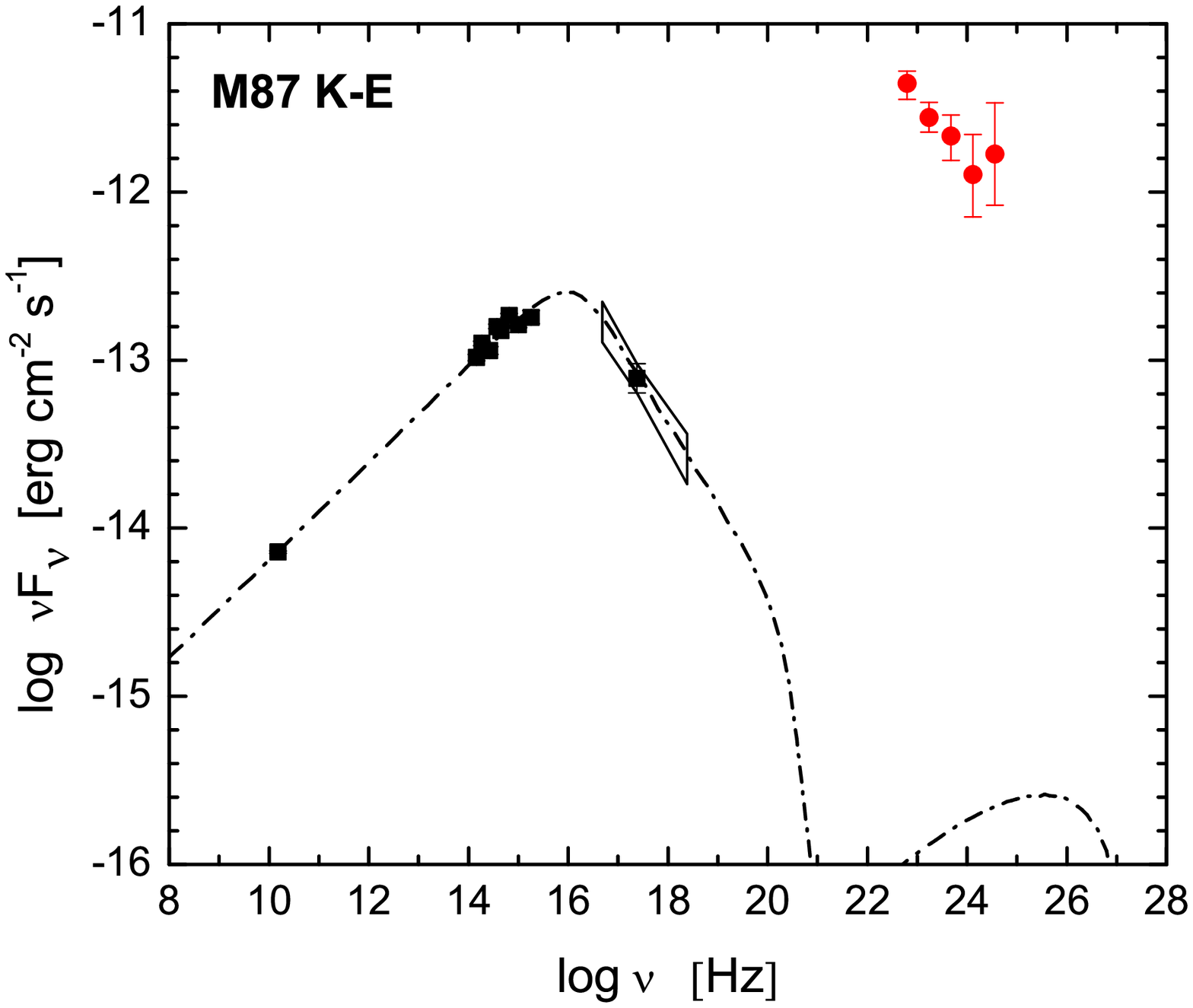}
\includegraphics[angle=0,scale=0.215]{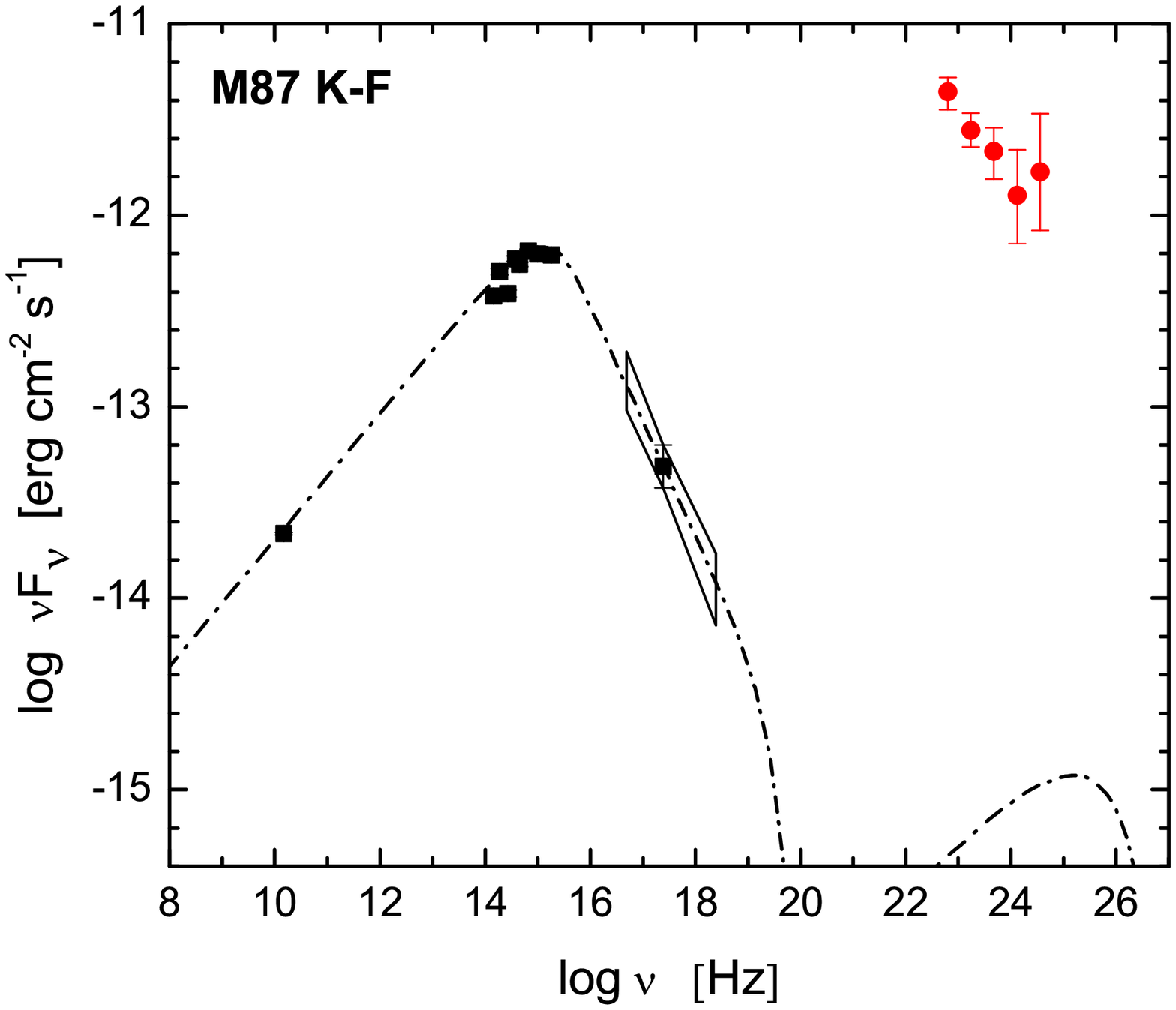}
\includegraphics[angle=0,scale=0.215]{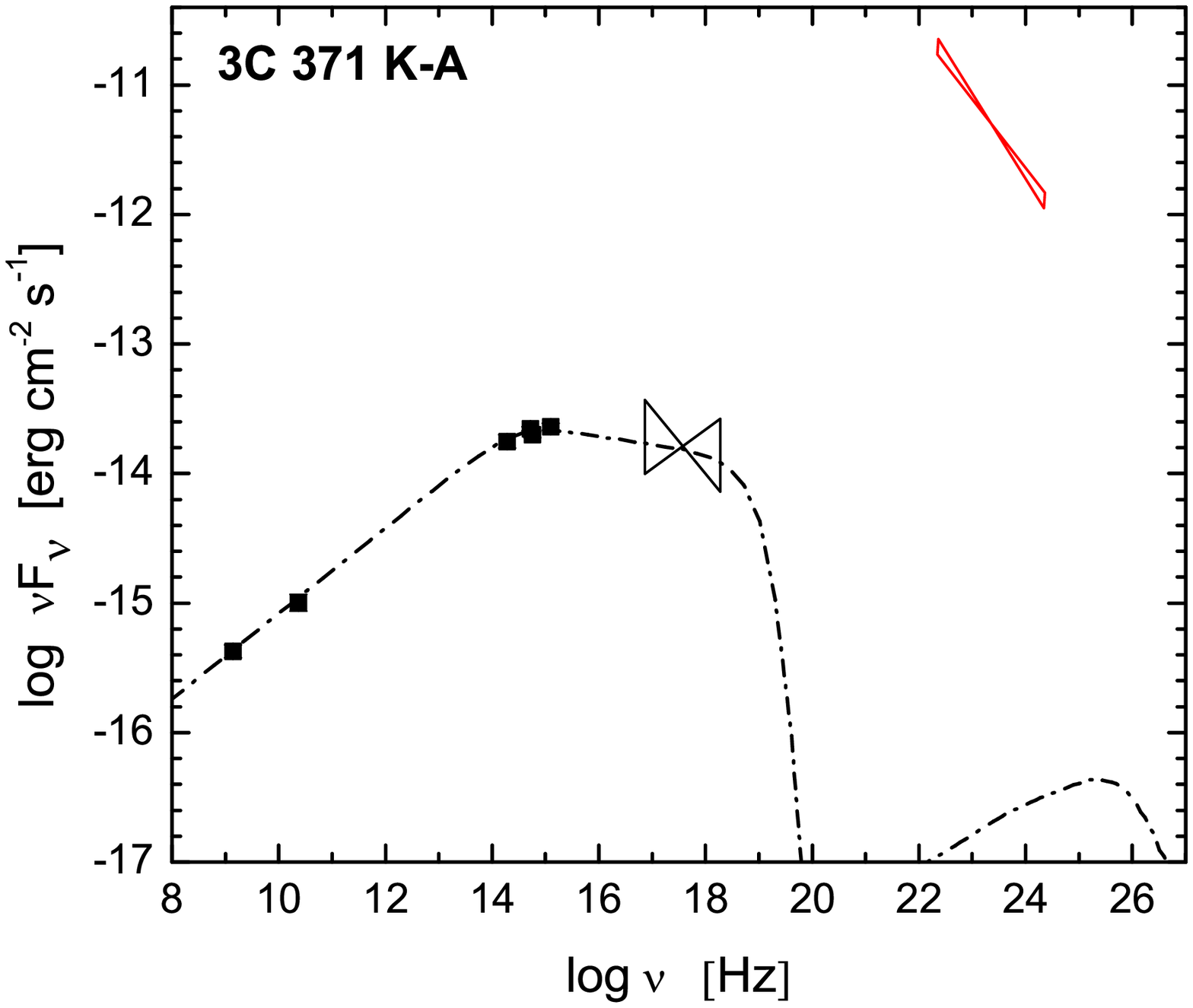}\\
\includegraphics[angle=0,scale=0.215]{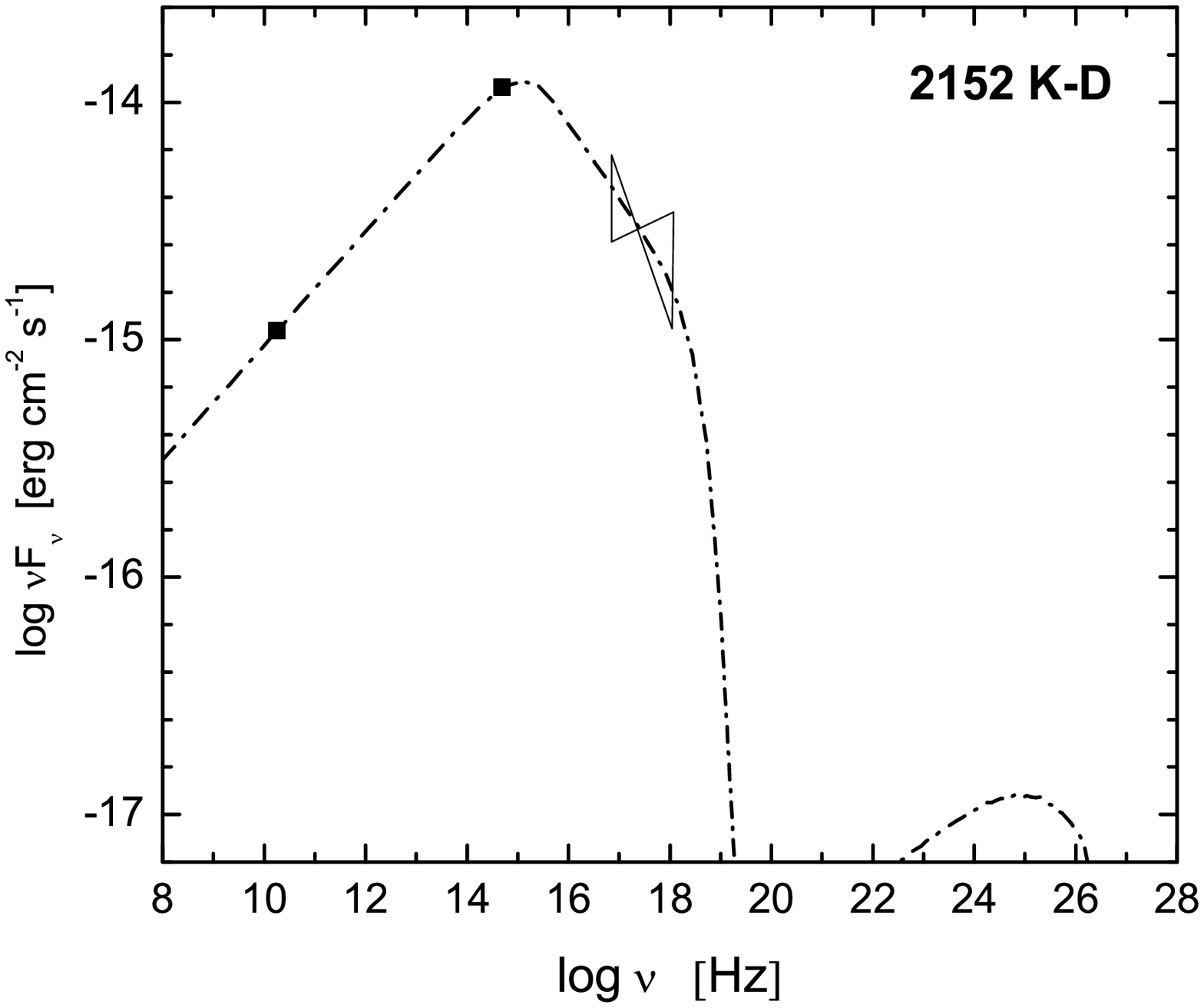}
\includegraphics[angle=0,scale=0.215]{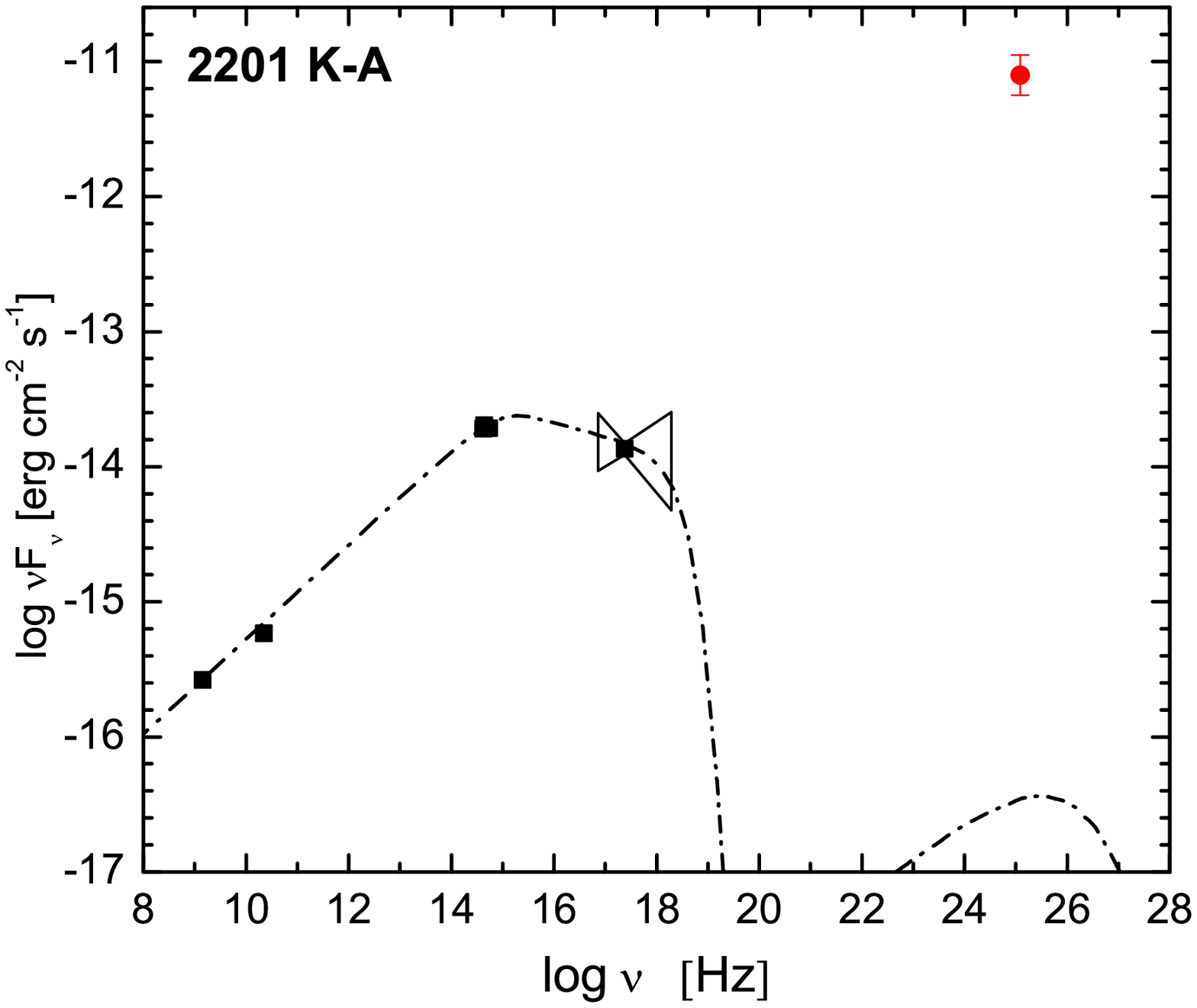}
\includegraphics[angle=0,scale=0.215]{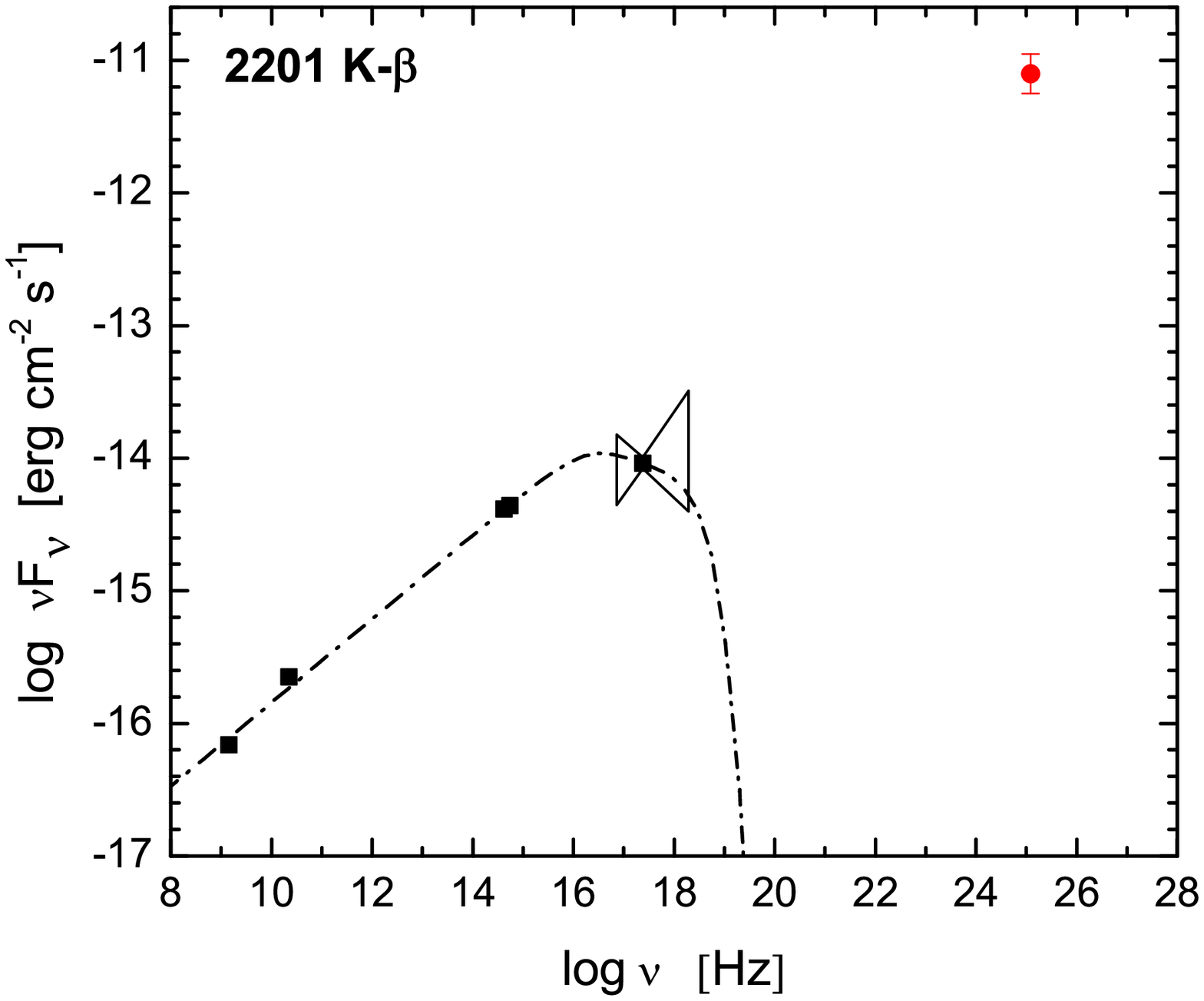}
\hfill\center{Fig.5---continued: the results of SED fitting with the synchrotron radiation (\emph{dash-dotted lines}) under the equipartition condition by assuming $\delta=1$.   }
\end{figure*}

\begin{figure*}
\includegraphics[angle=0,scale=0.215]{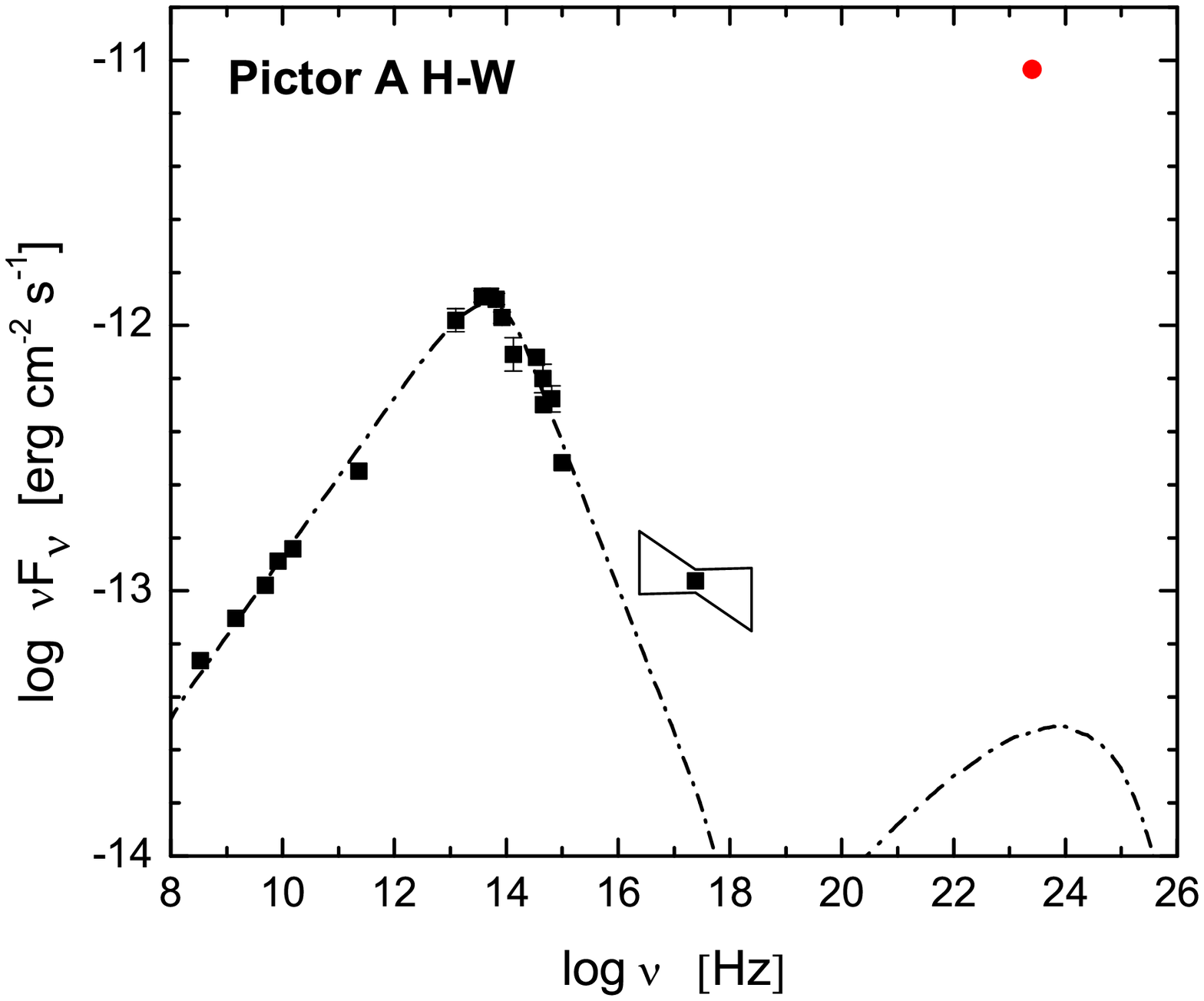}
\includegraphics[angle=0,scale=0.215]{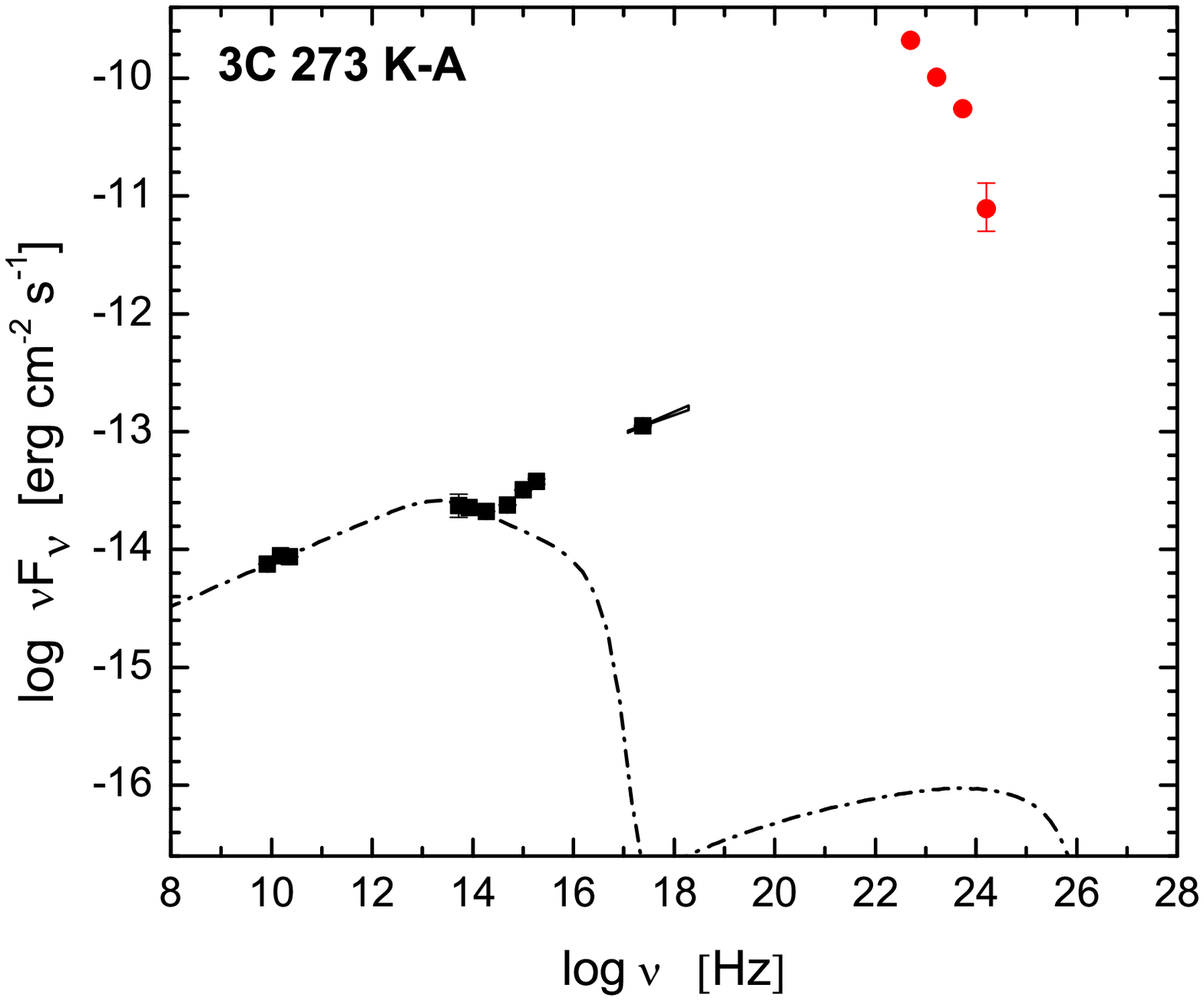}
\includegraphics[angle=0,scale=0.215]{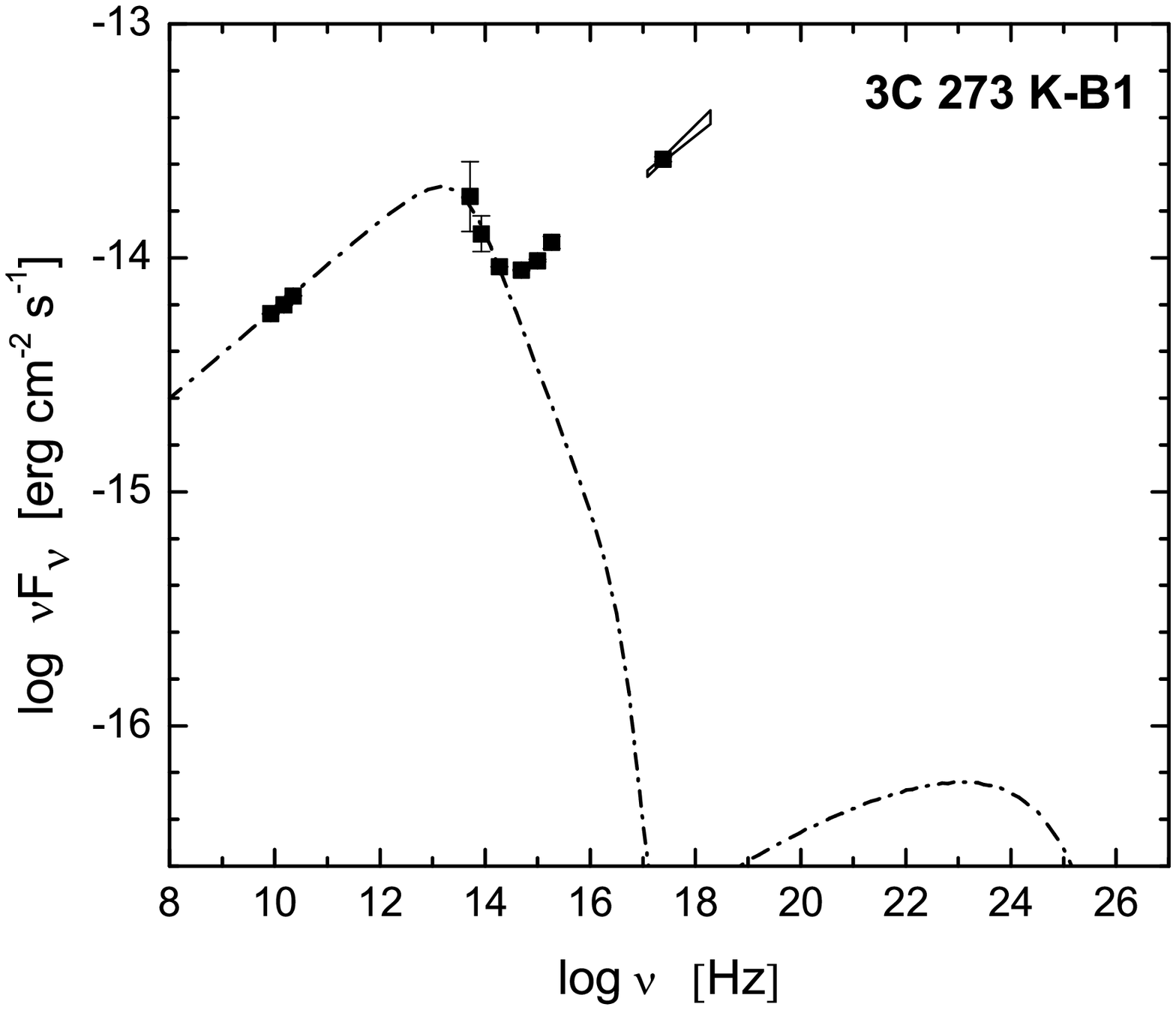}
\includegraphics[angle=0,scale=0.215]{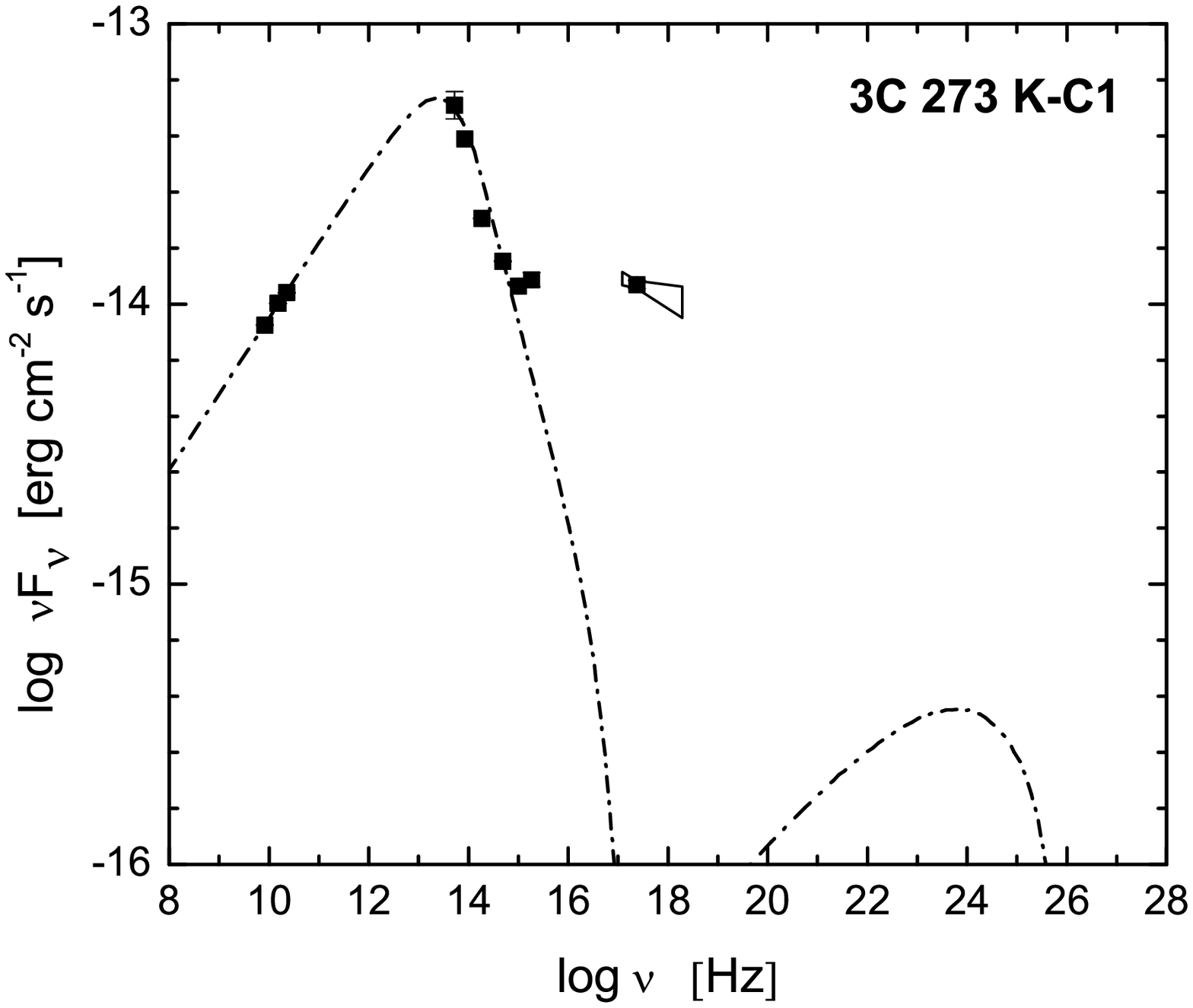}\\
\includegraphics[angle=0,scale=0.215]{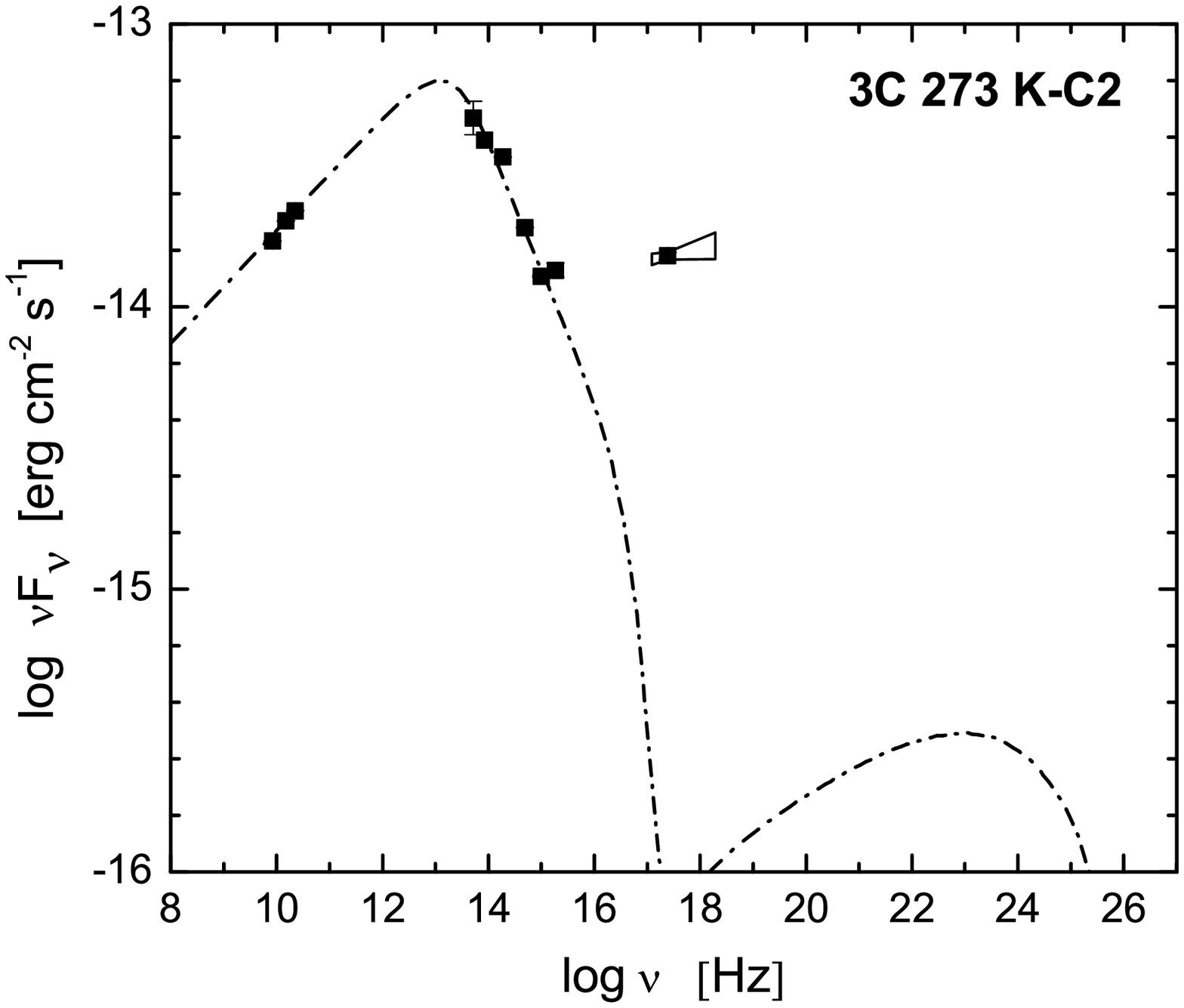}
\includegraphics[angle=0,scale=0.215]{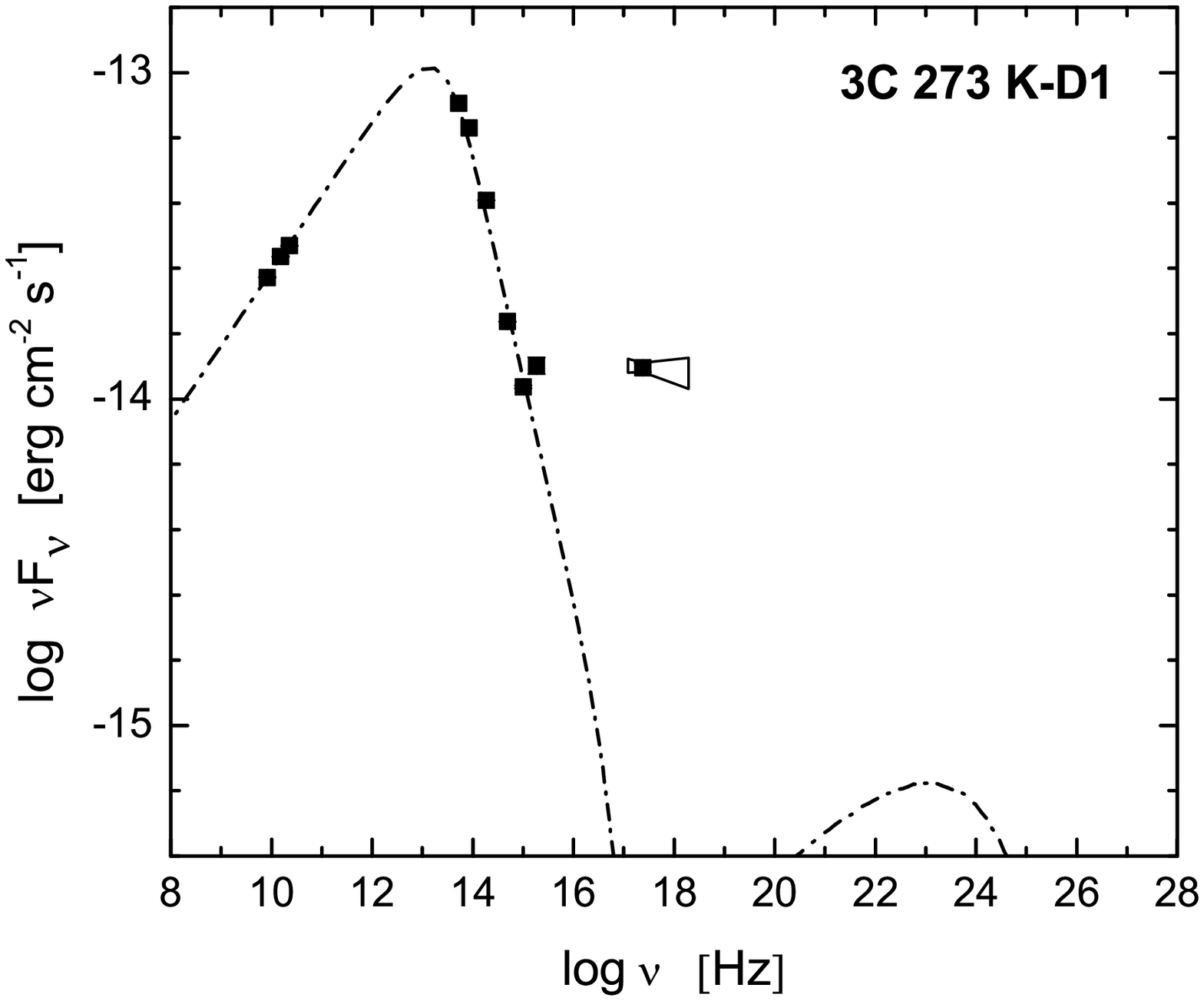}
\includegraphics[angle=0,scale=0.215]{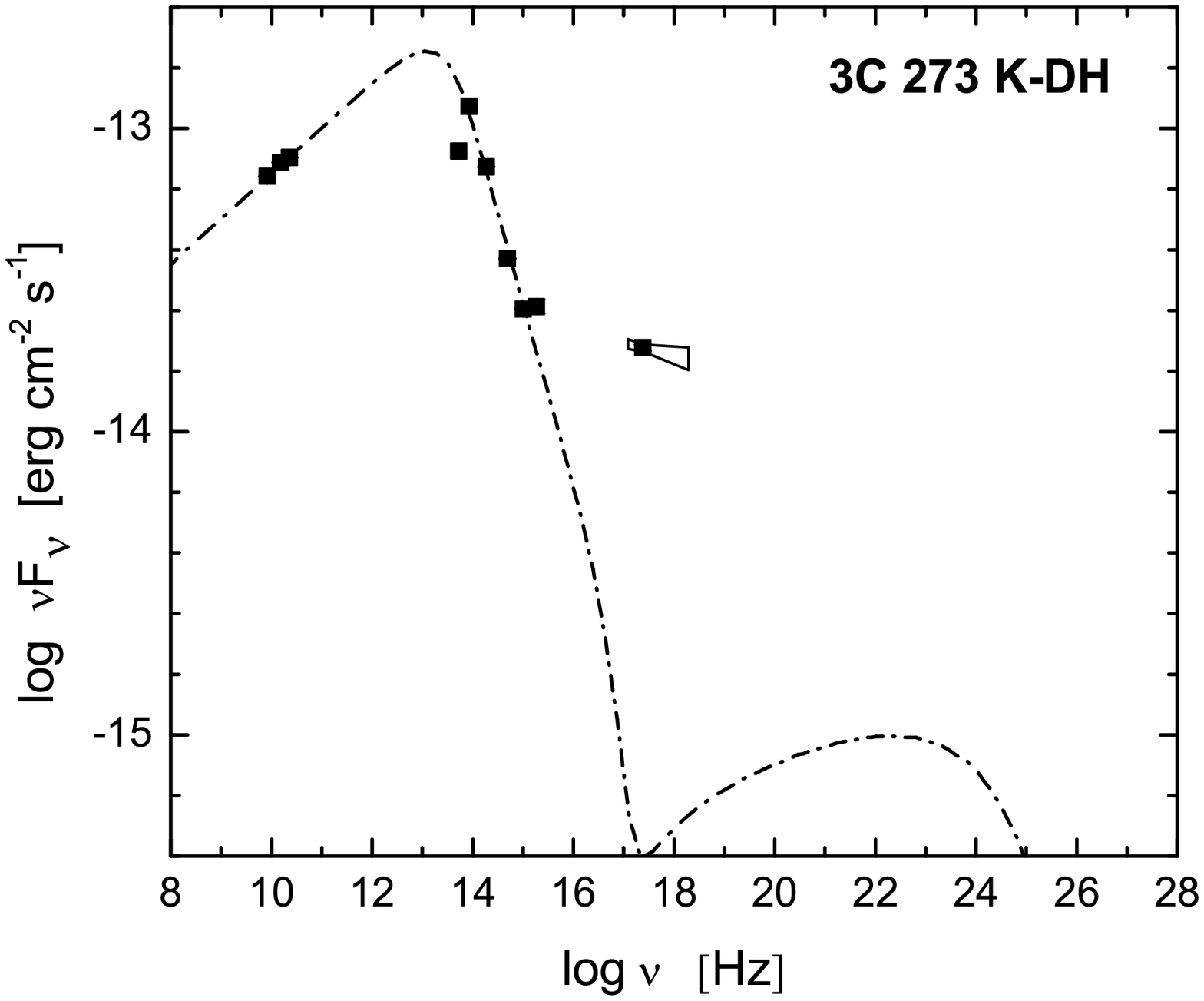}
\hfill\center{Fig.5---continued: the results of SED fitting with the synchrotron radiation (\emph{dash-dotted lines}) under the equipartition condition by assuming $\delta=1$ for the seven substructures. }
\end{figure*}

\clearpage

\begin{deluxetable}{lcccccccccccccccc}
\tabletypesize{\scriptsize} \rotate  \tablecolumns{15}\tablewidth{47pc} \tablecaption{Fitting Parameters and Derived Data for the Substructures with Leptonic Models }\tablenum{2} \tablehead{\colhead{Source}&\colhead{Comp}&\colhead{\tiny{$\gamma_{\rm min}$}}&\colhead{\tiny{$\log\gamma_{\rm b}$}}&\colhead{\tiny{$\log\gamma_{\rm max}$}}&\colhead{\tiny{$\log N_{0}$}}&\colhead{$p_{1}$}
&\colhead{$p_{2}$}& \colhead{$B$}&\colhead{$\delta$}&\colhead{$\log R$}&\colhead{\tiny{$\log P_{\rm e}$}}& \colhead{\tiny{$\log P_{\rm p}$}}& \colhead{\tiny{$\log P_{\rm B}$}} &\colhead{\tiny{$\log P_{\rm jet}$}}& \colhead{\tiny{$\log L_{\rm k}$}}\\
\colhead{}&\colhead{}&\colhead{}&\colhead{}&\colhead{}&\colhead{\tiny{[cm$^{-3}$]}}&\colhead{}
&\colhead{}&\colhead{\tiny{[$\mu$G]}}& \colhead{}&\colhead{[cm]}&\colhead{\tiny{[erg/s]}}& \colhead{\tiny{[erg/s]}}& \colhead{\tiny{[erg/s]}} & \colhead{\tiny{[erg/s]}}&\colhead{\tiny{[erg/s]}}}
\startdata
 \small{\textbf{IC/CMB}}&&&&&&&&&&&&&&&&\\
\hline
3C 15&K-C&35&5.2&6.6&-6.26&2.2&3.4&4.82&14&21.41&44.05&45.07&44.05&45.14&43.28\\
3C 17&K-S3.7&40&6&6.34&-3.78&2.74&3.4&17.26&8.5&21.54&44.99&46.28&44.98&46.32&44.27\\
&K-S11.3&140&5.2&6.2&-3.93&2.57&3.8&15.77&8&21.58&44.93&45.62&44.93&45.77&44.55\\
3C 31&K-R1&50&5.6&7&-7.15&2.17&3.6&1.87&9&21.72&43.47&44.3&43.47&44.41&41.98\\
&K-R2&100&5.74&6.74&-6.76&2.1&3.9&3.69&4.5&21.72&43.47&43.9&43.45&44.14&42.02\\
PKS 0208--512&K-K0&90&5.3&6.45&-5.24&2.48&3.8&5.3&8.8&22.3&45.51&46.34&45.51&46.45&44.95\\
&K-K1&80&5.6&6.4&-5.24&2.48&3.8&5.46&7&22.3&45.34&46.22&45.33&46.32&44.7\\
&K-K3&85&5&6.11&-5.05&2.47&3.8&6.88&5.7&22.6&45.96&46.81&45.96&46.92&45.51\\
3C 111&K-K22&160&5.3&7&-6.4&2.4&3.2&1.61&5.8&22.07&43.66&44.19&43.65&44.39&42.29\\
&K-K30&50&6.6&&-6.22&2.44&&2.23&7.2&22.07&44.13&45.19&44.12&45.26&42.75\\
&K-K61&50&5.48&6.6&-5.42&2.6&3.9&3.51&6.9&22.07&44.5&45.63&44.48&45.68&43.15\\
3C 120&K-K4&25&5.4&6.3&-5.84&2.37&3.8&4.86&20&21.15&43.85&45.16&43.84&45.2&42.89\\
&K-S2&110&6.41&7&-6.04&2.37&3.8&2.96&8.5&21.51&43.4&44.05&43.39&44.21&42.03\\
PKS 0637--752&K-K8.9&45&5&6.34&-5&2.47&3.8&8.44&7.5&22.33&45.85&46.96&45.83&47.02&45.34\\
&K-all&90&5.3&6.3&-4.76&2.57&3.8&6.93&4.2&22.93&46.37&47.24&46.36&47.34&45.86\\
3C 207&K-A&3&4.4&5.6&-6.81&1.97&4.4&5.9&13.3&22.04&45.47&47.22&45.45&47.24&45.23\\
PKS 1136--135&K-$\alpha$&4&4.78&&-5.76&2.52&&5.48&10&22.17&45.44&47.63&45.39&47.64&44.21\\
&K-A&2&4.78&&-6.04&2.46&&4.99&10&22.22&45.5&47.96&45.42&47.97&44.18\\
&K-B&3&4.6&4.6&-5.18&2.64&3.8&9.55&11&22.07&45.83&48.21&45.76&48.21&44.63\\
&K-C&100&5.65&6.08&-4.3&2.57&4.6&11.38&6.5&22.07&45.46&46.29&45.46&46.4&44.82\\
&K-D&140&5.7&6.3&-2.82&2.87&4.6&21.91&4.2&22.07&45.66&46.44&45.65&46.56&45.05\\
&K-E&130&4.81&6.7&-2.51&2.87&3.8&32.49&3.5&22.07&45.84&46.65&45.83&46.77&45.47\\
PKS 1150+497&K-B&3&4.6&5.6&-5.4&2.68&3.4&7.04&13.1&22.05&45.67&48.07&45.6&48.07&44.22\\
&K-C&70&5.38&7&-5.43&2.5&3.62&4.23&9&22.13&45&45.94&44.99&46.03&44.11\\
&K-D&3&4.78&5.3&-6.21&2.4&3.4&4.28&9.5&22.13&45.09&47.34&45.05&47.35&43.81\\
 &K-E&60&4.45&6.6&-5.74&2.36&3.3&4.65&9.4&22.02&44.9&45.85&44.9&45.94&44.1\\
 &K-IJ&85&5.26&6.28&-4.35&2.56&4.2&11.64&5&22.02&45.15&46.04&45.15&46.14&44.53\\
PKS 1229--021&K-A&120&5.3&6.03&-4.26&2.5&3.7&14.23&5.4&22.4&46.15&46.86&46.14&47&45.98\\
PKS 1354+195 &K-S4.0&85&4.6&6.1&-4.46&2.4&3.4&16.88&8.6&21.65&45.2&46.02&45.19&46.13&44.93\\
&K-S5.3&85&4.6&6.6&-4.48&2.4&3.4&16.39&7&21.65&45&45.81&44.99&45.93&44.59\\
3C 346&K-C&70&6.15&6.7&-4.26&2.56&5&13.62&5.8&21.88&45.14&46.12&45.13&46.2&44.48\\
3C 403&K-F1&45&5.66&6.08&-6.34&2.2&4.94&4.33&12.2&21.42&43.86&44.76&43.85&44.85&43.02\\
&K-F6&35&5.74&6.04&-6.03&2.28&5.6&4.85&13.5&21.42&44.05&45.13&44.04&45.2&43.17\\
PKS 2101--490&K-K6&200&5.28&6.6&-4.59&2.4&4&12.21&5.9&22.1&45.49&45.94&45.48&46.17&45.37\\
PKS 2209+080&K-E&70&5.34&6.48&-5.89&2.3&3.85&4.79&10&22.17&45.28&46.1&45.27&46.21&44.73\\
3C 454.3&K-A&80&5.48&6.18&-5.09&2.41&4.6&8.13&6.9&22.38&45.83&46.67&45.82&46.78&45.47\\
&K-B&5&5.13&5.85&-4.82&2.42&4.7&18.64&6.2&22.38&46.48&48.52&46.45&48.53&46.06\\
3C 33&H-S1&160&5&5.78&-3.4&2.4&3.8&50.36&3.5&21.25&44.56&45.1&44.56&45.3&44.19\\
&H-S2&150&5.11&5.85&-4.12&2.4&3.7&22.18&3&21.72&44.67&45.24&44.66&45.43&44.41\\
&H-N1&110&5.3&6&-4.84&2.4&3.8&10.44&4.5&21.64&44.21&44.91&44.2&45.05&43.51\\
&H-N2&100&5.15&6&-5.7&2.18&3.8&8.67&4.9&21.64&44.13&44.69&44.12&44.88&43.52\\
PKS 0405--123&H-N&200&4.78&6.08&-2.71&2.8&3.3&26.81&4&22.15&45.93&46.55&45.93&46.72&45.87\\
3C 105&H-S1&100&5.48&6.7&-6.64&2.07&5.8&4.63&11.2&21.66&44.34&44.77&44.32&45.01&43.41\\
&H-S3&200&5.43&5.83&-3.89&2.47&4.5&21.31&3.8&21.75&44.89&45.37&44.88&45.59&44.42\\
3C 111&H-N&180&5.9&6.3&-3.98&2.6&3.8&12.59&3.3&22.07&44.95&45.54&44.95&45.72&44.18\\
PKS 0836+299&H-B&68&6&6.9&-5.03&2.6&3.9&5.04&10.5&21.52&44.07&45.08&44.07&45.16&43.06\\
3C 228&H-S&103&5.18&6.15&-3.33&2.6&4.2&31.24&6.5&21.73&45.65&46.48&45.64&46.59&45.47\\
3C 245&H-D&200&4.7&5.9&-2.85&2.6&4&44.09&2.5&22.3&46.27&46.81&46.26&47.01&46.4\\
3C 263&H-K&140&5.32&5.7&-2.54&2.65&4.18&60.52&3.7&21.93&46.13&46.85&46.12&46.99&46.3\\
3C 275.1&H-N&110&5.4&6.95&-4.4&2.38&6&18.57&5.5&22.07&45.75&46.43&45.74&46.58&45.51\\
3C 295&H-NW&4&3.28&4.48&-4.14&2.2&3.8&63.07&13.5&21.25&45.96&47.97&45.92&47.97&46.32\\
PKS 1421--490&H-A&105&4.4&5.95&-3.25&2.4&4.4&63.61&8.8&21.71&46.5&47.24&46.5&47.38&46.88\\
3C 303&H-W&140&5.7&6.78&-3.89&2.66&3.4&12.43&6.5&21.89&45.17&45.88&45.16&46.02&44.67\\
3C 345&H-A&200&4.26&6.02&-4.46&2.1&3.2&41.95&2.1&22.09&45.65&45.95&45.64&46.25&45.4\\
3C 351&H-L&80&5.04&6.15&-4&2.5&3.9&21.08&5.3&22.1&45.88&46.77&45.87&46.87&45.62\\
&H-J&40&4.95&6&-4.33&2.4&4&22.99&14.3&21.4&45.42&46.55&45.41&46.61&45.21\\
3C 390.3&H-B&30&5.74&6&-5.58&2.4&4.2&5.83&11.6&21.52&44.29&45.54&44.28&45.58&43.26\\
Cygnus A&H-A&50&3.48&5.54&-4.89&2&4&34.96&9&21.52&45.61&46.52&45.61&46.61&45.41\\
&H-B&100&3.6&5.7&-4.21&2.24&3.9&33.23&6.5&21.52&45.29&46.03&45.28&46.16&44.7\\
&H-D&1&3.3&4.88&-5.4&1.9&3.2&31.5&11&21.52&45.74&47.92&45.69&47.93&45.62\\
PKS B2152--699&H-S&28&5&5.9&-4.26&2.5&3.3&20.45&11&20.84&43.97&45.31&43.96&45.35&43.04\\
&H-N&35&5.3&6.36&-4.6&2.5&3.5&13.07&9.6&21.02&43.81&45.06&43.8&45.1&42.7\\
3C 445&H-SW&30&5.23&6.48&-5.43&2.28&3.9&9.81&10&21.23&44.03&45.18&44.02&45.24&43.06\\
&H-SE&50&5.34&6.2&-4.91&2.41&3.6&10.97&6.3&21.53&44.32&45.36&44.31&45.43&43.41\\
\hline
 \small{\textbf{SSC}}&&&&&&&&&&&&&&&\\
\hline
3C 15&K-C&250&6.23&7.58&-0.77&2.2&3.4&0.66&1&21.41&47.09&47.24&40.03&47.48&\\
3C 17&K-S3.7&150&6.93&7.08&0.95&2.74&3.4&4.2&1&21.54&47.44&48.16&41.9&48.23&\\
&K-S113&500&5.81&6.85&0.02&2.57&3.8&7.9&1&21.58&46.77&46.9&42.52&47.14&\\
3C 31&K-R1&300&6.93&8.48&-1.02&2.17&3.6&0.04&1&21.72&47.59&47.6&38.13&47.9&\\
&K-R2&200&6.67&7.7&-2.46&2.1&3.9&0.2&1&21.72&46.46&46.56&39.62&46.82&\\
PKS 0208--512&K-K0&500&6.48&7.6&0.64&2.48&3.8&0.23&1&22.3&49.15&49.23&40.89&49.5&\\
&K-K1&200&6.78&7.48&0.38&2.48&3.8&0.21&1&22.3&49.08&49.56&40.81&49.68&\\
&K-K3&200&6&7.08&-0.24&2.47&3.8&0.49&1&22.6&49.09&49.57&42.15&49.69&\\
3C 111&K-K22&160&6.7&8.3&-0.32&2.4&3.2&0.02&1&22.07&48.22&48.75&38.31&48.86&\\
&K-K30&90&7.7&8.3&-0.05&2.44&3.6&0.04&1&22.07&48.48&49.28&38.92&49.34&\\
&K-K61&100&6.7&7.95&0.48&2.6&3.9&0.11&1&22.07&48.53&49.37&39.79&49.42&\\
3C 120&K-K4&100&6.6&7.3&0.62&2.37&3.8&0.49&1&21.15&47.5&48.19&39.25&48.27&\\
&K-S2&50&7.7&8.3&-0.1&2.37&3.8&0.09&1&21.51&47.61&48.61&38.54&48.65&\\
PKS 0637--752&K-K8.9&200&5.9&7.2&-0.22&2.47&3.8&1.1&1&22.33&48.57&49.04&42.31&49.17&\\
&K-all&200&6.18&7.18&-0.66&2.57&4&0.73&1&22.93&49.03&49.55&43.16&49.67&\\
3C 207&K-A&110&5.18&6.6&-2&1.97&4&1.5&1&22.04&47.99&48.27&42.01&48.45&\\
PKS 1136--135&K-$\alpha$&70&6.08&&0.62&2.52&&0.17&1&22.17&49.18&50.13&40.39&50.18&\\
&K-A&70&5.98&&0.34&2.46&&0.15&1&22.22&49.17&50.09&40.38&50.14&\\
&K-B&150&5.7&5.7&1.06&2.64&3.8&0.55&1&22.07&48.89&49.57&41.21&49.66&\\
&K-C&150&6.45&6.85&0.04&2.57&4.6&2.15&1&22.07&48.08&48.73&42.39&48.82&\\
&K-D&180&6.3&6.78&0.73&2.87&4.6&6.05&1&22.07&47.86&48.55&43.28&48.63&\\
&K-E&300&5.3&6.74&0.3&2.87&3.8&15&1&22.07&47.24&47.7&44.07&47.83&\\
PKS 1150+497&K-B&40&5.78&6.78&1&2.58&3.4&0.27&1&22.05&49.3&50.53&40.53&50.55&\\
&K-C&200&6.6&7.7&0.34&2.5&3.73&0.23&1&22.13&48.64&49.12&40.55&49.25&\\
&K-D&100&5.78&7.78&-0.1&2.4&3.8&0.14&1&22.13&48.64&49.37&40.15&49.44&\\
&K-E&200&5.85&7.3&-0.24&2.4&3.4&0.43&1&22.02&48.16&48.58&40.88&48.72&\\
&K-IJ&200&6&6.98&0&2.65&4&2.1&1&22.02&47.62&48.18&42.26&48.29&\\
PKS 1229--021&K-A&200&5.78&6.95&-0.4&2.5&3.7&3.3&1&22.4&48.44&48.92&43.41&49.05&\\
PKS 1354+195&K-S4.0&200&5.18&6.78&-0.34&2.4&3.4&6.9&1&21.65&47.3&47.74&42.55&47.88&\\
&K-S5.3&200&5.3&7.18&-0.47&2.4&3.4&4.9&1&21.65&47.18&47.62&42.25&47.75&\\
3C 346&K-C&80&6.72&7.4&-0.59&2.56&5&4.9&1&21.88&47.26&48.18&42.72&48.23&\\
3C 403&K-F1&100&6.65&7.08&-0.93&2.2&4.88&0.51&1&21.42&47.04&47.57&39.82&47.69&\\
&K-F6&30&6.78&7.08&-0.43&2.28&5.7&0.59&1&21.42&47.42&48.56&39.95&48.59&\\
PKS 2101--490&K-K6&200&5.95&7.3&-0.7&2.4&4&3&1&22.1&47.86&48.28&42.72&48.42&\\
PKS 2209+080&K-E&250&6.32&7.48&-0.6&2.3&3.85&0.5&1&22.17&48.41&48.66&41.31&48.85&\\
3C 454.3&K-A&150&6.32&7.04&-0.55&2.4&4.6&1.1&1&22.38&48.61&49.16&42.41&49.27&\\
&K-B&30&5.54&6.78&-1.12&2.42&4.7&6.8&1&22.38&48.28&49.53&43.99&49.55&\\
3C 33&H-S1&160&5.04&5.81&-2.15&2.4&3.8&145&1&21.25&44.72&45.26&44.39&45.41&\\
&H-S2&100&5.48&6.18&-2.15&2.4&3.8&16&1&21.72&45.76&46.5&43.43&46.57&\\
&H-N1&110&6&6.7&-1.24&2.4&3.8&2&1&21.64&46.51&47.19&41.46&47.27&\\
&H-N2&100&5.78&6.6&-2.3&2.18&3.8&2.3&1&21.64&46.18&46.71&41.58&46.82&\\
PKS 0405--123&H-N&100&5.28&6.45&0.01&2.8&3.3&17&1&22.15&47.7&48.61&44.33&48.66&\\
3C 105&H-S1&110&6.15&7&-1.82&2.07&4.5&0.9&1&21.66&47.1&47.44&40.8&47.6&\\
&H-S3&100&5.78&6.2&-1.52&2.47&4.5&16.5&1&21.75&46.24&47.02&43.5&47.08&\\
3C 111&H-N&50&6.41&6.85&-1.12&2.6&3.8&4&1&22.07&47.11&48.25&42.92&48.28&\\
PKS 0836+299&H-B&100&7.18&7.48&0.95&2.6&3.6&0.34&1&21.52&47.92&48.75&39.68&48.81&\\
3C 228&H-S&110&5.58&6.48&-0.26&2.6&4.2&32&1&21.73&47.08&47.88&44.04&47.94&\\
3C 245&H-D&300&5&6.18&-1.15&2.6&4&33.3&1&22.3&47.07&47.43&45.22&47.59&\\
3C 263&H-K&100&5.32&5.86&-0.96&2.66&4.04&135&1&21.93&46.64&47.51&45.69&47.57&\\
3C 275.1&H-N&150&5.86&6.85&-1.33&2.4&6&13.1&1&22.07&47.23&47.78&43.96&47.89&\\
3C 295&H-NW&650&3.52&5&-1.24&2.1&3.7&215&1&21.25&46.17&46.23&44.73&46.51&\\
PKS 1421--490&H-A&300&4.4&5.95&-1.35&2.4&4.4&550&1&21.71&46.32&46.62&46.48&46.97&\\
3C 303&H-W&100&6.32&7.41&0&2.66&3.4&4.7&1&21.89&47.53&48.39&42.69&48.45&\\
3C 345&H-A&200&5.4&6&-3.14&2.3&5&100&1&22.09&45.71&46.08&45.75&46.36&\\
3C 351&H-L&100&5.41&6.6&-1.15&2.5&3.9&18&1&22.1&47.24&48.03&44.28&48.09&\\
&H-J&140&5.48&6.48&-0.3&2.4&4&32&1&21.4&46.92&47.51&43.39&47.61&\\
3C 390.3&H-B&80&6.78&6.95&-0.11&2.4&4.2&0.74&1&21.52&47.45&48.28&40.36&48.34&\\
Cygnus A&H-A&200&3.6&5.6&-2.66&2&4&185&1&21.52&45.82&46.22&45.15&46.39&\\
&H-B&350&3.7&5.6&-2.66&2.05&3.8&95&1&21.52&45.62&45.82&44.57&46.05&\\
&H-D&300&3.48&5&-3.17&1.82&3.3&180&1&21.52&45.82&46.04&45.12&46.27&\\
PKS B2152--699&H-S&120&5.65&6.6&0.15&2.5&3.3&10.7&1&20.84&45.97&46.69&41.31&46.76&\\
&H-N&100&5.9&7.48&0.2&2.5&3.4&3&1&21.02&46.43&47.22&40.56&47.28&\\
3C 445&H-SW&100&6&7.3&-0.85&2.28&3.9&2.64&1&21.23&46.46&47.09&40.88&47.18&\\
&H-SE&130&6.02&7&-0.91&2.41&3.6&2.65&1&21.53&46.56&47.18&41.48&47.27&\\
\hline
 \small{\textbf{Synchrotron}}&&&&&&&&&&&&&&&\\
\hline
3C 66B&K-A&100&7.04&8.34&-3.46&2.44&3.2&46.29&1&20.97&&&&&\\
&K-B&100&5.93&8&-3.6&2.22&3.52&86.49&1&20.9&&&&&\\
3C 111&K-K9&100&6.32&8.4&-4.68&2.42&3.16&12.12&1&22.07&&&&&\\
&K-K14&100&6.48&8.4&-4.4&2.48&3.28&13.66&1&22.07&&&&&\\
&K-K40&100&6.3&8.4&-3.86&2.64&3.22&15.28&1&22.07&&&&&\\
&K-K45&100&7.41&8.48&-3.11&2.84&3.22&19.93&1&22.07&&&&&\\
&K-K51&100&6.32&8.4&-4&2.62&3.3&13.79&1&22.07&&&&&\\
&K-K97&100&6.32&8.78&-2.79&2.88&3.16&25.6&1&22.07&&&&&\\
&K-K107&100&6.34&8.78&-2.63&2.88&3.7&30.83&1&22.07&&&&&\\
&H-S&100&6.18&8.48&-2.38&2.88&3.76&41.21&1&22.07&&&&&\\
3C 120&K-K7&100&7.72&&-3.07&2.64&&38.01&1&21.48&&&&&\\
&K-S3&100&6.08&7.9&-4.26&2.38&3.26&22.49&1&21.51&&&&&\\
PKS 0521--365&K-K&100&6.18&8.18&-2.19&2.5&3.6&161.57&1&21.12&&&&&\\
M87&K-A&100&5.96&7.9&-2.19&2.3&4.2&324.81&1&20.33&&&&&\\
&K-B&100&5.96&7.9&-2.47&2.32&4.4&218.26&1&20.49&&&&&\\
&K-C1&100&5.78&8&-1.74&2.36&4.4&433.94&1&20.07&&&&&\\
&K-D&100&6.1&8&-1.82&2.38&3.86&372.25&1&19.98&&&&&\\
&K-E&100&6.65&8.7&-2.54&2.42&3.96&142.46&1&20.33&&&&&\\
&K-F&100&6.2&8.2&-2.74&2.34&4.2&148.3&1&20.46&&&&&\\
3C 371&K-A&100&6.18&8.3&-3.86&2.34&3.12&41.31&1&21.43&&&&&\\
PKS B2152--699&K-D&100&6.26&7.9&-2.52&2.52&3.62&103.66&1&20.84&&&&&\\
PKS 2201+044&K-A&100&6.3&7.95&-3.57&2.3&3.2&66.55&1&20.92&&&&&\\
&K-$\beta$&100&6.9&8&-3.27&2.37&3.2&73.42&1&20.69&&&&&\\
\enddata

\tablenotetext{1}{If no data are available in the columns of $\gamma_{\rm max}$ and $p_2$, that means the SEDs are fitted by a power-law spectrum, not a broken power-law spectrum. }
\end{deluxetable}

\end{document}